\documentclass{amsart}
\newcommand{\dee}{\mathrm{d}}
\newcommand{\tilpsi}{\hat{\psi}}
\newcommand{\tilPsi}{\hat{\Psi}}
\newcommand{\tilPhi}{\hat{\Phi}}
\newcommand{\tilphi}{\hat{\varphi}}
\newcommand{\tilxi}{\hat{\xi}}
\newcommand{\tilomega}{\hat{\omega}}
\newcommand{\che}{G}
\usepackage{bm}
\usepackage{amssymb}
\usepackage{graphicx}
\usepackage[usenames,dvips]{color}
\begin{document}
\pagenumbering{roman}
\begin{center}
\thispagestyle{empty}
{\large\bf Rapid Evaluation of Radiation Boundary Kernels
\\[1mm]
for Time--domain Wave Propagation on Blackholes${}^\dagger$}
\\[7.5mm]
Stephen~R.~Lau${}^\ddagger$
\\[5mm]
{\em Applied Mathematics Group,
Department of Mathematics\\
University of North Carolina, Chapel Hill,
NC 27599-3250 USA\\
Email address:} {\tt lau@email.unc.edu}
\\[5mm]
{\bf Abstract}
\end{center}
\noindent
For scalar, electromagnetic, or gravitational wave 
propagation on a background Schwarz\-schild blackhole,
we describe the exact nonlocal radiation outer boundary 
conditions ({\sc robc}) appropriate for a spherical outer 
boundary of finite radius enclosing the blackhole. 
Derivation of the {\sc robc} is based on Laplace and 
spherical--harmonic transformation of the 
{\em Regge--Wheeler equation}, the {\sc pde} governing the 
wave propagation, with the resulting radial {\sc ode} an 
incarnation of the {\em confluent Heun equation}. For a 
given angular integer $l$ the {\sc robc} feature integral 
convolution between a time--domain radiation boundary kernel
({\sc tdrk}) and each of the corresponding $2l+1$ 
spherical--harmonic modes of the radiating wave field. The 
{\sc tdrk} is the inverse Laplace transform of a 
frequency--domain radiation kernel ({\sc fdrk}) which is
essentially the logarithmic derivative of the asymptotically 
outgoing solution to the radial {\sc ode}. We 
numerically implement the {\sc robc} via a rapid 
algorithm involving approximation of the {\sc fdrk} by a 
rational function. Such an approximation is tailored to have 
relative error $\varepsilon$ uniformly along the axis of
imaginary Laplace frequency. Theoretically, $\varepsilon$ is 
also a long--time bound on the relative convolution error. Via 
study of one--dimensional radial evolutions, we demonstrate 
that the {\sc robc} capture the phenomena of quasinormal 
ringing and decay tails. Moreover, carrying 
out a numerical experiment in which a wave packet strikes 
the boundary at an angle, we find that the {\sc robc} 
yield accurate results in a three--dimensional setting.
Our work is a partial generalization to Schwarzschild 
wave propagation and Heun functions of the methods developed 
for flatspace wave propagation and Bessel functions by Alpert, 
Greengard, and Hagstrom ({\sc agh}), save for one key 
difference. Whereas {\sc agh} had the usual armamentarium 
of analytical results (asymptotics, order recursion relations, 
bispectrality) for Bessel functions at their disposal, what 
we need to know about Heun functions must be gathered 
numerically as relatively less is known about them. 
Therefore, unlike {\sc agh}, we are unable to offer an 
asymptotic analysis of our rapid implementation.
\vfill
\noindent
\underline{$\hspace{5cm}$}\\
${}^\dagger$ Based on Reference \cite{LauMathDiss}. Published as two
separate articles in J.~Comp.~Phys.~{\bf 199}, issue 1, 376-422 (2004) 
and Class.~and Quantum Grav.~{\bf 21}, 4147-4192 (2004).\\[2mm]
${}^\ddagger$ Now at Center for Gravitational Wave Astronomy, University of
Texas at Brownsville, 80 Fort Brown, Brownsville TX, 78520. Email address:
{\tt lau@phys.utb.edu}.
\newpage
\tableofcontents
\newpage
\addcontentsline{toc}{section}{List of main symbols}
\centerline{\bf List of main symbols}\label{listofsymbols}

\noindent
{\sc Section} \ref{sbsbSec:1.1.1}, 
$\dee s^2$\dotfill notation for generic 
line--element\\
$T$\dotfill static time coordinate\\
$r$\dotfill areal radius coordinate\\
$\theta,\phi$\dotfill standard angular coordinates\\
$\mathrm{m}$\dotfill geometrical mass of blackhole\\
$g_{\alpha\beta}$\dotfill metric tensor\\
$\tau$\dotfill dimensionless static time $T/(2\mathrm{m})$\\
$\rho$\dotfill dimensionless radius $r/(2\mathrm{m})$\\
$F$\dotfill ubiquitous metrical function 
$1-1/\rho = 1-2\mathrm{m}/r$\\
$\rho_*$\dotfill tortoise coordinate $\rho + \log(\rho -1)$\\
$\mu$\dotfill dimensionless retarded time $\tau - \rho_*$\\
$\nu$\dotfill dimensionless advanced time $\tau + 
\rho_*$\\
${\,}$\hfill(also used as Bessel order below)
\\[5mm]
{\sc Section} \ref{sbsbSec:1.1.2}, 
$\square$\dotfill d'Alembertian or wave operator\\
$\psi$\dotfill wave field in coordinates 
$(\tau,\rho,\theta,\phi)$ or $(T,r,\theta,\phi)$\\
$x^\alpha$\dotfill generic coordinate function\\
$\sqrt{-g}$\dotfill square root of (minus) 
determinant of metric\\
$g^{\alpha\beta}$\dotfill inverse metric tensor\\
$\jmath$\dotfill spin, takes values 0,1,2 only\\
$l$\dotfill primary harmonic angular index, runs from 
$\jmath$ to $\infty$\\
$m$\dotfill secondary harmonic angular index, runs 
from $-l$ to $l$\\
${\,}$\hfill (this is italic $m$, not plain $\mathrm{m}$ 
which is the mass)\\
$Y_{lm}(\theta,\phi)$\dotfill spherical harmonic\\ 
$\psi_{lm}(\tau,\rho)$\dotfill spherical--harmonic 
transform of $\psi$\\
$\psi_l$\dotfill $\psi_{lm}(\tau,\rho)$ with 
coordinates and index $m$ suppressed\\
$\Psi_l$\dotfill $\rho\psi_l$\\
$V(\rho)$\dotfill Regge--Wheeler potential\\
$\varphi_l$\dotfill $\psi_l(\mu+\rho_*,\rho)$, 
that is $\psi_l$ in retarded time coordinates\\
$\Phi_l$\dotfill $\rho\varphi_l$, same as 
$\Psi_l(\mu+\rho_*,\rho)$,\\
${\,}$\hfill that is $\Psi_l$ in retarded time 
coordinates
\\[5mm]
{\sc Section} \ref{sbsbSec:1.2.1}, 
$\mathcal{L}$\dotfill Laplace transform operation\\
$\sigma$\dotfill dimensionless Laplace frequency 
$2\mathrm{m}s$\\
$s$\dotfill Laplace frequency
\\[5mm]
{\sc Section} \ref{sbsbSec:1.2.2}, 
$\tilpsi_l$\dotfill Laplace transform (on $\tau$) of 
$\psi_l$\\
${\,}$\hfill and generic solution to radial {\sc ode}\\
$z$\dotfill the product $\sigma\rho$\\
$k_l(z)$, $i_l(z)$\dotfill modified spherical Bessel 
functions\\
$K_{l+1/2}(z)$, $I_{l+1/2}(z)$\dotfill modified 
cylindrical Bessel functions
\\[5mm]
{\sc Section} \ref{sbsbSec:1.2.3}, 
$\tilPsi_l$\dotfill Laplace transform (on $\tau$) of 
$\Psi_l$\\
$\rho_B$\dotfill dimensionless outer radius
\\[5mm]
{\sc Section} \ref{sbsbSec:1.3.1}, 
$\Theta_l$\dotfill generic solution to normal form of 
radial {\sc ode}\\
$\tilphi_l$\dotfill  Laplace transform (on
$\mu$) of $\varphi_l$, also $\Theta_l$ for 
$\jmath = 0$ case
\\[5mm]
{\sc Section} \ref{sbsbSec:1.3.2}, 
$\tilPhi_l$\dotfill Laplace transform (on $\mu$) of 
$\Phi_l$ and generic\\
${\,}$\hfill solution to normalized form of radial 
{\sc ode}\\
$W_l(\sigma\rho;\sigma)$ \dotfill
outgoing solution $\tilPhi_l^{+}(\rho;\sigma)$ to\\
${\,}$\hfill normalized form of 
radial 
{\sc ode}\\
$Z_l(\sigma\rho;\sigma)$ \dotfill
ingoing solution $\tilPhi_l^{-}(\rho;\sigma)$ to\\
${\,}$\hfill normalized form of radial {\sc 
ode}\\
$W_l(\sigma\rho)$, $Z_l(\sigma\rho)$\dotfill 
corresponding Bessel--type functions
\\[5mm]
{\sc Section} \ref{sbsbSec:1.3.3}, 
$\kappa$\dotfill $1-\jmath^2$, takes values 
$1,0,-3$ only\\
$d_n(\sigma)$\dotfill coefficients in asymptotic 
expansion for $W_l(z;\sigma)$\\
$c_n$\dotfill coefficients in asymptotic expansion for 
$W_l(z)$
\\[5mm]
{\sc Section} \ref{sbsbSec:1.4.1}, 
$\mathsf{N}$\dotfill $F^{1/2}$, temporal lapse for 
$(\tau,\rho)$ coordinates\\
$\mathsf{M}$\dotfill $F^{-1/2}$, radial lapse for 
$(\tau,\rho)$ coordinates\\
$*$\dotfill Laplace convolution\\
$\mathcal{L}^{-1}$\dotfill inverse Laplace 
transform operation\\
$\omega_l(\tau;\rho_B)$\dotfill time--domain 
radiation kernel ({\sc tdrk})\\
$\tilomega_l(\sigma;\rho_B)$\dotfill frequency--domain 
radiation kernel ({\sc fdrk})\\
$\omega(\tau)$, $\tilomega(\sigma)$\dotfill same 
objects with $l$ and $\rho_B$ suppressed
\\[5mm]
{\sc Section} \ref{sbsbSec:1.4.2}, $N_l(\rho_B)$\dotfill 
number of poles for $\tilomega_l(\sigma;\rho_B)$\\
$\sigma_{l,n}(\rho_B)$\dotfill pole location for 
$\tilomega_l(\sigma;\rho_B)$\\
$\alpha_{l,n}(\rho_B)$\dotfill pole strength for 
$\tilomega_l(\sigma;\rho_B)$\\
$\chi$\dotfill continuous variable on 
$[0,\infty)$ and coordinate for cut\\
$f_l(\chi;\rho_B)$\dotfill cut profile 
for  $\tilomega_l(\sigma;\rho_B)$\\
$N$, $\alpha_n$, $\sigma_n$, $f(\chi)$
\dotfill same variables with all $l$ and\\
${\,}$\hfill $\rho_B$ dependence suppressed
\\[5mm]
{\sc Section} \ref{sbsbSec:1.4.3}, 
$G_l(\rho_*,\rho_*',\tau)$\dotfill
time--domain Green's function.\\
$\hat{G}_l(\rho_*,\rho_*',\sigma)$\dotfill
frequency--domain Green's function.
\\[5mm]
{\sc Section} \ref{sbsbSec:1.4.4}, $\xi(\tau)$\dotfill 
approximate time--domain 
radiation kernel\\
$y$\dotfill $\mathrm{Im}\sigma$\\
$\hat{\xi}(\sigma)$\dotfill approximate frequency--domain
radiation kernel\\
$P(\sigma)$\dotfill polynomial in 
$\sigma$ of degree $d-1$\\
$Q(\sigma)$\dotfill polynomial in 
$\sigma$ of degree $d$\\
$\varepsilon$\dotfill relative error tolerance\\
$\Delta\tilomega(\sigma)$\dotfill numerical error 
in $\tilomega(\sigma)$
\\[5mm]
{\sc Section} \ref{sbsbSec:2.1.1}, 
$p$\dotfill integer appearing in cutoff $l-p$\\
${\,}$\hfill for asymptotic expansions
\\[5mm]
{\sc Section} \ref{sbsbSec:2.1.2}, 
$\theta$\dotfill an angle in $z$--plane\\
$I$, $J$, $N$, $M$\dotfill integer constants
\\[5mm]
{\sc Section} \ref{sbsbSec:2.1.3}, 
$k_{l,n}$\dotfill $n$th zero of the 
Bessel function $K_{l+1/2}(z)$\\
$\mathcal{C}$\dotfill asymptotic curve on which 
the $k_{l,n}$ accumulate\\
$P$\dotfill integer constant
\\[5mm]
Introduction for {\sc Section} \ref{sbSec:2.2}, 
$w_l(z)$\dotfill$z W_l'(z)/W_l(z)$\\
$w_l(z;\sigma)$\dotfill$z W_l'(z;\sigma)/W_l(z;\sigma)$
\\[5mm]
{\sc Section} \ref{sbsbSec:2.2.3}, 
$a_n$\dotfill coefficients used in closed--form\\
${\,}$\hfill expression for origin value of the kernel
\\[5mm]
{\sc Section} \ref{sbsbSec:2.2.4},
$u_l(z;\sigma)$\dotfill $\mathrm{Re}w_l(z;\sigma)$\\
$v_l(z;\sigma)$\dotfill $\mathrm{Im}w_l(z;\sigma)$\\
$u_l(z)$\dotfill $\mathrm{Re}w_l(z)$\\
$v_l(z)$\dotfill $\mathrm{Im}w_l(z)$
\\[5mm]
{\sc Section} \ref{sbsbSec:3.1.1}, $\mathbb{S}$\dotfill 
parameter space 
$\jmath=0,\rho_B\in[15,25],0\leq l \leq 10$
\\[5mm]
{\sc Section} \ref{sbsbSec:3.2.1}, $T_{l,n}(1/\rho_B)$\dotfill 
interpolating polynomial for pole 
locations
\\[5mm]
{\sc Section} \ref{sbsbSec:3.3.1}, $\nu$\dotfill Bessel order 
$l +1/2$, not advanced time\\
$x$\dotfill $\mathrm{Re}\sigma$\\
$d$\dotfill degree of polynomial $Q(\sigma)$
\\[5mm]
{\sc Section} \ref{sbsbSec:3.3.3},
$\beta_n$ \dotfill pole location for compressed kernel\\
$\gamma_n$\dotfill pole strength for compressed kernel
\\[5mm]
{\sc Section} \ref{sbsbSec:4.1.1}, $H$\dotfill horizon, inner 
two--surface at $r = 2\mathrm{m}$\\
$r_*$\dotfill physical tortoise coordinate $r + 2\mathrm{m}
\log(r/(2\mathrm{m})-1)$\\
$v$\dotfill physical advanced time $2\mathrm{m}\nu = T+r_*$\\
$u$\dotfill physical retarded time $2\mathrm{m}\mu = T-r_*$\\
$T_v$\dotfill time variable $T+2\mathrm{m}
\log(r/(2\mathrm{m})-1)$\\
${\,}$ \hfill closely related to $v = T_v + r$\\
$T_u$\dotfill time variable $T-2\mathrm{m}
\log(r/(2\mathrm{m})-1)$\\ 
${\,}$ \hfill closely related to $u = T_u - r$\\
$t$\dotfill time variable for numerical implementation,\\
${\,}$\hfill  $T_v$ shifted by suitable constant\\
$\Sigma$\dotfill spacetime foliation determined by\\ 
${\,}$\hfill $t$ and also
a generic level--$t$ slice\\
$N$\dotfill temporal lapse for $(t,r)$ 
coordinates,\\
$M$\dotfill radial lapse function for $(t,r)$
coordinates\\ 
$V^r$\dotfill shift vector for $(t,r)$ coordinates
\\[5mm]
{\sc Section} \ref{sbsbSec:4.1.2}, $U$\dotfill field 
$\psi_{lm}(\tau,\rho)$ 
from {\sc Section} \ref{sbsbSec:1.1.2} 
but now in terms of\\
${\,}$\hfill $(t,r)$ coordinates and with $l$ and $m$ suppressed
\\[5mm]   
{\sc Section} \ref{sbsbSec:4.1.3}, $e_\bot$\dotfill normal vector 
for a $\Sigma$ slice\\ 
$e_\vdash$ \dotfill normal vector for a round sphere in a 
$\Sigma$ slice\\
$e_{+}$\dotfill null vector $e_\bot + e_\vdash$\\
$X$\dotfill characteristic derivative $e_+[U]$,\\
${\,}$\hfill also with $l$ and $m$ indices suppressed
\\[5mm] 
{\sc Section} \ref{sbsbSec:4.1.4},  
$S(r,X,U,\partial_r U)$\dotfill source for $X$ evolution equation
\\[5mm]
{\sc Section} \ref{sbsbSec:4.2.1}, $r_B$\dotfill outer boundary radius 
$2\mathrm{m}\rho_B$\\
$Q$\dotfill number of subintervals for radial mesh\\
$r_q$\dotfill radial mesh point\\
$\Delta r$, $\Delta t$\dotfill radial and 
temporal discretization steps\\
$U^n_q$, $N_q$, $[X_-^M]_q$\dotfill 
examples of mesh functions\\
${\,}$ \hfill $U(t_n,r_q)$, $N(r_q)$, $X^M_{-}(r_q)$\\
$\bar{U}{}^{n+1}_q$, $\bar{X}{}^{n+1}_q$\dotfill 
predicted variables\\
$\widetilde{U}{}^{n+1}_q$, $\widetilde{X}{}^{n+1}_q$\dotfill 
corrected variables
\\[5mm]
{\sc Section} \ref{sbsbSec:4.2.2}, 
$r_{-1}$\dotfill ghost point $2\mathrm{m}-\Delta r$ 
inside horizon\\[5mm]
{\sc Section} \ref{sbsbSec:4.3.1}, $\mathsf{e}_+$\dotfill 
null vector 
$\mathsf{N}^{-1}\partial/\partial T + \mathsf{M}^{-1}
\partial/\partial r$\\
$\vartheta$\dotfill hyperbolic angle in the identity 
$\mathsf{e}_+ = e^\vartheta e_+$\\
$\mathsf{X}_l$\dotfill characteristic derivative 
$\mathsf{e}_+[\psi_l]$\\
$\Omega_l(T;r_B)$\dotfill physical time--domain 
radiation kernel
\\[5mm]
{\sc Section} \ref{sbsbSec:4.3.3}, 
$\beta_n^R,\beta_n^I$\dotfill real and imaginary parts 
of $\beta_n$ from 
{\sc Section} \ref{sbsbSec:3.3.3}\\
$\gamma_n^R,\gamma_n^I$\dotfill real and imaginary parts 
of $\gamma_n$ from
{\sc Section} \ref{sbsbSec:3.3.3}\\
$b_n^R +\mathrm{i}b_n^I$\dotfill$(\beta_n^R
+\mathrm{i}\beta_n^I)/(2\mathrm{m})$\\ 
$c_n^R +\mathrm{i}c_n^I$\dotfill$(\gamma_n^R
+\mathrm{i}\gamma_n^I)/(2\mathrm{m})$\\
$\Xi(t)$\dotfill approximate time--domain radiation kernel\\
$d_\mathrm{sing}$\dotfill number of approximating 
poles on negative real axis\\
$d_\mathrm{pair}$\dotfill number of conjugate 
pairs of approximating poles\\
$\mu_i$\dotfill reordered pole location on 
negative\\
${\,}$\hfill real axis, not retarded time\\
$\kappa_i$ \dotfill reordered pole strength on negative real axis\\
$m_j^R +\mathrm{i}m_j^I$\dotfill reordered pole strength\\
$k_j^R +\mathrm{i}k_j^I$\dotfill reordered pole location,\\
$H_i(t)$\dotfill cut piece of the kernel $\Xi(t)$ built with $\mu_i$ 
and $\kappa_i$\\
$G_j(t)$\dotfill ringing piece of kernel $\Xi(t)$ built\\
${\,}$ \hfill with 
$m_j^R +\mathrm{i}m_j^I$
and $k_j^R +\mathrm{i}k_j^I$\\
$F_j(t)$\dotfill auxiliary kernel  built with 
$m_j^R +\mathrm{i}m_j^I$ and $k_j^R +\mathrm{i}k_j^I$\\
${}_\mathtt{H}(\Xi*U)$, ${}_\mathtt{L}(\Xi*U)$\dotfill notation
for history and local\\
${\,}$\hfill parts of integral convolution\\[5mm]
{\sc Section} \ref{sbsbSec:4.3.4}, 
$\mathsf{N}_B$\dotfill $\mathsf{N}(r_B)$\\
${}_\mathtt{L}\overline{(\Xi*U)}{}^{n+1}$\dotfill 
predicted local part of convolution\\        
${}_\mathtt{L}\widetilde{(\Xi*U)}{}^{n+1}$\dotfill
corrected local part of convolution
\\[5mm]
{\sc Section} \ref{sbsbSec:5.1.1}, 
$f$\dotfill one--dimensional bump function\\
$A$, $a$, $b$, $\delta$\dotfill parameters specifying $f$\\
$U^\mathrm{free}$\dotfill free solution\\
$U^\mathrm{ROBC}$\dotfill {\sc robc} solution\\
$U^\mathrm{SOBC}$\dotfill Sommerfeld solution
\\[5mm]
{\sc Section} \ref{sbsbSec:5.2.1}, 
$f$\dotfill three--dimensional bump function\\
$\mathbf{r}$, $\mathbf{v}$, $\mathbf{r}_0$,
$v_x$, $v_y$, $v_z$,
$x_0$, $y_0$, $z_0$\dotfill parameters specifying $f$\\
$U_{lm}$, $X_{lm}$\dotfill $U$ and $X$, but now with $l$ and $m$ 
indices\\
$(r_q,\theta_i$, $\phi_j)$\dotfill spherical polar grid
\\[5mm]
{\sc Section} \ref{sbsbSec:5.2.2}, 
$\mathrm{vol}(\Sigma)$\dotfill volume of spatial slice $\Sigma$
\newpage
\clearpage
\pagenumbering{arabic}

\setcounter{section}{-1}
%
%
\section{Introduction}
\subsection{Background}
Consider the Cauchy problem\footnote{We use {\em Cauchy 
problem} in lieu of {\em initial--value problem} in order to 
reserve the latter for the process of generating initial 
data, one that requires the solution of elliptic {\sc pde} 
for theories involving constraints such as general 
relativity or fluid flow.} for the scalar wave equation,
\begin{equation}
- \partial^2_tU + \Delta U = 0\, ,
\label{flatspacewaveeq}
\end{equation}
on $[t_0,t_F]\times\mathbb{E}^3$, the Cartesian product of a 
closed time interval and Euclidean three--space. First, we 
specify suitable initial--value or canonical data 
$U|_{t_\mathrm{0}}$ and $\partial_tU|_{t_0}$ on 
$\mathbb{E}^3$ at the initial time $t_0$. Next, using the 
rule (\ref{flatspacewaveeq}), we evolve the data until the 
final time $t_F$, along the way generating the solution 
$U$ throughout the temporally bounded but spatially 
unbounded domain 
$[t_0,t_F] \times \mathbb{E}^3$. Provided physically 
reasonable initial data, this problem is 
well--posed; however, it is not the evolution problem 
one typically encounters in numerical wave simulation. 
Usually the numerical mesh covers only a {\em finite} 
portion of $\mathbb{E}^3$.

\begin{figure}
\scalebox{0.7}{\includegraphics{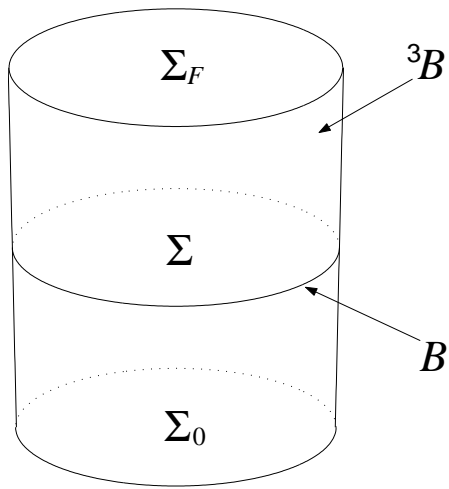}}
\caption{{\sc Finite spacetime domain $\mathcal{M}\subset
\mathbb{E}^3$ with boundary $\partial\mathcal{M}
=\Sigma_0\cup\Sigma_F\cup{}^3\!B$.} Respectively, $\Sigma_0$
and $\Sigma_F$ denote the solid round ball $\Sigma$ at the  
initial time $t_0$ and the final time $t_F$. Radiation
boundary conditions are given on the three--dimensional 
timelike cylinder ${}^3\!B$. Our geometric perspective on
the ``quasilocal'' spacetime region $\mathcal{M}$ comes  
from Refs.~\cite{BrownYork} and
\cite{BrownLauYork}.}
\label{introFigure1}
\end{figure}

With the finiteness of numerical meshes in mind, 
consider the following more realistic evolution problem. 
Let $\Sigma\subset\mathbb{E}^3$ be a round, solid, 
three--dimensional ball determined by  $r \leq r_B$, with 
$r_B$ a fixed outer radius, on which we specify compactly
supported initial data $U|_{t_0}$ and $\partial_tU|_{t_0}$ 
at $t = t_0$. Again, the goal is to evolve the data, although 
now generating the solution $U$ on the finite domain 
$\mathcal{M} = [t_0,t_F] \times \Sigma$ depicted in 
{\sc Fig}.~\ref{introFigure1}. Respectively, let $\Sigma_0$ 
and $\Sigma_F$ denote the ball $\Sigma$ at $t = t_0$ and 
$t=t_F$. One element of the boundary $\partial\mathcal{M} 
= \Sigma_0\cup\Sigma_F\cup{}^3\!B$ is a timelike 
three--dimensional cylinder ${}^3\!B$ determined by $t_0
\leq t \leq t_F$ and $r = r_B$. Note that ${}^3\!B$ is 
the history in time of the spherical spatial boundary $B = 
\partial\Sigma$. As it stands, such an evolution problem is 
not well--posed, since $\mathcal{M}$ is larger than the 
future domain of dependence of $\Sigma_0$. Indeed, 
$U|_{t_0}$ and $\partial_tU|_{t_0}$ are {\em free data}, and 
we have no control over data on $\mathbb{E}^3/\Sigma_0$, the 
region exterior to the initial $\Sigma$ ball. Data on this 
exterior region may contain so--called ingoing radiation 
which will impinge upon ${}^3\!B$ at later times, 
affecting the solution $U$ within $\mathcal{M}$. Most often
in numerical wave simulation the goal is to forbid such 
ingoing radiation by the choice of {\em radiation boundary 
conditions}, that is explicit rules governing the behavior 
of $U$ and $\partial_tU$ on ${}^3\!B$. Often referred 
to as {\em nonreflecting boundary conditions} ({\sc nrbc}), 
for the described problem such conditions ideally specify 
that the spherical boundary $B$ is completely transparent. 
Due to the free nature of the initial data, {\em exact} 
{\sc nrbc} are inherently nonlocal in both space and time. 
With {\sc nrbc} specified along ${}^3\!B$, we may refer 
to such an evolution as a {\em mixed Cauchy--boundary value 
problem}.

More generally, one might consider radiation boundary 
conditions associated with some other {\sc pde} and/or 
different type of $B$ boundary, say cubical or irregularly 
shaped. Refs.~\cite{Lindman} through \cite{Hagstrom2} 
pertain either to the
described spherical problem or to more general radiation 
boundary conditions. This is certainly not an exhaustive 
list, and we point the reader to review articles 
\cite{Givoli,Hagstrom1,Hagstrom2} for more comprehensive 
listings. Although we do not attempt an extensive literature 
review, we make mention of a few approaches to radiation 
boundary conditions in order to put our work in some 
context. Two pioneering early works are those of Engquist 
and Majda \cite{EngMajda1,EngMajda2} and Bayliss and Turkel 
\cite{BayTurkel}. Each develops a hierarchy of {\em local} 
differential conditions of increasing complexity. Engquist 
and Majda's work is based on exact radiation boundary 
conditions as expressed within the theory of 
pseudo--differential operators, and their approach is not 
necessarily tied to a spherical geometry nor to the ordinary 
wave equation. Also considering more than just our problem 
above, Bayliss and Turkel base their approach on asymptotic 
expansions about infinity, for example the standard 
multipole expansion for a radiating field obeying the 
ordinary wave equation (\ref{flatspacewaveeq}). Another 
approach to radiation boundary conditions relies on the 
introduction of absorbing layers, and as an example we 
mention Ref.~\cite{Kosloff}. In the introduction of his 
review article \cite{Hagstrom1}, Hagstrom describes the 
main advances made in the 1990s on several fronts related 
to radiation boundary conditions: (i) improved 
implementations of hierarchies, such as the ones mentioned, 
(ii) new absorbing layer techniques exhibiting 
reflectionless interfaces, and (iii) efficient algorithms 
for evaluation of exact nonlocal boundary operators. 
Results \cite{GroteKeller1,GroteKeller2,GroteKeller3} of
Grote and Keller fall within this first category. An 
advance on the second front was the introduction of {\em 
perfectly matched layers} by B\'{e}renger \cite{Beren}, 
while a key advance on the third was the rapid 
implementation of {\sc nrbc} for spherical boundaries by 
Alpert, Greengard, and Hagstrom \cite{AGH1,AGH2}. See also 
related work by Sofronov \cite{Sofronov}. Hagstrom discusses 
the state of the art for both fronts (ii) and (iii) in his 
second review article \cite{Hagstrom2}.

\subsection{Alpert, Greengard, Hagstrom nonreflecting boundary 
conditions} Since our investigation will follow the approach of 
Alpert, Greengard, and Hag\-strom ({\sc agh}) which belongs to 
category (iii) in the last paragraph, let us describe it in more
detail. With $U_{lm}(t,r)$ 
denoting a single spherical harmonic mode of the wave field 
($l$ and $m$ are the standard orbital and azimuthal integers), we 
consider the reduced ordinary $3+1$ wave equation,
\begin{equation}
\frac{\partial^2U_l}{\partial t^2} - 
\frac{\partial^2U_l}{\partial r^2} -
\frac{2}{r}\frac{\partial U_l}{\partial r} 
+ \frac{l(l+1)}{r^2}U_l = 0,
\end{equation}
stemming from the (\ref{flatspacewaveeq}).
We have dropped the index $m$ on $U_l$ since it does not appear
in the {\sc pde}. Let $\hat{U}_l(s,r) = \mathcal{L}[U_l](s,r)$
denote the Laplace transform of the mode. Provided that we assume 
both compactly supported initial data and a large enough radius $r$, 
the transform $\hat{U}_l(s,r)$ obeys a homogeneous {\sc ode},
\begin{equation}
      z^2\frac{\dee^2\hat{U}_l}{\dee z^2}
     +2z\frac{\dee\hat{U}_l}{\dee z}
     -\left[
      z^2
     +l(l+1)
      \right]
      \hat{U}_l = 0\, ,
\label{msbe1}
\end{equation}
known as the {\em modified spherical Bessel equation}. Here 
$z = s r$ 
is shorthand for the product of Laplace frequency $s$ and radius
$r$. In terms of the standard half--integer order {\em MacDonald 
function} \cite{Watson} $K_{l+1/2}(z)$, we define an associated 
function $W_l(z) = (\pi z/2)^{1/2} \exp(z) K_{l+1/2}(z)$ more 
closely related to a 
confluent hypergeometric function. The function $z^{-1} \exp(-z) 
W_l(z)$ is the outgoing solution to (\ref{msbe1}). Here $W_{l}(z)$ 
has been chosen so that $W_l(z) \sim 1$ as $z \rightarrow \infty$, 
that is ``normalized at infinity.''  Moreover, it turns out that 
$W_l(z)$ is a polynomial of degree $l$ in inverse $z$. For
example, $W_0(z) = 1$, $W_1(z) = 1+z^{-1}$, $W_2(z) = 1+ 3z^{-1}
+3z^{-2}$.

{\sc agh} introduce a {\em nonreflecting boundary kernel}, here 
called a {\em time--domain radiation kernel} ({\sc tdrk}) as 
follows. Again assuming compactly supported initial data and a 
large enough radial value, the Laplace transform of the mode
$U_l(t,r)$ takes the form
\begin{equation}
\hat{U}_l(s,r) = a_l(s) r^{-1} \exp(-sr) W_l(sr)\, ,
\end{equation}
where $a_l(s)$ is an analytic function of $s$ depending on the 
details of the initial data. Differentiation of the last 
equation with respect to $r$ leads to the identity (the 
prime denotes differentiation in argument)
\begin{equation}
s\hat{U}_l(s,r) + \partial_r\hat{U}_l(s,r) + r^{-1}
\hat{U}_l(s,r) = 
r^{-1} \big[sr W_l'(sr)/W_l(sr)\big]\hat{U}_l(s,r)\, .
\label{fdident}
\end{equation}
This identity is of course valid at the fixed outer boundary radius 
$r_B$, provided that the outer two--sphere boundary $B$ lies beyond 
the support of the initial data. A well--known classical result,
$W_l(z)$ has $l$ simple zeros $\{k_{l,n}: n = 1,\cdots, l\}$ which 
lie in the lefthalf plane. The $k_{l,n}$ are also the zeros of 
$K_{l+1/2}(z)$, as discussed in \cite{AGH1} and below. As a result, 
the object $\tilomega_l(s;r_B) = s r_B 
W_l'(sr_B)/W_l(sr_B)$, a {\em frequency--domain radiation kernel} 
({\sc fdrk}), is a sum of $l$ simple poles in the lefthalf $s$--plane, 
\begin{equation}
\tilomega_l(s;r_B) = \sum_{n = 1}^l 
                     \frac{k_{l,n}/r_B}{s-k_{l,n}/r_B}\, .
\end{equation}
Whence its inverse Laplace transform $\omega_l(t;r_B)$, the 
{\sc tdrk} of {\sc agh}, is a corresponding sum of 
exponentials,
\begin{equation}
\omega_l(t;r_B) =  \sum_{n=1}^l 
(k_{l,n}/r_B)\exp\big(k_{l,n}t/r_B\big) \, . 
\end{equation}
We find the following for the first three such kernels:
\begin{align}
\omega_0(t;r_B) & = 0\, ,
\\
\omega_1(t;r_B) & = -r_B^{-1}\exp\big(-t\big/r_B\big)\, ,
\nonumber
\\
\omega_2(t;r_B) & =
-r_B^{-1} \exp\big(-{\textstyle \frac{3}{2}}t/r_B\big)
  \big[3\cos\big({\textstyle \frac{1}{2}}\sqrt{3}t/r_B\big)
+\sqrt{3}\sin\big({\textstyle \frac{1}{2}}\sqrt{3}t/r_B\big)
\big]\, .
\nonumber
\end{align}
By elementary properties of the Laplace transform, including the
Laplace convolution theorem, inverse transformation of the
identity (\ref{fdident}) yields the following {\sc robc} (a 
condition of complete transparency):\footnote{Domain
reduction appears in the early work \cite{GustaKreiss79} of
Gustafsson and Kreiss, although domain reduction via Laplace
convolution appears shortly thereafter in the work \cite{Hagstrom0}
of Hagstrom. In Ref.~\cite{Friedlander} Friedlander considered
essentially the same convolution kernel but in a different
context.}
\begin{equation}
\left.
\left[
\frac{\partial U_{l}}{\partial t}+
\frac{\partial U_{l}}{\partial r}+ 
\frac{U_{l}}{r}\right]\right|_{r = r_B}
= r^{-1}_B \int_0^t \omega_l(t-t';r_B)
U_{l}(t',r_B){\rm d}t'\, .
\label{nrbc}
\end{equation}
From a numerical standpoint, this boundary condition becomes expensive 
for high--order modes, since $\omega_l(t;r_B)$ is made up of $l$ 
exponential factors. However, elementary exponential identities do 
afford an efficient recursive evaluation of the convolution \cite{AGH1}. 
For wave propagation on flat 2+1 spacetime, the analogous cylindrical 
{\sc tdrk} for a given angular mode is again a sum of exponentials, but 
now the sum is over a set including a continuous sector \cite{AGH1}. 
Remarkably, this is a feature shared by blackhole kernels. Such 
continuous distributions are primarily relevant for the low--order modes 
and quite expensive to evaluate.

{\sc agh} describe an algorithm for kernel compression, by
which we mean approximation of the {\sc fdrk} by a rational
function, where the approximation is of specified relative
error $\varepsilon$ uniformly for $\mathrm{Re}s \geq 0$.
The resulting rational function, itself a sum of simple
poles, typically involves far fewer terms. As a result, in
the {\sc agh} approach the cost of updating the numerical 
solution at the outer boundary $B$ is minimized subject to the 
prescribed tolerance $\varepsilon$ (which also turns out to be 
a relevant error measure in the time--domain). Therefore, we 
may describe their implementation of {\sc robc} for the
ordinary wave equation as {\em rapid}. {\sc agh} have given a 
formidable asymptotic analysis of their rapid implementation, 
proving in particular that the number $d$ of poles needed to 
approximate the {\sc fdrk} to within the specified tolerance 
scales like
\begin{equation}
d \sim O\big(\log\nu\log(1/\varepsilon)
+\log^2\nu+\nu^{-1}\log^2(1/\varepsilon)\big)
\label{AGHscaling}
\end{equation}
as $\nu = l+1/2 \rightarrow \infty$ and $\varepsilon
\rightarrow 0^{+}$ \cite{AGH1}. Increased performance as $\nu
\rightarrow\infty$ is also seen for rational approximation
of the cylindrical kernels relevant for 2+1 wave propagation.
However, as remarked upon later in 
{\sc Section}~\ref{sbsbSec:3.3.1}, for both the 2+1 and 
blackhole scenarios rational approximation of low--order kernels 
(associated with costly continuous sectors) also leads to savings.

\subsection{Problem statement}
Now consider the evolution problem for the scalar wave 
equation,
\begin{equation}
\sum_{\mu,\nu = 0}^3 
\frac{1}{\sqrt{-g}}\frac{\partial}{\partial x^\mu} 
\left(\sqrt{-g}g^{\mu\nu}
\frac{\partial U}{\partial x^\nu}\right)
 = 0\, ,
\label{Schwarzschildwaveeq}
\end{equation}
describing a field $U$ propagating on a Schwarzschild 
blackhole determined by $g_{\mu\nu}$ metric functions. A 
slight modification of (\ref{Schwarzschildwaveeq}) yields a 
wave equation flexible enough to also describe propagation 
of electromagnetic or gravitational waves on a Schwarzschild
blackhole \cite{Wheeler,ReggeWheeler,Zerilli,ChandraDet}.
As a model of gravitational wave propagation, the problem 
has applications in relativistic astrophysics: 
non--spherical gravitational collapse and stellar 
perturbations, among others. Gravitational wave propagation 
is also of considerable theoretical interest in general 
relativity. With 
numerical wave simulation on a finite mesh in mind, we again 
choose a finite domain $\Sigma$, now a round, 
three--dimensional, thick shell also bounded internally by 
the blackhole horizon $H$. The outer boundary $B$, one 
element of $\partial\Sigma = B\cup H$, is again specified 
by $r = r_B$, while the inner boundary $H$ corresponds to 
$r = 2{\rm m}$ (twice the geometrical mass of the 
blackhole). Let us set the task of evolving data 
$U|_{t_0}$ and $\partial_tU|_{t_0}$ given on $\Sigma_0$, in 
order to generate the solution $U$ on the finite domain 
$\mathcal{M} = [t_0,t_F] \times \Sigma$ with boundary 
$\partial\mathcal{M} = \Sigma_0\cup\Sigma_F\cup{}^3\!B
\cup{}^3\!H$.\footnote{Although it hardly needs to be
noted for this simple introduction, our time variable
$t$ is closely related to the advanced 
Eddington--Finkelstein coordinate $v$ discussed 
in {\sc Section} \ref{sbsbSec:4.1.1}.} 
Here ${}^3\!H$, a portion of the {\em future event 
horizon}, is the three--dimensional characteristic history 
$[t_0,t_F]\times H$ of $H$. To accomplish the task, we need 
explicit outer boundary conditions on ${}^3\!B = [t_0,t_F]
\times B$, ones stemming from the assumption of trivial initial 
data on the outer spatial region exterior to $\Sigma_0$. We 
refer to these as {\em radiation outer boundary conditions} 
({\sc robc}).\footnote{We include the adjective ``outer'' 
in our acronym in order to distinguish between boundary 
conditions at $B$ and those at $H$. As shown later, setting 
appropriate boundary conditions at $H$ is not nearly so 
difficult for our problem, heuristically since $H$ acts as 
a one--way membrane of sorts. However, for dynamical 
spacetimes the issue of inner boundary conditions (at 
``apparent horizons'') is a difficult problem in its own 
right.} The {\sc robc} corresponding to 
(\ref{Schwarzschildwaveeq}) are more subtle than simple 
nonreflection, in part due to the back--scattering of waves 
off of curvature. 

In discussing {\em finite} outer boundary conditions, and
{\sc robc} in particular, for relativity, we should first
make a distinction between general relativity, in which the
dynamics of spacetime is governed by the full nonlinear
Einstein equations, and its perturbation theory, in which
the dynamics of disturbances on a fixed background
solution to the Einstein equations is governed by a linear
{\sc pde} similar to (\ref{Schwarzschildwaveeq}). In this
second paradigm one examines the propagation of weak
gravitational waves on a fixed background spacetime (which
may or may not be curved). York's survey article \cite{York}
on the dynamics and kinematics of general relativity is the
best jumping off point for a study of the literature we now
mention. Within the context of a mixed Cauchy--boundary
value problem, Friedrich and Nagy have made theoretical
progress towards solving the full Einstein equations on a
bounded domain \cite{FriedrichNagy}; however, their results
do not appear suited for numerical work. For the most part,
approaches towards numerical outer boundary conditions in
the full theory have relied either on matching Cauchy
domains to characteristic surfaces (see
\cite{Bishopetal1,Bishopetal2} and references therein) or
ensuring that the outer boundary is at a large enough
distance so that perturbation theory can be brought into
play (see \cite{Rupright,Rezzolla} and reference therein).
In this latter approach, the relevant perturbative
wave equation is essentially (\ref{Schwarzschildwaveeq});
however, the corresponding exact nonlocal {\sc robc} are
not used. A very different approach towards theoretical and
numerical outer boundary conditions has been given in
\cite{Lau}. However, it would seem of limited practical use,
since it relies on the ``many--fingered'' nature of time in
general relativity to completely freeze the flow of time at
the outer boundary. In terms of harmonic coordinates
Szil\'{a}gyi, Schmidt, and Winicour have theoretically and
numerically studied mixed Cauchy--boundary value problems
for the Einstein equations linearized about flat Minkowski
spacetime \cite{SziSchWin}. Attempts towards implementation
of {\sc robc} in numerical relativity have mostly relied on
improved versions of the well--known Sommerfeld condition,
although Novak and Bonazzola have considered more general
nonreflecting boundary conditions with relativistic
applications in mind \cite{NovBon}.
The Sommerfeld condition is a local boundary condition which
is exact for a spherically symmetric outgoing wave in
flatspace. Recently, Allen, Buckmiller, Burko, and Price 
have discussed the effect of such approximate boundary 
conditions on long--time numerical simulation of waves on 
the Schwarzschild geometry \cite{ABBP} (we consider one of 
their numerical experiments below). For comments on such 
approaches as well as other remarks on numerical relativity, 
see the review article by Cook and Teukolsky \cite{CookTeu}. 
To date, there seems to have been no truly systematic analysis 
of algorithm error for treatments of {\sc robc} in numerical 
relativity.

We use frequency--domain methods and gather results which
resemble those used and found in seminal work \cite{Leaver3}
by Leaver and also in related work \cite{Andersson} by
Andersson. Despite this resemblance, neither Leaver nor
Andersson considered {\em boundary} integral kernels belonging
to the finite timelike cylinder $^3\!B$. The starting point
for these authors was the exact solution to the Cauchy problem
as expressed via an integral
(Green's function) representation involving $\Sigma_0$
{\em spatial} convolution with initial data (actually Leaver
also considered more general driving source terms beyond just
initial data). As Leaver noted in footnote 21 of
\cite{Leaver3}, the seeds of this approach are found in Morse
and Feshbach's 1953 treatment \cite{MorFesh} of the ordinary
flatspace wave equation, although they have origins in the
19th century. Many authors have since used frequency--domain
and complex--analytic methods to examine the Cauchy problem
for perturbations on the Schwarzschild geometry from this
Green's function perspective, and Andersson's article
\cite{Andersson} is a salient recent example. However, we
stress up--front that our problem of imposing {\sc robc} via
domain reduction is not the same as Leaver and Andersson's
problem, and our work has quite a different focus on $^3\!B$
{\em temporal} integral convolution. Moreover, the methods
---those of {\sc agh}--- that we describe and use in this
article and its follow--up were only fully developed for the
ordinary wave equation in the late 1990s. {\sc Section}
\ref{sbsbSec:1.4.3} further compares and contrasts our
theoretical analysis with that of Leaver and Andersson.

\subsection{Overview of results}
In this article we describe both the exact {\sc robc} 
for (\ref{Schwarzschildwaveeq}) and an algorithm for their 
rapid numerical implementation. As mentioned, our approach 
to the problem follows {\sc agh} quite closely. The equation 
(\ref{Schwarzschildwaveeq}) is linear, but necessarily with 
variable coefficients. Nevertheless, exploiting its time and 
rotational symmetries, we may likewise use Laplace and 
spherical--harmonic transformation in order to obtain a
second--order radial {\sc ode} which turns out to be an
incarnation of the {\em confluent Heun equation} 
\cite{CHE1,CHE2}, also related to the {\em generalized 
spheroidal wave equation}\footnote{The ordinary {\em 
spheroidal wave equation} stems from variable separation of 
the ordinary wave equation (\ref{flatspacewaveeq}) in 
oblate or prolate spheroidal coordinates \cite{MorFesh}.} 
discussed in some literature \cite{Wilson,Leaver1}. Following 
{\sc agh}, we may formally introduce the {\sc tdrk} as 
the inverse Laplace transform of the homogeneous logarithmic 
derivative of the asymptotically outgoing solution. Analytically, 
the {\sc fdrk}, the logarithmic derivative in question, is a sum 
of poles, although now the sum is over both a discrete set
and a continuous set (similar to the situation for the 
flatspace wave equation
in $2+1$ rather than 3+1 dimensions \cite{AGH1}).

Employing both direct numerical construction of blackhole 
{\sc fdrk}'s as well as their compression along the lines of 
{\sc agh}, we numerically implement the exact {\sc robc} for 
waves on Schwarzschild. In principle, we may implement the 
conditions to arbitrary numerical accuracy even for long--time 
simulations. Via study of one--dimensional radial 
evolutions, we demonstrate that the described {\sc robc} 
capture the long--studied phenomena of both quasinormal 
ringing and late--time decay tails. Since our implementation
is based on the exact nonlocal {\em history--dependent} 
{\sc robc}, we sidestep issues raised by Allen, Buckmiller, 
Burko, and Price \cite{ABBP} and are indeed able to capture 
decay tails with our boundary conditions. We demonstrate 
this below with an example considered in \cite{ABBP}. 
We also consider a three--dimensional evolution based on a 
spectral code, one 
showing that the {\sc robc} yield accurate results for the 
scenario of a wave packet striking the boundary at an angle. 
Our work is a partial generalization 
to Schwarzschild wave propagation and Heun functions of the 
methods developed for flatspace wave propagation 
and Bessel functions by {\sc agh}, save for one key 
difference. Whereas {\sc agh} had the usual armamentarium 
of analytical results (asymptotics, order recursion 
relations, bispectrality) for Bessel functions at their 
disposal, what we need to know about Heun functions must 
be gathered numerically as relatively less is known about 
them. 

Due to this key difference, we are, unfortunately, unable
to offer a rigorous asymptotic analysis of our rapid
implementation. However, our numerical work suggests that
the number $d$ of approximating poles grows at a rate not   
at odds with the one mentioned above. Indeed, as seen
in Table \ref{threeTable2} from {\sc Section} \ref{sbsbSec:3.3.1}, for 
$\varepsilon = 10^{-10}$ and $r_B = 30\mathrm{m}$ we have 
found that $d = 20$ is sufficient for all $l \leq 64$. With
the same $r_B$ but $\varepsilon = 10^{-6}$ instead, $d = 14$ 
is sufficient for all $l \leq 256$. These $d$ values gives us
some idea of the cost associated with our rapid implementation.
Let us focus attention on a single spherical harmonic mode,
ignoring the cost of performing a (numerical) harmonic
transform. Depending on the algorithm used for wave simulation,
such a transform may or may not be performed at each numerical
time step. For the range of $l$ values and $\varepsilon$  
accuracies we consider, $d$ is roughly on the order of 10.  
Therefore, at each numerical time step we expect a boundary
operation count which is some multiple $10p$ of this $d$
value. (Actually poles whose locations are properly complex
are more costly than those which lie on the real axis. This
affects $p$, but no matter.) A practical radial discretization
of $[2\mathrm{m},30\mathrm{m}]$ will have some multiple $1000q$
of a thousand mesh points, whence $p/q$ is a straightforward
estimate for the {\em percentage} cost (at each numerical
time--step) of updating the solution at the edge of the
computational domain relative to the cost of updating on the
interior. Note that while $p$ remains fixed, $q$ increases with
mesh resolution, so that the cost of our {\sc robc} relative to
interior cost becomes accordingly negligible. Let us support
this assertion with a concrete example. For the one--dimensional
radial evolution described in {\sc Section} \ref{sbsbSec:5.1.1},
we have kept track of the total {\sc cpu} time spent both on 
interior work and the {\sc robc}. We list the ratio of these 
times as percentages in Table \ref{introTable1}. Such 
percentages belie the true savings of the implementation. 
Without some form of {\sc robc}, one would be forced to consider 
the free evolution of waves on a domain {\em larger} than 
$[2\mathrm{m},30\mathrm{m}]$, one large enough to ensure that
the waves would not reach the larger
outer boundary during the simulation.\footnote{Or at least
large enough so that waves reflected off the outer
boundary would not disturb the smaller computational domain of
interest.} The cost of using a larger domain as ``boundary
conditions'' is usually some {\em multiple} of the interior
cost. Therefore, our {\sc robc} typically cost less than a 
percent of what evolution on such a larger domain costs.   
In {\sc Section} \ref{sbsbSec:4.3.3} we discuss memory and 
storage issues relevant to our implementation of {\sc robc}.
\begin{table}
\begin{center}
\begin{tabular}{|l||c|c|c|c|c|}
\hline
\multicolumn{1}{|c||}{} &
\multicolumn{5}{c|}{Number of radial mesh points}\\
\cline{2-6}
$l$ & 1024 & 2048 & 4096 & 8182 & 16384 \\
\hline
\hline
$0$ & 2.24 & 1.49 & 1.00 & 0.39 & 0.21 \\
\hline
$33$ & 2.41 & 1.76 & 1.31 & 0.50 & 0.26 \\
\hline
$64$ & 2.55 & 1.81 & 1.37 & 0.51 & 0.27 \\
\hline
\end{tabular}
\end{center}
\vskip 2mm
\caption{{\sc Relative cost of $\varepsilon = 10^{-10}$ robc.} For
the one--dimensional evolution described in {\sc Section} 
\ref{sbsbSec:5.1.1}, we list percentage values for {\sc cpu} time spent 
on {\sc robc} relative to {\sc cpu} time spent on interior work. Whence 
each value is essentially the
aforementioned ratio $p/q$ for a particular $q$.} 
\label{introTable1}
\end{table}

\subsection{Summary}

{\sc Section} \ref{Sec:1} 
discusses variable transformations, various resulting forms 
of (\ref{Schwarzschildwaveeq}), and the exact {\sc robc}.
We start off by defining dimensionless coordinates for 
time $\tau$, radius $\rho$, and Laplace frequency $\sigma$.
For example, $r = 2{\rm m}\rho$ and $s = \sigma/(2{\rm m})$.
The outer boundary $B$ is determined $\rho = \rho_B$. With these 
coordinates we introduce the asymptotically outgoing solution 
$W_l(\sigma\rho;\sigma)$ to the radial equation, the one 
corresponding to the Bessel--type function $W_l(sr) = 
W_l(\sigma\rho)$ above. For a given angular index $l$
the {\sc tdrk} $\omega_l(\tau;\rho_B)$ is the inverse Laplace 
transform of the {\sc fdrk} $\tilomega_l(\sigma;\rho_B) 
= \sigma \rho_B W_l'(\sigma\rho_B;\sigma)/
W_l(\sigma\rho_B;\sigma)$. We then write 
the {\sc robc} as an integral convolution between 
$\omega_l(\tau;\rho_B)$ and each of the corresponding $2l+1$ 
modes $\psi_{lm}(\tau,\rho_B)$ of the 
radiating field $\psi(\tau,\rho,\theta,\phi)$, 
where $\psi$ is $U$ from above but expressed in terms of different 
coordinates. Afterwards, we describe the key representation of 
$\tilomega_l(\sigma;\rho_B)$ as a (continuous and discrete) 
sum of poles. 
{\sc Section} \ref{Sec:1} ends with the derivation of an 
estimate for 
the relative error $\varepsilon$ associated with approximating 
$\omega_l(\tau;\rho_B)$ by a numerical kernel 
$\xi_l(\tau;\rho_B)$.

{\sc Section} \ref{Sec:2} describes numerical evaluation of
both $W_l(\sigma\rho_B;\sigma)$ and $\tilomega_l(\sigma;\rho_B$), 
with the former considered as a function of complex Laplace 
frequency $\sigma$ (mostly lying in the lefthalf plane) and the 
latter as a function of purely imaginary $\sigma = {\rm i}y$. 
Both types of evaluation rely on numerical integration over 
certain paths in the complex plane. We consider several numerical 
methods, but the main ones involve path integration in terms 
of a complex variable $z =\sigma\rho$. While numerical 
evaluation of $W_l(\sigma\rho_B;\sigma)$ is important 
insofar as studying the analytic structure of this 
function is concerned, implementation of {\sc robc} 
mainly requires that we are able to accurately evaluate 
$\tilomega_l({\rm i}y;\rho_B)$ for any $y \in \mathbb{R}$.
In this section we also discuss in detail the
accuracy of our numerical methods.

{\sc Section} \ref{Sec:3} focuses on the 
sum--of--poles representations of $\tilomega_l(\sigma;\rho_B)$. 
The first subsection is a qualitative description of the 
analytic structure of $W_l(\sigma\rho_B;\sigma)$ and its 
relevance for the exact sum--of--poles representation. This 
subsection examines the zeros
in frequency $\sigma$ of $W_l(\sigma\rho_B;\sigma)$ which
correspond to poles of $\tilomega_l(\sigma;\rho_B)$. It
also studies the branch behavior of 
$W_l(\sigma\rho_B;\sigma)$ along the negative Re$\sigma$ axis, 
behavior that gives rise to a continuous 
pole distribution (these are not really poles in the sense of
complex analysis). This 
distribution appears in the 
exact sum--of--poles representation, and we graphically 
examine it. The second subsection presents our direct numerical 
construction of $\tilomega_l(\sigma;\rho_B)$ for
$0\leq l \leq 10$ and $\rho_B \in [15,25]$. We discuss several 
numerical accuracy checks of our direct construction. 
The third describes kernel compression, and the
resulting approximation of $\tilomega_l(\sigma;\rho_B)$ by a 
rational function which is itself a sum of poles. In this
third subsection we consider the bandwidth $0\leq l \leq 64$. 

{\sc Section} \ref{Sec:4} presents the details of
our implementation of {\sc robc}. As mentioned, our implementation 
is designed around the MacCormack predictor--corrector algorithm 
\cite{LeVeque}. The first subsection describes both our choice of 
spacetime foliation and the associated first--order evolution system 
of {\sc pde}, while the second reviews the MacCormack algorithm in 
the context of this system. As an aside, the second subsection also 
addresses the issue of inner boundary conditions at the horizon $H$.
The third subsection describes the implementation, showing how
{\sc robc} fit within the framework of the interior prediction
and correction. In this final subsection we also
address some memory issues relevant to our implementation.

{\sc Section} \ref{Sec:5}
documents the results of several numerical
tests of our implementation of {\sc robc}. Throughout this
section, we consider a blackhole of mass $\mathrm{m} = 2$
enclosed within an outer boundary of radius $r_B = 60$ so
that the horizon is located at $2\mathrm{m} = 4$ and $\rho_B
= 15$. The choice $\mathrm{m} = 2$ is not particularly
special, and has been made only to provide an example in
which the mass is neither $\frac{1}{2}$ nor $1$. The bulk of this
section focuses on one--dimensional radial evolutions of   
single spherical--harmonic modes. However, in the last 
subsection we consider a three--dimensional test.

A final sections offers some conclusions
and also discusses applications and extensions of our results.

In {\sc Appendix}~\ref{Sec:A} we discuss a certain modification 
of the MacCormack algorithm which is featured in our implementation 
of {\sc robc}. In {\sc Appendix}~\ref{Sec:B} we provide numerical
tables for several compressed kernels.

Starting on page~\pageref{listofsymbols} we have listed our 
{\em main} symbols in order of appearance, with the section 
number given for where each symbol first appears. Symbols not 
listed there are defined and used locally. Although some 
symbols in this work have multiple meanings, within a given 
section a symbol's meaning does not change. We point out that 
as a complex variable $\sigma = x + \mathrm{i}y$, and, 
therefore, for the complex variable $z = \sigma\rho$ we 
always write $z = \mathrm{Re}z + \mathrm{i}\mathrm{Im}z$. 
%
%
\section{Wave equation and radiation outer boundary 
conditions}\label{Sec:1}
This section sets up the theoretical framework on which 
the subsequent sections rest. We here discuss various {\sc pde}
and {\sc ode} relevant for wave propagation on the 
Schwarzschild geometry. We then derive the exact
and nonlocal radiation outer boundary conditions ({\sc robc})
appropriate for asymptotically outgoing fields, thereby
paving the way for their numerical implementation in later
sections.

\subsection{Wave equation on Schwarzschild background}\label{sbSec:1.1}

\subsubsection{Line--element}\label{sbsbSec:1.1.1}
Consider the diagonal line--element describing a static, 
spherically symmetric, vacuum blackhole of mass $\mathrm{m}$,
\begin{equation}
\dee s^2 = - F \dee T^2
           + F^{-1} \dee r^2
             + r^2 \dee \theta^2
             + r^2\sin^2\theta \dee \phi^2\, ,
\label{staticelement}
\end{equation}
written with respect to the standard static time $T$ and areal radius
$r$ \cite{MTW,Taylor}. We use uppercase $T$ here to save
lowercase $t$ for a different time coordinate needed later. 
Note that the metric coefficient $F(r) = 1 - 2\mathrm{m}/r$ 
vanishes ---and so $F^{-1}$ is singular--- as $r\rightarrow
2\mathrm{m}$. As is well known, $r = 2\mathrm{m}$ does not 
represent a physical singularity,
rather the coordinate system is degenerate for this value of the radius.
In these coordinates the round sphere determined by $r = 2\mathrm{m}$ is 
the bifurcate cross--section of the event horizon of the blackhole. 
In this work, we are chiefly interested in the ``exterior region'' 
defined by $2\mathrm{m} < r < \infty$.

It will prove convenient to pass to and work with dimensionless
coordinates $(\tau,\rho,\theta,\phi)$ defined by
\begin{equation}
2\mathrm{m}\tau = T \, , \quad 2\mathrm{m}\rho = r
\end{equation}
($\theta$ and $\phi$ are already dimensionless). After the
rescaling $\dee s^2 \mapsto \dee s^2/(4\mathrm{m}^2)$, we may rewrite the
line--element (\ref{staticelement}) in the dimensionless form
\begin{equation}
\dee s^2 = - F \dee \tau^2
           + F^{-1} \dee \rho^2
             + \rho^2 \dee \theta^2
             + \rho^2\sin^2\theta \dee \phi^2\, ,
\label{dlesselement}
\end{equation}
where now $F(\rho) = 1-1/\rho$ so the unphysical singularity is located
at $\rho = 1$.

We also consider the outgoing and ingoing systems of 
Eddington--Finkelstein coordinates \cite{MTW,Taylor},
here in dimensionless form. To construct them, first 
introduce the Regge--Wheeler {\em
tortoise coordinate} \cite{ReggeWheeler,MTW}
\begin{equation}
\rho_{*} = \rho + \log(\rho-1)\, .
\end{equation}
Recall that this transformation is valid for $1 < \rho < \infty$
which corresponds to $-\infty < \rho_{*} < \infty$, and that
$\rho \rightarrow 1^{+}$ corresponds to $\rho_{*}
\rightarrow - \infty$. In terms of the tortoise coordinate 
we write (\ref{dlesselement}) as
\begin{equation}
\dee s^2 = F \big(-\dee \tau^2
           + \dee \rho_*^2\big)
             + \rho^2 \dee \theta^2
             + \rho^2\sin^2\theta \dee \phi^2\, .
\label{dlesselementstar}
\end{equation}  
The characteristic coordinate $\mu = \tau - \rho_{*}$ is the 
{\em retarded time}, 
and the set $(\mu,\rho,\theta,\phi)$ is the outgoing 
Eddington--Finkelstein system. With respect to it, the 
line--element takes the form
\begin{equation}
\dee s^2 = - F \dee \mu^2
           - 2\dee\mu\dee\rho
             + \rho^2 \dee \theta^2
             + \rho^2\sin^2\theta \dee \phi^2\, .
\label{oneEFelement}
\end{equation}
In the $(\mu,\rho,\theta,\phi)$ system $g_{\rho\rho} = 0$, so 
that the vector field $\partial/\partial\rho$ is characteristic 
or {\em null}, whereas in the $(\tau,\rho,\theta,\phi)$ system 
$g_{\rho\rho} = F^{-1}(\rho)$, so that $\partial/\partial\rho$ 
is spacelike on the exterior region. Level--$\mu$ hypersurfaces are 
characteristic and outgoing (cones which open up towards the future) 
with $\partial/\partial\rho$ as their outgoing generator. In 
{\sc Section} \ref{sbsbSec:4.1.1} we consider the ingoing 
Eddington--Finkelstein coordinate system which is based on the
characteristic coordinate $\nu = \tau + \rho_*$ known as 
{\em advanced time}.

\subsubsection{Wave equation}\label{sbsbSec:1.1.2}
The covariant d'Alembertian or wave equation associated
with the diagonal line--element (\ref{dlesselement}) is the following:
\begin{equation}
       \left(1-\frac{1}{\rho}\right)^{-1}
       \frac{\partial^2 \psi}{\partial\tau^2}
      -\frac{1}{\rho^2}\frac{\partial}{\partial\rho}
       \left[\rho^2\left(1-\frac{1}{\rho}\right)
       \frac{\partial\psi}{\partial\rho}
       \right]
      - \frac{\Delta_{S^2}\psi}{\rho^2} = 0\, ,
\label{scalarwave}
\end{equation}
where $\Delta_{S^2}$ is the Laplace operator (with 
negative eigenvalues) belonging to the unit--radius round sphere
$S^2$. Notice that we use $\psi$ for the wave field associated with 
the static time coordinate $\tau$ (or associated with its counterpart 
$T$) introduced above, whereas in the introduction we have used $U$ 
for the wave field. Later we use $U$ for the wave field associated with 
a certain time variable $t$ related to ingoing Eddington--Finkelstein 
coordinates. Our numerical work is based on $t$ (which is why $U$ and 
$t$, rather than $\psi$ and $T$, appear in the introduction). For 
flat spacetime $T$ and $t$ are the same, and so $\psi$ and $U$ are 
also formally the same for $\mathrm{m} = 0$.

Introducing the standard set
$Y_{lm}(\theta,\phi)$ of basis
functions for square--integrable functions on $S^2$, we consider
an appropriate expansion
\begin{equation}
\psi(\tau,\rho,\theta,\phi) =
\sum_{l,m}
\psi_{lm}(\tau,\rho) Y_{lm}(\theta,\phi)
\end{equation}
of the field $\psi$ in terms of spherical--harmonic modes
$\psi_{lm}$. The spherical--harmonic transform of
(\ref{scalarwave}),
\begin{equation}
       \left(1-\frac{1}{\rho}\right)^{-1}
       \frac{\partial^2 \psi_l}{\partial\tau^2}
      -\frac{1}{\rho^2}\frac{\partial}{\partial\rho}
       \left[\rho^2\left(1-\frac{1}{\rho}\right)
       \frac{\partial\psi_l}{\partial\rho}
       \right]
      + \frac{l(l+1)\psi_l}{\rho^2}
      = 0\, ,
\label{maineq}
\end{equation}
is the {\sc pde} governing the evolution of a generic mode     
$\psi_{l}$. On $\psi_{l}$ we have suppressed the $m$, since 
it does not appear in the {\sc pde}.

Addition of a single simple term to (\ref{maineq}) yields a
modified wave equation flexible enough to describe either the
mode evolution of an electromagnetic field $A_\beta$ or the 
mode evolution of small gravitational perturbations 
$\delta g_{\alpha\beta}$ on the Schwarzschild background. 
The modified equation is
\begin{equation}
       \left(1-\frac{1}{\rho}\right)^{-1}
       \frac{\partial^2 \psi_l}{\partial\tau^2}
      -\frac{1}{\rho^2}\frac{\partial}{\partial\rho}
       \left[\rho^2\left(1-\frac{1}{\rho}\right)
       \frac{\partial\psi_l}{\partial\rho}
       \right]
      + \frac{l(l+1)\psi_l}{\rho^2}
      - \frac{\jmath^2\psi_l}{\rho^3}
       = 0
\label{maineq2}
\end{equation}
with the {\em spin} $\jmath = 0,1,2$ corresponding to scalar,
electromagnetic, and gravitational radiation. We review the 
history of this correspondence in the next paragraph. We may 
cast (\ref{maineq2}) in a particularly simple form via
simultaneous transformation of the independent and dependent
variables. Indeed, setting $\Psi_l = \rho\psi_l$ and here 
viewing $\partial/\partial\rho_{*}$ as shorthand for
$(1-\rho^{-1})\partial/\partial\rho$, we rewrite (\ref{maineq2})
as follows:
\begin{equation}
\frac{\partial^2\Psi_{l}}{\partial\tau^2} -
\frac{\partial^2\Psi_{l}}{\partial\rho^2_{*}} + V(\rho)\Psi_l =
0\, . \label{maineq3}
\end{equation}
The {\em Regge--Wheeler potential}
\begin{equation}
V(\rho) = \left(1-\frac{1}{\rho}
\right)\left[\frac{l(l+1)}{\rho^2} +
  \frac{1-\jmath^2}{\rho^3}\right]
\label{onepotential}
\end{equation}
would depend only implicitly on $\rho_{*}$ were we using
$\rho_{*}$ as the independent variable. As we will see in
{\sc Section} \ref{sbsbSec:1.2.3}, the Laplace transform of
(\ref{maineq3}) is important theoretically, since it
elucidates the role of Laplace frequency as a spectral 
parameter.

Wheeler derived the $j=1$ version of 
(\ref{maineq3},\ref{onepotential}) 
in 1955 \cite{Wheeler}, showing that each of the two polarization
states for an electromagnetic field on the Schwarzschild geometry
is described by one copy of the equation. Regge and Wheeler then 
derived the $\jmath = 2$ equation for odd--parity (or axially) 
gravitational perturbations in 1957 \cite{ReggeWheeler}, and Zerilli 
introduced a similar equation describing even--parity (or polar) 
gravitational perturbations in 1970 \cite{Zerilli}. In the 1970s 
Chandrasekhar and Detweiler demonstrated that the Zerilli equation 
can be derived from   
(\ref{maineq3}), although the derivation involves differential
operations (see \cite{ChandraDet} and references therein).

Adopting $\mu$ as the time coordinate, we define $\varphi_l 
(\mu,\rho) = \psi_l(\mu+\rho_{*},\rho)$ and write 
(\ref{maineq2}) as
\begin{equation}
       2\frac{\partial^2 \varphi_l}{\partial\mu\partial\rho}
      +\frac{2}{\rho}\frac{\partial\varphi_l}{\partial\mu}
      -\frac{1}{\rho^2}\frac{\partial}{\partial\rho}
       \left[\rho^2\left(1-\frac{1}{\rho}\right)
       \frac{\partial\varphi_l}{\partial\rho}
       \right]
      + \frac{l(l+1)\varphi_l}{\rho^2}
      - \frac{\jmath^2\varphi_l}{\rho^3}
       = 0\, .
\label{maineq2EF}
\end{equation}
Another way to obtain this wave equation is to form the
d'Alembertian associated with (\ref{oneEFelement}) and then
implement a spherical--harmonic transformation. 
Similar to above, we may either set $\Phi_l = \rho\varphi_l$ or
$\Phi_l(\mu,\rho) = \Psi_l(\mu+\rho_{*},\rho)$, thereby
expressing (\ref{maineq3}) as
\begin{equation}
2\frac{\partial^2\Phi_{l}}{\partial\mu\partial\rho_{*}} -
\frac{\partial^2\Phi_{l}}{\partial\rho^2_{*}} + V(\rho)\Phi_l =
0\, , \label{maineq3EF}
\end{equation}
again viewing $\partial/\partial\rho_{*}$ as a shorthand. As we
will see, for a given $l$ the {\sc fdrk} is built from the 
outgoing solution to the formal Laplace transform of 
(\ref{maineq3EF}). Table \ref{oneTable1} lists the wave fields 
we have introduced, and it also briefly describes the 
theoretical importance of each field's Laplace transform. 
The statements made in the table are 
explained in {\sc Sections} \ref{sbSec:1.2} and \ref{sbSec:1.3}.
\begin{table}\begin{center}
\small
\begin{tabular}{|c|l||c|l|}\hline
\multicolumn{2}{|c||}{Static time coordinate system 
$(\tau,\rho)$} &
\multicolumn{2}{c|}{Retarded time coordinate system 
$(\mu,\rho)$}
\\
\hline
$\psi_l(\tau,\rho)$ 
&
{\sc ode} for L.t.~is analogous to
&
$\varphi_l(\mu,\rho)$  
&
$\jmath = 0$ {\sc ode} for L.t.~is directly 
\\
&
the spherical Bessel equation.
& 
&
the confluent Heun equation.
\\
\hline
$\Psi_l(\tau,\rho)$ 
& 
{\sc ode} for L.t.~elucidates role
&
$\Phi_l(\mu,\rho)$
&
{\sc ode} for L.t.~has outgoing
\\
&
of $\sigma$ as a spectral parameter.
& 
&
solution normalized at $\infty$.
\\
\hline
\end{tabular}
\end{center}
\vskip 2mm
\caption{{\sc Wave fields and their relevance.}
L.t.~stands for {\em Laplace transform}. As we discuss in 
{\sc Section} \ref{sbSec:1.2}, via the Laplace transform we 
trade a {\sc pde} for an {\sc ode}.}
\label{oneTable1}
\end{table}

\subsection{Laplace transform and radial wave 
equation}\label{sbSec:1.2}
\subsubsection{Laplace transform}\label{sbsbSec:1.2.1}
The Schwarzschild geometry is static,\footnote{More precisely,
$\partial/\partial\tau$ is a hypersurface--orthogonal vector field
which satisfies the Killing equation $\pounds_{\partial/\partial\tau}
g_{\mu\nu} = 0$, where $\pounds$ denotes Lie differentiation.}
and with respect to the chosen coordinates we indeed see that the
components of the metric tensor are
$\tau$--independent. In turn, the variable coefficients of the linear
wave equation described in the last subsection do not depend on time,
a scenario permitting study of the equation via the technique of
Laplace transform. The description of the technique in Chapter 12 
of the textbook \cite{Greenberg} by Greenberg is suitable for our
purposes. Let $\mathcal L$ denote the transform operation,
\begin{equation}
{\mathcal L}[g](\sigma) = \int^{\infty}_{0} e^{-\sigma\tau}
                          g(\tau){\rm d}\tau\, .
\end{equation}
Here we use $\sigma$ for the variable dual to $\tau$ with respect
to the Laplace transform. We may define a physical variable $s =
\sigma/(2\mathrm{m})$, with dimensions of inverse length, which is dual to
$t$ and satisfies $st = \sigma\tau$. We may also define a formal 
Laplace transformation on the retarded time $\mu$ by replacing 
$\tau$ with $\mu$ in the last equation. For the time being we 
proceed with the transformation on $\tau$.

\subsubsection{Laplace transform of the wave equation}\label{sbsbSec:1.2.2}
Let us formally compute the Laplace transform of (\ref{maineq2}),
in order to get an {\sc ode} in the radial variable $\rho$.
With $\tilpsi_l = {\mathcal L}[\psi_l]$ and a dot denoting $\tau$
differentiation, we have
\begin{align}
      {\mathcal L}[\dot{\psi}_l](\sigma,\rho)  & = 
                                          \sigma\tilpsi_l(\sigma,\rho)
                                        - \psi_l(0,\rho)\, ,
                                          \nonumber \\
      {\mathcal L}[\ddot{\psi}_l](\sigma,\rho) & = 
                                          \sigma^2\tilpsi_l(\sigma,\rho)
                                         - \sigma\psi_l(0,\rho)
                                         - \dot{\psi}_l(0,\rho)\, ,
\end{align}
provided $\mathrm{Re}\sigma > 0$.
We assume that the initial data $\psi_l(0,\rho)$ and
$\dot{\psi}_l(0,\rho)$ vanish in a neighborhood of $\rho$,
as is true for compactly supported data so long as we
choose $\rho$ large enough. This 
assumption ensures formally that upon Laplace transformation we may 
replace $\tau$ partial differentiation by $\sigma$ multiplication. 
Whence, after some simple algebra we find
\begin{equation}
      \frac{\dee^2\tilpsi_l}{\dee\rho^2}
     +\frac{2\rho-1}{\rho(\rho-1)}
      \frac{\dee\tilpsi_l}{\dee\rho}
     +\left[
      \frac{-\sigma^2\rho^2}{(\rho-1)^2}
     -\frac{l(l+1)}{\rho(\rho-1)}
     +\frac{\jmath^2}{\rho^2(\rho-1)}
      \right]
      \tilpsi_l = 0
\label{maineq4}
\end{equation}
for the Laplace transform of (\ref{maineq2}).

It is instructive to see what happens to (\ref{maineq4}) in the
$\mathrm{m} \rightarrow 0^{+}$ limit. Before taking the limit, 
first recall that $\rho = r/(2\mathrm{m})$ and 
$\sigma = 2 \mathrm{m} s$, so that the product 
$z = \sigma\rho = sr$ is independent of $\mathrm{m}$. 
With this in mind, we divide the overall equation by a factor 
of $\sigma^2$ and find
\begin{equation}
      \frac{\dee^2\tilpsi_l}{\dee z^2}
     +\frac{2z-\sigma}{z(z-\sigma)}
      \frac{\dee\tilpsi_l}{\dee z}
     +\left[
      \frac{-z^2}{(z-\sigma)^2}
     -\frac{l(l+1)}{z(z-\sigma)}
     +\frac{\sigma\jmath^2}{z^2(z-\sigma)}
      \right]
      \tilpsi_l = 0\, ,
\label{maineq4z}
\end{equation}
where $\sigma$ is now shorthand for $2\mathrm{m}s$. 
Formally then, the $\mathrm{m}
\rightarrow 0^{+}$ limit along with multiplication by $z^2$ sends 
the last equation into the {\em
modified spherical Bessel equation} ({\sc msbe}) \cite{MorFesh}:
\begin{equation}
      z^2\frac{\dee^2\tilpsi_l}{\dee z^2}
     +2z\frac{\dee\tilpsi_l}{\dee z}
     -\left[
      z^2
     +l(l+1)
      \right]
      \tilpsi_l = 0\, .
\label{msbe}
\end{equation}
As linearly independent solutions of the {\sc msbe} we may 
take
\begin{equation}
k_l(z) = \sqrt{\frac{\pi}{2z}}K_{l+1/2}(z) \, , \quad i_l(z) =
\sqrt{\frac{\pi}{2z}}I_{l+1/2}(z)\, , \label{sphericalbessels}
\end{equation}
where $K_{l+1/2}(z)$, MacDonald's function, and 
$I_{l+1/2}(z)$ are standard modified (cylindrical) Bessel 
functions of half--integer order \cite{Watson}. Later we 
emphasize the close
parallels between these Bessel functions and the corresponding 
solutions to (\ref{maineq4z}).

\subsubsection{Laplace frequency as a spectral 
parameter}\label{sbsbSec:1.2.3}
We observe that
\begin{equation}
      \frac{\dee^2\tilPsi_{l}}{\dee\rho^2_{*}}
     -V(\rho)\tilPsi_l
     =\sigma^2\tilPsi_l\, ,
\label{notselfadjoint}
\end{equation}
is the formal Laplace transform of (\ref{maineq3}). This is
a remarkable form of the radial {\sc ode} for several reasons.
First, as might be expected from the suggestive form of the 
equation, we could consider it in the context of an eigenvalue 
problem, although one in which the operator on the {\sc lhs} is 
not self--adjoint. More precisely, suppose we seek solutions to 
(\ref{notselfadjoint}) which vanish at $\rho = \rho_B$ (a fixed 
constant) and are also asymptotically 
outgoing, that is behave as $e^{-\sigma\rho}$ for large 
$\rho$. We do then (numerically) find solutions corresponding 
to a discrete (but finite) set of $\sigma$ values, but these turn 
out to be values in the lefthalf plane.\footnote{The results we 
describe in subsequent sections justify this statement, although 
in what follows we work with a different form of the {\sc ode} 
stemming from yet another transformation $\tilPsi_l = 
\exp(-\sigma\rho_{*})\tilPhi_l$ of the dependent variable. 
See {\sc Section} \ref{sbsbSec:1.3.2} and 
what follows.} For such
$\sigma$ the term $e^{-\sigma\rho}$ blows up as $\rho$ gets large, 
spoiling any possible self--adjointness for ${\rm d}^2/{\rm 
d}\rho^2_{*} - V(\rho)$ on $[\rho_B,\infty)$ with these boundary 
conditions. 

Let us also consider the Bessel analog of
(\ref{notselfadjoint}). Namely,
\begin{equation}
      \frac{\dee^2\tilPsi_{l}}{\dee\rho^2}
     -\frac{l(l+1)}{\rho^2}\tilPsi_l
     =\sigma^2\tilPsi_l\, .
\label{Besselnotselfadjoint}
\end{equation}  
We reach this equation by first passing to $z =\sigma\rho$ as 
above, taking the $\mathrm{m}\rightarrow
0^{+}$ limit, and then passing back to $\rho = z/\sigma$. 
For the type of eigenvalue problem mentioned above, the operator 
${\rm d}^2/{\rm d}\rho^2 - l(l+1)/\rho^2$ is again not 
self--adjoint; however, this fact is not our prime concern 
now. The discussion in {\sc Section} \ref{sbsbSec:1.2.2} shows that 
(\ref{Besselnotselfadjoint}) has solutions, such as 
$(\sigma\rho)^{1/2} K_{l+1/2}(\sigma\rho)$, of a special form. 
Indeed, they simultaneously solve an {\sc ode} in the 
spectral parameter $\sigma$ \cite{bispec},
\begin{equation}
      \frac{\dee^2\tilPsi_{l}}{\dee\sigma^2}
     -\frac{l(l+1)}{\sigma^2}\tilPsi_l
     =\rho^2\tilPsi_l\, .
\label{Besselnotselfadjoint2}
\end{equation}
Accordingly, we describe solutions to 
(\ref{Besselnotselfadjoint}) as {\em bispectral}. 
Unfortunately, solutions to the more complicated 
{\sc ode} (\ref{notselfadjoint}) are not
bispectral in this sense, and the lack of an associated
differential equation in the spectral parameter $\sigma$ 
complicates our numerical investigations. More comments on 
this point follow in {\sc Section} \ref{sbsbSec:2.1.2}.

\subsection{Normal and normalized form of the radial wave 
equation}\label{sbSec:1.3}
\subsubsection{Normal form}\label{sbsbSec:1.3.1}
Standard analysis 
\cite{BenOrs,Chattopadhyay} of 
the {\sc ode} (\ref{maineq4}) shows that $\rho = 0$ and
$\rho=1$ are regular singular points, corresponding respectively to
indicial exponents $\pm\jmath$ and
$\pm\sigma$, whereas $\rho = \infty$ is an irregular singular 
point \cite{Chattopadhyay}. To put (\ref{maineq4}) in a 
``normal form,'' we transform the
dependent variable $\tilpsi_l$ in order to (i) set one indicial 
exponent to zero at each singular point and (ii) ``peel--off'' the 
essential singularity at infinity as best we can. To this end, 
let us set
\begin{equation}
      \tilpsi_l = \rho^{\jmath}
                  (\rho - 1)^{-\sigma}
                  e^{-\sigma\rho}
                  \Theta_l
                =
                  \rho^{\jmath}
                  e^{-\sigma\rho_{*}}
                  \Theta_l\, ,
\label{normtrans}
\end{equation}
noting that such a transformation clearly achieves condition (i).
Moreover, the large--$\rho$ behavior of (\ref{maineq4})
suggests that in order to achieve condition (ii) we should peel
off either the factor $\exp(-\sigma\rho)$ or $\exp(\sigma\rho)$. Our
choice of peeling off $\exp(-\sigma\rho)$ indicates our intention
to examine asymptotically outgoing radiation fields. 
In (\ref{normtrans}) we have peeled off a factor of
$(\rho-1)^{-\sigma}$ rather than $(\rho - 1)^{\sigma}$ in order
that the tortoise coordinate appears in the argument of the 
exponential factor in the transformation. Under
our transformation Eq.~(\ref{maineq4}) becomes
\begin{equation}
\frac{\dee^2\Theta_l}{\dee\rho^2} 
+\left[-2\sigma + \frac{1+2\jmath}{\rho}
               +\frac{1-2\sigma}{\rho-1}
                \right]
               \frac{\dee\Theta_l}{\dee\rho}
               +\left[\frac{-2\sigma (1+\jmath)}{\rho-1}
               +\frac{\jmath (\jmath + 1) - l(l+1)}{\rho (\rho -1)}
                \right]\Theta_l = 0\, .
\label{maineq5}
\end{equation}
We remark that one may also obtain the $\jmath = 0$ version of 
(\ref{maineq5}) directly from (\ref{maineq2EF}) via formal 
Laplace transform on the retarded time $\mu$, i.~e.~for $\jmath = 0$ 
we can say $\Theta_l = \tilphi_l$.

Eq.~(\ref{maineq5}) is a realization of the {\em (singly) confluent 
Heun equation} \cite{CHE1,CHE2}
\begin{equation}
\frac{\dee^2\che}{\dee\rho^2}
               +\left[\beta + \frac{\gamma}{\rho}
               +\frac{\delta}{\rho-1} \right]
                \frac{\dee\che}{\dee\rho}
               +\left[\frac{\alpha\beta}{\rho-1}
               +\frac{q}{\rho (\rho-1)}
                \right]\che = 0\, ,
\label{oneCHE}
\end{equation}
which has the generalized Riemann scheme \cite{CHE2}
\begin{equation}
\left[\begin{array}{cccc}
1        & 1        & 2   &                         \\
0        & 1        & \infty                & ; \rho\\
0        & 0        & \alpha                & ; q   \\
1-\gamma & 1-\delta & \gamma+\delta-\alpha  &       \\
         &          & 0                     &       \\
         &          & -\beta                 &      \\
\end{array}
\right]\, .
\label{grs}
\end{equation}
The first three columns of the scheme's second row indicate
singular--point locations, while the corresponding columns of the first
row indicate their types. That is to say, we have regular
singular points at $\rho = 0$ and $\rho = 1$ and an irregular singular
point at $\infty$ which arises as the confluence of {\em two} regular 
singular points (the 2 in the third column of the first row indicates 
this confluence).\footnote{{\sc Appendix} B of \cite{LauMathDiss} shows 
how the confluent Heun equation arises from the {\em Heun equation}, an 
{\sc ode} similar 
to the familiar hypergeometric equation, although possessing four rather 
than three regular singular points.} The remaining information in the 
first two columns specifies the indicial exponents at the regular singular 
points, while the remaining information in the third column specifies 
the Thom\'{e} exponents corresponding to the two 
{\em normal solutions} about the point at $\infty$. These 
solutions have the asymptotic behavior
\begin{equation}
\che^{+} \sim \rho^{-\alpha} e^{0\cdot\rho}\, ,
\qquad
\che^{-} \sim \rho^{-\gamma-\delta+\alpha} e^{-\beta\rho}\, ,
\label{oneThomesolutions}
\end{equation}
as $\rho\rightarrow\infty$ (in some sector which we discuss later). 
Finally, in the fourth column we have the independent variable $\rho$ as 
well as the {\em accessory parameter} $q$. An {\sc ode} with the 
singularity structure of the confluent Heun equation is determined by 
the indicial exponents belonging to the regular singularities along with 
the Thom\'{e} exponents only up to a free parameter $q$. {\sc Appendix}
B of \cite{LauMathDiss} discusses this point in more detail.

\subsubsection{Normalized form at infinity}\label{sbsbSec:1.3.2}
Viewed as the confluent Heun equation, we see that our radial wave
equation (\ref{maineq5}) has the following generalized Riemann scheme:
\begin{equation}
\left[\begin{array}{cccl}
1        & 1        & 2   &                 \\
0        & 1        & \infty           & ; \rho\\
0        & 0        & 1+\jmath         & ; \jmath(\jmath+1)-l(l+1)\\
-2\jmath & 2\sigma  & 1+\jmath-2\sigma &    \\
         &          & 0                &    \\
         &          & 2\sigma          &    \\
\end{array}
\right]\, ,
\label{grs2}
\end{equation}
showing that the normal solutions to (\ref{maineq5})
obey
\begin{equation}
\Theta^{+}_l \sim \rho^{-1-\jmath} e^{0\cdot\rho}\, ,\qquad  
\Theta^{-}_l \sim \rho^{-1-\jmath+2\sigma}e^{2\sigma\rho}\, ,
\end{equation}
as $\rho\rightarrow\infty$. We may also write
$\Theta^{-}_l \sim \rho^{-1-\jmath}\exp(2\sigma\rho_{*})$
for the large--$\rho$ behavior of the second solution. The scheme 
(\ref{grs}) shows that confluent Heun functions are generally 
specified by five parameters. However, our specific scheme (\ref{grs2}) 
corresponds to a two--parameter family of functions (those 
parameters being $\sigma$ and $l$, with $\jmath$ viewed as 
fixed). This is comparable to the situation regarding the 
flatspace radial wave equation and the associated one--parameter 
family of Bessel functions (that parameter being the Bessel 
order $l+1/2$). Bessel functions (suitably transformed) 
are a one--parameter family within the larger two--parameter 
class of confluent hypergeometric functions (which may also be 
represented as either Whittaker functions or Coulomb wave 
functions) \cite{Slater,Curtis,Thompson}.

Numerical considerations below dictate that we work instead
with an outgoing solution which is ``normalized at infinity,''
that is to say approaches unity for large $\rho$. Therefore, we
now enact the transformation
\begin{equation}
\Theta_l = \rho^{-1-\jmath}\,
           \tilPhi_l\, ,
\end{equation}
or in terms of the original field $\tilpsi_l = \rho^{-1}
\exp(-\sigma\rho_{*})\tilPhi_l$, whereupon we find
\begin{equation}
                \frac{\dee^2\tilPhi_l}{\dee\rho^2}
                +\left[-2\sigma - \frac{1}{\rho}
                +\frac{1-2\sigma}{\rho-1} \right]
                 \frac{\dee\tilPhi_l}{\dee\rho}
                +\left[\frac{1-\jmath^2}{\rho^2} -
                 \frac{1-\jmath^2 + l(l+1)}{\rho (\rho-1)}
                 \right]\tilPhi_l = 0
\label{maineq6}
\end{equation}
as the {\sc ode} satisfied by $\tilPhi_l$. Remarkably, this
equation agrees with that obtained directly from
(\ref{maineq3EF}) via formal Laplace transform on the retarded
time $\mu$, whence our choice $\tilPhi_l$ with a hat for the 
dependent variable here. We emphasize that this statement is 
true for all possible spin values ($\jmath = 0,1,2$), whereas
the identification $\Theta_l = \tilphi_l$ mentioned before is 
valid only for $j = 0$. 

We again set $z = \sigma\rho = sr$ (independent of $\mathrm{m}$) and
divide (\ref{maineq6}) by an overall factor of $\sigma^2$,
thereby reaching the following particularly useful form of the 
radial wave equation:
\begin{equation}
                \frac{\dee^2\tilPhi_l}{\dee z^2}
                +\left[-2 - \frac{1}{z}
                +\frac{1-2\sigma}{z-\sigma} \right]
                 \frac{\dee\tilPhi_l}{\dee z}
                +\left[\frac{1-\jmath^2}{z^2} -
                 \frac{1-\jmath^2 + l(l+1)}{z (z-\sigma)}
                 \right]\tilPhi_l = 0\, .
\label{maineq7}
\end{equation}
With $W_l(z;\sigma)$ and $Z_l(z;\sigma)$ respectively denoting 
the outgoing and ingoing solutions to this {\sc ode}, the 
corresponding solutions to (\ref{maineq6}) are 
$W_l(\sigma\rho;\sigma)$ and $Z_l(\sigma\rho;\sigma)$ with 
$\sigma$ here viewed as fixed. As $\rho\rightarrow\infty$ these 
obey
\begin{equation}
W_l(\sigma\rho;\sigma) \sim 1\, , \qquad Z_l(\sigma\rho;\sigma)
\sim e^{2\sigma\rho_*}\, ,
\end{equation}
as shown by the material presented above. Respectively, we 
might also denote $W_l$ and $Z_l$ by 
$\tilPhi^+_l$ and $\tilPhi^-_l$.
We will often refer to $W_l(\sigma\rho;\sigma)$ as a confluent Heun 
function (or just Heun function), even though it differs from a Heun 
function by the $\rho^{-1-\jmath}$ factor (as shown above). This
terminology streamlines our presentation, allowing us to draw the
flatspace/Schwarzschild distinction via the Bessel/Heun modifiers.

Recall that $\sigma \rightarrow 0$ as $\mathrm{m}\rightarrow 0^+$, 
whereas the product $z = \sigma\rho$ remains fixed in the said limit.
Therefore, in the $\mathrm{m}\rightarrow 0^+$ limit 
Eq.~(\ref{maineq7}) becomes
an {\sc ode}
\begin{equation}
\frac{\dee^2\tilPhi_l}{\dee z^2} - 2\frac{\dee\tilPhi_l}{\dee z}
-\frac{l(l+1)}{z^2}\tilPhi_l = 0 \label{nbesseq}
\end{equation}
which could also be obtained straight from the {\sc msbe}
(\ref{msbe}) via the transformation $\tilpsi_l =
z^{-1}e^{-z}\tilPhi_l$. In terms of the two--parameter
functions introduced above, $W_l(z;0)$ and $Z_l(z;0)$
are respectively the formal outgoing and ingoing solutions to 
(\ref{nbesseq}). We shall also write these as simply
$W_l(z)$ and $Z_l(z)$ when there is no cause for confusion.
A few examples may be illuminating. The first three outgoing
solutions to the {\sc msbe} are the following {\em spherical 
MacDonald functions}:
\begin{equation}
k_{0}(z) = \frac{e^{-z}}{z}\, , \quad
k_{1}(z) = \frac{e^{-z}}{z}\left(1+\frac{1}{z}
                 \right)\, ,\quad
k_{2}(z) = \frac{e^{-z}}{z}\left(1+\frac{3}{z}+
               \frac{3}{z^2}\right)\, .
\end{equation}
Now consider the following polynomials in inverse $z$:
\begin{equation}
W_{0}(z) = 1\, , \quad W_{1}(z) = 1+\frac{1}{z}\, ,\quad W_{2}(z)
= 1+\frac{3}{z}+
               \frac{3}{z^2}\, .
\end{equation}
From the discussion above we see that these are outgoing solutions
to (\ref{nbesseq}), and clearly ones which are normalized at
infinity. We shall see that outgoing solutions $W_l(z;\sigma)$ to
(\ref{maineq7}) are similar, albeit not simple
polynomials in inverse $z=\sigma\rho$.

\subsubsection{Asymptotic expansion for normalized form}\label{sbsbSec:1.3.3}
For our purposes Eq.~(\ref{maineq7}) will prove the most useful
form of the frequency--space radial wave equation, so let us
describe its outgoing solution $W_l(z;\sigma)$ as a formal asymptotic
series. Our discussion in {\sc Section} \ref{sbsbSec:1.3.2} 
focused on the variable 
$\rho$, but the same normalization issues are pertinent for $z$.
First, for convenience we set $\kappa = 1-\jmath^2$; hence
$\kappa$ takes the values $1,0,-3$ for scalar, electromagnetic, and
gravitational cases respectively. We will often work with $\kappa$ 
instead of $\jmath$.  Here and in what follows, we suppress the
solution's $\kappa$ dependence.

Assume a solution to (\ref{maineq7}) taking the form
\begin{equation}
W_l(z;\sigma) \sim \sum_{n=0}^\infty d_n(\sigma)z^{-n}\, ,
\label{seriesatinfty}
\end{equation}
demanding that $d_0(\sigma) = 1$. Of course the remaining
$d_n(\sigma)$ will in
general also depend on $l$ and $\kappa$, but we suppress
this dependence here. Standard calculations then determine both
$d_1(\sigma) = l(l+1)/2$ and the following three--term recursion
relation:
\begin{equation}
d_{n+1}(\sigma)
= \frac{[l(l+1)-n(n+1)]d_n(\sigma)
+\sigma (n^2 + \kappa - 1)
d_{n-1}(\sigma)}{2(n+1)}\, .
\label{bthreeterm}
\end{equation}
A dominant balance argument shows the $d_{n+1}(\sigma)/d_n(\sigma)
= O(n)$, whence the series (\ref{seriesatinfty}) is generally
divergent and only summable in the sense of an asymptotic expansion.
Olver shows that the sector of validity for this asymptotic
expansion includes the entire $z$--plane (see Chapter 7 of 
\cite{Olver}).

Set $c_n = d_n(0)$. Sending $\sigma \mapsto 0$ in
(\ref{bthreeterm}) then yields the simple two--term recursion
relation
\begin{equation}
c_{n+1} = \frac{[l(l+1)-n(n+1)]}{2(n+1)}\,c_n\, ,
\label{ctwoterm}
\end{equation}
with solution (see Ref.~\cite{Watson}, p.~202)
\begin{equation}
c_{n} = \frac{\Gamma (l+n+1)}{2^n n!\Gamma (l-n+1)}\, .
\label{bfromGamma}
\end{equation}
When $l$ is an integer, as is the case here, the series
$\sum_{n=0}^\infty c_n z^{-n}$ truncates, showing for all $l$ that
the solution $W_l(z)$ is a polynomial of degree $l$ in inverse
$z$. All coefficients $c_n$ are positive and nonzero, and
we can ultimately conclude that all zeros of $W_l(z)$ lie in the
lefthalf $z$--plane. Furthermore, the last nonzero coefficient is
\begin{equation}
c_{l} = \frac{\Gamma (2l+1)}{2^l l!}\, ,
\end{equation}
and from this formula we may appeal to the asymptotic
behavior of the gamma function (see Ref.~\cite{MorFesh}, p.~486) 
in order to show
\begin{equation}
c_l \sim \Gamma(l)2^l\sqrt{l/\pi}
\label{largecl}
\end{equation}
as $l$ becomes large.

\subsection{Radiation outer boundary conditions}\label{sbSec:1.4}
This subsection derives and discusses {\sc robc} for a single
spherical--harmonic mode $\psi_{lm}(\tau,\rho)$, but we
continue the practice of everywhere suppressing the subscript $m$.
This subsection's formulae are valid for all possible spin values 
$\jmath = 0,1,2$; however, for concrete examples we choose $\jmath 
= 0$.

\subsubsection{Derivation of the radiation kernel}\label{sbsbSec:1.4.1}
\begin{figure}[t]
\scalebox{0.70}{\includegraphics{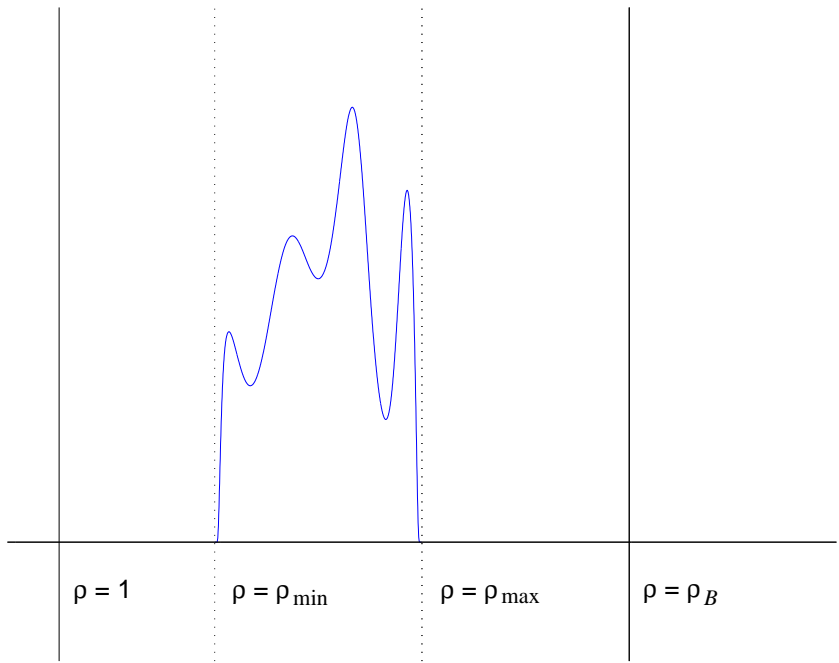}}
\caption{{\sc Initial wave--packet configuration in the 
frequency domain.}
The undulations represent the initial data which is compactly
supported on $[\rho_\mathrm{min}, \rho_\mathrm{max}]$. The
domain $\mathsf{S}_0 = (\rho_\mathrm{max},\infty)$ lies to
the right of the data.}
\label{oneFigure1}
\end{figure}

Although we will now derive exact equations, let us define 
the radial computational domain to be $2\mathrm{m}$ times the $\rho$ 
interval $[1,\rho_B]$. The radial numerical mesh will 
be a discretization of this domain. Now consider an infinite 
radial domain ${\mathsf S}$ defined by $\rho > 
\rho_\mathrm{max}$, with $\rho_\mathrm{max} < \rho_B$. 
Let ${\mathsf S}_0$ denote the intersection of 
${\mathsf S}\times[0,\infty)$ with an initial $\tau = 0$ Cauchy 
surface. Assume that
the initial data $\psi_{l}(0,\rho)$ and $\dot{\psi}_{l}(0,\rho)$ 
is of compact support and, moreover, vanishes on ${\mathsf S}_0$. 
The condition $\rho_\mathrm{max} < \rho_B$ ensures that the computational 
domain edge $\rho_B$ does not intersect the support of the initial 
data (see {\sc Fig.}~\ref{oneFigure1}). Then the analysis of the 
last subsection establishes the formal expression\footnote{Our notation 
here is not strict, since of course the {\sc rhs} has no 
$\sigma$ dependence. A strict notation would replace each $\sigma$ 
on the {\sc rhs} with a blank $\square$ or dot $\cdot$ to indicate 
that this dependence is lost in the integration.}
\begin{equation}
\psi_l(\tau,\rho)
   = {\mathcal L}^{-1}
     \left[a_{l}(\sigma) \frac{e^{-\sigma\rho_{*}}
           W_{l}(\sigma\rho;\sigma)}{\rho}
         + b_{l}(\sigma) \frac{e^{-\sigma\rho_{*}}
           Z_{l}(\sigma\rho;\sigma)}{\rho}
     \right](\tau)
\end{equation}
as the general solution to the wave equation (\ref{maineq2}) on 
the history ${\mathsf S}\times[0,\infty)$ of ${\mathsf S}_0$. 
Here the coefficients $a_{l}(\sigma)$ and $b_{l}(\sigma)$ are 
arbitrary functions analytic in the righthalf $\sigma$--plane. 
Now, $Z_{l}(\sigma\rho;\sigma) \sim \exp(2\sigma\rho_{*})$ as $\rho
\rightarrow\infty$, showing that $b_{l}(\sigma)$ must be zero 
(for otherwise the solution is not asymptotically outgoing 
as expected for an initial ``wave packet'' of compact support). 

The solution within the computational domain is also 
obtained as above via inverse Laplace transform. However, now 
the relevant frequency--space radial function solves the 
inhomogeneous version of (\ref{maineq4}), in which case the
source
\begin{equation}
(\rho^2-\rho)^{-1}
J_{l}(\rho;\sigma) = -\rho^2(\rho-1)^{-2}
                  \big[\dot{\psi}_{l}(0,\rho)
                 +\sigma\psi_{l}(0,\rho)\big]
\end{equation}
replaces zero on the {\sc rhs} of the equation, as is necessary 
for non--trivial initial data. This solution 
appropriately matches $a_{l}(\sigma)\rho^{-1} e^{-\sigma\rho_{*}} 
W_{l}(\sigma\rho;\sigma)$ at the largest radius $\rho_\mathrm{max}$
on which the data is supported. Let us further assume that the 
initial data is supported only on 
$[\rho_\mathrm{min},\rho_\mathrm{max}]$, where $\rho_\mathrm{min}
> 1$ (again, see {\sc Fig.}~\ref{oneFigure1}). In this case,
taking the second solution
$\rho^{-1} \exp(-\sigma\rho_*)Z_l(\sigma\rho;\sigma)$ as ingoing
at the horizon and using well--known methods
associated with one--dimensional
Green's functions, one can show that
\begin{equation}
a_{l}(\sigma) \propto
\int^{\rho_{\mathrm{max}}}_{\rho_{\mathrm{min}}}
\exp(-\sigma\rho)(\rho-1)^{-\sigma}\rho^{-1}
           Z_{l}(\sigma\rho;\sigma) 
           J_{l}(\rho;\sigma){\rm d}\rho\, ,
\label{inandaround}
\end{equation}
where the proportionality constant is determined by 
calculating the Wronskian of the two chosen linearly independent 
solutions to the homogeneous equation \cite{Chattopadhyay}.

Let us now derive the explicit form of the radiation 
kernel, assuming that we now work in the region $\rho > \rho_B$.
Consistent with our presentation thus far, we denote by 
$\psi_l(\tau,\rho)$ the function satisfying
\begin{equation}
{\mathcal L}[\psi_l](\sigma,\rho) = \tilpsi_l(\sigma,\rho) =
a_{l}(\sigma) \rho^{-1}e^{-\sigma\rho_{*}}W_l(\sigma\rho;\sigma)\, .
\end{equation}
Differentiation of this formula by $\rho$ then gives
\begin{equation}
\partial_\rho\tilpsi_l(\sigma,\rho) =
\left[\sigma\frac{W_l'(\sigma\rho;\sigma)}{W_l(\sigma\rho;\sigma)}
-\sigma - \frac{\sigma}{\rho-1} -\frac{1}{\rho}\right]
\tilpsi_l(\sigma,\rho)\, ,
\end{equation}
with the prime denoting differentiation with respect to the first
slot of $W_l(z;\sigma)$. Next, we rearrange terms and introduce some
new symbols, thereby arriving at
\begin{equation}
\frac{\sigma\tilpsi_l(\sigma,\rho)}{{\sf N}(\rho)} +
\frac{\partial_\rho\tilpsi_l(\sigma,\rho)}{{\sf M}(\rho)} +
\frac{\tilpsi_l(\sigma,\rho)}{{\sf M}(\rho)\rho} =\rho^{-1}
{\sf N}(\rho)\tilpsi_l(\sigma,\rho)
\left[\sigma\rho\,
\frac{W_l'(\sigma\rho;\sigma)}{W_l(\sigma\rho;\sigma)}
\right]\, ,
\end{equation}
where in terms of the function $F(\rho)$ appearing in the
line--element (\ref{dlesselement}) we have introduced the 
temporal lapse function ${\sf N}(\rho)
= F^{1/2}(\rho)$ and the radial lapse function ${\sf M}(\rho) =
F^{-1/2}(\rho)$. The metrical function ${\sf N}$ 
describes the proper--time separation between neighboring 
three--dimensional level--$\tau$ hypersurfaces, whereas, in a 
given such three--surface, ${\sf M}$ describes the proper--radial 
separation between neighboring concentric two--spheres \cite{York}. 
Upon inverse Laplace transformation, the last equation becomes
\begin{equation}
\frac{1}{{\sf N}}\frac{\partial\psi_l}{\partial\tau} +
\frac{1}{{\sf M}}\frac{\partial\psi_l}{\partial\rho} +
\frac{\psi_l}{{\sf M}\rho} = \rho^{-1}
{\sf N}(\rho)\psi_l(\tau,\rho)*
{\mathcal L}^{-1}
\left[\sigma\rho_B\,
\frac{W_l'(\sigma\rho_B;\sigma)}{W_l(\sigma\rho_B;\sigma)}
\right](\tau)\, ,
\end{equation}
with $*$ here indicating Laplace convolution (defined just below).
On this equation we remark that the direction 
${\sf N}^{-1}\partial/\partial\tau + {\sf M}^{-1}\partial/\partial\rho$
is null and outgoing, whence the derivative of the field appearing 
on the {\sc lhs} is along a characteristic. The last equation 
holds in particular at $\rho_B$, and as our {\sc robc} we adopt the 
following:
\begin{equation}
\left.\left[\frac{1}{{\sf N}}\frac{\partial\psi_l}{\partial\tau} +
\frac{1}{{\sf M}}\frac{\partial\psi_l}{\partial\rho} +
\frac{\psi_l}{{\sf M}\rho}\right]\right|_{\rho=\rho_B} =
\rho_B^{-1}{\sf N}(\rho_B)\int^\tau_0
\omega_l(\tau-\tau';\rho_B)\psi_l(\tau',\rho_B){\rm d}\tau'\, ,
\label{ROBC}
\end{equation}
where we have introduced the {\em time--domain radiation kernel}
({\sc tdrk})
\begin{equation}
\omega_l(\tau;\rho_B) = {\mathcal L}^{-1}
\left[\sigma\rho_B\,
\frac{W_l'(\sigma\rho_B;\sigma)}{W_l(\sigma\rho_B;\sigma)}
\right](\tau)\, .
\end{equation}
We refer to that appearing within the square brackets on the 
{\sc rhs} as the {\em frequency--domain radiation kernel}
({\sc fdrk}), and we also denote it by 
\begin{equation}
\tilomega_l(\sigma;\rho_B) = \sigma\rho_B\,
\frac{W_l'(\sigma\rho_B;\sigma)}{W_l(\sigma\rho_B;\sigma)}\, . 
\label{formalfdrk}
\end{equation}
We assume that the {\sc fdrk} 
$\tilomega_l(\sigma;\rho_B)$ has the appropriate $\sigma$--decay 
necessary for a well--defined $\omega_l(\tau;\rho_B) = {\mathcal 
L}^{-1}[\tilomega_l(\sigma;\rho_B)](\tau)$.
Both $\omega_l(\tau;\rho_B)$ and $\tilomega_l(\sigma;\rho_B)$
do of course depend on the values of $l$ and $\rho_B$ (and on
the choice of spin $\jmath$), but to avoid 
clutter we will sometimes suppress this dependence and write
simply $\omega(\tau)$ and $\tilomega(\sigma)$. Finally, we note
that the {\sc robc} can be written simply as
\begin{equation}
\left.\left[\frac{1}{{\sf N}}\frac{\partial\Psi_l}{\partial\tau} +
\frac{1}{{\sf M}}\frac{\partial\Psi_l}{\partial\rho}
\right]\right|_{\rho=\rho_B} =
\rho_B^{-1}{\sf N}(\rho_B)\int^\tau_0
\omega_l(\tau-\tau';\rho_B)\Psi_l(\tau',\rho_B){\rm d}\tau'
\label{ROBCb}
\end{equation}
in terms of the field $\Psi_l = \rho\psi_l$ appearing in 
(\ref{maineq3}).

\subsubsection{Representation of the kernel}\label{sbsbSec:1.4.2}
In {\sc Section} \ref{Sec:3} we undertake a fairly thorough 
numerical investigation 
of the analytic behavior of both $W_l(\sigma\rho_B;\sigma)$ and
$\tilomega_l(\sigma;\rho_B)$ as functions of the complex variable 
$\sigma$. As a result of our investigation, we shall make the following
conjectures regarding the {\sc fdrk} 
$\tilomega_l(\sigma;\rho_B)$. First, for $l$ fixed 
$\tilomega_l(\sigma;\rho_B)$ is analytic on 
${\mathbb C}\backslash(-\infty,0]$, 
save for $N_l = N_l(\rho_B) \in {\mathbb Z}_{\geq 0}$ simple poles with
locations $\{\sigma_{l,n} = \sigma_{l,n}(\rho_B) : n = 1,\cdots,N_l\}$ 
lying in the lefthalf $\sigma$--plane. Second, 
$\tilomega_l(\sigma;\rho_B)$ is bounded in a neighborhood of the 
origin $\sigma = 0$. Third, $\mathrm{Re}\tilomega_l(\sigma;\rho_B)$ is 
continuous and $\mathrm{Im}\tilomega_l(\sigma;\rho_B)$ jumps by a sign 
across the branch cut along the negative Re$\sigma$ axis. The 
integer $N_l(\rho_B)$ is constant over sizable regions of the $\rho_B$ 
parameter space. However, the pole locations $\sigma_{l,n}$ do vary 
smoothly with respect to changes of $\rho_B$, apparently subject to
\begin{equation}
\sigma_{l,n}(\rho_B) \sim \sum_{k=1}^\infty 
                     \sigma_{l,n,k}\rho_B^{-k}\, ,
\label{izerolocations}
\end{equation}
where the $\sigma_{l,n,k}$ are constants. This series is perhaps only 
summable in the sense of an asymptotic expansion, and we have only
numerically observed the first two terms.
\begin{figure}[t]
\scalebox{0.70}{\includegraphics{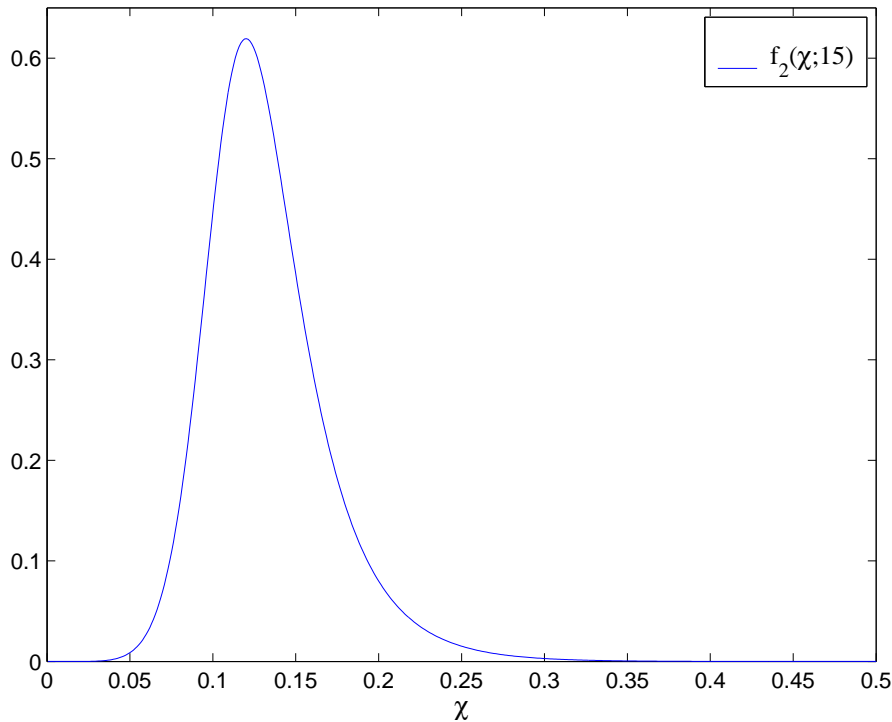}}
\caption{{\sc Typical cut profile}. For 
this plot $l=2$, $\rho_B = 15$, and $\jmath = 0$.
\label{oneFigure2}}
\end{figure}

Let us now define the $n$th {\em pole strength} and a {\em cut profile}
respectively via the formulae
\begin{equation} \alpha_{l,n}(\rho_B) = -\rho_B
\sigma_{l,n}'(\rho_B)\, ,\quad f_l(\chi;\rho_B) = 
\mathrm{Im}\tilomega_l(\chi e^{{\rm i}\pi};\rho_B)\, ,
\end{equation}
with $\chi \geq 0$ and the prime here standing for 
$\partial/\partial\rho_B$
differentiation. Like the pole locations, both of these objects 
also vary with respect to changes of $\rho_B$ as indicated. 
As can be inferred from the third conjecture of the last paragraph, 
it is the case that 
$\mathrm{Im}\tilomega_l(\chi e^{-{\rm i}\pi};\rho_B)
= -f_l(\chi;\rho_B)$. To give a concrete example, 
we choose $\jmath = 0$, $l=2$, and $\rho_B=15$, in which case 
we have numerically found that $N_2(15) = 2$, 
$\sigma_{2,n}(15) \simeq -0.0969 \pm {\rm i} 0.0612$, 
and $\alpha_{2,n}(15) \simeq -0.0936 \pm {\rm i} 0.0647$, for $n = 1(+)$
and $2(-)$. For these parameter values the corresponding cut profile is 
shown in {\sc Fig.}~\ref{oneFigure2}. The plot is typical in the sense 
that for all $l$ and $\rho_B$ considered here,
$f_l(\chi;\rho_B)$ decays sharply in the $\chi\rightarrow 0^+$ and
$\chi\rightarrow \infty$ limits (except for $l = 0$ where the 
decay in the $\chi\rightarrow 0^+$ limit is not as sharp). 
However, the shape of the 
profile can be qualitatively different for other parameter values. 
Moreover, for certain exceptional values of the parameters, the 
profile can even blow up at a particular $\chi$ point, in which 
case numerical evidence suggests that the integral in 
(\ref{poleandcut}) is defined in the sense of a Cauchy Principal 
Value. We discuss all of these issues in {\sc Section} 
\ref{sbSec:3.1}.

In terms of the pole locations and strengths and the cut profile, we 
claim that the {\sc fdrk} has the following representation 
(suppressing $\rho_B$ dependence for now):
\begin{equation}
\tilomega_l(\sigma) = \sum_{n = 1}^{N_l}
\frac{\alpha_{l,n}}{\sigma-\sigma_{l,n}} - 
\frac{1}{\pi}\int^\infty_0
\frac{f_l(\chi)}{\sigma+\chi}\,{\rm d}\chi\, ,
\label{poleandcut}
\end{equation}
for all $\sigma$ not equal to a $\sigma_{l,n}$ and not lying on
$(-\infty,0]$. Despite the right closed bracket  here, we shall
also evaluate this representation at $\sigma = 0$. Given the
described structure of the $\sigma$--function 
$\tilomega_l(\sigma)$,
the derivation of such a representation amounts to a simple 
exercise
involving the Residue Theorem and a ``keyhole'' contour. 
Although we will often describe the second term on the {\sc rhs}
of (\ref{poleandcut}) as corresponding to a continuous set of
poles, these are not really poles in the sense of complex
analysis (which are properly isolated singularities). 
Formally, we compute the inverse Laplace transform of
(\ref{poleandcut}), with result
\begin{equation}
\omega_l(\tau) = \sum_{n = 1}^{N_l} \alpha_{l,n}
\exp(\sigma_{l,n}\tau) -
\frac{1}{\pi}\int^\infty_0 f_l(\chi)\exp(-\chi\tau){\rm d}\chi\, .
\end{equation}
Evidently then, direct numerical construction of the {\sc fdrk}
would amount to numerical computation of the pole locations and
strengths and the cut profile. For $l \leq 10$ we consider such
a direct construction in {\sc Section} \ref{sbSec:3.2}, although
we show in {\sc Section} \ref{sbSec:3.3} how this brute--force 
approach may be bypassed insofar as implementation of {\sc robc} 
is concerned.

\subsubsection{Comparison with the Green's function method}
\label{sbsbSec:1.4.3}
Any temptation to identify the pole locations
$\sigma_{l,n}(\rho_B)$ in the representation (\ref{poleandcut})
with so--called quasinormal modes \cite{KokkotasSchmidt} should
be resisted. For a given $l$ there are an infinite number of
quasinormal modes \cite{KokkotasSchmidt}, fixed numerical values
intrinsic to the blackhole geometry and certainly insensitive to
any particular choice of outer boundary radius $\rho_B$. However,
for $\rho_B > 1$ the poles now under examination, that is the zeros
in $\sigma$ of the Heun--type function $W_l(\sigma\rho_B;\sigma)$,
are finite in number, and they do depend on $\rho_B$. Moreover, the
boundary value problem associated with these Heun zeros is different
than the usual one associated with quasinormal modes. This usual
boundary value problem was considered in the pioneering work
\cite{Leaver3} of Leaver, and more recently in a careful study by
Andersson \cite{Andersson}. The goal of both authors was to examine
a given multipole field $\psi_l(\tau,\rho) = \rho^{-1}
\Psi_l(\tau,\rho)$ in terms of a Green's function representation
involving initial data. Andersson refers to this as the initial
value problem for the scalar field (or, more generally, for
electromagnetic or gravitational perturbations), although when
we mentioned that name in the first paragraph of the introduction
we did not have this Green's function approach in mind. In
Eq.~(6) of \cite{Andersson}, Andersson expresses the scalar
field as
\begin{equation}
\Psi_l(\tau,\rho_*)
= \int G_l(\rho_*,\rho_*';\tau)
\partial_\tau\Psi_l(0,\rho_*) {\rm d}\rho_*'
+ \int \partial_\tau G_l(\rho_*,\rho_*';\tau)
\Psi_l(0,\rho_*) {\rm d}\rho_*'\, .
\label{andersoneq}
\end{equation}
In this equation we view the field $\Psi_l$ introduced in
(\ref{maineq7}) as depending on $\rho_*$ (as Andersson does), and
we have also slightly modified Andersson's notations to suit our
own. The appropriate limits of $\rho_*$ integration in
(\ref{andersoneq}) are discussed in \cite{Andersson}. This problem
perhaps resembles our own; however, as we now demonstrate, it is
different both in concept and detail.

Both Leaver and Andersson considered the (here Laplace) transform
of the Green's function in (\ref{andersoneq}), a frequency--domain
Green's function $\hat{G}_l(\rho_*,\rho_*';\sigma)$ associated
with the following boundary value problem. The solution is pure
ingoing at the horizon [$\tilPsi^-_l(\sigma,\rho_*) \sim
\exp(\sigma\rho_*)$ as $\rho_* \rightarrow -\infty$] and outgoing
at infinity [$\tilPsi^+_l(\sigma,\rho_*) \sim \exp(-\sigma\rho_*)$
as $\rho_*\rightarrow\infty$]. In fact, we briefly considered
$\hat{G}_l(\rho_*,\rho_*';\sigma)$ in and around (\ref{inandaround}), 
although we
shall make no further use of it in this article or the follow--up
article. Leaver and Andersson's approach was essentially to examine
the value $\Psi_l(\tau,\rho_*)$, as expressed by (\ref{andersoneq}),
via a careful analysis of $\hat{G}_l(\rho_*,\rho_*';\sigma)$. (Of
no concern here, Leaver further considered more general driving
source terms beyond just the initial data.) When
$\hat{G}_l(\rho_*,\rho_*';\sigma)$ is considered as an analytic
function of complex $\sigma$ and continued into the lefthalf plane,
its pole locations are the quasinormal modes and there is also an
associated branch cut along the negative Re$\sigma$ axis
\cite{Leaver3,Andersson,NollertSchmidt}. These complex analytic
features play a prominent role in describing the physical behavior
of the field (see \cite{Leaver3,Andersson} and references therein).

Our key representation (\ref{poleandcut}) stems from continuation
into the lefthalf $\sigma$--plane of the {\sc fdrk}
$\tilomega_l(\sigma;\rho_B)$, the expression (\ref{formalfdrk})
involving the logarithmic derivative of $W_l(\sigma\rho;\sigma) =
\exp(\sigma\rho_*) \tilPsi^+_l(\sigma,\rho)$. Here we again view
$\tilPsi_l$ as depending on $\rho$, rather than $\rho_*$ as in
the last paragraph. We stress that the {\sc fdrk}
$\tilomega_l(\sigma;\rho_B)$ is not the Green's function
$\hat{G}_l(\rho_*,\rho_*';\sigma)$ considered by both Leaver and
Andersson. Indeed, $\tilomega_l(\sigma;\rho_B)$ is a {\em boundary}
integral kernel. Moreover, it is built solely with the outgoing
solution $\tilPsi^+_l$ to the homogeneous {\sc ode}, whereas
construction of the Green's function requires two linearly
independent homogeneous solutions, $\tilPsi^-_l$ and $\tilPsi^+_l$.
As mentioned, the pole locations associated
with $\tilomega_l(\sigma;\rho_B)$ are not the quasinormal modes,
rather the special frequencies, finite in number, for which the
outgoing solution $W_l(\sigma\rho;\sigma)$ also vanishes at
$\rho_B$. Despite the fact that $\hat{G}_l(\rho_*,\rho_*';\sigma)$
and $\tilomega_l(\sigma;\rho_B)$ are different integral kernels,
we remark that they share the same {\em qualitative} features
in the lefthalf $\sigma$--plane (each has poles and a branch cut).

On top of these technical differences between our work and those
of Leaver and Andersson, we point out that our overall goal is very
different. As mentioned, their goal was to examine the actual
value of the field via the representation (\ref{andersoneq})
based on {\em spatial} convolution. On the contrary, our
{\em boundary} kernel $\omega_l(\tau;\rho_B)$ is associated with
{\em temporal} convolution, with the goal being to impose exact
radiation boundary conditions at a given outer sphere $B$. That is
to say, our goal is domain reduction via the introduction of
integral convolution over the history $^3\!B$ of the boundary $B$.
With this distinction in mind, compare our key Eq.~(\ref{ROBC})
with Andersson's key equation, as we have written it in
(\ref{andersoneq}). Perhaps the approach of Andersson and Leaver
could also be used to numerically implement exact radiation
boundary conditions in an alternative way [by setting $\rho_* =
\rho_B + \log(\rho_B-1)$ in (\ref{andersoneq})], but they did not
address this question {\em per se}. Moreover, such an approach
would necessarily relate {\sc robc} to the details of the data
on the initial surface, which would seem awkward from a numerical
standpoint.\footnote{We are definitely not critical of the most
excellent works of Leaver and Andersson. Indeed, just as their
Green's function technique would seem not the best way to implement
{\sc robc}, we do not believe we could directly reproduce their
results with our boundary kernel technique.} Even were such an
implementation carried out, memory and speed issues would
inevitably arise. Besides developing exact {\sc robc} via domain
reduction, we also intend to provide an efficient and rapid
implementation of these conditions.

\subsubsection{Approximation of the kernel}\label{sbsbSec:1.4.4}
Our numerical implementation of {\sc robc} rests on 
approximation of the exact kernel $\omega(\tau)$ (now
suppressing $l$ as well as $\rho_B$) by a compressed 
kernel $\xi(\tau)$. We explain this terminology later, but here 
collect an estimate needed to address the relative error 
associated with such an approximation. We start with the 
Laplace inversion formula,
\begin{equation}
\psi(\tau) = \frac{1}{2\pi{\rm i}}
\int^{{\rm i}\infty}_{-{\rm i}\infty}
\tilpsi(\sigma) e^{\sigma\tau}{\rm d}\sigma\, ,
\end{equation}
where we are assuming that any singularities of
$\tilpsi(\sigma)$ lie in the lefthalf plane Re$\sigma < 0$. A
change of variables casts the inversion formula into the form
\begin{equation}
\psi(\tau) =
\frac{1}{2\pi}
\int^{\infty}_{-\infty}
\hat{\psi}({\rm i}y) e^{{\rm i}y\tau}{\rm d}y\, ,
\end{equation}
thereby introducing the Fourier transform
$\widetilde{\psi}(y) = \tilpsi({\rm i}y)$ of
$\psi(\tau)$. It
then follows that
\begin{equation}
\big\|\psi\big\|_{L_2(0,\infty)} =
\big\|\psi\big\|_{L_2(\mathbb{R})} =
\big\|\widetilde{\psi}\big\|_{L_2(\mathbb{R})}\, .
\end{equation}
The first equality follows subject to the
assumption that $\psi(\tau)$ vanishes for $\tau \leq 0$,
while the second is Parseval's identity.
                                                                                               
Now suppose $\xi(\tau)$, with Fourier transform $\widetilde{\xi}(y)$,
is an approximation to the kernel $\omega(\tau)$. Later we shall have
its Laplace transform $\tilxi(\sigma)$ as a rational
function $P(\sigma)/Q(\sigma)$, and then
$\widetilde{\xi}(y) = P({\rm i}y)/Q({\rm i}y)$.
Also introduce the Fourier transform $\widetilde{\omega}(y) =
\tilomega({\rm i}y)$ of $\omega(\tau)$. In terms of these variables
we have
\begin{align}
\big\|\xi*\psi-\omega*\psi\big\|_{L_2(0,\infty)}
& = 
\big\|\widetilde{\xi}\widetilde{\psi}
-\widetilde{\omega}\widetilde{\psi}\big\|_{L_2(\mathbb{R})}
\label{oneestimates} \\
& \leq  {\rm sup}_{y\in\mathbb{R}}
\frac{|\widetilde{\xi}(y)
-\widetilde{\omega}(y)|}{|\widetilde{\omega}(y)|}
\times \big\|\widetilde{\omega}
\widetilde{\psi}\big\|_{L_2(\mathbb{R})}
\nonumber \\
& =  {\rm sup}_{\sigma\in {\rm i}\mathbb{R}}
\frac{|\tilxi(\sigma)-\tilomega(\sigma)|}{|\tilomega(\sigma)|}
\times \big\|\omega*\psi\big\|_{L_2(0,\infty)}
\nonumber
\end{align}
as our basic estimate. Because of this estimate, we focus on 
finding approximations $\tilxi(\sigma)$ to $\tilomega(\sigma)$ 
which have small relative supremum error along the imaginary 
axis.

Finally, suppose that we do not quite know 
$\tilomega({\rm i}y)$. Rather, as is the case, we must 
generate $\tilomega({\rm i}y)$ itself numerically. Then,
instead of the relative error 
\begin{equation}
{\rm sup}_{\sigma\in {\rm i}\mathbb{R}}
\frac{|\tilxi(\sigma)-\tilomega(\sigma)|}{|\tilomega(\sigma)|}\, ,
\label{oneerror1}
\end{equation}
we should consider an expression like
\begin{equation}
{\rm sup}_{\sigma\in {\rm i}\mathbb{R}}
\frac{|\tilxi(\sigma)-\tilomega(\sigma)|}{|\tilomega(\sigma)|}
+ {\rm sup}_{\sigma\in {\rm i}\mathbb{R}}
\frac{|\Delta\tilomega(\sigma)|}{|\tilomega(\sigma)|}\, ,
\label{oneerror2}
\end{equation}
where the final term is an estimate of the supremum relative 
error in our knowledge of $\tilomega({\rm i}y)$. For the methods 
we develop to generate $\tilomega({\rm i}y)$, this second term is 
negligible with respect to the first one.
%
%
\section{Numerical evaluation of the outgoing solution and
kernel}\label{Sec:2}
This section describes the handful of numerical methods used 
in this work. The first 
subsection describes a numerical method for evaluating the outgoing 
solution $W_l(\sigma\rho_B;\sigma)$ at a given complex $\sigma$,
and this method allows us to numerically study the analytic 
structure of $W_l(\sigma\rho_B;\sigma)$ as a function of Laplace
frequency. The lefthalf $\sigma$--plane is the domain of interest,
and a study of $W_l(\sigma\rho_B;\sigma)$ on this domain, carried 
out in {\sc Section} \ref{sbSec:3.1}, justifies the key representation 
(\ref{poleandcut}). As we indicated in {\sc Section} 
\ref{sbsbSec:1.4.4}, our numerical approximations to 
the {\sc fdrk} $\tilomega_l(\sigma;\rho_B)$ are tailored to have
small relative supremum error along the Im$\sigma$ axis. 
Therefore, insofar as 
implementation of {\sc robc} is 
concerned, we primarily need numerical methods for obtaining 
accurate numerical profiles for Re$\tilomega_l({\rm i}y;\rho_B)$ 
and Im$\tilomega_l({\rm i}y;\rho_B)$ with $y\in\mathbb{R}$.
The second subsection describes two such methods.

Before describing our numerical methods, we note that Leaver has
analytically represented a solution to the Regge--Wheeler equation
(more generally to the generalized spheroidal wave equation) as an
infinite series in Coulomb wave functions, where the expansion
coefficients obey a three--term recursion relation \cite{Leaver1}.
Such a series can alternatively be viewed as a sum of confluent
hypergeometric functions. One approach towards our goal of numerically
evaluating the outgoing solution would be to use the appropriate Leaver
series. However, beyond the issue of numerically solving the relevant
three--term recursion relation, numerical evaluation of Coloumb wave
functions (for complex arguments) is already somewhat tricky
\cite{Thompson}. Here we describe far simpler methods, which are
nevertheless extremely accurate. Although simple, our methods are
very accurate only for a limited range of frequencies (which happen
to be precisely the frequencies we are interested in). The Leaver
series is valid over the whole frequency plane. Although we have
not compared our methods with the Leaver series, we believe they
are better suited for our purposes.

\subsection{Numerical evaluation of the outgoing 
solution}\label{sbSec:2.1}
From now on let us simply refer to $W_l(\sigma\rho_B;\sigma)$
as a Heun function and $W_l(\sigma\rho_B)$ as a Bessel function.
This is not quite correct since, as discussed in {\sc Section}
\ref{sbsbSec:1.3.2}, 
$W_l(\sigma\rho_B;\sigma)$ and $W_l(\sigma\rho_B)$ respectively
differ from Heun and Bessel functions by transformations on the
dependent variable; however, this terminology will streamline
our presentation. We now present numerical methods for computing 
the complex value $W_l(\sigma\rho_B;\sigma)$. The methods have been 
designed to successfully compute the similar value 
$W_l(\sigma\rho_B)$, formally $W_l(\sigma\rho_{\rm
B};0)$ in our notation. $W_l(\sigma\rho)$ solves
the {\sc ode}
\begin{equation}
\frac{\dee^2\tilPhi_l}{\dee\rho^2}
-2\sigma\frac{\dee\tilPhi}{\dee\rho} - \frac{l(l+1)}{\rho^2}
\tilPhi = 0\label{nbesseq2}
\end{equation}
obtained directly from (\ref{nbesseq}) via the substitution $z =
\sigma\rho$. {\sc Section}
\ref{sbsbSec:1.3.3} noted that $W_l(\sigma\rho)$
is a polynomial $\sum_{n=0}^l c_n (\sigma\rho)^{-n}$ of degree
$l$ in inverse $\sigma\rho$, with coefficients $c_n$ given in
(\ref{bfromGamma}). With the {\em exact} form of $W_l(\sigma\rho)$
we could in principle compute the value
$W_l(\sigma\rho_B)$ directly.\footnote{Due to the growth 
(\ref{largecl}) of the Bessel coefficients, such direct 
computation is plagued by increasing 
loss of accuracy as $l$ grows.} Nevertheless, if we pretend 
that the exact form of $W_l(\sigma\rho)$ is not at our disposal, 
then the task of numerically computing $W_l(\sigma\rho_B)$ shares 
essential features with our ultimate task of computing 
$W_l(\sigma\rho_B;\sigma)$. The task of computing
$W_l(\sigma\rho_B)$ has been an invaluable model, and for
ease of presentation we mostly focus on it here.

\subsubsection{Numerical integration}\label{sbsbSec:2.1.1}
Focusing on the $\rho$--dependence of the solution, we write 
$W_l(\sigma\rho) = \sum_{n=0}^{l} (c_n\sigma^{-n}) \rho^{-n}$.
Since we shall not allow ourselves to evaluate 
$W_l(\sigma\rho_B)$ as $\sum_{n=0}^{l} (c_n\sigma^{-n}) 
\rho^{-n}_B$, we truncate the series after some fixed number 
$l-p$ of terms, assuming that
\begin{equation}
W_l(\sigma\rho) \sim
\sum_{n=0}^{l-p} (c_n\sigma^{-n}) \rho^{-n}
\label{aexpansion}
\end{equation}
is at our disposal.
Truncation by hand of this already finite series serves as a 
model for the scenario involving $W_l(\sigma\rho;\sigma)$, where 
only a divergent formal series, such as the one specified by
(\ref{seriesatinfty}) and (\ref{bthreeterm}), is at our disposal.
With our truncated series we can still generate an accurate
approximation to the value $W_l(\sigma\rho_\infty)$, so long as
$\rho_\infty$ is large enough. Let us set $\rho_\infty = {\tt
scale}*\rho_B$, with ${\tt scale}$ a large number.
Evaluation of the truncated sum and its $\rho$ derivative at
$\rho_\infty$ then generates initial data for the {\sc ode}.
Moreover, the generated data is approximate to the exact data
$\{W_l(\sigma\rho_\infty),W^\rho_l(\sigma\rho_\infty)\}$ giving
rise to $W_l(\sigma\rho)$. Here the superscript $\rho$ denotes
$\partial/\partial\rho$ differentiation, whereas a prime $'$ 
would denote differentiation in argument. We stress that
our approximation to the exact data can be rendered arbitrarily
accurate by choosing $\rho_\infty$ large enough. Finally, we
{\em numerically} integrate (\ref{nbesseq2}) in $\rho$ from 
$\rho_\infty$ all the way down to $\rho_B$, thereby computing a 
candidate for the value $W_l(\sigma\rho_B)$. 
\begin{figure}[t]
\scalebox{0.70}{\includegraphics{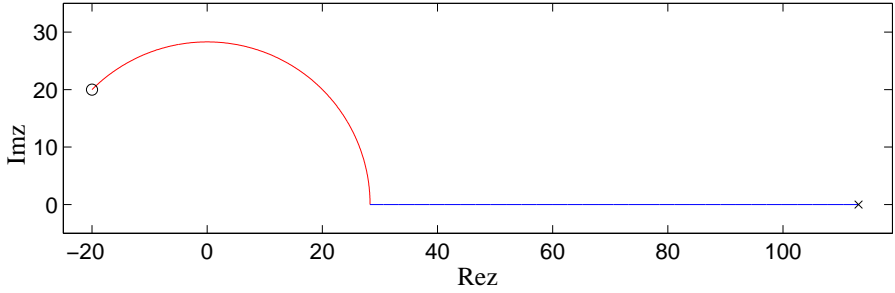}}
\caption{{\sc Two--component path in $z$--plane.}
The figure depicts the ray--and--arc path described in the text.
Here we have $\rho_B = 20$, $\sigma = -1+{\rm i}$, and
${\tt scale} = 4$, so that $z_B = -20 + {\rm i} 20$ (marked by
an o) and $z_\infty = 4*20*\sqrt{2} \simeq 113$ (marked by an x).
Typically {\tt scale} will be much larger, but the value here makes
for a good figure.
\label{2path}}
\end{figure}

As it stands, the description in the last paragraph is an
outline for a stable numerical method, provided Re$\sigma
>0$. However, for the case Re$\sigma < 0$ of interest
the described method is not stable. To see why,
consider the $l = 1$ outgoing solution $W_1(\sigma\rho) = 
1+(\sigma\rho)^{-1}$ to (\ref{nbesseq2}). As a second linearly 
independent solution to the $l=1$ {\sc ode}, take the ingoing 
solution $Z_1(\sigma\rho) = \exp(2\sigma\rho)[1-(\sigma\rho)^{-1}]$. 
Further, suppose that initial data for the {\sc ode} is obtained 
from a truncated sum as described above, with $\{1,0\}$ in 
place of exact data $\{1+(\sigma\rho_\infty)^{-1}, -\sigma 
(\sigma\rho_\infty)^{-2}\}$. The initial data $\{1,0\}$ 
corresponds to a linear combination 
$a W_1(\sigma\rho) + b Z_1(\sigma\rho)$ 
with $a \simeq 1$ and $b$ such that 
$b Z_1(\sigma\rho_\infty) 
\simeq 0$. To fix some realistic numbers, let $\rho_B = 20$, 
${\tt scale} = 250$ so $\rho_\infty = 5000$, and $\sigma = 
-0.05$. Then we compute $a \simeq 1.0040$ and 
$b Z_1(\sigma\rho_\infty) \simeq 8.0320\times 10^{-6}$, where 
$b \simeq 1.1229\times 10^{212}$. The exact value we wish 
to calculate is $W_1(-1) = 0$. However, with the chosen
initial data, even an {\em exact} integration of 
(\ref{nbesseq2}) from $\rho_\infty = 5000$ to $\rho_B = 20$ 
yields the value $3.0393 \times 10^{211}$. Since error in the 
initial conditions is exponentially enhanced, the second 
solution $Z_1(-0.05\rho)$ becomes dominant as $\rho$ is 
decreased.
\begin{figure}
\scalebox{0.70}{\includegraphics{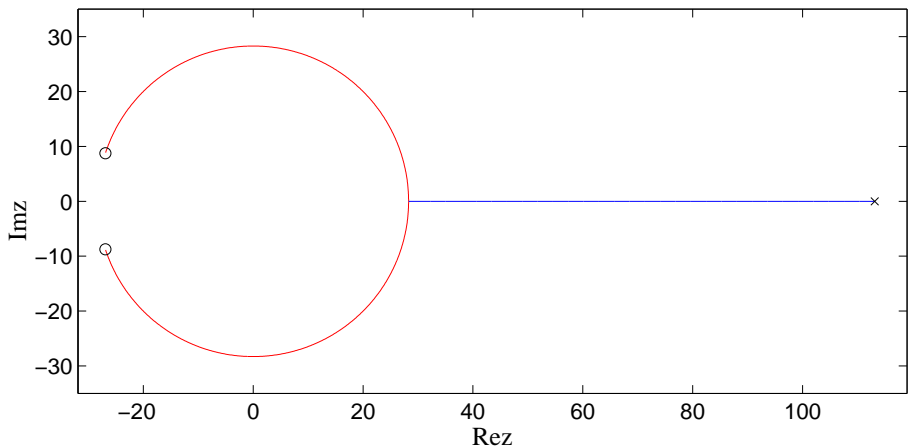}}
\caption{
{\sc Mirror image paths.} The figure depicts the
two globally different paths connecting $z_\infty$
to $z$--points above and below the negative Re$z$
axis. Use of such different paths puts a
branch cut on the negative Re$z$ axis for
Heun functions and for Bessel functions not of
half--integer order.
\label{ab2path}}
\end{figure}

\subsubsection{Two--component path integration}\label{sbsbSec:2.1.2}
The simple discussion at hand suggests that we should complexify
the variable $\rho$, rotating $\rho_\infty$ off the real
axis by an angle $\theta$ large enough to ensure that the
product $\sigma\rho_\infty$ lies in the righthalf plane.
Then integration along a ray in the complex $\rho$--plane
from $\rho_\infty$ towards the complex point
$\exp({\rm i}\theta)\rho_B$ (with $\rho_B$ still real
here) would exponentially {\em suppress} error in the
initial conditions. At the end of such a ray integration,
a second integration over an arc of $\theta$ radians would
be needed to undo the phase of $\exp({\rm i}\theta)\rho_B$.
We effect such a rotation of the $\rho$ coordinate as follows.
We choose to work with the variable $z = \sigma\rho$, the
solution $W_l(z)$, and the truncated series
$\sum_{n=0}^{l-p} c_n z^{-n}$. Our integration will now be
carried out in the complex $z$--plane rather than the
$\rho$--plane, although the strategy is essentially the
same. We define $z_\infty$ to be a large
real number ${\tt scale}*|\sigma\rho_B|$, and obtain
initial data approximate to $\{W_l(z_\infty),W_l'(z_\infty)\}$ 
via evaluation of the truncated series and its $z$ derivative at 
$z_\infty$. Even for large $l$ we have typically chosen 
$l-p = 5$ terms to define the truncated series. Then to
compute $W_l(\sigma\rho_B)$, we must numerically integrate the 
{\sc ode} (\ref{nbesseq}) from $z_\infty$ to $z_B = \sigma\rho_B$ 
along some path in the complex $z$--plane. A possible two--component
path is shown in {\sc Fig}.~\ref{2path}. It is composed of a
straight ray followed by a circular arc, with the
terminal point of the ray being the real $z$--point
$|\sigma\rho_B|$. The arc subtends an angle equal to
the argument of $\sigma$. If $\sigma$ happens to lie in the
third quadrant, then the relevant two--component path looks
like the one in {\sc Fig}.~\ref{2path} except reflected across the
Re$z$ axis. 
\begin{figure}
\scalebox{0.70}{\includegraphics{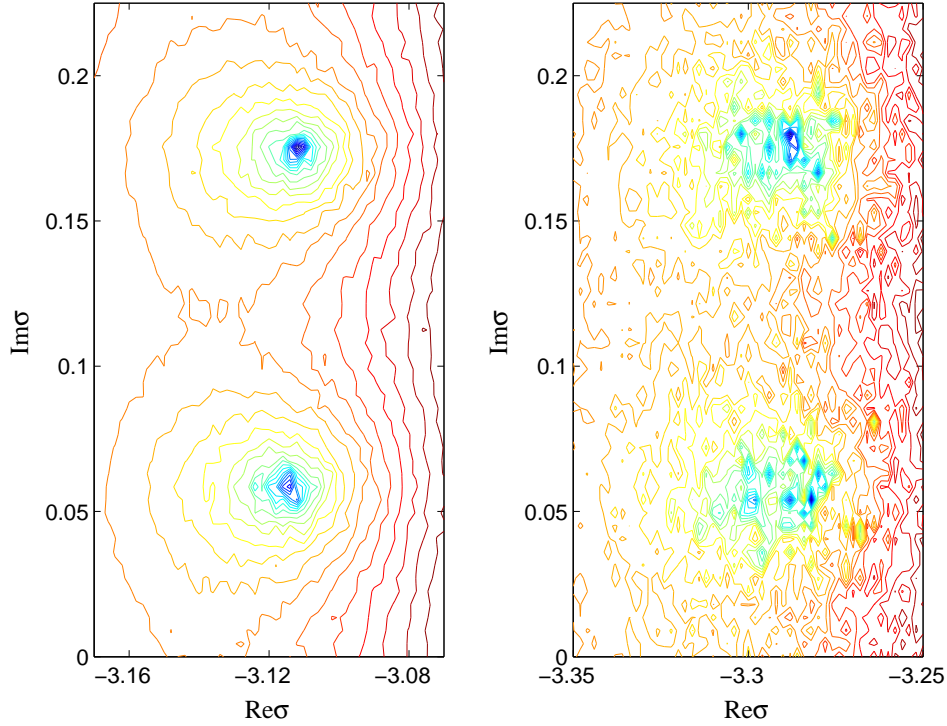}}   
\caption{
{\sc Contour lines of $\log |W_{l}(15\sigma)|$ on $\sigma$--plane
for $l = 70$ and $74$.}
On the {\sc lhs} we plot $\log |W_{70}(15\sigma)|$. The method
used to generate this plot is the one based on two--component
paths, and the integration scheme along each path component is
fifth--order Runga--Kutta--Fehlberg \cite{Atkinson}. Relevant
parameters here are $I=50$, $J=50$, $N=49152$, $M=49152$,
$l= 70$, $\rho_B = 15$, ${\tt scale} = 500$, $p=65$, and
$\kappa = 1$. $I$ and $J$ respectively specify the vertical
and horizontal discretization of the $\sigma$--plane. $N$ and
$M$ are respectively the number of integration steps along the
ray and arc. Other parameters are described in the text. On the
{\sc rhs} $l=74$ rather than $70$ and $p=69$ rather than $65$ 
(the initial condition is still determined by $5 = l - p$ terms).
All other parameters are the same as for the {\sc lhs} plot.
\label{twoFigure3}}
\end{figure}

Evaluation of the Heun function $W_l(\sigma\rho_B;\sigma)$ features 
numerical integration of the {\sc ode} (\ref{maineq7}) from 
$z_\infty$ to $z_B = \sigma\rho_B$ along the same two--component 
path. Although $z$ of course changes along the integration path, the 
$\sigma$ in $W_{l}(z;\sigma)$ remains fixed 
throughout the integration. We are then integrating a 
{\em different} {\sc ode} for each value of $\sigma$. Since
these Heun functions are not bispectral (see the discussion in 
{\sc Section} \ref{sbsbSec:1.2.3}), there would seem no way
around such a cumbersome approach. Were we only interested in 
$W_l(\sigma\rho_B)$, and not $W_l(\sigma\rho_B;\sigma)$ as well,
such an approach would be unnecessary (for then we could  
integrate with respect to frequency $\sigma$). In 
essence our two--component path method for evaluation of either 
$W_l(\sigma\rho_B)$ or $W_l(\sigma\rho_B;\sigma)$ is an 
integration with respect to radius rather than frequency. 
Indeed, even for the Bessel case, we connect each $z_B$ to the 
point $z_\infty$ by its own integration path, and during the 
integration do not record values for $W_l(z)$ along the path. 
Recording such values throughout the integration would be a 
more efficient way of mapping out the $\sigma$ dependence of 
$W_l(\sigma\rho_B)$. 
\begin{figure}[t]   
\scalebox{0.70}{\includegraphics{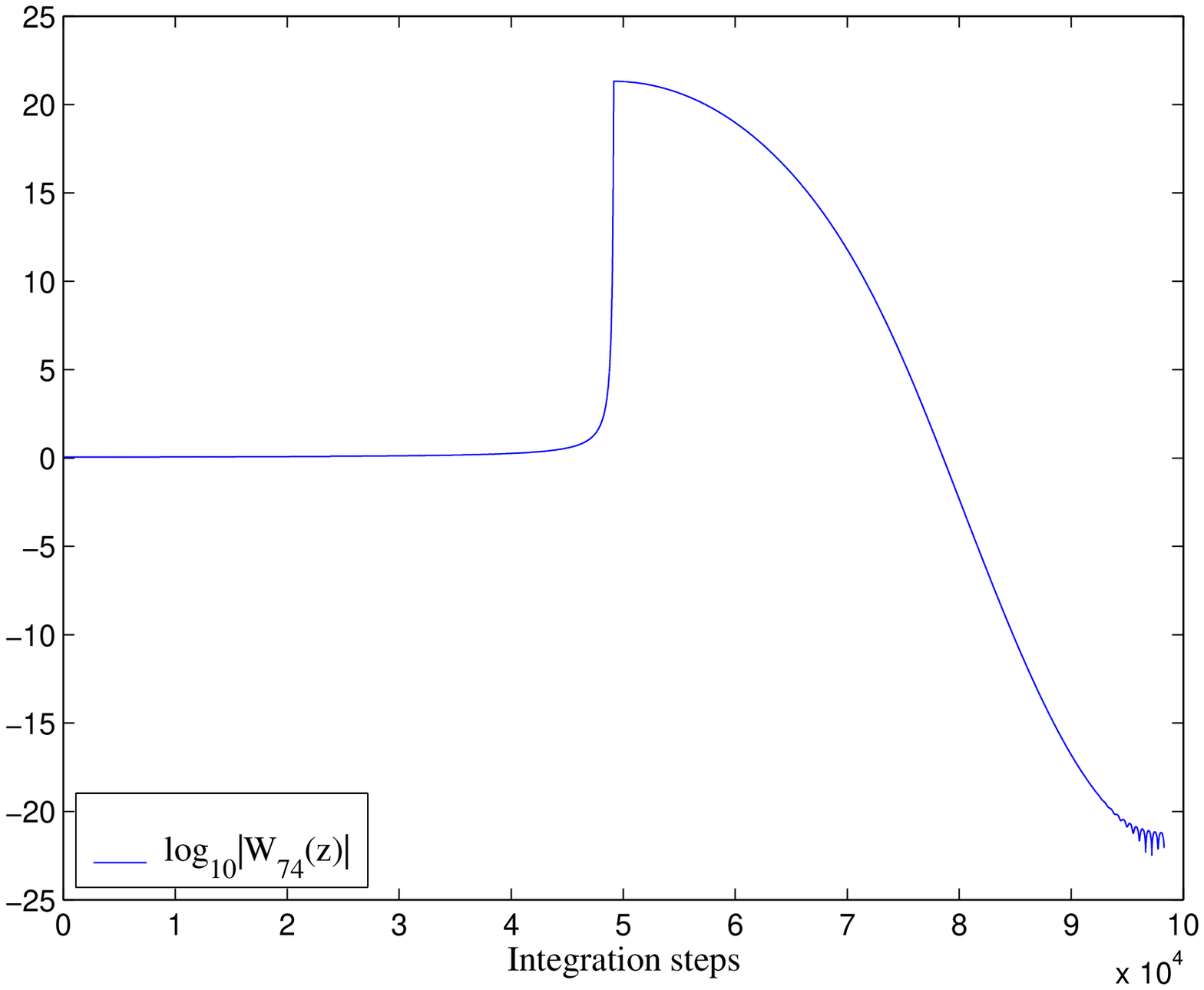}}
\caption{{\sc Value of $\log_{10}|W_{74}(z)|$ along
two--component path from $z_\infty$ to $z_B = 15\sigma$
for $\sigma = -3.29128 + {\rm i}0.05785$}. The point
$-3.29128 + {\rm i}0.05785$ is close to a zero of
$W_{74}(15\sigma)$ and lies in the region of the
$\sigma$--plane shown in Fig.~\ref{twoFigure3}.
The parameters here are the same as those listed in the
caption of Fig.~\ref{twoFigure3}.
Along the horizontal axis we have the
$N + M =98304$
integration steps. The function is of order unity at
$z_\infty$ and close to zero at the terminal point $z_B$.
However, note that the modulus of $W_{74}(z)$ gets larger
than $10^{20}$ during the integration.}
\label{twoFigure4}
\end{figure}

Let us note two key features of two--component paths.
First, for any choice of $z_B \neq 0$ the associated path avoids 
the origin where the function $W_l(z)$ is singular. Second,
considering two terminal points, one $z^\mathsf{a}_B$ just above
and the other $z^\mathsf{b}_B$ just below the negative real axis, we 
note that the respective two--component paths connecting them to
$z_\infty$ are mirror images, as depicted in {\sc Fig}.~\ref{ab2path}.
Therefore, the path leading from $z_\infty$ to $z^\mathsf{a}_B$ is
globally different than the path leading from $z_\infty$ to
$z^\mathsf{b}_B$, this being true despite the fact that $z^\mathsf{a}_B$
and $z^\mathsf{b}_B$ may lie arbitrarily close to each other in the
$z$--plane. For $l\in\mathbb{Z}_{\geq 0}$ the functions $W_l(z)$ 
are clearly analytic on the punctured $z$--plane, 
so that $W_l(z^\mathsf{a}_B) = W_l(z^\mathsf{b}_B)$ in
the limit that these points meet on the negative Re$z$ axis.
However, for Heun functions we shall find 
$W_l(z^\mathsf{a}_B;\sigma^\mathsf{a}) 
\neq W_l(z^\mathsf{b}_B;\sigma^\mathsf{b})$, and in turn
$W_l(\sigma^\mathsf{a}\rho_B;\sigma^\mathsf{a}) 
\neq W_l(\sigma^\mathsf{b}\rho_B;\sigma^\mathsf{b})$, 
for corresponding $\sigma^\mathsf{a} = z_B^\mathsf{a}/\rho_B$ and 
$\sigma^\mathsf{b} = z_B^\mathsf{b}/\rho_B$. Therefore, the negative
Re$\sigma$ axis is a branch cut for $W_l(\sigma\rho_B;\sigma)$
as a function of $\sigma$. Our path choices for connecting points
in the second and third quadrants to $z_\infty$ have been made
precisely to put this branch cut on the negative Re$\sigma$ axis.
\begin{figure}
\scalebox{0.70}{\includegraphics{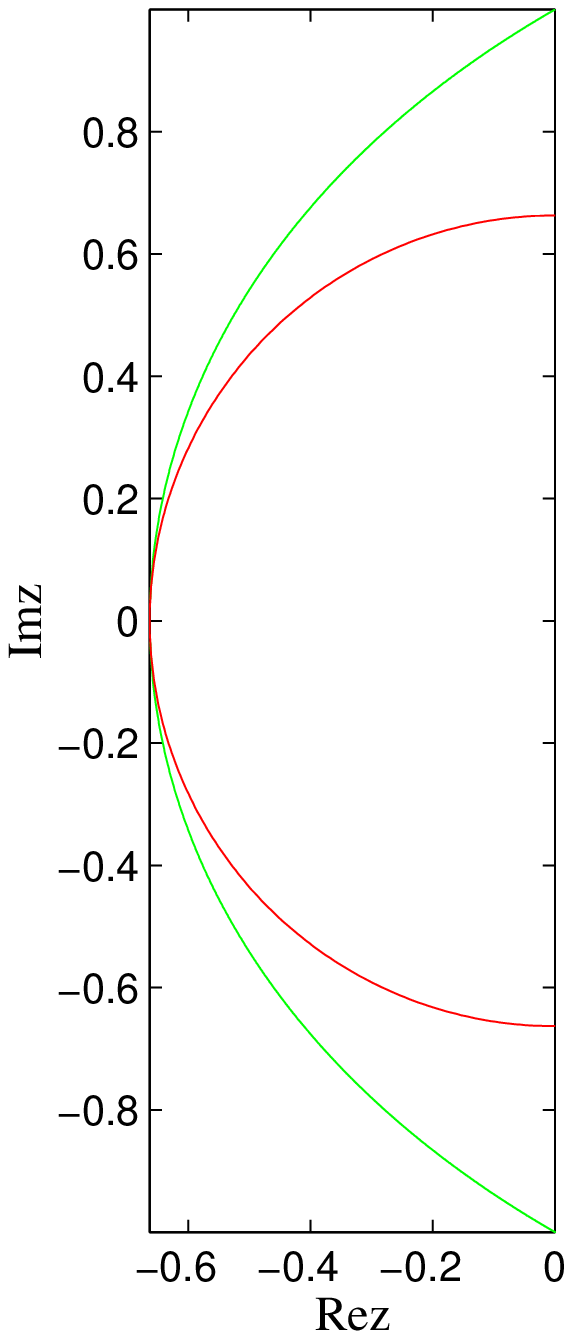}}
\caption{
{\sc Asymptotic curve $\mathcal{C}$.} The large
curve $\mathcal{C}$ is the one described in the
text and near which the scaled zeros of the MacDonald
function lie. It has parametric form
$z(\lambda) = -\sqrt{\lambda^2-\lambda\tanh\lambda}\pm {\rm i}    
\sqrt{\lambda\coth\lambda-\lambda^2}$ for $\lambda$
in the domain $[0,\lambda_0]$ with $\lambda_0
\simeq 1.1997$ such that $\tanh\lambda_0 = 1/\lambda_0$.
The small curve is a circle tangent to the $\mathcal{C}$
point $(x_0,0)$, where $x_0 = -\sqrt{\lambda_0^2-1}
\simeq -0.6627$.}
\label{twoFigure5}
\end{figure}

As we demonstrate below, the described two--component path 
method is quite accurate for low $l$. However, for large $l$ 
and some values of $\sigma$ there is a considerable loss of 
precision associated with evaluating $W_l(\sigma\rho_B)$ by 
this method (this is true no matter what integration scheme
is used along the path components). Therefore, we shortly 
introduce a more accurate method based on one--component paths.
Before turning to the improved method,
let us first heuristically describe the trouble the two--component
method can run into for large $l$. In {\sc Fig}.~\ref{twoFigure3} we
graphically demonstrate the breakdown in the method which occurs
(for the specified parameter values) when $l$ gets beyond $70$.
The relevant task under consideration is to obtain $W_l(\sigma\rho_B)$
in a region around those zeros of $W_l(\sigma\rho_B)$ which have large 
negative real parts. On the {\sc lhs} we plot $\log |W_{70}(15\sigma)|$,
using the logarithm to distribute contour lines more evenly. For the 
portion of the $\sigma$--plane shown only two of seventy zero locations 
are evident. Note the onset of degradation in the numerical solution. 
On the {\sc rhs} we plot 
$\log |W_{74}(15\sigma)|$, and in the plot two of seventy--four zero 
locations are somewhat evident, despite significant degradation. This 
degradation stems from the following phenomenon. Although they do avoid 
the origin, two--component integration paths, especially those which 
terminate near a zero with large negative real part, tend to pass 
through a region near the origin where the solution is quite large. 
The phenomenon becomes more pronounced as $l$ grows. Two--component
paths connect $z_\infty$ (where the solution is of order unity) to
$z_B$ (which might be at or near a zero of the solution in question),
and at each of these points the solution is in some sense small.
Therefore, loss of accuracy is an issue if the connecting path indeed
passes through a large--solution region. We document an instance
of this situation in {\sc Fig}.~\ref{twoFigure4}.
\begin{figure}
\scalebox{0.70}{\includegraphics{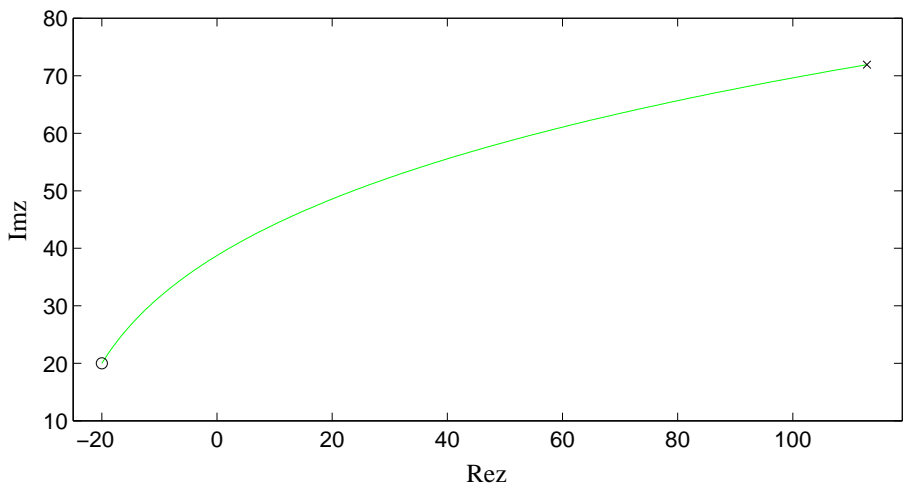}}
\caption{{\sc One--component path.}
To generate the depicted curve we have set $R \simeq 38.7479$
and chosen $\eta$ from the range $0.5162 \lesssim \eta \lesssim
1.8650$, ensuring that the terminal point $z_B = 20 +
{\rm i}20$ and the initial point $z_\infty \simeq 113
+{\rm i} 72$ (comparable with the analogous point in
Fig.~\ref{2path}). For one--component paths the meaning of
the {\tt scale} variable is a little different. For such
paths the real component of $z_\infty$ is set by
$R*{\tt scale}$.
\label{twoFigure6}}
\end{figure}

\subsubsection{One--component path integration}\label{sbsbSec:2.1.3}
We now describe an alternative class of integration paths tailored
to mitigate the problem of passing through regions where the solution
is large. Members of this alternative class are one--component paths, 
and this new class yields an improved version of the integration method 
based the two--component paths. As the new method will be more accurate, 
we will use it to quantify the accuracy of the two--component method.

The one--component paths of interest are essentially dilations of
a certain curve $\mathcal{C}$ depicted in
{\sc Fig}.~\ref{twoFigure5}. A parametric description of $\mathcal{C}$
in terms of transcendental functions is given in the figure caption.
The curve $\mathcal{C}$ is intimately related to the zeros of
$W_l(z)$, also the zeros of the MacDonald function $K_{l+1/2}(z)$.
As a degree--$l$ polynomial in inverse $z$, the function $W_l(z)$
has $l$ zeros. Let $n\in\mathbb{Z}_{\geq 0}$ run from $1$ to $l$ 
(with $n = 0$ if $l=0$) and $k_{l,n}$ denote the zeros of $W_l(z)$. 
It is known that the
scaled zeros $(l+1/2)^{-1}k_{l,n}$ lie arbitrarily close to $\mathcal{C}$
as $l$ becomes large (see results listed or summarized in 
Refs.~\cite{Olver,AGH1,Jiang,Watson}). 
A simple numerical experiment performed in {\sc Section} 
\ref{sbsbSec:3.1.1} confirms this assertion even for small $l$. 
Therefore, for a given $l$, dilation of $\mathcal{C}$ 
by $l + 1/2$ yields a curve on
which the solution $W_l(z)$ tends to remain small. Our one--component
integration paths are quartic approximations to (dilations of)
$\mathcal{C}$, and an example is depicted in 
{\sc Fig}.~\ref{twoFigure6}. The approximation is given parametrically
by $R(g(\eta),\eta)$, where $R$ is fixed and 
$g(\eta) = a\eta^4 + b\eta^2 + c$
is a quartic polynomial such that upon multiplication by $R$
the $\mathcal{C}$ points $(0, \pm 1)$, $(x_1,\pm y_1)$, and
$(x_0,0)$ all lie on the parametric approximation.
We have $x_1 = -\sqrt{2/(e^2+1)}$,
$y_1 = \sqrt{2/(e^2-1)}$, and $x_0 = -\sqrt{\lambda_0^2-1}$
with $\lambda_0 \simeq 1.1997$ obeying $\tanh(\lambda_0) =
1/\lambda_0$. From the
parametric description of $\mathcal{C}$ given in the caption
of {\sc Fig}.~\ref{twoFigure5}, one may verify that each of these
points indeed lies on $\mathcal{C}$. Numerically then
$a \simeq 0.1534$, $b \simeq 0.5093$, and $c \simeq -0.6627$.
\begin{figure}[t]
\scalebox{0.70}{\includegraphics{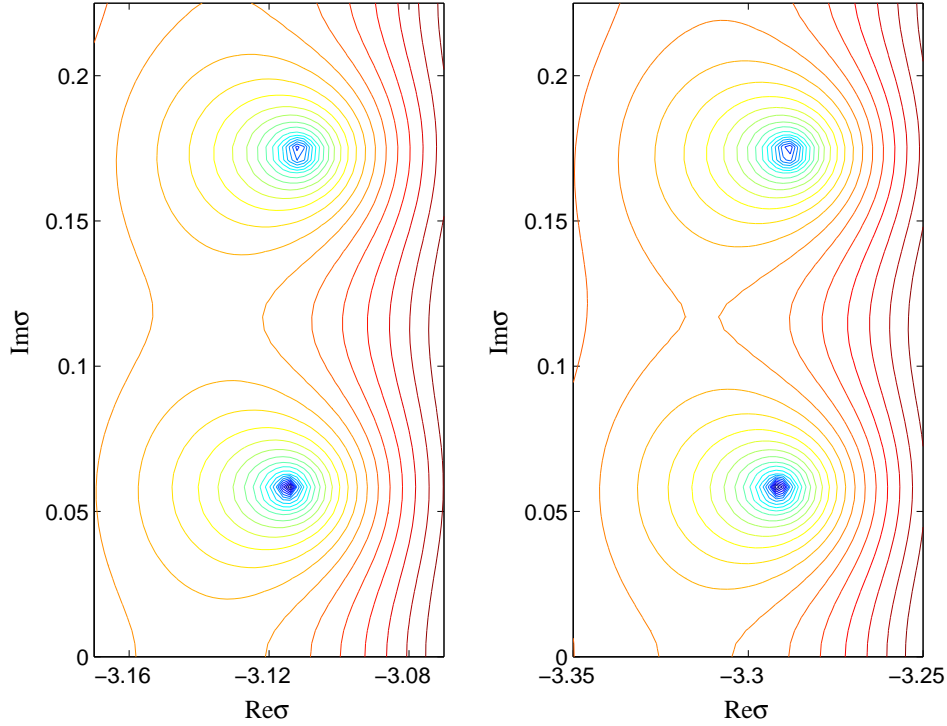}}
\caption{{\sc Contour lines of $\log |W_{l}(15\sigma)|$
on $\sigma$--plane for $l = 70$ and $74$.}
On the {\sc lhs} we have plotted $\log |W_{70}(15\sigma)|$.
The method used to generate the plot is the one based on the
one--component paths, again with Runga--Kutta--Fehlberg
integration. Relevant parameters here are $I=50$, $J=50$,
$P = 98304$, $l = 70$, $\rho_B = 15$, ${\tt scale}
= 500$, $p=65$, and $\kappa = 1$. $P$ is the number of
subintervals for the numerical integration. Other
parameters are described in the text or the caption for
Fig.~\ref{twoFigure3}. The plot on the {\sc rhs} is nearly
the same, except that now $l=74$ and $p=69$ (i.e. the initial
condition still determined by $5 = l - p$ terms). All
other parameters are the same as for the {\sc lhs} plot.
\label{twoFigure7}}
\end{figure}
\begin{figure}[t]
\scalebox{0.70}{\includegraphics{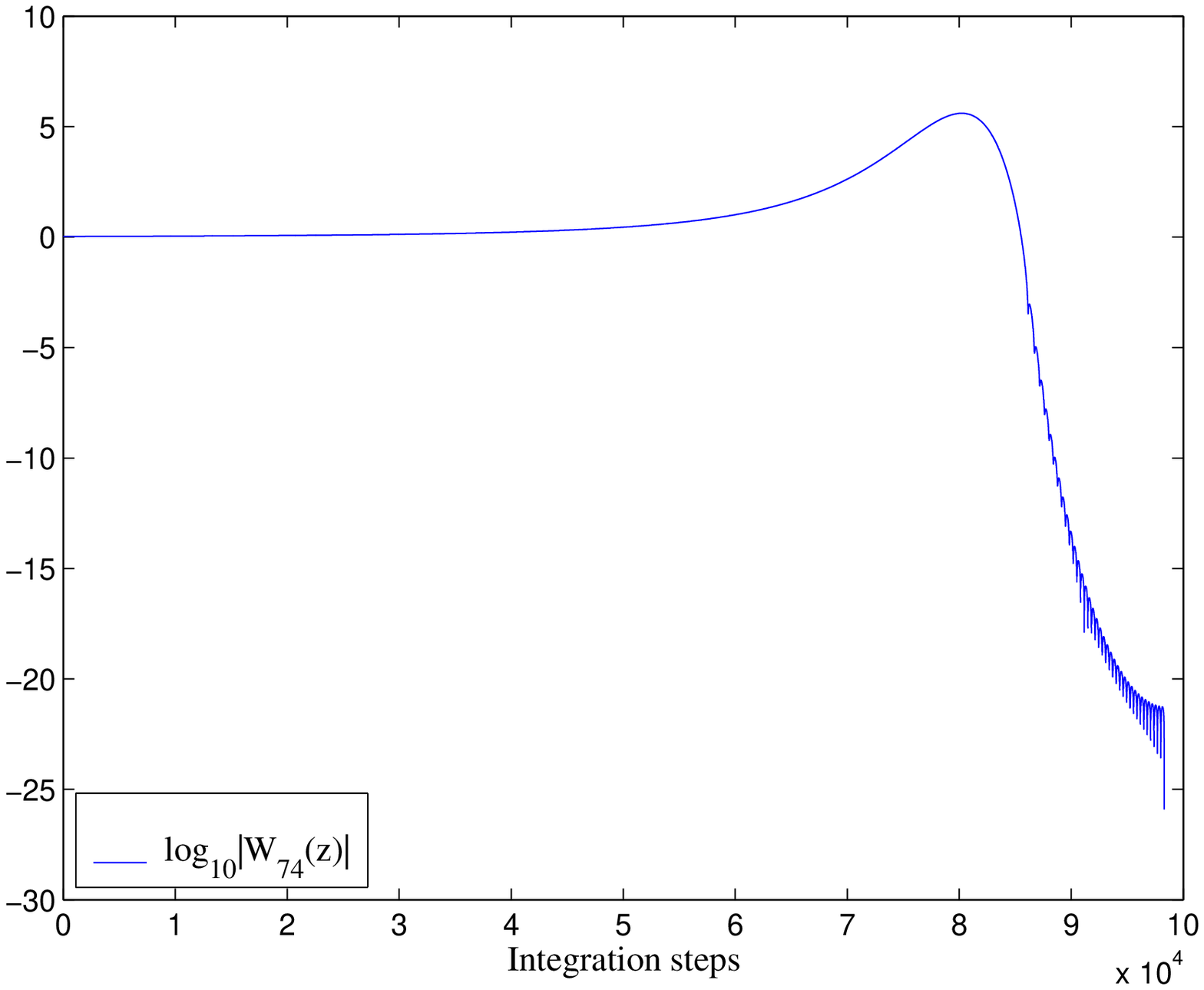}}
\caption{
{\sc Value of $\log_{10}|W_{74}(z)|$ along
one--component path from $z_\infty$ to $z_B = 15\sigma$
for $\sigma = -3.29128 + {\rm i}0.05785$}.
The point $-3.29128 + {\rm i}0.05785$ is
close to a zero of $W_{74}(15\sigma)$ and lies
in the region of the $\sigma$--plane shown in
Fig.~\ref{twoFigure7}. Along the horizontal axis we
have the $P = 98304$ integration steps. Note that the
modulus of $W_{74}(z)$ never gets so large as it does in
Fig.~\ref{twoFigure4}.}
\label{twoFigure8}
\end{figure}

We repeat the graphical investigation described and carried out 
at the end of {\sc Section} 
\ref{sbsbSec:2.1.2}, but now with the one--component 
method. The relevant contour plots of $\log |W_{70}(15\sigma)|$ and 
$\log |W_{74}(15\sigma)|$ are shown in {\sc Fig}.~\ref{twoFigure7}. 
Comparing this set of plots with the corresponding set in 
{\sc Fig}.~\ref{twoFigure3}, we see significantly less degradation in
former set. Moreover, we again investigate the size of the solution 
along an integration path in {\sc Fig}.~\ref{twoFigure8}. These figures 
and their captions argue that the method based on one--component 
paths is more accurate than the one based on two--component paths,
at least insofar as zero--finding is concerned.

\subsubsection{Accuracy of the numerical evaluation}\label{sbsbSec:2.1.4}
To check the accuracy of our methods for evaluating either
$W_l(\sigma\rho_B;\sigma)$ or $W_l(\sigma\rho_B)$ we may 
compare values obtained independently from one--component path and 
two--component path integration. For the evaluation of 
$W_l(\sigma\rho_B)$ there are other checks. First, numerical values for 
$W_l(\sigma\rho_B)$ can be checked against direct evaluations of 
$\sum_{n=0}^l c_n(\sigma\rho_B)^{-n}$.
However, as we have seen in (\ref{largecl}), for large $l$ the final
coefficient $c_l$ and the ones just before it quickly become too large
to faithfully evaluate this exact expression. One can use extended
precision (say in {\sc Mathematica}) to get around this problem. Another 
check, useful for large values of $l$ even without extended precision, 
involves the known continued fraction expansion
\begin{equation}
z \frac{W_l'(z)}{W_l(z)} = -\frac{l(l+1)}{2(z+1)+}\,
\frac{(l-1)(l+2)}{2(z+2)+} \cdots
\frac{2(2l-1)}{2(z+l-1)+}\, \frac{2l}{2(z+l)}\, .
\label{continuedfraction}
\end{equation}
This formula follows from recurrence relations obeyed by MacDonald 
functions \cite{Watson}.
It remains valid for non--integer $l$; however, in this case the 
{\sc rhs} of the equation is an infinite continued fraction.
Lenz's method may be used to evaluate this continued fraction for any
$l$ (see the appendix of Ref.~\cite{Thompson}). Now, both of
our integration methods also return the derivative $W_l'(z_B)$
in addition to $W_l(z_B)$. To see why, let $W = U+{\rm i}V$
(suppressing the argument and $l$). In order to integrate the 
second--order {\sc ode} (\ref{nbesseq}) for the complex variable $W$, 
we switch to a first--order system of {\sc ode} for the real vector 
$(U^\star,U,V,V^\star)$. The $\star$ denotes differentiation with 
respect to any relevant path parameter. With knowledge of $U^\star$
and $V^\star$ and the Cauchy--Riemann equations, one can recover
$W'$. Therefore, both the one--component and two--component 
path methods may also be used to evaluate $z_B W_l'(z_B)/W_l(z_B)$, 
and this value can then be checked against the continued fraction 
(\ref{continuedfraction}) evaluated at $z_B$. In this context,
notice that the zeros in $\sigma$ of the reciprocal
$W_l(\sigma\rho_B)/(\sigma\rho_B W'_l(\sigma\rho_B))$ are also
the zeros of $W_l(\sigma\rho_B)$, owing to the fact that the zeros 
of the MacDonald function are simple \cite{Watson}. Appealing to the 
above checks, we find that even the inferior numerical method 
based on two--component paths is quite accurate for $l \leq 10$; 
and we offer the following concrete investigations to sharpen
this statement.\footnote{We remark that these accuracy checks
test our methods where we need them most, that is on those tasks 
necessary for a numerical construction the kernel via the 
representation (\ref{poleandcut}).}

{\em Accuracy in zero--finding.}
The function $W_{10}(15\sigma)$ has ten
zeros, which come in five complex--conjugate pairs. Using the
secant algorithm, we compute the five zeros with positive imaginary
parts via our two independent methods. Note that whether the
one--component or two--component path method is used, each 
function call in the secant algorithm involves a numerical 
integration. The results, listed in Table \ref{twoTable1}, 
indicate that for low $l$ the two--component path method is 
associated with absolute errors equal to or better than 
$10^{-12}$, at least insofar as zero--finding is 
concerned.\footnote{Computing the same zeros in extended precision 
with {\sc Mathematica}, we have checked that the one--component 
path method yields the zeros with absolute errors near 
$10^{-15}$.} We reach the same conclusion upon computing the 
zeros of the Heun function $W_{10}(15\sigma;\sigma)$ via the two 
methods. This is remarkable in that there is no {\em a priori} 
relationship between the asymptotic curve $\mathcal{C}$ and the
zeros in $\sigma$ of the Heun function $W_l(\sigma\rho_B;\sigma)$.
However, carrying out the same graphical experiments for Heun 
functions that we carried out for Bessel functions and documented 
in {\sc Figs}.~\ref{twoFigure4} and \ref{twoFigure8}, we again find that 
the one--component path method is better than the two--component
path method at keeping the solution small during the integration.
Numerical experiments described in {\sc Section} 
\ref{sbsbSec:3.1.1} further
clarify this issue. 
\begin{table}
\begin{center}
\begin{tabular}{|l|}
\hline
$-0.461469660361894 + {\rm i} 0.057844346363414$
\\
\hline
$-0.441019397698505 + {\rm i} 0.174104528053357$
\\
\hline
$-0.397835221905874 + {\rm i} 0.292329812596140$
\\
\hline
$-0.325747971123938 + {\rm i} 0.414999032164771$
\\
\hline
$-0.207261082243274 + {\rm i} 0.548846630604906$
\\
\hline
\end{tabular}
\end{center}
\vskip 2mm
\begin{center}
\begin{tabular}{|l|}
\hline
$-0.461469660361817 + {\rm i} 0.057844346363415$
\\
\hline
$-0.441019397698458 + {\rm i} 0.174104528053339$
\\
\hline
$-0.397835221905853 + {\rm i} 0.292329812596129$
\\
\hline
$-0.325747971123933 + {\rm i} 0.414999032164771$
\\
\hline
$-0.207261082243273 + {\rm i} 0.548846630604906$
\\
\hline
\end{tabular}
\end{center}
\vskip 2mm
\caption{{\sc Zeros of $W_{10}(15\sigma)$ computed via 
two different methods.} In the top table we list
five of the ten zeros (the other five are complex   
conjugates). These have been found using the two--component
path method in tandem with the secant algorithm. In the second
table we list the same zeros, although now found using the
one--component path method with the secant algorithm.
\label{twoTable1}}
\end{table}

\begin{figure}[t]
\scalebox{0.70}{\includegraphics{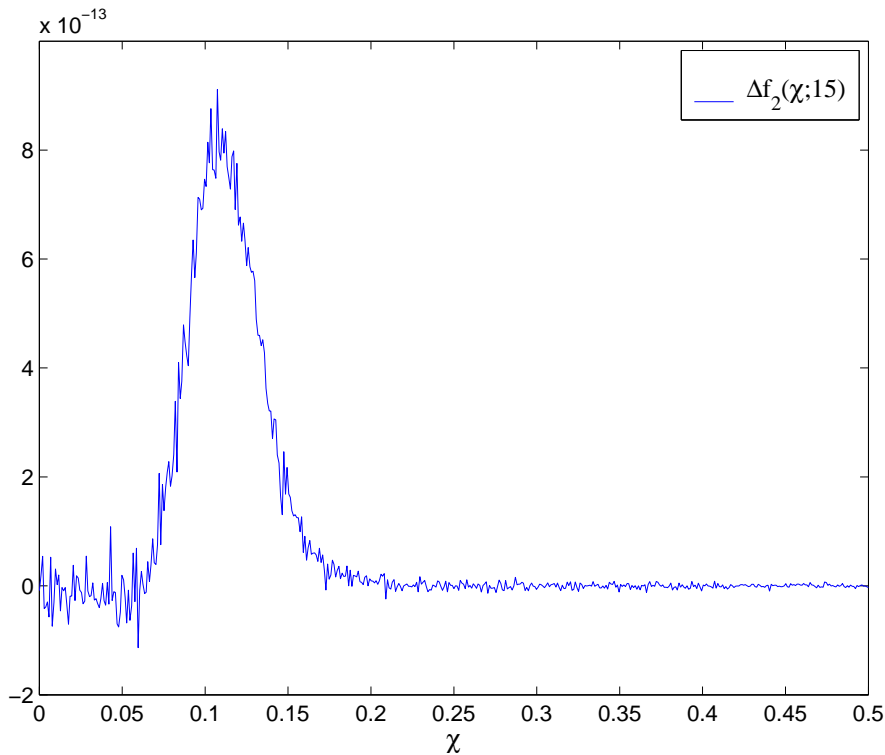}}
\caption{{\sc Absolute error in the cut profile $f_2(\chi;15)$.}
Here we plot the difference $\Delta f_2(\chi;15)$ of two
numerical computations of $f_2(\chi;15)$, one based on the
two--component path method and the other on the one--component
path method. For the two--component method we have $N = 131072
= M$, while for the one--component method $P = 262144$.
Parameters common to both computations are $\kappa = 1$,
${\tt scale} = 1000$ and $p = -3$ so $l-p = 5$. There are
512 $\chi$--subintervals in the plot.
\label{twoFigure9}}
\end{figure}

{\em Accuracy in the cut profile.} As applied to the
Heun case, both the two--component and one--component path methods 
also return $W_l'(\sigma\rho_B;\sigma)$. This can 
be seen via argumentation similar to that given above in the 
context of the real vector $(U^\star,U,V,V^\star)$. Therefore, we 
have two independent methods for calculating 
$\sigma\rho_B W_l'(\sigma\rho_B;\sigma)/W_l(\sigma\rho_B;\sigma)$,
where $\sigma$ may be chosen pure real and negative (say with 
the convention that all paths approach the negative 
Re$\sigma$ axis running through the second quadrant). That is to say,
each of our methods may be used to evaluate the cut profile
\begin{equation}
f_l(\chi;\rho_B) = 
{\rm Im}\left[
e^{{\rm i}\pi}\chi\rho_B 
W_l'(e^{{\rm i}\pi}\chi\rho_B;e^{{\rm i}\pi}\chi)
/W_l(e^{{\rm i}\pi}\chi\rho_B;e^{{\rm i}\pi}\chi)\right]\, .
\label{twocutprofile}
\end{equation}
Using each method to obtain its own numerical graph for the profile
$f_2(\chi;15)$ shown in {\sc Figure} \ref{oneFigure2} of 
{\sc Section} \ref{sbsbSec:1.4.2}, we then plot the difference of 
these graphs in {\sc Fig}.~\ref{twoFigure9}.
Similar graphs for other values of $l\leq 10$ indicate that the 
two--component path method evaluates the maximum value of 
$|f_l(\chi;15)|$ with an absolute error better than $10^{-10}$. 
For the following reasons we believe that the one--component path 
method computes this maximum with an even smaller absolute error. 
The essential support of $|f_l(\chi;15)|$ corresponds to a region of 
the $\sigma$ plane near those zeros of $W_l(\sigma\rho_B;\sigma)$ 
with largest negative real part. Therefore, in connecting $z_\infty$ 
to a purely real $z_B = \exp({\rm i}\pi)\chi\rho_B$ on the cut, a
one--component path runs all the way near (a dilation of) 
$\mathcal{C}$, indicating that the numerical solution along 
such a one--component path again tends to remain small. Experiments 
like those documented in {\sc Figs.}~\ref{twoFigure4}
and \ref{twoFigure8} confirm this expectation.

We will mainly use the described integration methods for small 
$l\leq 10$. However, we note that via comparison with both the 
continued fraction expression and extended precision calculations
in {\sc Mathematica}, we believe that our one--component path
method maintains single precision accuracy up to about $l = 50$, 
at least insofar as zero--finding is concerned. Finally, we 
mention that we have carried out all integration using the 
Runga--Kutta--Fehlburg scheme with fixed step--size along 
individual path components. In light of the sufficient accuracy 
noted here and the next subsection, we have not 
found it necessary to introduce any sort of adaptive integration.
Furthermore, we have not found the local truncation error 
estimate (stemming from comparison between the fourth and 
fifth--order integration schemes) provided by
Runga--Kutta--Fehlburg to be a useful diagnostic for our 
purposes. Relying on our own accuracy checks, we have simply 
used the straight explicit fifth--order scheme.

\subsection{Numerical evaluation of the radiation kernel}\label{sbSec:2.2}
The numerical methods discussed in the last subsection work well for
values of $l \leq 10$, and via (\ref{poleandcut}) will allow
us to directly construct sufficiently accurate sum--of--poles
representations of the {\sc fdrk}. Moreover, even for
moderately large $l$ these methods prove useful in qualitative
investigations of the outgoing solution's analytic structure.
However, when it comes to building an accurate sum--of--pole
representation of the {\sc fdrk} for high $l$, the described methods
lack the necessary accuracy.

In this subsection we describe different methods for direct
evaluation of the radiation kernel itself along the
Im$\sigma$ axis, ones sufficiently accurate even for 
high $l$. Given accurate profiles for the real and imaginary parts
of the radiation kernel along this axis, we may then extract an
accurate sum--of--poles representation via a method described
in {\sc Section} \ref{sbSec:3.3} and due to Ref.~\cite{AGH1}. 
Here we described two methods for evaluating 
$\tilomega_l(\sigma;\rho_B)$ when
$\sigma$ is pure imaginary, one accurate so long as 
$|\sigma| \gg 0$ and the other so long as $0 \neq |\sigma| \lesssim 1$. 
There is some interval of overlap on the Im$\sigma$ axis 
on which both methods are accurate and may be compared. We warn the 
reader that we also use the notation $\tilomega_l(\sigma;\rho_B)$ 
for the Bessel {\sc fdrk} 
$\sigma\rho_B W'_l(\sigma\rho_B)/W_l(\sigma\rho_B)$.
In order to avoid the confusion which might arise from
this dual meaning of the symbol $\tilomega_l(\sigma;\rho_B)$,
in this subsection we sometimes adopt the following notation. For the 
product of $z$ with the Heun logarithmic 
derivative we may use
\begin{equation}
w_l(z;\sigma) = z W'_l(z;\sigma)/W_l(z;\sigma)\, ,
\end{equation}
while for the corresponding Bessel object we may use 
\begin{equation}
w_l(z) = z W'_l(z)/W_l(z)\, .
\label{Besselflatspacekernel}
\end{equation}
Formally $w_l(z) = w_l(z;0)$, in parallel with the conventions
of {\sc Section} \ref{sbsbSec:1.3.2}. For the Heun case 
$\tilomega_l(\sigma;\rho_B) = w_l(\sigma\rho_B;\sigma)$, while for
the Bessel case $\tilomega_l(\sigma;\rho_B) = 
w_l(\sigma\rho_B)$.
\begin{figure}[t]
\scalebox{0.70}{\includegraphics{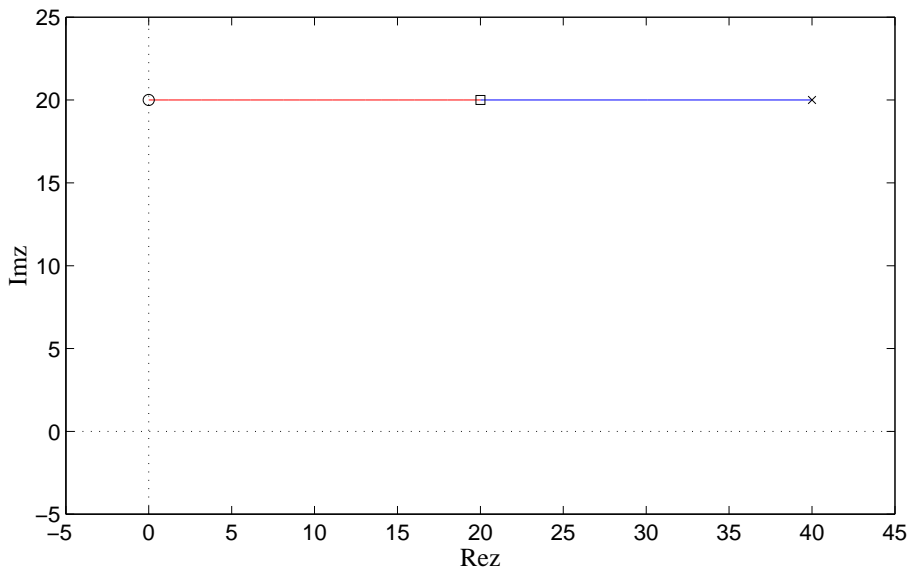}} 
\caption{{\sc
Two--component path in $z$--plane.} The figure depicts a straight
path of the type described in the text. Here we have $\rho_B =
20$, $\sigma = {\rm i}$, ${\tt scale}_1 = 20$, and ${\tt scale}_2
= 40$, so that the points $z_1$, $z_2$ and $z_B$ respectively
correspond to the marked cross, square, and circle. Typically
${\tt scale}_1$ and ${\tt scale}_2$ will be much larger, but the
values here make for a good figure. \label{wpath2}}
\end{figure}

\subsubsection{Evaluation of the kernel for large imaginary
frequencies} \label{sbsbSec:2.2.1}
We turn first to the evaluation of
$\tilomega_l(\sigma;\rho_B)$ for $\sigma\in {\rm i}\mathbb{R}$ 
and $|\sigma| \gg 0$. For the remainder of this subsection
$\sigma = {\rm i}y$ for real $y$. In this scenario we find it 
useful to again work with the complex
variable $z = \sigma \rho$. As before, for a given $\sigma$ the 
terminal evaluation point will be denoted by 
$z_B = \sigma \rho_B$, and it lies on the Im$z$ axis.
Consider two positive real numbers ${\tt scale}_1 > {\tt scale}_2$ 
and associated $z$--points $z_1 = {\tt scale}_1 + z_B$ and $z_2 = {\tt
scale}_2 + z_B$. The point $z_1$ is analogous to the point
$z_\infty$ introduced before. Further consider a straight path
like the one shown in {\sc Fig}.~\ref{wpath2} running through all of
these points. Let us now outline the method for obtaining the
value $\tilomega_l(z_B/\rho_B;\rho_B)$, mostly considering only
the model Bessel case to streamline the presentation. First, using
the truncated series $\sum_{n=0}^{l-p} c_n z_1^{-n}$, we
compute initial values for the {\sc ode} (\ref{nbesseq}). Next, we
integrate the {\sc ode} along the straight path from $z_1$ to
$z_2$ (the first portion of the path in {\sc Fig}.~\ref{wpath2}). As
Re$z > 0$ along this path, we again have exponential suppression
of errors both in the initial conditions and due to roundoff. The
result of this integration is accurate numerical values for $W_l(z_2)$ 
and $W'_l(z_2)$, from which we can directly build a numerical value
for the kernel $z_2 W'_l(z_2)/W_l(z_2)$ at this intermediary point. 
The assumption here is that $z_2$ is still large enough in modulus 
to ensure that the solution $W_l(z_2)$ is not too large. Finally, we
integrate the radiation kernel itself along the straight path from
$z_2$ to $z_B$, carrying this out as follows.

Whether we are working with (\ref{maineq7}) or
(\ref{nbesseq}), we have an {\sc ode} of the form
\begin{equation}
\frac{{\rm d}^2\tilPhi_l}{{\rm d} z^2}
+ R(z;\sigma)\frac{{\rm d}\tilPhi_l}{{\rm d} z}
+ S(z;\sigma)\tilPhi_l = 0\, ,
\end{equation}
so that both $w_l(z)$ and $w_l(z;\sigma)$ obey a first--order
nonlinear {\sc ode} of the form
\begin{equation}
\frac{{\rm d}w_l}{{\rm d}z}
- \frac{w_l}{z}
+ \frac{w_l^2}{z}
+ R(z;\sigma) w_l + z S(z;\sigma) = 0\, .
\label{firstorderode}
\end{equation}
We use the same symbol $w_l$ as the dependent variable here, 
since as a first--order {\sc ode} there is only one linearly
independent solution. With the accurate value for $w_l(z_2)$ 
at our disposal at the end of the first
integration leg, we integrate this last {\sc ode} from $z_2$ to
the terminal value $z_B$ (the second portion of the path in
{\sc Fig}.~\ref{wpath2}) in order to obtain the desired complex value
$w_l(z_B) = z_B W'_l(z_B)/W_l(z_B)$. For the
Heun case Eq.~(\ref{firstorderode}) is an {\sc ode} for $w_l(z;\sigma)
= z W'_l(z;\sigma)/W_l(z;\sigma)$, and it is again integrated from
$z_2$ to $z_B$, given an accurate value for $w_l(z_2;\sigma)$.

The method just described can be unstable if the terminal point
$z_B$ lies too close to the origin. Indeed, notice that some of
the terms in the {\sc ode} (\ref{firstorderode}) are singular at
the origin, showing that finiteness of the kernel derivative
at the origin depends on exact cancellation of singular terms.
Round off error will spoil any such exact cancellation in an
integration towards a terminal $z_B$ equal to or near zero. Later
we demonstrate that the value $\tilomega_l(0;\rho_B)$ of the
kernel at the origin can be computed in closed form for both the
Bessel and Heun cases. This raises the possibility that the value
$w_l(z_B)$ ---or in the Heun case the value $w_l(z_B;\sigma)$---
corresponding to a non--zero $|z_B| \ll 1$ might be numerically 
computed via integration of
(\ref{firstorderode}) out from the origin. However, our numerical
experiments suggest that this is not a viable approach. Moreover,
the following analytical reasoning would also seem to dash this
possibility. For the Bessel $l = 2$ case we have the exact
expression
\begin{equation}
z W'_2(z)/W_2(z) = -\frac{3z+6}{z^2 + 3z + 3}\, .
\end{equation}
A linearization stability analysis of the Bessel--case {\sc ode}
(\ref{firstorderode}) about this solution indicates that small
perturbations of $z W'_2(z)/W_2(z)$ grow exponentially on paths
running away from the origin along the Im$z$ axis.
\begin{figure}[t]
\scalebox{0.70}{\includegraphics{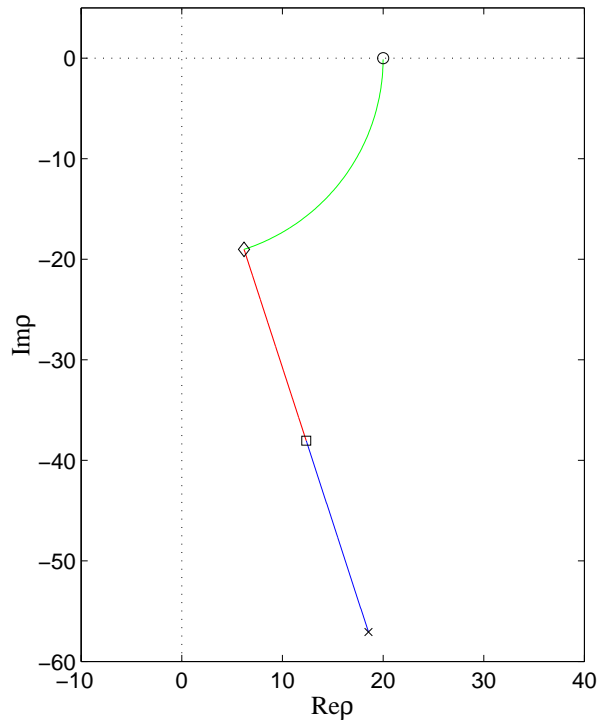}}
\caption{{\sc
Three--component path in $\rho$--plane.} The figure depicts an   
example three--component path of the type described in the text. Here
we have $\rho_B = 20$, ${\tt scale}_1 = 60$, ${\tt scale}_2 = 40$,
and $\theta = - 2\pi/5$ so that the points $\rho_1$, $\rho_2$, 
$\rho_3$ and $\rho_B$ respectively correspond to $60\exp(-{\rm
i}2\pi/5)$ (cross), $40\exp(-{\rm i}2\pi/5)$ (square),
$20\exp({-\rm i}2\pi/5)$ (diamond), and 20 (circle). The depicted
path corresponds to a $\sigma$ value lying on the positive
imaginary axis, so that between $\rho_1$ and $\rho_3$ we have
Re$(\sigma\rho) > 0$. Better suppression of error would be had for 
$\theta = -\pi/2$, in which case the portion of the path between
$\rho_1$ and $\rho_3$ would lie on the negative imaginary axis.
However, the final integration would then be over a longer
arc, and on this final arc we do not expect to have error
suppression. There seems to be some trade--off here, which is why
we have kept $\theta$ as a parameter. In any case, typically ${\tt
scale}_1$ and ${\tt scale}_2$ will be much larger, but the values
here make for a good figure.\label{twoFigure11}}
\end{figure}

\subsubsection{Evaluation of the kernel for small imaginary
frequencies} \label{sbsbSec:2.2.2}
For the reasons just laid down, we use a different
method for small {\em non--zero} imaginary $\sigma$. The new
method employs integration in the complex $\rho$--plane rather
than the $z$--plane. We introduce new positive real numbers ${\tt
scale}_1 > {\tt scale}_2 > \rho_B$, a phase factor $\exp({\rm
i}\theta)$, and the following associated $\rho$--points (all in 
polar form):
$\rho_1 = {\tt scale}_1*\exp({\rm i}\theta)$, $\rho_2 = {\tt
scale}_2*\exp({\rm i}\theta)$, $\rho_3 = \rho_B\exp({\rm
i}\theta)$, and $\rho_B$. These points define a three--component
path in the $\rho$--plane such as the one shown in
{\sc Fig}.~\ref{twoFigure11}. Let us now outline the new method for 
computing the value $\tilomega_l(\sigma;\rho_B)$, again mostly 
considering only the model Bessel case to streamline the 
presentation. First,
using the asymptotic expansion (\ref{aexpansion}), we compute
initial values for the {\sc ode} (\ref{nbesseq2}). Next, we
integrate (\ref{nbesseq2}) along the straight ray from $\rho_1$ 
to $\rho_2$ (the first portion of the path in 
{\sc Fig}.~\ref{twoFigure11}). We
choose the angle $\theta$ such that Re$(\sigma\rho) > 0$ along
this path, ensuring exponential suppression of errors both in the
initial conditions and due to roundoff. The result of this
integration is accurate numerical values for $W_l(\sigma\rho_2)$ 
and $W^\rho_l(\sigma\rho_2)$, from which we can directly build a 
numerical value for $w_l(\sigma\rho_2)$ at
the intermediary point $\rho_2$. Similar to before, the assumption
is that $\rho_2$ is large enough in modulus to ensure that
the solution $W_l(\sigma\rho_2)$ is not too large. Finally, we
integrate $w_l(\sigma\rho)$ itself along a two--component
ray--and--arc path from $\rho_2$ to the real point $\rho_B$, by
way of an intermediate point $\rho_3$. Such a remaining
two--component path is depicted in {\sc Fig}.~\ref{twoFigure11} as the 
final two portions of the curve connecting $\rho_2$ to $\rho_B$.
This integration is carried out as follows.

Whether we are working with (\ref{maineq6}) or
(\ref{nbesseq2}),
we have an {\sc ode} of the form
\begin{equation}
\frac{{\rm d}^2\tilPhi_l}{{\rm d} \rho^2} +
\mathcal{R}(\rho;\sigma)
\frac{{\rm d}\tilPhi_l}{{\rm d} \rho}
+ \mathcal{S}(\rho;\sigma) \tilPhi_l = 0\, ,
\label{pqrhoode}
\end{equation}
whence both $w_l(\sigma\rho;\sigma)$ and $w_l(\sigma\rho)$ 
obey a first--order nonlinear {\sc ode} of the form
\begin{equation}
\frac{{\rm d}w_l}{{\rm d}\rho}
- \frac{w_l}{\rho}
+ \frac{w_l^2}{\rho}
+ \mathcal{R}(\rho;\sigma) w_l + \rho \mathcal{S}(\rho;\sigma)
= 0 \, .
\label{wpqrhoode}
\end{equation}
To reach this equation, we have used, for example, 
$w_l(\rho\sigma) = \rho W_l^{\rho}(\rho\sigma)/W_l(\rho\sigma)$.
Given the initial value for $w_l(\sigma\rho_2)$ obtained 
from the first leg of the
integration in the last paragraph, we first integrate
(\ref{wpqrhoode}) along the straight ray from $\rho_2$ to 
$\rho_3$ which has the same modulus as the terminal point $\rho_B$. The
final leg, a rotation back to the Re$\rho$ axis, is an
integration of (\ref{wpqrhoode}) along an arc from $\rho_3$ to
$\rho_B$.
\begin{figure}[t]
\scalebox{0.70}{\includegraphics{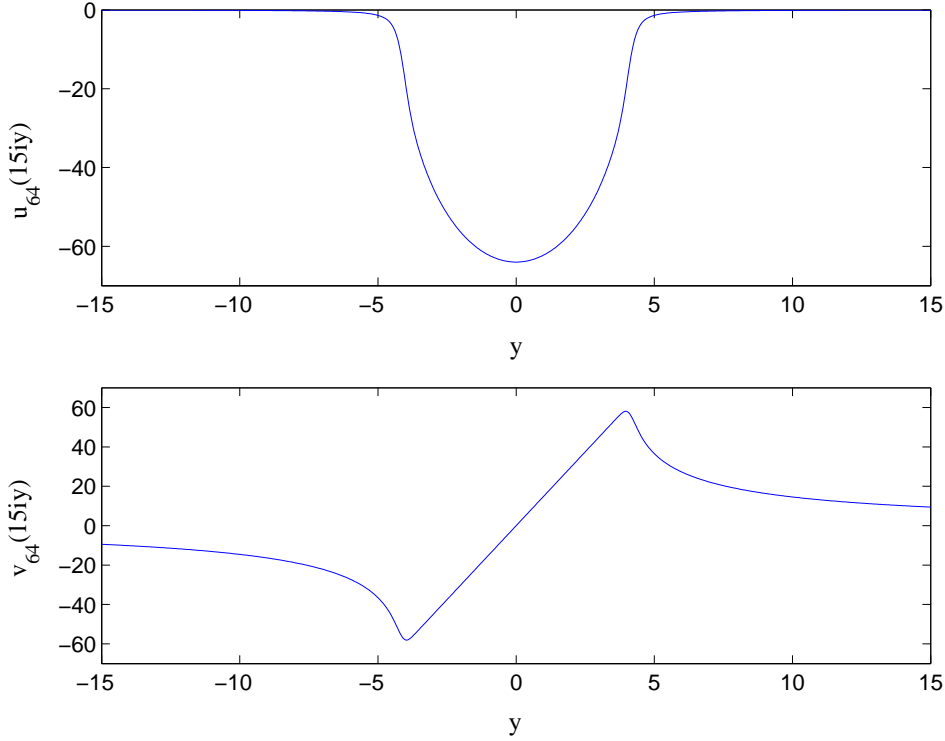}}
\caption{{\sc
Bessel fdrk $\tilomega_{64}({\rm i}y;15) = w_{64}(15{\rm i}y)$.}
Here we plot the functions $u_{64}(15{\rm i}y) = \mathrm{Re}w_{64}(15{\rm
i}y)$ and $v_{64}(15{\rm i}y) = \mathrm{Im}w_{64}(15{\rm i}y)$,
with the $y$ axis split into 512 subintervals.
For $|y| > {\tt break} = 1$ we have evaluated
$\tilomega_{51}({\rm i}y;15)$ using two--component integration 
with the following parameter values: $N = 131072$,
$M =131072$, ${\tt scale}_1 = 1000$, and ${\tt scale}_2
= 100$. $N$ and $M$ are respectively the number of integration   
steps taken along the first and second components of the path.   
For $|y| \leq {\tt break}$ we have evaluated $\tilomega_{51}({\rm
i}y;15)$ using three--component integration with the
parameter values $N = 131072$, $M =131072$, $P =
2048$, $\theta = \pi/4$, ${\tt scale}_1 = 1000$, and ${\tt
scale}_2 = 100$. $N$, $M$, and $P$ are respectively the number of 
integration steps taken along the first, second, and third
components of the path. For both integration methods $\kappa = 1$
and $p = 59$. Typically, we have chosen ${\tt break}$ smaller,
but now have ${\tt break} = 1$ to demonstrate the three--component
method.
\label{twoFigure12}}
\end{figure}

\subsubsection{Value of the kernel at the origin}\label{sbsbSec:2.2.3}
For the Bessel case the origin value $\tilomega_l(0;\rho_B)$
of the radiation kernel is the limit $\lim_{\sigma\rightarrow0}
w_l(\sigma\rho_B)$, while for the
Heun case the value $\tilomega_l(0;\rho_B)$ is the limit
$\lim_{\sigma\rightarrow0} w_l(\sigma\rho_B;\sigma)$. Whether
considering the Bessel or Heun case, we may derive an exact
expression for the value $\tilomega_l(0;\rho_B)$. Turn first to
the Bessel case, where $W_l(\sigma\rho) = \sum^l_{n=0} c_n
(\sigma\rho)^{-n}$ is of course singular at $\sigma = 0$. However,
with this exact expression it is easy to check that
\begin{equation}
\lim_{\sigma\rightarrow 0} w_l(\sigma\rho_B) = - l\, .
\end{equation}
We stress that this calculation of $\tilomega_l(0;\rho_B)$ makes
use of the exact form of the outgoing solution, which is not at
our disposal in the Heun case. A separate recipe for getting
this value, one without appeal to the exact form of
$W_l(\sigma\rho_B)$, goes as follows. Set $\sigma = 0$ in
(\ref{nbesseq2}), thereby reaching an {\sc ode}
\begin{equation}
\frac{{\rm d}^2\tilPhi_l}{{\rm d}\rho^2} -
\frac{l(l+1)}{\rho^2}\tilPhi_l = 0\
\end{equation}
with solutions $\rho^{l+1}$ and $\rho^{-l}$. We now use 
$[\rho \partial_\rho \log \rho^{-l}]|_{\rho = \rho_B}$ as the 
origin value $\tilomega_l(0;\rho_B)$, again finding $-l$.
\begin{figure}
\scalebox{0.70}{\includegraphics{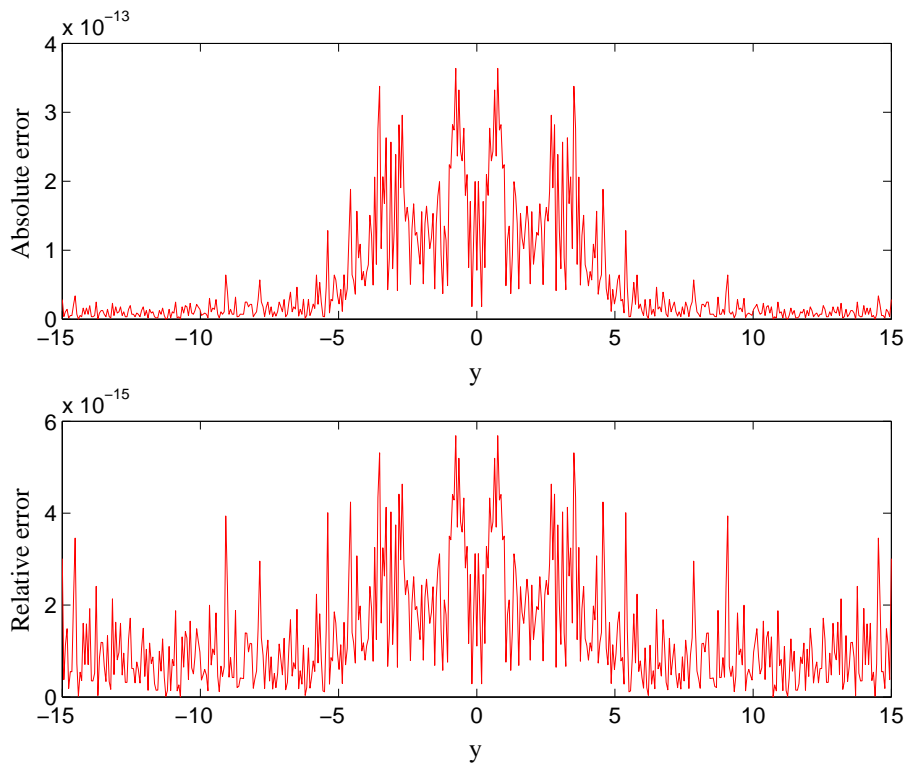}}
\caption{{\sc Error in Bessel fdrk
$\tilomega_{64}({\rm i}y;15) = w_{64}(15{\rm i}y)$.}
Here we plot the absolute error
$|\Delta w_{64}(15{\rm i}y)|$ as well as the relative
error $|\Delta w_{64}(15{\rm i}y)|/|w_{64}(15{\rm i}y)|$.
These errors have been computed against the ``exact''
$w_{64}(15{\rm i}y)$ generated with the continued
fraction expression (\ref{continuedfraction}).
Parameters are the same as in {\sc Fig}.~\ref{twoFigure12}.
\label{twoFigure13}}
\end{figure}

Let us turn to the Heun case and follow this recipe for
getting the value $\tilomega_l(0;\rho_B)$. We set $\sigma = 0$ in
(\ref{maineq6}), obtaining the following {\sc ode}:
\begin{equation}
\frac{{\rm d}^2\tilPhi_l}{{\rm d}\rho^2} +
\left[-\frac{1}{\rho}+\frac{1}{\rho-1}\right] \frac{{\rm
d}\tilPhi_l}{{\rm d}\rho} + \left[\frac{\kappa}{\rho^2}-
\frac{\kappa+l(l+1)}{\rho(\rho-1)}\right]\tilPhi_l = 0\, ,
\end{equation}
Both solutions to this equation may be expressed in terms of
infinite series in inverse $\rho$. The one corresponding to
$\rho^{-l}$ above has the form
\begin{equation}
\sum_{n=0}^{\infty} a_n\rho^{-(l+n)}\, ,
\end{equation}
where $a_0 = 1$ and
\begin{equation}
a_{n+1} = \frac{(l+n)(l+n+2)+\kappa}{(l+n+1)(l+n+2)-l(l+1)}\,
a_n\, .
\end{equation}
The series is positive and absolutely convergent for
all $\rho > 1$. We then have
\begin{equation}
\tilomega_l(0;\rho_B) = -\sum^{\infty}_{n=0} (l+n) a_n
\rho^{-n}_B\Big/\sum^{\infty}_{n=0} a_n \rho^{-n}_B
\label{originkernel}
\end{equation}
as our concrete expression for the value in question. Notice that
this value approaches $-l$ in the $\rho_B \rightarrow \infty$
limit as expected.
\begin{figure}
\scalebox{0.70}{\includegraphics{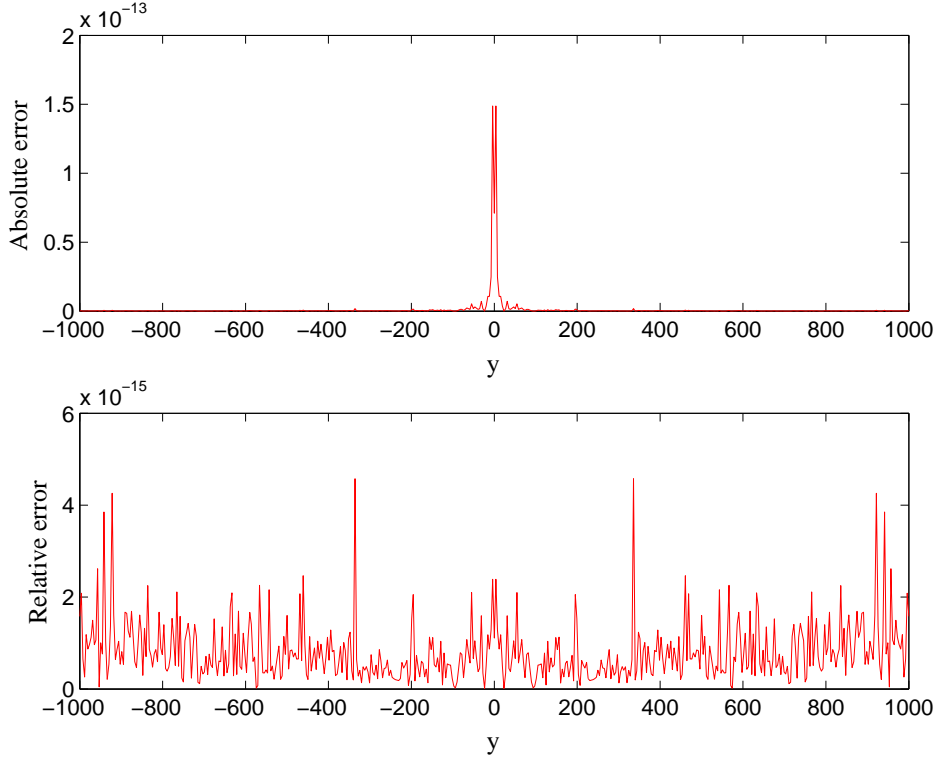}} 
\caption{
{\sc Error in Bessel fdrk
$\tilomega_{64}({\rm i}y;15)$.}
These are essentially the same plots as
those in {\sc Fig}.~\ref{twoFigure13}, save that here we
have a larger $y$--interval.
\label{twoFigure14}}
\end{figure}

\subsubsection{Accuracy of the numerical evaluation}\label{sbsbSec:2.2.4}
{\sc Fig}.~\ref{twoFigure12} depicts the real part 
$u_{64}(15{\rm i}y)$ and the imaginary part $v_{64}(15{\rm i}y)$ of 
the Bessel {\sc fdrk} $\tilomega_{64}({\rm i}y;15) 
= w_{64}(15{\rm i}y)$ along the Im$\sigma$ axis for 
$l = 64$ and $\rho_B = 15$. We have generated these plots using 
the methods described in this subsection, and have listed other 
parameters set while obtaining them in the figure caption. To 
examine the accuracy of these numerical profiles, we may compare 
them with corresponding profiles obtained via the continued 
fraction expansion (\ref{continuedfraction}). We consider the 
profiles stemming from the continued fraction expansion as the 
``exact'' ones. With the two sets of profiles, one may compute 
corresponding absolute and relative error measures. We plot these
errors in {\sc Fig}.~\ref{twoFigure13} and 
{\sc Fig}.~\ref{twoFigure14} (the second being a pull--back of 
the first). From these figures we conclude that our numerical
methods evaluate $w_{64}(15{\rm i}y)$ with an absolute supremum 
error less than $10^{-12}$ and a relative supremum error less than 
$10^{-14}$, at least for $|y| < 1000$. For the Bessel case at 
hand we have found comparable error bounds associated with
all other values of $l\in\{10,11,\dots,64\}$, although we note that
the corresponding $y$--interval needs to shrink by as much as
an order of magnitude to maintain these bounds for $l=10$.
\begin{figure}[t]
\scalebox{0.70}{\includegraphics{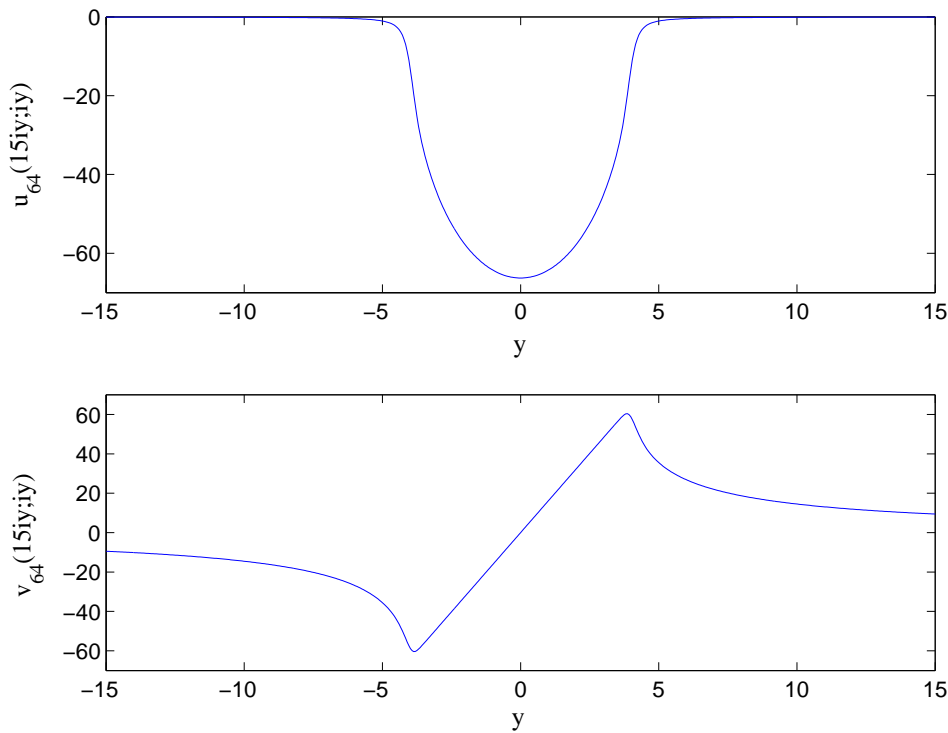}}
\caption{
{\sc Heun fdrk $\tilomega_{64}({\rm i}y;15)
= w_{64}(15{\rm i}y;{\rm i}y)$.}
Here we plot the functions
$u_{64}(15{\rm i}y;{\rm i}y)
= \mathrm{Re}w_{64}(15{\rm i}y;{\rm i}y)$ and
$v_{64}(15{\rm i}y;{\rm i}y)
= \mathrm{Im}w_{64}(15{\rm i}y;{\rm i}y)$.
All parameters in these plots match those
listed in {\sc Fig}.~\ref{twoFigure12} depicting the Bessel
{\sc fdrk}.\label{twoFigure15}}
\end{figure}

\begin{figure}[t]
\scalebox{0.70}{\includegraphics{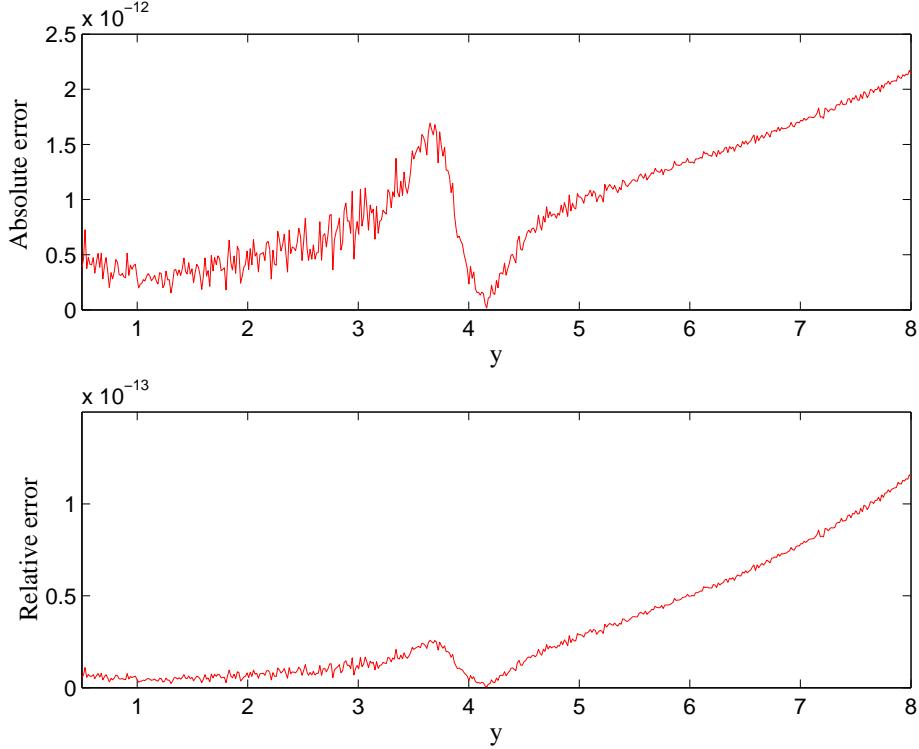}}
\caption{{\sc
Error in Heun fdrk $\tilomega_{64}({\rm i}y;15)
= w_{64}(15{\rm i}y;{\rm i}y)$.}
Here we plot the absolute error measure
$|\Delta w_{64}(15{\rm i}y;{\rm i}y)|$ and the relative error
measure $|\Delta w_{64}(15{\rm i}y;{\rm i}y)|/
|w_{64}(15{\rm i}y;{\rm i}y)|$ described in the text. We have
$y\in[0.5,8]$ for both. All other parameters set in
generating these plots are the same as those listed in the
caption of {\sc Fig}.~\ref{twoFigure12}.
\label{twoFigure16}}
\end{figure}

{\sc Fig}.~\ref{twoFigure15} depicts the real part
$u_{64}(15{\rm i}y;{\rm i}y)$ and the imaginary part 
$v_{64}(15{\rm i}y;{\rm i}y)$ of the Heun {\sc fdrk} 
$\tilomega_{64}({\rm i}y;15) = w_{64}(15{\rm i}y;{\rm i}y)$ 
along the Im$\sigma$ axis. We have again chosen the 
representative case $l = 64$ and $\rho_B = 15$, setting the 
rest of the parameters to the same values used to generate the 
Bessel profiles depicted in {\sc Fig}.~\ref{twoFigure12}.
Note that the two sets of profiles are qualitatively very similar. 
However, they are different. In particular, now the real part 
has a minimum value of $-66.2816976576098$ rather than $-64$.
For the Heun case at hand we have no analog of the continued 
fraction expansion with which to check the accuracy of the 
profiles. Nevertheless, at least for $y$ values of order unity, 
we can perform an accuracy check by comparing the two--component
and three--component path methods for evaluating the kernel. 
Such a comparison is shown in 
{\sc Fig}.~\ref{twoFigure16}. With the two numerically obtained 
kernels we form an absolute 
error measure $|\Delta w_{64}(15{\rm i}y;{\rm i}y)|$
and also a relative error measure
$|\Delta w_{64}(15{\rm i}y;{\rm i}y)|/
|w_{64}(15{\rm i}y;{\rm i}y)|$, over $y\in[0.5,8]$ for both.
In the denominator of the relative
error, we happen to have used the kernel stemming from the
three--component method. {\sc Fig}.~\ref{twoFigure16} 
displays plots of both error measures. Note that poor 
performance for the three--component method is evident in the 
right portions of the plots. For the three--component method 
the length of the third and final integration path grows with $y$.
Therefore, for large $y$ one expects a corresponding loss of 
precision for the three--component method. 
%
%
\section{Sum--of--poles representation of the 
radiation kernel}\label{Sec:3}
In this section we focus on both exact and approximate 
representation of the {\sc fdrk} $\tilomega_l(\sigma;\rho_B)$ 
as a sum of poles. In the first subsection we qualitatively discuss 
the exact representation (\ref{poleandcut}) of the {\sc fdrk} as 
a (continuous and discrete) sum of poles, highlighting what we 
believe to be its main features. In the second subsection we document 
our particular numerical construction of the {\sc fdrk} as a 
sum of poles, and give an analysis of its numerical error. 
We stress that from a theoretical standpoint we are {\em conjecturing} 
that the Schwarzschild {\sc fdrk} ---built from the Heun function 
$W_l(z;\sigma)$--- admits the representation (\ref{poleandcut}),
although we do provide compelling numerical evidence for a 
representation of this form. This is in contrast to the case of the 
flatspace {\sc fdrk} ---built from the Bessel function $W_l(z)$--- 
which we theoretically know admits such a representation \cite{AGH1}. 
Since for us the representation (\ref{poleandcut}) of the 
Schwarzschild {\sc fdrk} is ultimately conjecture, there is more 
need to painstakingly justify it numerically, and we do so in the
second subsection. In the third subsection we turn to kernel 
compression, by which we mean approximation of the {\sc fdrk} by a 
proper rational function $P(\sigma)/Q(\sigma)$ which is itself a 
sum of poles. In the first and second subsections we consider 
only the $\jmath = 0$ ($\kappa = 1$) case, but also consider the 
$\jmath = 2$ ($\kappa = -3$) case in subsection three. As yet, we 
have not examined the $j = 1$ ($\kappa = 0$) case corresponding to
electromagnetic radiation.

In this section and in the simulations we describe later, 
we have almost exclusively worked with an outer boundary 
radius $\rho_B \in [15,25]$, corresponding to a physical outer 
boundary radius $r_B \in [30\mathrm{m},50\mathrm{m}]$. Therefore, 
were we considering a more general isolated source of gravitational 
radiation, one with gravitational radius $2\mathrm{m}$, then the 
boundary two--sphere $B$ would be located outside of the
{\em strong--field region} as defined by Thorne \cite{Thorne}.
Moreover, for wave simulation on a fixed background as we consider 
here, the location of $B$ corresponds to a metric coefficient 
$F(\rho_B)$ from (\ref{dlesselement}) in the range $0.9\overline{33} 
\leq F(\rho_B) \leq 0.96$. Whence $B$ lies in a region where the 
Schwarzschild metric is flat up to small correction. The {\sc robc} 
described in {\sc Section} \ref{sbSec:1.4} are not tied to the 
weak--field region. However, for this region our numerical methods 
for examining/constructing the {\sc fdrk} are accurate.
\begin{figure}
\scalebox{0.70}{\includegraphics{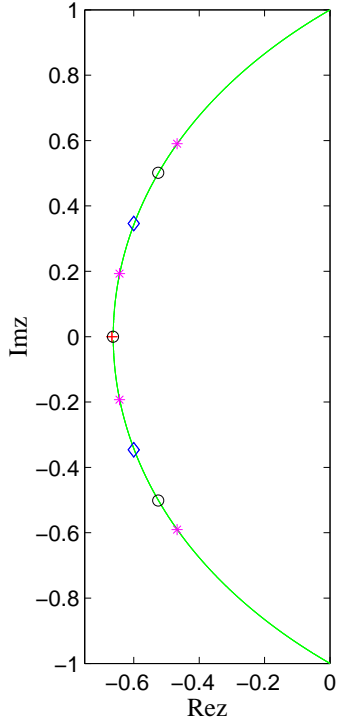}}
\caption{{\sc Scaled zeros of MacDonald functions.}
Here we plot scaled zeros $(l+1/2)^{-1}k_{l,n}$ for
$l = 1,2,3,4$. The cross is the scaled zero of
$K_{1/2}(z)$, the diamonds are the scaled zeros of
$K_{3/2}(z)$, the circles are the scaled zeros of
$K_{5/2}(z)$, and the stars are the scaled zeros
of $K_{7/2}(z)$. To the eye these zeros, corresponding as
they do to small $l$ values, already lie close to the
curve $\mathcal{C}$ shown and described in both
the text and the caption of {\sc Figure} \ref{twoFigure5}
in {\sc Section} \ref{sbsbSec:2.1.3}.
As $l$ gets large the scaled zeros $(l+1/2)^{-1}k_{l,n}$
lie closer and closer to $\mathcal{C}$.
\label{threeFigure1}}
\end{figure}
\begin{figure}
\scalebox{0.70}{\includegraphics{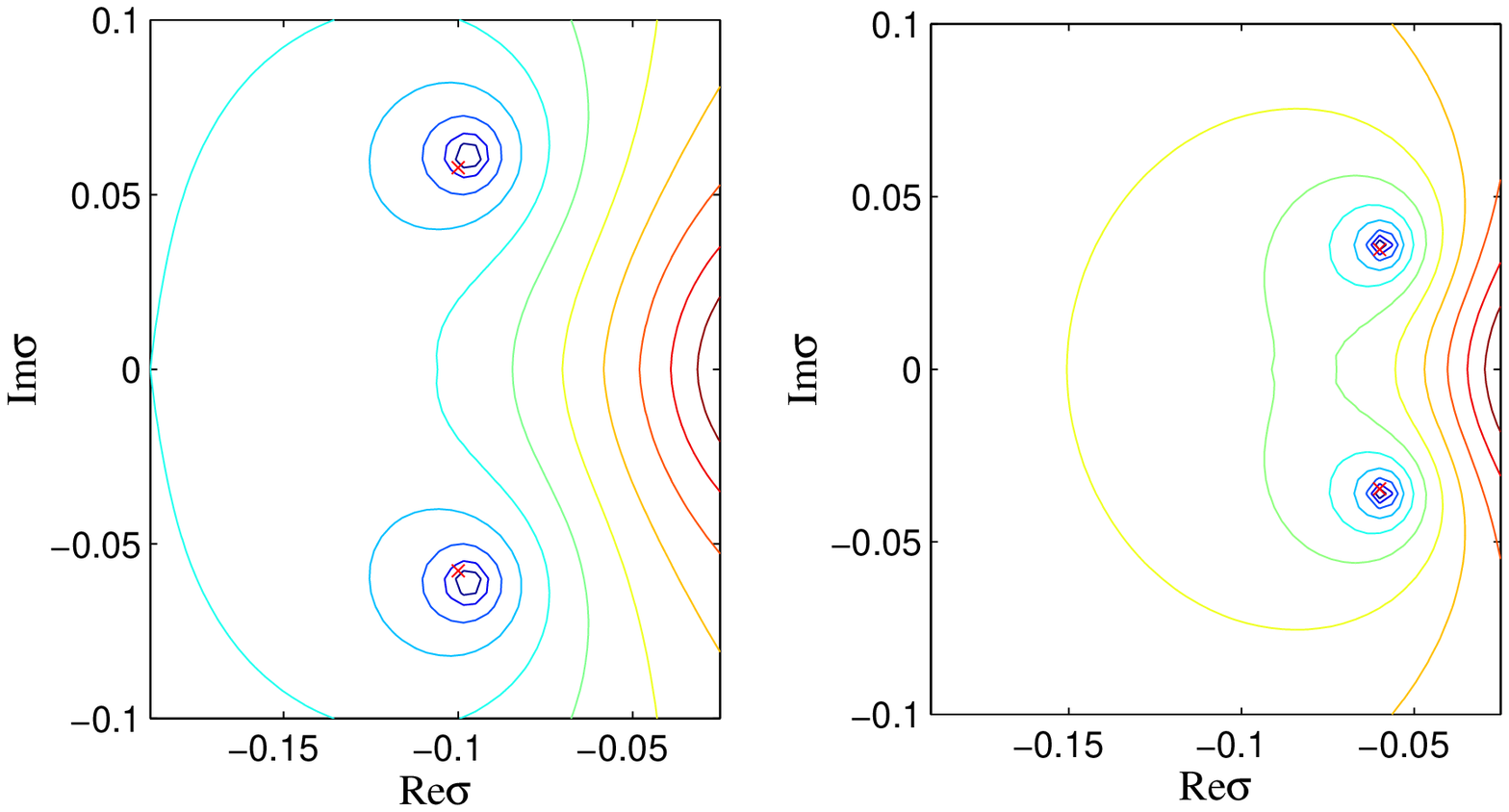}}
\caption{{\sc Zeros of $W_2(15\sigma;\sigma)$ and
                       $W_2(25\sigma;\sigma)$.}
The lefthand plot shows contour lines of $\log
|W_2(15\sigma;\sigma)|$, and the righthand one shows contour lines
of $\log |W_2(25\sigma;\sigma)|$. The logarithm mollifies the
singularity at the origin, distributing contour lines more evenly.
Notice that the zero locations are closer to the origin in the
righthand plot. In the lefthand plot and the righthand plot we have 
also respectively marked as crosses the zeros of the Bessel
functions $W_2(15\sigma) = 1 + 3(15\sigma)^{-1} +
3(15\sigma)^{-2}$ and $W_2(25\sigma) = 1 + 3(25\sigma)^{-1} +
3(25\sigma)^{-2}$. Perhaps evident even to the eye, the Bessel and
Heun zeros lie closer to each other in the righthand plot
(corresponding to the larger value of $\rho_B$).
\label{threeFigure2}}
\end{figure}

\subsection{Qualitative study of pole locations and 
cut profile}\label{sbSec:3.1}
Using the one--component path method described in 
{\sc Section} \ref{sbsbSec:2.1.3}, 
we first turn to the analytic structure of $W_l(\sigma\rho_B;\sigma)$
as a function of frequency $\sigma$ in the lefthalf plane,
assuming that a zero of this function corresponds to a pole
location appearing in (\ref{poleandcut}). Using the same method,
we then draw some quick observations concerning the cut profile
$f_l(\chi;\rho_B)$ in (\ref{poleandcut}). We mainly focus on the
restricted parameter space $\mathbb{S}$ determined by $\rho_B
\in [15,25]$ and $0\leq l \leq 10$, but also make mention of 
some remarkable features which crop up for other parameter 
values outside of this space. Our parameter space $\mathbb{S}$
has been chosen with the following reasons in mind. First, 
its $\rho_B$ interval is as discussed in the last paragraph. 
Second, it includes the first few values of $l$, which we want
to single out for special attention. Third (and related to the 
first two), it avoids by design the aforementioned remarkable 
features. We stress that our discussion in this first subsection is 
mostly qualitative and amounts to a collection of conjectures without 
substantial numerical or
analytical proof. Although we are bypassing a truly thorough study of
some interesting phenomena, we do not believe these phenomena to
be directly relevant for numerical implementation of {\sc robc} 
(further remarks on this point to follow).
\begin{figure}
\scalebox{0.70}{\includegraphics{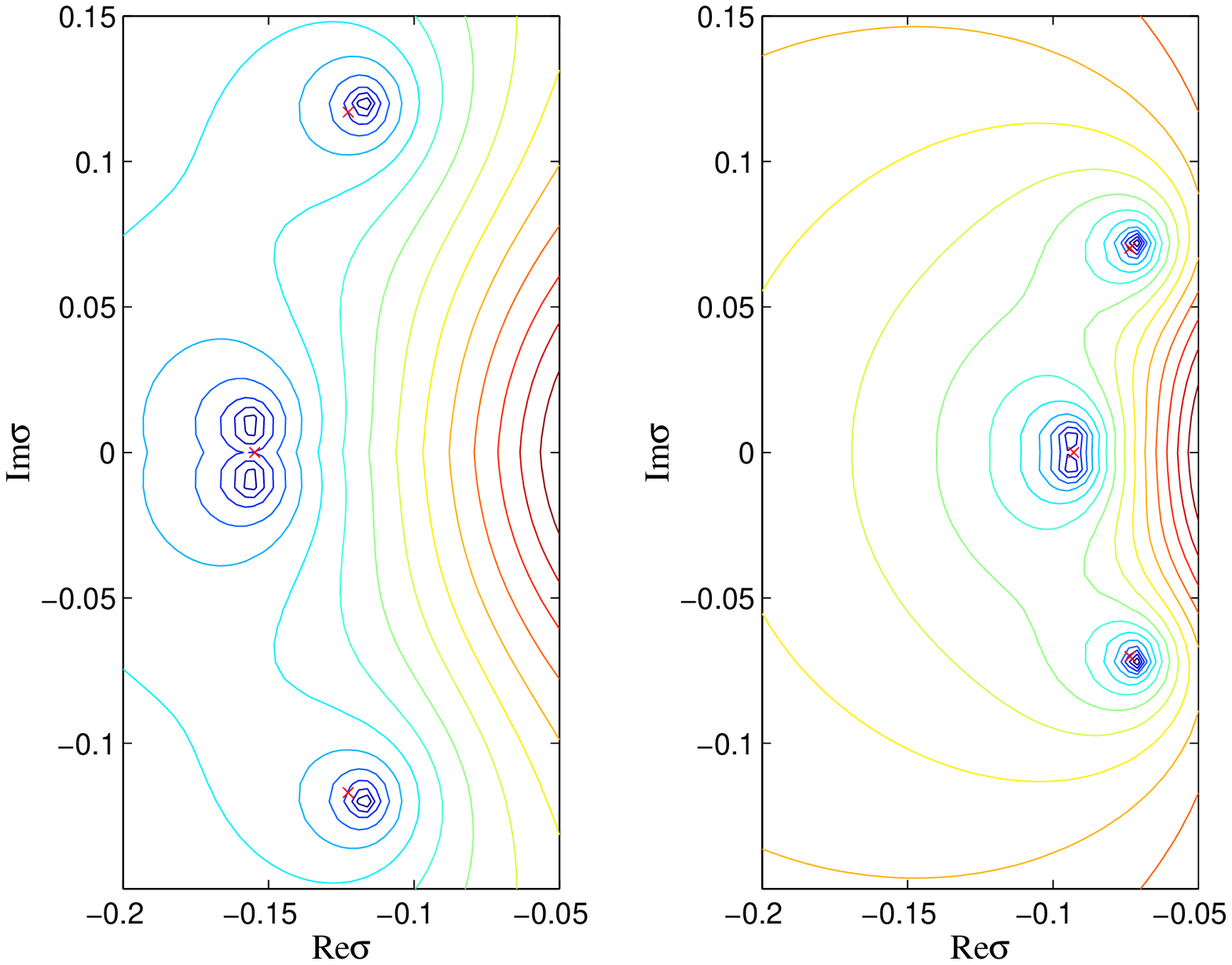}}
\caption{{\sc Zeros of $W_3(15\sigma;\sigma)$ and
$W_3(25\sigma;\sigma)$.} The lefthand plot shows contour lines of
$\log |W_3(15\sigma;\sigma)|$, and the righthand one shows lines
of $\log |W_3(25\sigma;\sigma)|$. Again, crosses mark the
corresponding Bessel zeros. Notice in the lefthand
plot that the single real Bessel zero corresponds to a pair of
zeros in the Heun case. Actually, in the righthand plot there are
also two distinct Heun zeros, each one near the Bessel zero lying
on the real axis, but the resolution is almost too low to see them
as distinct. \label{threeFigure3}}
\end{figure}
\begin{figure}[t]
\scalebox{0.70}{\includegraphics{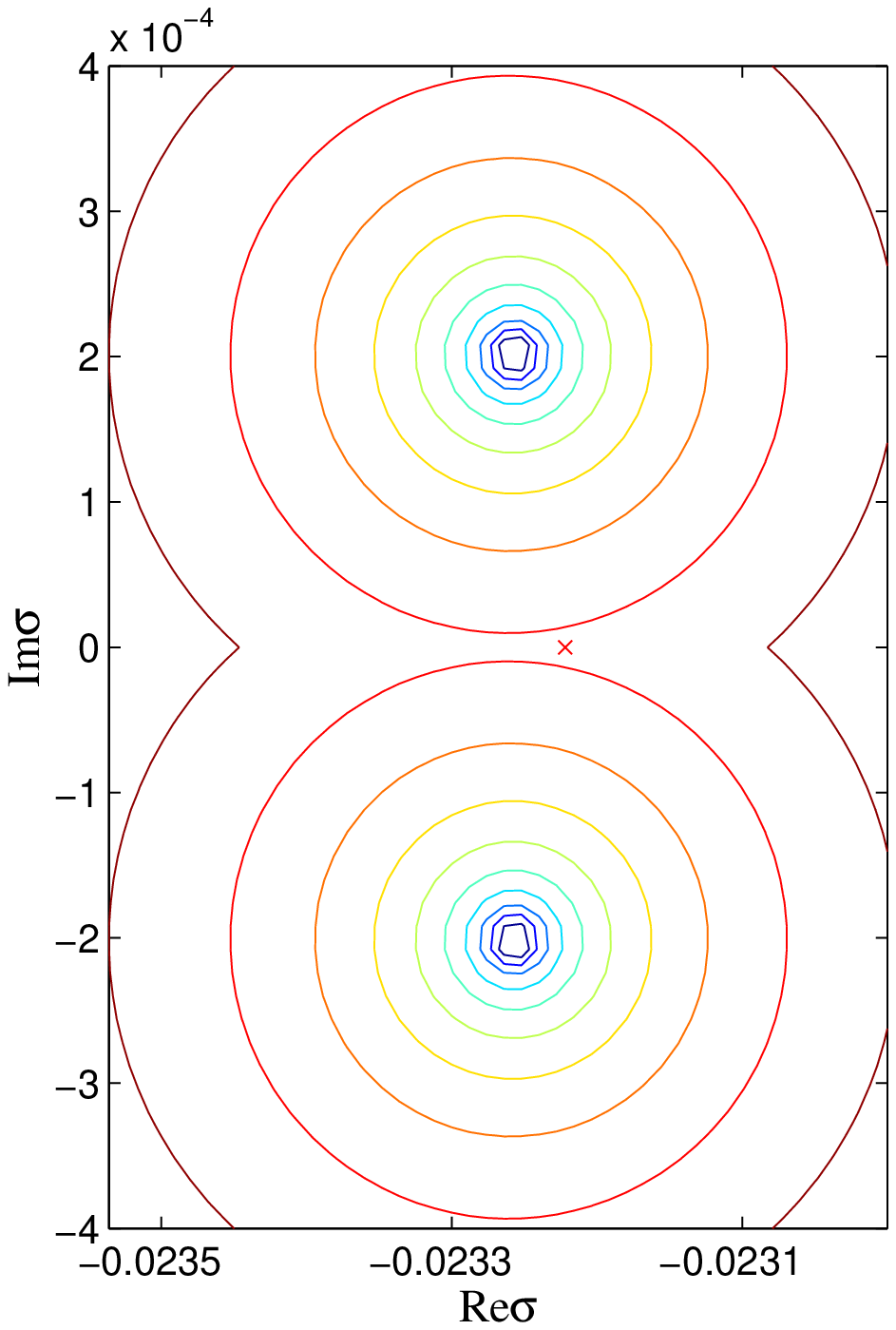}}
\caption{{\sc Zeros of $W_3(100\sigma;\sigma)$.} Here we plot
contour lines of $\log |W_3(100\sigma;\sigma)|$. Note that two
zeros of $W_3(100\sigma;\sigma)$ lie outside of the plot, as we
focus on the pair of zeros closest to the real axis. The red cross
is a zero of $W_3(100\sigma)$. We might gather from this plot that
the feature associated with odd $l$ and discussed in the text
persists as $\rho_B$ gets large. \label{threeFigure4}}
\end{figure}

\subsubsection{Zeros of the outgoing solution as a 
function of $\sigma$}\label{sbsbSec:3.1.1}
Recall that in {\sc Section} \ref{sbsbSec:2.1.3} we denoted by
$\{k_{l,n}: n = 1,\cdots,l\}$ the zero set of the MacDonald
function $K_{l+1/2}(z)$ which is also the zero set of $W_l(z)$.
With this notation the zeros in $\sigma$ of $W_l(\sigma\rho_B)$ are 
then simply the $k_{l,n}/\rho_B$. Let us collect several facts
concerning such sets, summarizing results derived or listed in
Refs.~\cite{Olver,AGH1,Jiang,Watson}. First, for even $l$ these zeros
come in complex--conjugate pairs, while for odd $l$ they again come
in complex--conjugate pairs save for a lone zero which lies on the
negative Re$\sigma$ axis. Second, the scaled zeros 
$(l+1/2)^{-1}k_{l,n}$ lie close to the asymptotic curve
$\mathcal{C}$ introduced in {\sc Section} \ref{sbsbSec:2.1.3}. See
{\sc Fig}.~\ref{threeFigure1} for a graphical demonstration of this claim.
Hence, for each $l$
one may imagine the zeros distributed in a crescent pattern in the
lefthalf $\sigma$--plane. As concrete examples, the zeros of
$W_2(15\sigma)$ are approximately $-0.1000\pm {\rm i}0.0577$,
while those of $W_3(15\sigma)$ are approximately $-0.1226 \pm
{\rm i}0.1170$ and $-0.1548+{\rm i}0$. Respectively, these zero
sets are marked by crosses in the lefthand plots of 
{\sc Figs}.~\ref{threeFigure2} and \ref{threeFigure3}.
\begin{figure}[t]
\scalebox{0.70}{\includegraphics{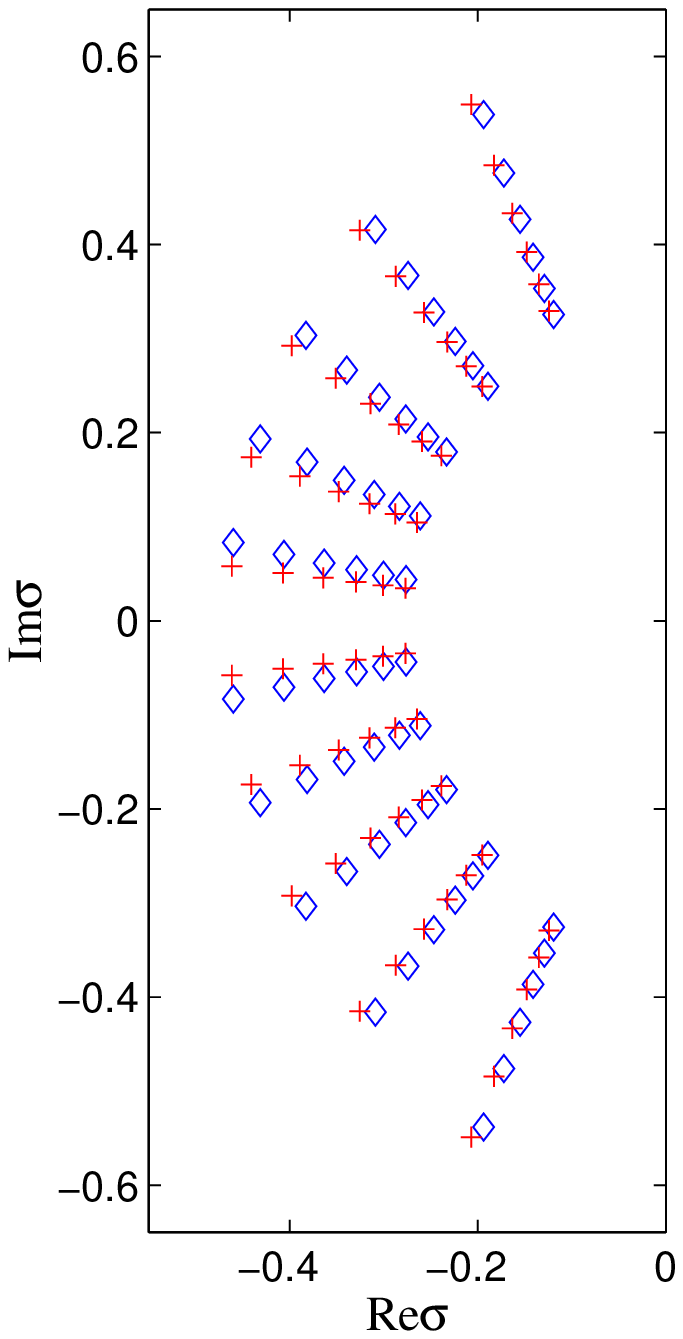}}
\caption{{\sc Zeros of $W_{10}(\sigma\rho_B;\sigma)$ and
$W_{10}(\sigma\rho_B)$.}
Here we plot zeros of these functions for $\rho_B
= 15,17,19,21,23,25$. The outermost crescent of
diamonds are the zeros of $W_{10}(15\sigma;\sigma)$,
while the innermost crescent of diamonds are the zeros
of $W_{10}(25\sigma;\sigma)$. The outermost crescent of
crosses are the zeros of $W_{10}(15\sigma)$, while the
innermost crescent of crosses are the zeros of
$W_{10}(25\sigma)$.
\label{threeFigure5}}
\end{figure}
\begin{figure}[t]
\scalebox{0.70}{\includegraphics{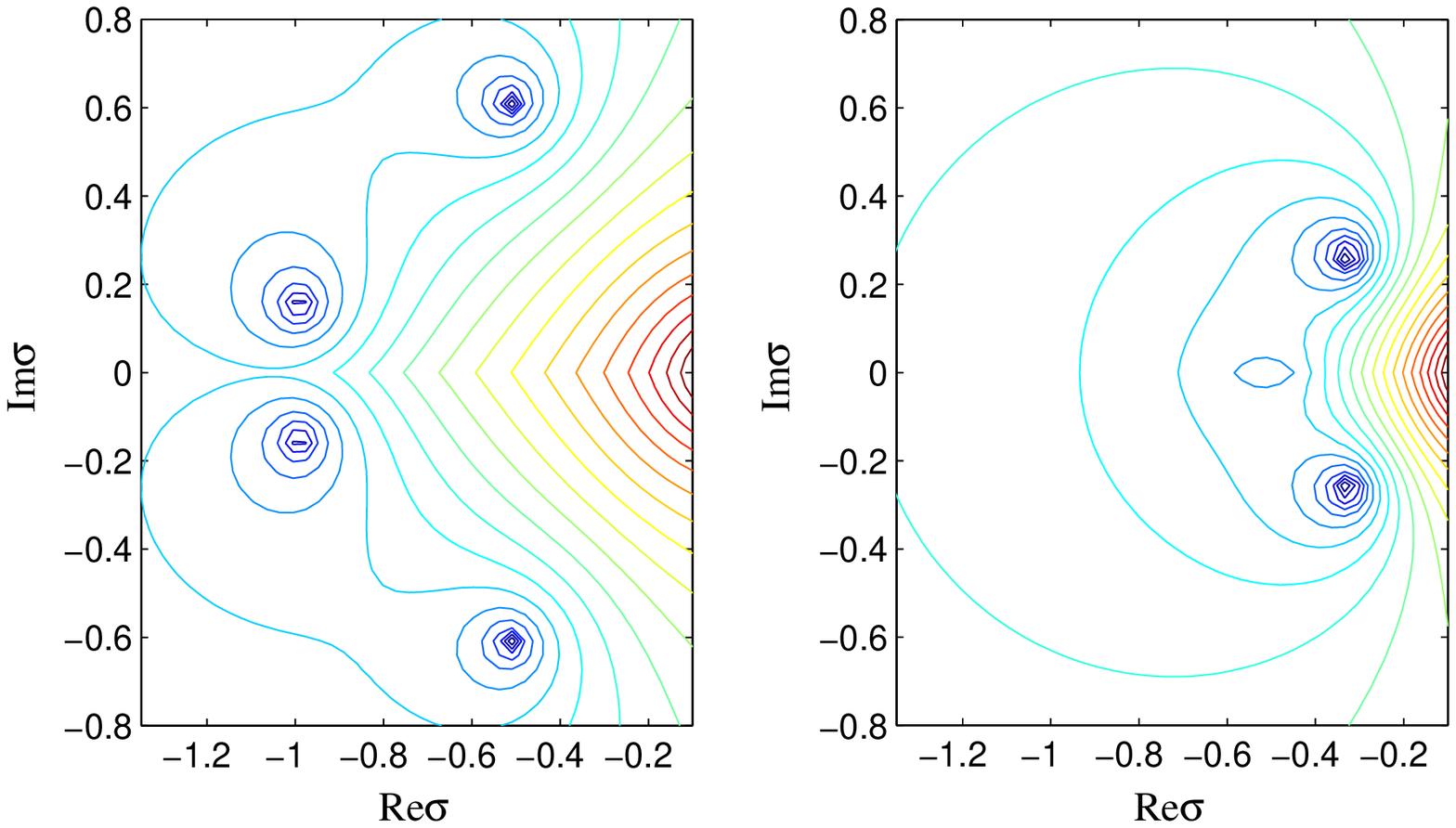}}
\caption{{\sc Zeros of $W_2(\sigma\rho_B;\sigma)$.} On the left we
have a plot of $\log|W_2(2\sigma;\sigma)|$ and on the
right one of $\log|W_2(4\sigma;\sigma)|$, demonstrating that a new
pair of zeros is created between $\rho_B = 4$ and $\rho_B = 2$.
The initial stages of this creation are perhaps evident in the
rightmost plot. \label{threeFigure6}}
\end{figure}

{\em Zeros for chosen parameter space.}
Dealing with $W_l(\sigma\rho_B;\sigma)$ as a function of $\sigma$
in the Heun scenario, we have denoted the zeros of this function by
$\sigma_{l,n} = \sigma_{l,n}(\rho_B)$. Over the range $\mathbb{S}$ 
of parameters mentioned above, the zeros $\sigma_{l,n}$ behave
qualitatively similar to the zeros $k_{l,n}/\rho_B$ of
$W_l(\sigma\rho_B)$, save for one key difference associated with
odd $l$. {\sc Fig}.~\ref{threeFigure2} displays contour plots showing
zero locations for $W_2(15\sigma;\sigma)$ and
$W_2(25\sigma;\sigma)$. Likewise, {\sc Fig}.~\ref{threeFigure3}
displays contour plots showing zero locations for
$W_3(15\sigma;\sigma)$ and $W_3(25\sigma;\sigma)$. These plots
exhibit the main features associated with the zeros of
$W_l(\sigma\rho_B;\sigma)$ over $\mathbb{S}$. To the
eye, apparent zero locations in these plots nearly match zero
locations (marked as crosses) for the corresponding
Bessel functions. However, as shown in {\sc
Fig}.~\ref{threeFigure3} and also noted in the figure caption, there is
a key difference associated with odd $l$. To appreciate the
difference, compare the zeros of $W_3(15\sigma)$ with those of
$W_3(15\sigma;\sigma)$. Notice that the single zero of
$W_3(15\sigma)$ lying on the negative Re$\sigma$ axis
corresponds to two zeros of $W_3(15\sigma;\sigma)$, one lying just
above and the other just below the negative Re$\sigma$ axis.
This is a generic feature belonging to all odd values of $l
\in \{1,3,5,7,9\}$ and $\rho_B \in [15,25]$ considered here. 
Therefore, for the Heun scenario and the parameter space 
$\mathbb{S}$ at hand there are an
even number of zeros whether $l$ is even or odd. Moreover, we
believe that this feature (of a pair of complex--conjugate zeros
lying close to the real axis and together corresponding to a
single Bessel zero) persists as $\rho_B$ gets
large, as evidenced by {\sc Fig}.~\ref{threeFigure4} and its caption.
More precisely, if $N_l = N_l(\rho_B)$ denotes the number of zeros
belonging to $W_l(\sigma\rho_B;\sigma)$, then for each $l \in 
\{0,\dots,10\}$
we observe that $N_l$ is constant on $[15,25]$ with $N_0 = 0$,
$N_1 = 2 = N_2$, $N_3 = 4 = N_4$, $N_5 = 6 = N_6$, $N_7 = 8 =
N_8$, and $N_9 = 10 = N_{10}$. Our conjecture is that each 
$N_l$ is also the same constant on $[15,\infty)$.
\begin{figure}[t]
\scalebox{0.70}{\includegraphics{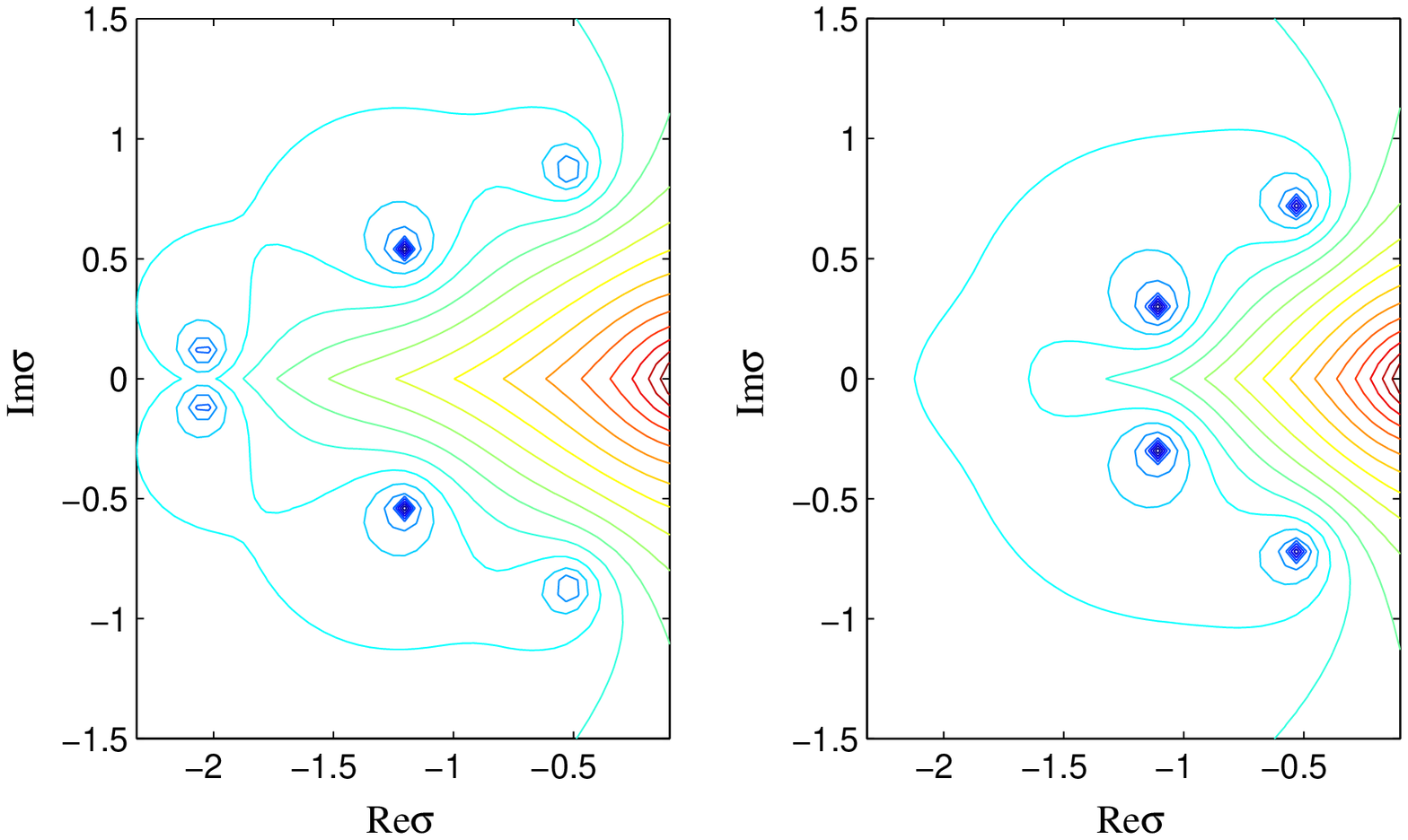}}
\caption{{\sc Zeros of $W_2(\sigma\rho_B;\sigma)$.} On the left we
have a plot of $\log|W_2(1.5\sigma;\sigma)|$ and on the
right one of $\log|W_2(1.75\sigma;\sigma)|$, demonstrating that a
second new pair of zeros is created between $\rho_B = 1.75$ and
$\rho_B = 1.5$. As $\rho_B \rightarrow 1^{+}$ more and more zero
pairs are created. \label{threeFigure7}}
\end{figure}

{\em Asymptotics of zeros.} We discuss two asymptotic
regimes for zero locations: one large $\rho_B$ at fixed $l$ and
the other large $l$ at fixed $\rho_B$. Turning to the first
regime, we conjecture that the zeros $\sigma_{l,n}(\rho_B)$
approach the Bessel zeros $k_{l,n}/\rho_B$ as $\rho_B$
becomes large. That is to say, the first term in the asymptotic
expansion (\ref{izerolocations}) is
\begin{equation}
\sigma_{l,n,1} = k_{l,n}.
\end{equation}
{\sc Fig}.~\ref{threeFigure5} is a typical piece of graphical evidence
indicating such behavior. Using the zero locations shown in this figure,
we have confirmed for each $n$ that $\big|\sigma_{10,n}(\rho_B) - 
k_{10,n}/\rho_B\big| = O(\rho_B^{-2})$ over $[15,25]$, 
in parallel with the first two terms in (\ref{izerolocations}).
\begin{figure}[t]
\scalebox{0.70}{\includegraphics{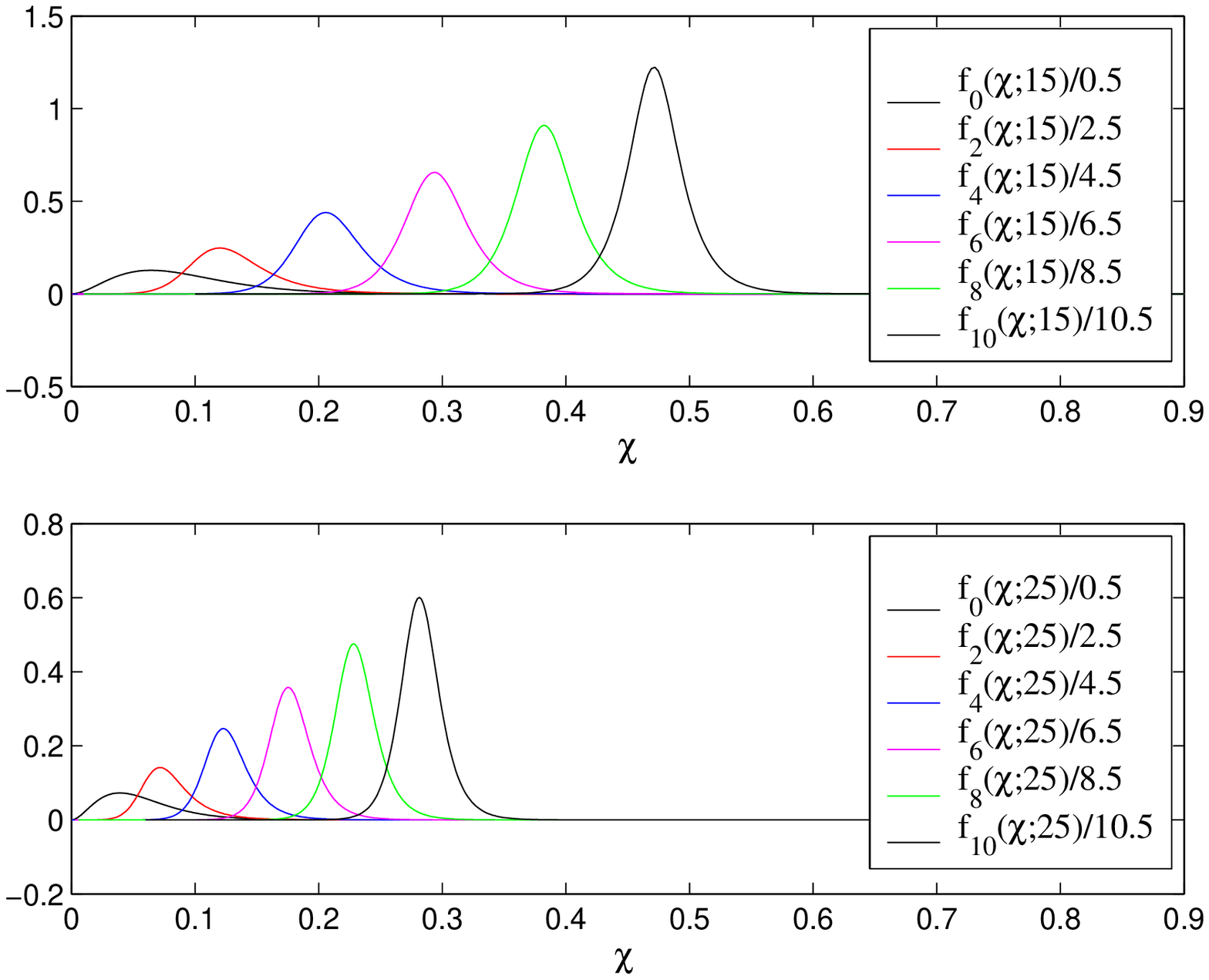}}
\caption{{\sc Even cut profiles scaled by order for $\rho_B = 15$
and $25$.} Here, for example, $f_0(\chi;15)/0.5$ is on the far left, 
and $f_{10}(\chi;15)/10.5$
on the far right.
\label{threeFigure8}}
\end{figure}

Now turning to the second asymptotic regime, we remark on the
order scaling of Heun zeros as $l$ becomes large. We have observed
that the scaled zeros $(l+1/2)^{-1}\sigma_{l,n}$ tend to
accumulate on the same fixed curve $\mathcal{C}_{\rho_B}$ as $l$
becomes large. For example, using the 20 scaled zeros
$\sigma_{20,n}(15)/20.5$ to hint at a candidate $\mathcal{C}_{15}$,
one finds that ---to the eye at least--- all scaled zeros for $l <
20$ lie on this curve. Of course $\mathcal{C}_{\rho_B}$ is of a
different shape than the dilated curve $\mathcal{C}/\rho_B$,
although choosing a still larger fixed $\rho_B$ value yields
better agreement between the two curves. That is to say, consistent 
with the first type of asymptotic behavior discussed, we have
$\rho_B\mathcal{C}_{\rho_B} \rightarrow \mathcal{C}$ holding 
pointwise as $\rho_B \rightarrow \infty$. We are not confident that 
this order scaling is robust for very large $l$.
\begin{figure}[t]
\scalebox{0.70}{\includegraphics{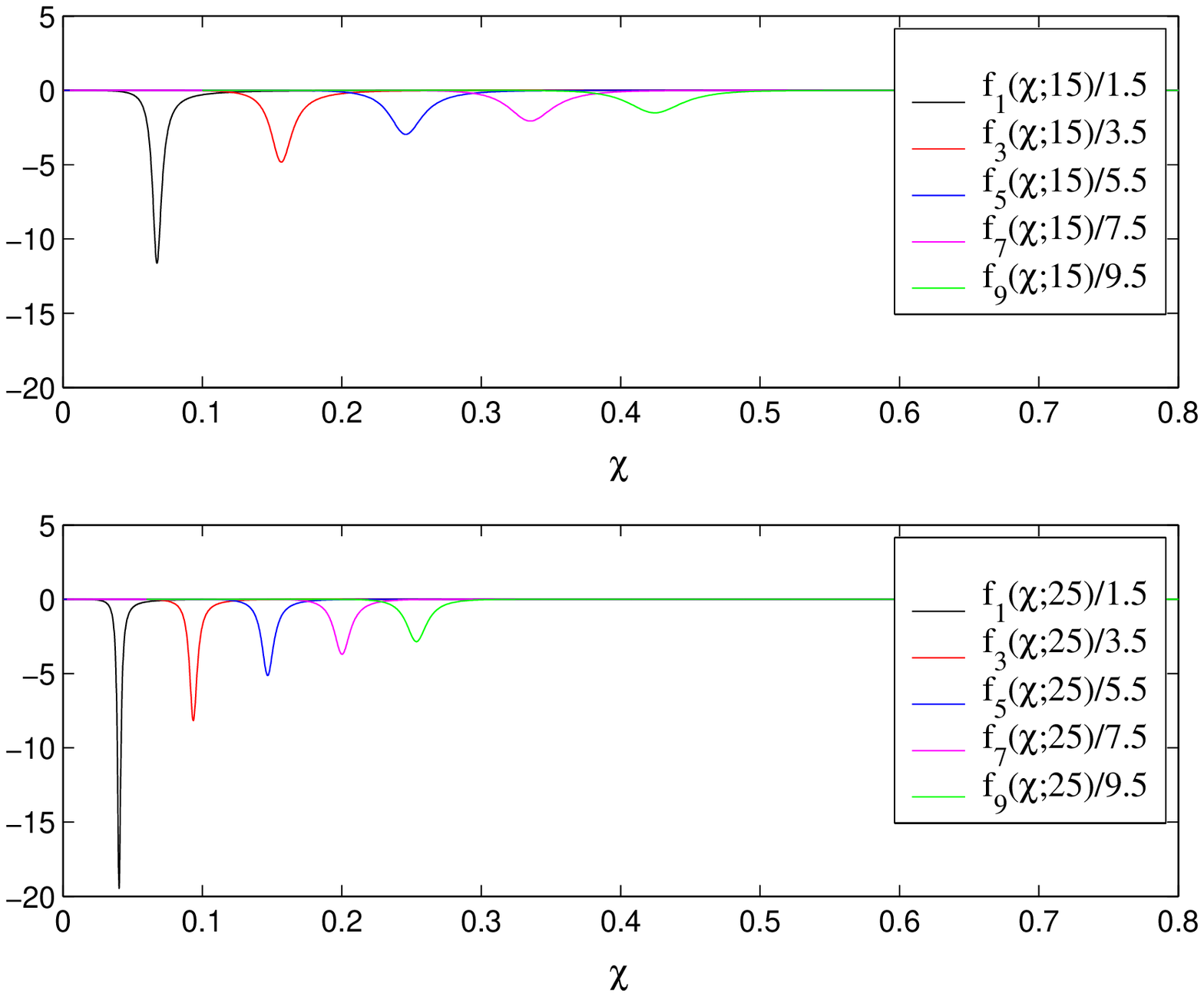}}
\caption{{\sc Odd cut profiles scaled by order for $\rho_B = 15$
and $25$.} Here, for example, $f_1(\chi;15)/1.5$ is on the far left, 
and $f_{9}(\chi;15)/9.5$
on the far right.
\label{threeFigure9}}
\end{figure}
\begin{figure}[t]
\scalebox{0.70}{\includegraphics{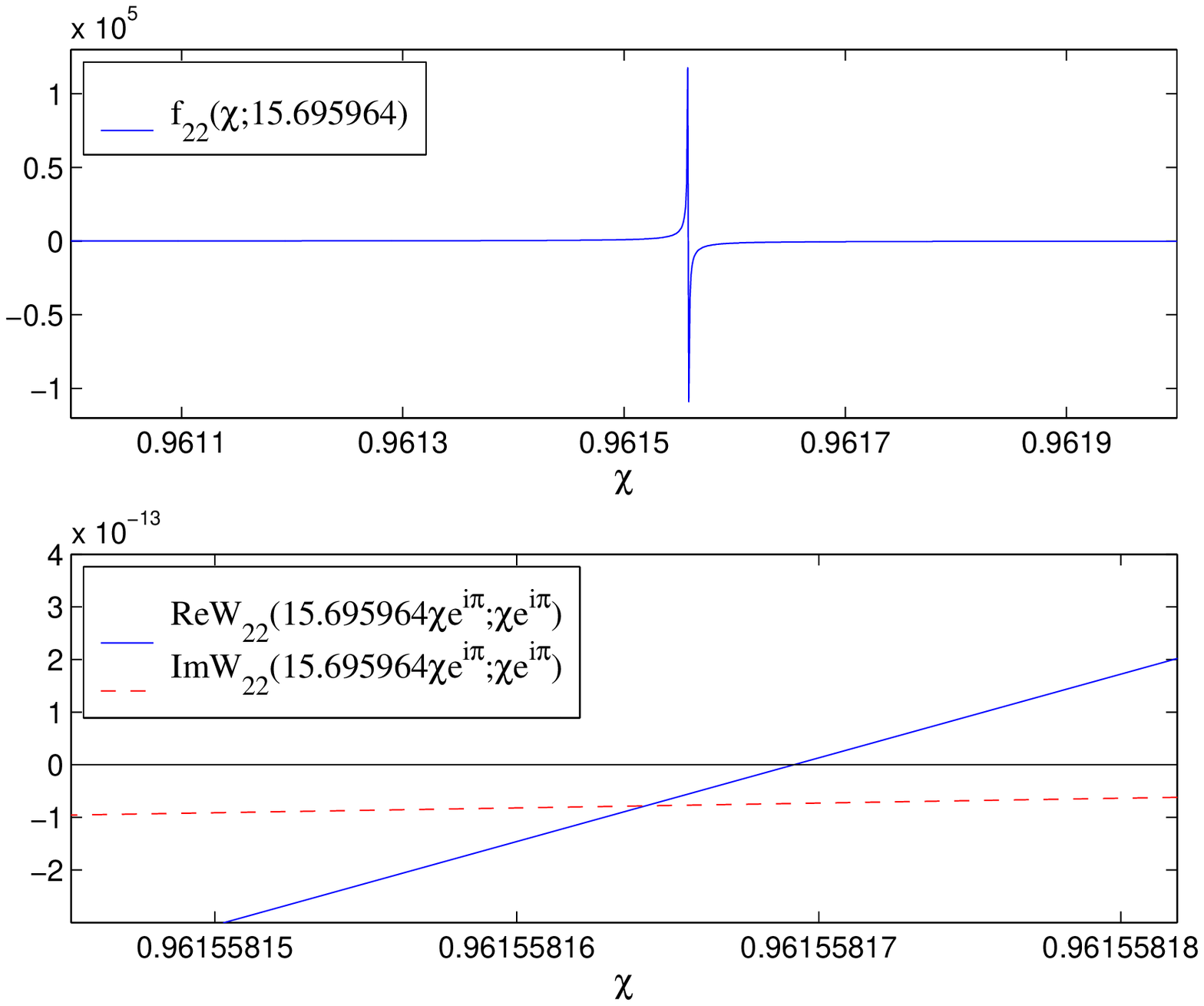}}
\caption{{\sc Blow--up in cut profile near a critical
value.} The value $15.695964$ is close to the critical value
$\rho_B^{c_1}$ for $l = 22$. See the text for more description.
\label{threeFigure10}}
\end{figure}

{\em Zero pair creation as $\rho_B$ approaches unity.}
We have claimed that over the $\rho_B$ interval $[15,25]$ Heun
zero behavior is similar to Bessel zero behavior, save for the
aforementioned curious feature concerning odd $l$. As mentioned,
we believe our claim remains true over $[15,\infty)$ for all low $l$ 
here of interest. However, as we now argue, the number $N_l$ of 
zeros for a given $l$ is not conserved as $\rho_B \rightarrow 1^+$. 
The behavior we have noticed is the following.\footnote{The Heun
zeros under consideration are analogous to the ``flatspace quasinormal
modes'' discussed in the introduction of \cite{NollertSchmidt}
by Nollert and Schmidt. Nevertheless, despite the infinite 
number of zeros as $\rho_B \rightarrow 1^{+}$, they are 
apparently not what physicists call ``quasinormal modes.''}

For each value $l \in \{0,\dots,10\}$ there is a critical value
$\rho_B^{c_1}$ of $\rho_B$ (less than $15$ of course) for which a
new pair of zeros is created on the negative Re$\sigma$ axis,
the branch cut. Below that there is yet another critical value
$\rho_B^{c_2}$ for which yet another new pair of zeros is created
on the negative Re$\sigma$ axis, and so on. Therefore, it would 
seem that as a function $N_l(\rho_B)$ is step--like and blows
up as $\rho_B\rightarrow 1^{+}$. Let us remark on the nature of
the zero lying on the branch cut for a critical value
$\rho_B^{c_j}$. As $\rho_B$ is increased past $\rho_B^{c_1}$, say,
two zeros appear to collide on the branch cut. However, they do
not merge into a double zero, rather they pass ``over and under
each other," with each zero remaining on its own analytic
neighborhood continued across the branch cut.

In {\sc Figs}.~\ref{threeFigure6} and \ref{threeFigure7} we document the
creation of the first and second new zero pairs for
$W_2(\sigma\rho_B;\sigma)$ associated with decreasing $\rho_B$
from $4$ down to $1.5$. In Table \ref{critical} we have listed 
approximations to the critical values $\rho_B^{c_1}$ for 
$l \in \{0,\dots,10\}$. 
Using $\sigma = \chi \exp (\mathrm{i}\pi)$ along the negative
Re$\sigma$ axis, the table also lists approximate values for the
location $\chi^{c_1}$ of each zero--pair creation. We have obtained
these numbers using the method discussed below in reference to
{\sc Fig}.~\ref{threeFigure10}. All of the 
approximate critical values in the table are well below our $\rho_B$ 
interval $[15,25]$; however, for higher $l$ values creation of a zero 
pair can occur in our interval. For instance, in what follows we 
determine that creation of the first new zero pair for $l = 22$ 
occurs for $\rho^{c_1}_B \simeq 15.70$ and $\chi^{c_1} \simeq 0.96$.
\begin{table}\begin{center}
\begin{tabular}{|c||c|c|c|c|c|c|c|c|c|c|c|}
\hline $l$ & 0 & 1 & 2 & 3& 4 & 5 & 6 & 7 & 8 & 9 & 10
\\
\hline $\rho^{c_1}_B$ 
                      & $1.49$
                      & $1.52$ 
                      & $2.52$ 
                      & $2.06$ 
                      & $3.81$ 
                      & $2.67$ 
                      & $5.11$ 
                      & $3.31$ 
                      & $6.44$ 
                      & $3.97$ 
                      & $7.75$ 
\\
\hline
$\chi^{c_1}$
                      & $0.63$
                      & $1.31$
                      & $0.79$
                      & $1.49$
                      & $0.85$
                      & $1.60$
                      & $0.89$
                      & $1.67$
                      & $0.91$
                      & $1.71$
                      & $0.92$
\\
\hline
\end{tabular}
\end{center}
%
%
%
\vskip 2mm
\caption{{\sc Approximate critical values of $\rho_B$}. Here 
we list rough values corresponding to the creation of the 
first new zero pair for $W_l(\sigma\rho_B;\sigma)$. For example, 
as $\rho_B$ is lowered from $1.5$ to $1.48$, the number $N_0$ of 
zeros for $W_0(\sigma\rho_B;\sigma)$ jumps from 0 to 2. We also 
list approximate values for the  location $-\chi^{c_1}$ of each 
zero--pair creation, again with $\pm 0.01$ error bounds.
\label{critical}}
\end{table}

\subsubsection{Parameter dependence of the cut 
profile}\label{sbsbSec:3.1.2}
Let us now discuss the cut profile $f_l(\chi;\rho_B)$ appearing 
in the representation (\ref{poleandcut}) of the {\sc fdrk}. We 
first remark on the behavior of the profile over the chosen 
parameter range $\mathbb{S}$, and then turn to exceptional behavior 
associated with critical parameter values lying outside 
$\mathbb{S}$.

{\em Behavior over the chosen parameter range.} For
$0 \leq l\leq 10$ and for $\rho_B = 15$ and $25$ we 
plot scaled even profiles 
in {\sc Fig}.~\ref{threeFigure8} and scaled odd profiles in 
{\sc Fig}.~\ref{threeFigure9}. The scaling allows us to view all 
profiles on the same plot. Notice that the order--scaling is
different for even and odd cases. As $\rho_B$ is increased
towards $25$, the other endpoint of our interval, all of these
profiles retain their shape; however, both their maximum value
(in absolute value) and the essential window of their support
vary. 

{\em Cauchy principal value.}
The chosen parameter space
$\mathbb{S}$  has been carefully tailored to avoid
the exceptional situation where a zero lies on the branch cut.
However, over our $\rho_B$ interval $[15,25]$, we shall of course
be interested in $l$ values higher than $10$, and on this interval 
zero pair creation is an issue for such $l$. A glance at the form
(\ref{twocutprofile}) of the cut profile $f_l(\chi;\rho_B)$ 
given earlier indicates that a negative real zero
$\sigma = \exp({\rm i}\pi)\chi$ of $W_l(\sigma\rho_B;\sigma)$ 
should give rise to a singular cut profile. 

For the aforementioned exceptional case $l = 22$ and 
$\rho_B^{c_1} \simeq 15.70$, we depict the profile blow--up in 
{\sc Fig}.~\ref{threeFigure10}. In the top plot we have the 
cut profile $f_{22}(\chi;15.695964)$, where $15.695964$ is 
approximately the critical value $\rho_B^{c_1}$ of $\rho_B$ 
corresponding to the creation of the first new zero pair for the 
function $W_{22}(\sigma\rho_B;\sigma)$. For $\rho_B$ values 
larger than $\rho_B^{c_1}$ the function has $22$ zeros, but as 
$\rho_B$ is lowered below $\rho_B^{c_1}$ a new pair of zeros 
appears from the branch cut. The lower plot depicts 
Re$W_{22}(15.695964\sigma;\sigma)$ (solid line) and
Im$W_{22}(15.695964\sigma;\sigma)$ (dotted line) as well as
their intersection point below the zero line. For $\rho_B =
15.695962$ this intersection point lies above zero, while for
$\rho_B = 15.695966$ it lies below zero. As this intersection
point appears to move smoothly with varying $\rho_B$, we
conjecture the existence of a zero on the branch cut for a
critical $\rho_B^{c_1}$ between $\rho_B = 15.695962$ and $\rho_B =
15.695966$ (actually we know it lies between $\rho_B = 15.695962$
and $\rho_B = 15.695964$). Our guess at the value, $\rho_B =
15.695964$, should be within $2\times 10^{-6}$ of the true
$\rho_B^{c_1}$. Furthermore, we note that for $\rho_B$ slightly
above the critical value the profile $f_{22}(\chi;\rho_B)$ is a
positive peak like one in {\sc Fig}.~\ref{threeFigure8}, but as
$\rho_B$ is lowered past $\rho_B^{c_1}$ the profile transitions to
a negative (and sharper) peak like one in {\sc
Fig}.~\ref{threeFigure9}. 

Despite the blow--up discussed in the last paragraph, we emphasize 
that along the Im$\sigma$ axis, the {\sc fdrk} 
$\tilomega_{22}({\rm i}y;\rho_B)$ itself changes smoothly as 
$\rho_B$ varys across $\rho_B^{c_1}$. Indeed, the pieces 
Re$\tilomega_{22}({\rm i}y;\rho_B)$ and
Im$\tilomega_{22}({\rm i}y;\rho_B)$ may be computed either via the 
representation (\ref{poleandcut}) or numerically via the methods
outlined in {\sc Section} \ref{sbSec:2.2}. 
Using the latter methods, we observe 
that both pieces vary smoothly as $\rho_B$ varys across the critical 
value $\rho_B^{c_1}$. We therefore offer the following
conjecture. Although the cut profile $f_{22}(\chi;\rho_B)$ is 
singular at a particular point $\chi_{c_1} \simeq 0.96$ when 
$\rho_B = \rho_B^{c_1} \simeq 15.70$, the corresponding integral 
contribution
\begin{equation}
-\frac{1}{\pi}\int^{\infty}_{0} \frac{f_{22}(\chi;\rho_B)}{{\rm i}y
+\chi}{\rm d}\chi
\end{equation}
to $\tilomega_{22}({\rm i}y;\rho_B)$ varys smoothly as $\rho_B$ 
varys across $\rho_B^{c_1}$. This would seem to indicate that  
$f_{22}(\chi;\rho_B^{c_1})/({\rm i}y +\chi)$ has a distributional 
interpretation, and we believe its integral to be defined in the 
sense of Cauchy Principal Value. The nearly antisymmetrical 
blow--up in the top plot of {\sc Fig.}~\ref{threeFigure10} is 
in accord with this conjecture.

\subsection{Numerical construction of the radiation 
kernel}\label{sbSec:3.2}
We now document our numerical construction of the representation
(\ref{poleandcut}) over the chosen parameter space $\mathbb{S}$,
also discussing in detail the accuracy of the construction.

\subsubsection{Construction for chosen parameter 
space}\label{sbsbSec:3.2.1}
To obtain the numerical kernel, we have used the one--component
path method described in {\sc Section}
\ref{sbsbSec:2.1.3} both to obtain pole
locations and strengths as well as the cut profile for certain
parameter choices. Let us first discuss our treatment of the 
poles.

{\em Construction of pole locations and strengths.}
Let a choice of $l \in \{0,\dots,10\}$ remain fixed 
throughout this paragraph. We choose nine Chebyshev points 
$\{1/\rho_B^k : k = 0,1,\dots, 8\}$ on the interval $[1/25,1/15]$. 
That is to say, the formula
\begin{equation}
  2\xi_B^k = 0.7+0.04+(0.7-0.04)\cos[\pi(2k+1)/(2n+2)]\, .
\end{equation}
determines the nine numbers $\xi^k_B = 1/\rho^k_B$. Our choice of 
nine Chebyshev points suffices for our purposes, although we
make no claim that nine is the optimal number of points. Next,
for each $k$ we have used the secant algorithm to find the zero set
$\{\sigma_{l,n}(\rho_B^k)\}$ of $W_l(\sigma\rho_B^k;\sigma)$. 
Then, at fixed $l$ and $n$ we interpolate each function 
$\sigma_{l,n}(\rho_B)$ by an eighth degree Chebyshev polynomial 
$T_{l,n}(1/\rho_B)$ in inverse $\rho_B$, so that this polynomial 
approximates $\sigma_{l,n}(\rho_B)$ on the interval $[15,25]$. 
On the same interval, we approximate the pole strengths 
$\alpha_{l,n}(\rho_B)$ by $T'_{l,n}(1/\rho_B)/\rho_B$, 
where here the prime $'$ denotes ${\rm d}/{\rm d}\xi_B$ 
differentiation.

{\em Construction of the cut profile.} Given any
small positive $\eta$, say $\simeq 10^{-12}$, we assume 
the existence of corresponding finite integration limits, 
$\chi_{\rm min}$ (which may or may not be $0$) and 
$\chi_{\rm max}$, such that the integral 
\begin{equation}
-\frac{1}{\pi}\int_{\chi_{\rm min}}^{\chi_{\rm max}}
\frac{f_{l}(\chi;\rho_B)}{{\rm i}y + \chi}{\rm d}\chi
\label{windowintegral}
\end{equation}
approximates the true value
\begin{equation}
-\frac{1}{\pi}\int_{0}^{\infty}
\frac{f_{l}(\chi;\rho_B)}{{\rm i}y + \chi}{\rm d}\chi
\label{trueintegralforcut}
\end{equation}
to within relative error $\eta$ uniformly for $y \in 
\mathbb{R}$. We stress that this is an assumption, although one 
apparently true for the analogous cut profiles stemming from 
integral order ($\nu = l+1/2 = n$) Bessel functions 
$W_{(n-1/2)}(\sigma\rho_B)$, as shown in the fourth section 
of \cite{AGH1}. In practice we have ``eyeballed'' the 
integration window $[\chi_{\rm min},\chi_{\rm max}]$, for 
example referring to {\sc Figure} \ref{oneFigure2} of 
{\sc Section} \ref{sbsbSec:1.4.2}, we have chosen 
$[0.0005, 1.125]$ for $l = 2$ and $\rho_B = 15$. The 
correctness of our guess will be confirmed when we later 
examine the accuracy of the kernel. Finally, to obtain the 
cut contribution to the value $\tilomega_l({\rm i}y;\rho_B)$ 
for a given $y$, we discretize the integral 
(\ref{windowintegral}) via the Simpson rule. For the profile 
in the aforementioned figure we have used 1024 subintervals. 
Values $f_{l}(\chi_j;\rho_B)$ belonging to nodes $\chi_j$ 
in the corresponding discrete sum are computed 
with the one--component path method. Since for our chosen
parameter range the considered profiles appear to be of 
definite sign, we expect that these sums are not plagued by
cancellation error.
\begin{figure}[t]
\scalebox{0.70}{\includegraphics{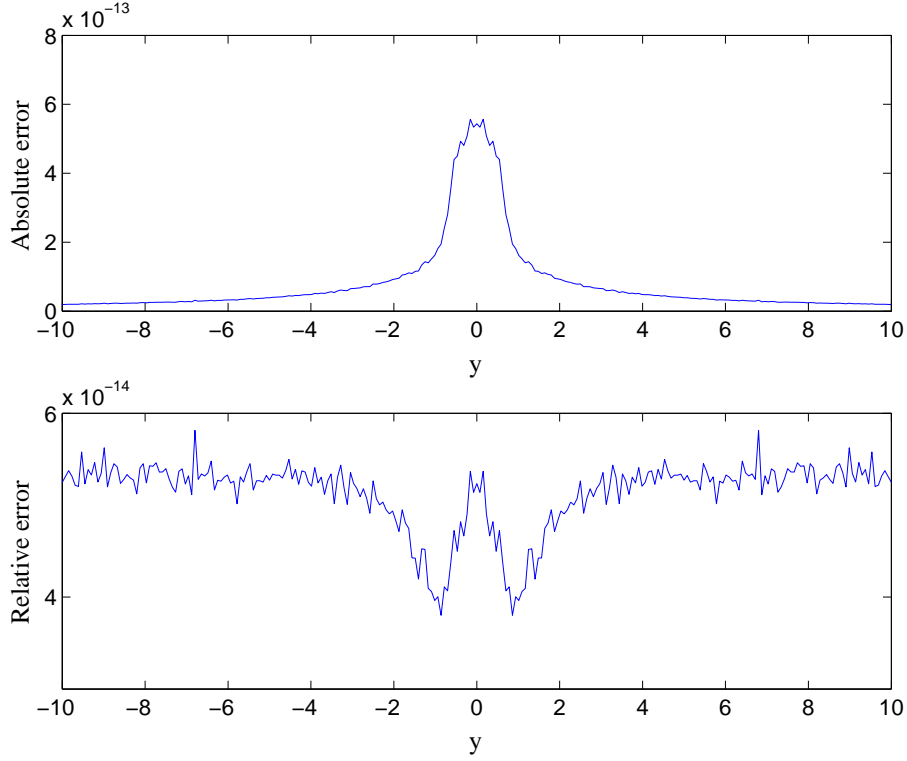}}
\caption{{\sc Error in Heun {\sc fdrk}
$\tilomega_{10}({\rm i}y;15) = w_{10}(15{\rm i}y;{\rm i}y)$.}
With $256$ subintervals of the $y$ direction, we plot the
absolute error $|\Delta w_{10}(15{\rm i}y;{\rm i}y)|$
and a relative error $|\Delta w_{10}(15{\rm i}y;{\rm i}y)|
/|w_{10}(15{\rm i}y;{\rm i}y)|$
described in the text.
\label{threeFigure11}}
\end{figure}

The foregoing construction of the cut contribution is applicable
for a fixed value of $\rho_B$, whereas our construction for poles
yielded locations and strengths over the whole $\rho_B$ 
interval $[15,25]$. In order to handle the cut contribution 
over the whole interval, we again introduce the nine Chebyshev 
points $\{1/\rho_B^k : k = 0,1,\dots, 8\}$, and ---for each 
$\chi_j$ integration node--- construct an eighth degree 
polynomial in $1/\rho_B$ which interpolates 
$f_l(15\chi_j/\rho_B;\rho_B)$. Notice that we are also scaling
the integration nodes $\chi_j$ associated with $\rho_B = 15$ 
[and determined by the choice of $\chi_{\rm min}$ and 
$\chi_{\rm max}$ as well as the number of subintervals chosen 
to evaluate the integral (\ref{windowintegral}) via Simpson's rule].
\begin{figure}[t]
\scalebox{0.70}{\includegraphics{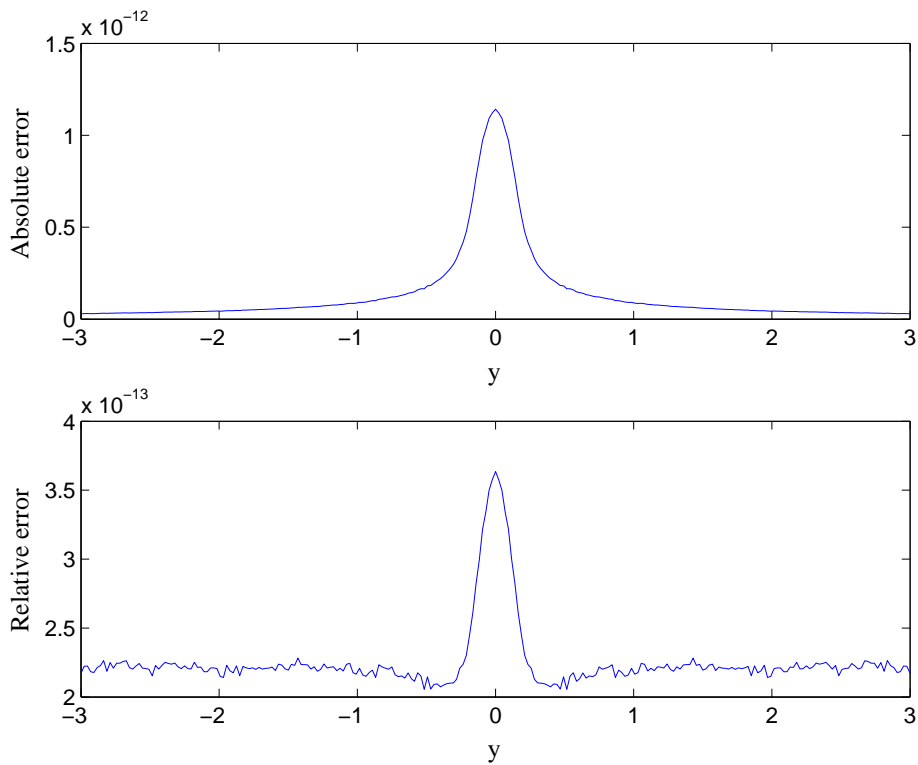}}
\caption{{\sc Error in Heun {\sc fdrk}
$\tilomega_{3}({\rm i}y;15) = w_{3}(15{\rm i}y;{\rm i}y)$.}
Again with $256$ subintervals of $y$, we plot the absolute error
$|\Delta w_{3}(15{\rm i}y;{\rm i}y)|$
and a relative error $|\Delta w_{3}(15{\rm i}y;{\rm i}y)|     
/|w_{3}(15{\rm i}y;{\rm i}y)|$
described in the text.
\label{threeFigure12}}
\end{figure}

\subsubsection{Accuracy of the construction}\label{sbsbSec:3.2.2} 
We check the accuracy of our numerical kernel in two ways,
and as one result provide compelling numerical evidence that
the {\sc fdrk} indeed admits the sum--of--poles representation 
(\ref{poleandcut}).

{\em Value of the kernel at the origin.} Building the
pole and cut contributions to the kernel as described, we may
compute a numerical value for $\tilomega_l(0;\rho_B)$ and check
it against the accurate series (\ref{originkernel}). We find 
that for {\em any} choice of $l$ and $\rho_B$ in $\mathbb{S}$
the numerical value for $\tilomega_l(0;\rho_B)$ has 
absolute error less than $10^{-11}$ (in fact on the order of 
$10^{-12}$ or better). We stress that this level of accuracy 
holds even when $1/\rho_B$ is not a Chebyshev node, in which 
case the pole and the cut contributions to the kernel are 
obtained via interpolation. 

{\em Direct check.}
For $1\leq l\leq 10$ we now have two independent numerical methods
for evaluating the Heun {\sc fdrk} $\tilomega_l({\rm i}y;\rho_B)$. The 
first is evaluation of the numerical kernel directly constructed via 
the representation (\ref{poleandcut}) as described in this subsection. The 
second is evaluation using path integration as described in 
{\sc Section}
\ref{sbSec:2.2}. As a final and perhaps most convincing accuracy 
check, we compare these two methods. In {\sc Fig}.~\ref{threeFigure11} 
we have such a comparison for 
$\tilomega_{10}({\rm i}y;15) = w_{10}(15{\rm i}y;{\rm i}y)$. With 
the two numerically obtained kernels we form an absolute error 
measure $|\Delta w_{10}(15{\rm i}y;{\rm i}y)|$ 
and a relative error measure 
$|\Delta w_{10}(15{\rm i}y;{\rm i}y)|/|w_{10}(15{\rm i}y;{\rm i}y)|$, 
where in forming the denominator of the relative error we happen to 
have used the directly constructed kernel. We plot these error 
measures in the figure. {\sc Fig}.~\ref{threeFigure12} depicts 
similar plots for $|\Delta w_{3}(15{\rm i}y;{\rm i}y)|$ and 
$|\Delta w_{10}(15{\rm i}y;{\rm i}y)|/|w_{3}(15{\rm i}y;{\rm i}y)|$.
Over all of $\mathbb{S}$, save for $l = 0$ cases, this check indicates
that we have relative and absolute errors better than $10^{-11}$.
As for $l = 0$, we believe the integration methods of 
{\sc Section} \ref{sbSec:2.2} to be less reliable than the directly 
constructed kernel. Indeed, $\tilomega_0(\mathrm{i}y;\rho_B)$ is 
quite concentrated around the origin, and very small values of $y$ 
negate the exponential error suppression built into the 
three--component integration method of {\sc Section} \ref{sbsbSec:2.2.2}.
In any case, for $l = 0$ the first accuracy check of 
$\tilomega_0(0;\rho_B)$ should be convincing in and of itself. 
Indeed, in the absence of a pole contribution to the kernel, one 
expects the largest error at the $y$--origin. Moreover, the largest
error is indeed concentrated near the origin in other small--$l$ 
plots such as those shown in {\sc Fig}.~\ref{threeFigure12}. Via
comparison with the series (\ref{originkernel}), we have found that 
our directly constructed numerical kernel yields a value for 
$\tilomega_0(0;15)$ with an absolute error better than 
$7.5\times10^{-13}$ and a relative error better than
$2.2\times10^{-11}$.

\subsection{Approximation of the kernel by rational 
functions}\label{sbSec:3.3}
Throughout this subsection we suppress
all $l$ and $\rho_B$ dependence (and we continue to
suppress $\kappa = 1 - \jmath^2$ spin dependence as always).

\subsubsection{Overview and results}\label{sbsbSec:3.3.1}
We have assumed that the {\sc fdrk} $\tilomega(\sigma)$
admits a (continuous and discrete) sum--of--poles 
representation (\ref{poleandcut}), 
and the numerical investigations outlined in {\sc Sections}
\ref{sbSec:3.1} and \ref{sbSec:3.2} (especially the direct check
in {\sc Section} \ref{sbsbSec:3.2.2}) indicate that 
this assumption is valid. This 
sum--of--poles representation is quite similar in form to the one
associated with integer--order Bessel functions and appropriate 
for wave propagation on a flat $2+1$ dimensional spacetime. 
As such, we believe that {\sc Lemmas} 3.4 and 3.5 in 
Ref.~\cite{AGH1} pertain to our representation 
(\ref{poleandcut}) of the Heun {\sc fdrk}, demonstrating that 
for Re$\sigma \geq 0$ we may accurately approximate the kernel 
by a rational function,
\begin{equation}
\frac{P(\sigma)}{Q(\sigma)} =
\sum_{n=1}^{d} \frac{\gamma_n}
                {\sigma-\beta_n}\, ,
\label{dpolesum}
\end{equation}
which is itself a sum of $d$ simple poles. That these lemmas are 
indeed pertinent is an assumption. The approximating pole 
locations $\beta_n$ and strengths $\gamma_n$ in (\ref{dpolesum}) 
will be computed numerically.

For our rational approximations $P(\sigma)$ and $Q(\sigma)$ 
are polynomials with ${\rm deg}(Q) = d = {\rm deg}(P)+1$. Both 
$P(\sigma)$ and $Q(\sigma)$ depend on $l$, $\rho_B$, and $\kappa$.
As mentioned, we suppress this dependence throughout. Our 
choice $P(\sigma)/Q(\sigma)$ will be tailored to ensure that 
the relative supremum error
\begin{equation}
{\rm sup}_{y\in\mathbb{R}}\frac{\big|\tilomega(x+{\rm
i}y)-P(x+{\rm i}y)/Q(x+{\rm i}y)\big|}{\big|\tilomega(x+{\rm
i}y)\big|}
\label{superror}
\end{equation}
is smaller than a prescribed tolerance $\varepsilon$. In 
practice, we set $\varepsilon$ to $10^{-6}$ or $10^{-10}$. 
Eq.~(\ref{oneestimates}) has motivated our interest in this 
error measure. We note that for the $\varepsilon$ considered, the
second error term in (\ref{oneerror2}) may be neglected so
long as we aim for slightly better than the desired tolerance
(for example, aiming for $7.5\times 10^{-11}$ rather than 
$10^{-10}$). As for the veracity of this statement, we offer the
studies carried out in {\sc Sections} \ref{sbsbSec:2.2.4} and 
\ref{sbsbSec:3.2.2} of the relative error 
$|\Delta\tilomega({\rm i}y)|/
|\tilomega({\rm i}y)|$ in our knowledge of the true kernel.

Throughout what 
follows, $x \geq 0$ is a fixed constant, and we have always set
$x = 0$ in our work; however, for generality we retain $x$ in the 
analysis. Our thinking here is that the kernel 
$\tilomega(\sigma)$ should be analytic in the righthalf 
$\sigma$--plane; hence $Q(\sigma)$ should have zeros only in 
the lefthalf $\sigma$--plane (for simplicity, let us assume that
$P(\sigma)$ and $Q(\sigma)$ share no common zeros). If we set 
$x = 0$, then the error above measures how well $P(\sigma)/Q(\sigma)$ 
approximates $\tilomega(\sigma)$ along the Im$\sigma$ axis,
precisely the contour of integration appearing in the inverse 
Laplace transform. 

To find the desired rational approximation to the kernel, we solve
a least--squares problem of the form
\begin{equation}  
       {\rm min}_{P,Q}
       \int^\infty_{-\infty}
       \left|
       \frac{P(x+{\rm i}y)}{Q(x+{\rm i}y)}
      -\tilomega(x + {\rm i}y)
       \right|^2
       {\rm d}y\, ,
\label{leastsqr}
\end{equation}
that is we minimize the integral over the space of
polynomials $P(\sigma)$ and $Q(\sigma)$ such that ${\rm deg}(Q) = 
d = {\rm deg}(P)+1$. After we have numerically solved this problem 
and found $P(\sigma)$ and $Q(\sigma)$, we compute the relative 
supremum error (\ref{superror}) via numerical evaluations on a fine mesh. 
If this error is larger than the set tolerance $\varepsilon$, then 
we increment $d$ and try again. The upcoming
{\sc Section} \ref{sbsbSec:3.3.2} 
describes the algorithm we have used to solve (\ref{leastsqr}). 
We admit that for us the algorithm is half ``black box.'' That is to 
say, although we shall describe it in some detail, we provide no hard 
analytical proof as to why the algorithm actually converges to a 
solution of the least--squares problem. Moreover, we do not 
discuss why minimization of this least--square error is 
associated with small relative supremum error. Nevertheless, we
can check the accuracy of the final output and are able to achieve 
the desired tolerance.
\begin{table}
\begin{center}
\begin{tabular}{|l|l||l|l|}
\hline
\multicolumn{2}{|c||}{$\varepsilon = 10^{-6}$} & 
\multicolumn{2}{c|}{$\varepsilon = 10^{-10}$}
\\
\hline
  $l$ & $d$ & $l$ & $d$ 
\\
\hline
       &    & 0      & 20 
\\
\hline
       &    & 1      & 14 
\\
\hline
       &    & 2      & 10 
\\
\hline
       &    & 3--5   & 9 
\\
\hline
       &    & 6--7   & 8  
\\
\hline
0      & 12 & 8--9   & 9  
\\
\hline
1      & 6  & 10--11 & 10 
\\
\hline
2--5   & 5  & 12--14 & 11 
\\
\hline
6--7   & 6  & 15--18 & 12 
\\
\hline
8--10  & 7  & 19--24 & 13 
\\
\hline
11--17 & 8  & 25--32 & 14 
\\
\hline
18--27 & 9  & 33--42 & 15 
\\
\hline
28--44 & 10 & 43--56 & 16 
\\
\hline
45--64 & 11 & 57--64 & 17
\\
\hline 
\end{tabular}
\end{center}
\vskip 2mm
\caption{{\sc Number $d$ of poles for given $\varepsilon$.} 
Number of poles needed to approximate the {\sc fdrk} for
$\rho_B = 15$ with relative 
supremum error $\varepsilon$. We find these numbers to be
valid for both $\jmath = 0,2$. However, for the case 
$\jmath = 2$ (gravitational radiation), one must assume 
$l \geq 2$. We have not yet examined the case $\jmath = 1$ 
(electromagnetic radiation).}
\label{threeTable2}
\end{table}
\begin{figure}[t]
\scalebox{0.70}{\includegraphics{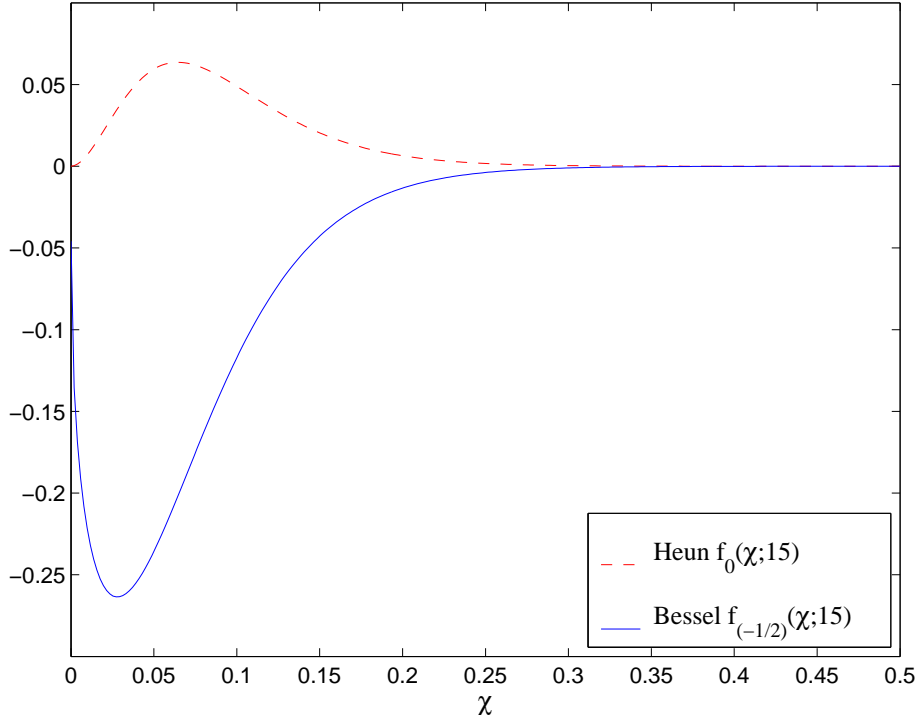}}
\caption{{\sc Cut profiles for lowest--order kernels.}
Here we plot the (positive) Heun profile $f_0(\chi;15)$ and
the (negative) Bessel profile $f_{(-1/2)}(\chi;15)$, both
for $\rho_B = 15$. To generate both we have used the
one--component path method described in {\sc Section}
\ref{sbsbSec:2.1.3}.
The Bessel {\sc fdrk} $\tilomega_{(-1/2)}(\sigma;15)$ associated
with $f_{(-1/2)}(\chi;15)$ is relevant for wave propagation in
flat 2+1 dimensional spacetime.
\label{threeFigure13}}
\end{figure}

Before turning to a description of the algorithm, let us first 
give a quick overview of our results. In Table \ref{threeTable2} 
we list the required number $d$ of poles $\beta_n$ in (\ref{dpolesum}) 
corresponding to $\rho_B = 15$ and 
two choices of the error tolerance $\varepsilon$. We 
find these pole numbers to suffice for both the $\jmath=0$ and $\jmath=2$ 
cases. Notice that $d$ and $l$ appear to grow at different rates, 
indicating increased performance in the large $l$ limit. However, 
as we here consider only the relatively small bandwidth 
$0\leq l\leq 64$, we make this observation with some skepticism. We 
have extended the $\varepsilon = 10^{-6}$ side of the table and
found that $d = 14$ is sufficient for $l = 256$. {\sc agh} have 
rigorously shown that for Bessel functions and flatspace wave 
propagation the number of approximating poles scales like
\begin{equation}
d \sim O\big(\log\nu\log(1/\varepsilon) + 
\log^2\nu+\nu^{-1}\log^2(1/\varepsilon)\big)
\end{equation}
as $\nu = l+1/2 \rightarrow \infty$ and $\varepsilon \rightarrow 
0^{+}$. While it is beyond our mathematical ability to analytically 
establish this or similar growth for Heun functions, the growth 
indicated in Table \ref{threeTable2} is not at odds with this
result.  Compression is also economical for low--order
kernels which are dominated by their continuous sectors. Table
\ref{threeTable2} lists $d = 20$ as sufficient for $\varepsilon
= 10^{-10}$ and $l = 0$. A numerical quadrature of the integral
appearing in (\ref{poleandcut}) would require considerably more
nodes to achieve this tolerance.

As mentioned above, we have
set $x=0$ in all cases (so that the least--squares problem is
associated with the Im$\sigma$ axis), and in particular we
have done so for $l=0$. This is perhaps remarkable in light of
the fact that {\sc agh} found it necessary to choose $x$ 
nonzero while approximating the zero--order Bessel {\sc fdrk},
a case for which the estimate (\ref{oneestimates}) is not
applicable. We stress that one need only deal with this 
zero--order Bessel kernel (a circular rather than spherical
boundary kernel) while considering wave propagation
on flat 2+1 dimensional spacetime. To emphasize the difference 
between the $l = 0$ Heun {\sc fdrk} and the $\nu = l + 1/2 = 0$ 
Bessel {\sc fdrk},
we plot their cut profiles in {\sc Fig}.~\ref{threeFigure13}.
Note that for both cases the kernel, as represented by 
(\ref{poleandcut}), is determined solely by its cut profile.
As documented by {\sc agh} in \cite{AGH1}, loss of 
differentiability at the origin for the Bessel profile precludes 
a rational approximation of the zero--order Bessel {\sc fdrk} by 
the methods described here, that is unless one chooses $x$ positive 
and non--zero. For a non--zero positive $x$ a more complicate 
estimate, valid only for finite time, must be used in place of 
(\ref{oneestimates}) \cite{AGH1}. {\sc Fig}.~\ref{threeFigure13} 
suggests that we may work with $x=0$ even for the lowest--order 
($l= 0$) Heun {\sc fdrk}.

\subsubsection{Linear least--squares problem}\label{sbsbSec:3.3.2} 
To bypass the 
nonlinearity of the problem (\ref{leastsqr}), we switch to the 
linear iterative procedure
\begin{equation}
       {\rm min}_{P^{(k+1)},Q^{(k+1)}}
       \int^\infty_{-\infty}
       \left|
       \frac{P^{(k+1)}(x+{\rm i}y)}{Q^{(k)}(x+{\rm i}y)}
      -\frac{Q^{(k+1)}(x+{\rm i}y)}{Q^{(k)}(x+{\rm i}y)}\,
       \tilomega(x+{\rm i}y)
       \right|^2
       {\rm d}y\, ,
\label{kthenergy}
\end{equation}
where given $Q^{(k)}(\sigma)$ the task is to find the minimizing 
polynomials $P^{(k+1)}(\sigma)$ and $Q^{(k+1)}(\sigma)$. 
To commence the iteration, we need a degree $d$ polynomial $Q^{(0)}(\sigma)$ 
as an initial guess, and we describe the construction of initial
polynomials in the upcoming {\sc Section} \ref{sbsbSec:3.3.3}. 

Assuming that we have $Q^{(k)}(\sigma)$ at our disposal, 
let us now discuss the solution to the least--squares problem 
(\ref{kthenergy}) at the $k$th level. To do so, we introduce 
the inner product
\begin{equation}
\left\langle h, g\right\rangle_k = 
\int_{-\infty}^{\infty} h(y)\bar{g}(y) m_k(y){\rm d}y\, ,
\end{equation}
where $h(y)$ and $g(y)$ are suitable functions and $m_k(y) 
= 1/|Q^{(k)}(x+{\rm i}y)|^2$ is the $k$th weight function. 
\vskip 4pt
\noindent
{\sc Claim:} 
Set $\mathcal{P}(y) =
P^{(k+1)}(x + {\rm i}y)$, $\mathcal{Q}(y) =
Q^{(k+1)}(x + {\rm i}y)$, $\mathcal{W}(y) =
\tilomega(x + {\rm i}y)$, and
\begin{equation}
h_n(y) =
\left\{
\begin{array}{ll}
(x+{\rm i}y)^{n/2-1}
&
{\rm for}\,\,n = 2,4,6,\dots , 2d
\\
&
\\
(x+{\rm i}y)^{(n-1)/2}\,\mathcal{W}(y)
&
{\rm for}\,\,n = 1,3,5,\dots ,2d-1.
\end{array}\right.
\end{equation}
Minimization of the integral (\ref{kthenergy}) is equivalent to
\begin{equation}
\left\langle 
-\mathcal{P} + \mathcal{Q}\mathcal{W}, 
h_n\right\rangle_k = 0
\label{orthog}
\end{equation}
for $n = 1,\cdots,2d$. 

To verify the claim, first introduce the expansions 
$\mathcal{P}(y) = \sum_{j=0}^{d-1}p_j (x+{\rm i}y)^j$ and 
$\mathcal{Q}(y) = \sum_{j=0}^d q_j (x+{\rm i}y)^j$, assuming 
$q_d = 1$. Next, consider a small perturbation $\delta p_m$ of the $m$th 
coefficient $p_m$.  Vanishment of the induced first--order variation 
of the integral (\ref{kthenergy}) may then be written as
\begin{equation}
\left\langle -\mathcal{P} + \mathcal{Q}\mathcal{W},
h_{2m+2}\right\rangle_k \overline{\delta p_m} + 
\overline{\left\langle
-\mathcal{P} + \mathcal{Q}\mathcal{W},
h_{2m+2}\right\rangle_k} \delta p_m = 0.
\end{equation}
Finally, exploit the freedom that $\delta p_m$ can be either real 
or imaginary, thereby reaching (\ref{orthog}) for even $n = 2m+2$.  
Eq.~(\ref{orthog}) for odd $n$ follows similarly upon introduction 
of $\delta q_m$ perturbations. $\Box$

With the claim (\ref{orthog}) in mind, we orthogonalize the 
$2d+1$ functions
\begin{eqnarray}
\lefteqn{\mathcal{W}(y),\,1,\, (x+{\rm i}y)\mathcal{W}(y),\,
(x+{\rm i}y),\,
\dots\, ,} & & \label{listoffs} \\
& & (x+{\rm i}y)^{d-1}\mathcal{W}(y),\,
(x+{\rm i}y)^{d-1},\, 
(x+{\rm i}y)^d\mathcal{W}(y)\, .
\nonumber
\end{eqnarray}
The result is a family $\{g_n(y):n=1,\dots,2d+1\}$ of
orthogonal functions, with the last member $g_{2d+1}(y)$ 
the sought for solution $-\mathcal{P}(y) 
+ \mathcal{Q}(y)\mathcal{W}(y)$
in (\ref{orthog}). We iteratively compute the functions in this 
family via the formulae
\begin{equation}
g_n(y) =
\left\{\begin{array}{ll}
\mathcal{W}(y)
& {\rm for}\,\,n = 1\\ 
& \\
1 - c_{21}\mathcal{W}(y) &
{\rm for}\,\,n = 2\\
& \\
(x+{\rm i}y)g_{n-2}(y)
-\sum_{j=1}^{{\rm min}\{4,n-1\}}c_{nj}g_{n-j}(y)
& 
\begin{array}{l}
{\rm for}\,\,n = 3,\\
\dots,2d+1\, ,
\end{array}
\end{array}\right.
\end{equation}
where the real constants $c_{nj}$ are given by
\begin{equation}
c_{nj} =
\left\{\begin{array}{ll} \displaystyle
\frac{\left\langle 1,\mathcal{W}\right\rangle_k}{\left\langle 
\mathcal{W},\mathcal{W}
\right\rangle_k} & {\rm for}\,\, 
n = 2\,\,{\rm and}\,\,j=1 \\
& \\
\displaystyle
\frac{\left\langle (\sigma g_{n-2}),g_{n-j}\right\rangle_k}{\left\langle
g_{n-j},g_{n-j}\right\rangle_k} &
\begin{array}{l}
{\rm for}\,\, n = 3,\dots,2d+1\\
{\rm and}\,\,j = 1,\dots,{\rm min}\{4,n-1\}\, .
\end{array}
\end{array}\right.
\end{equation}
Here we use the notation $(\sigma g_{n-2})(y)$ for the function 
$(x+{\rm i}y)g_{n-2}(y)$. 

A few key observations confirm the correctness of these 
formulae. As suggested in {\sc Figure} \ref{twoFigure15}
of {\sc Section} \ref{sbsbSec:2.2.4},
the profiles Re$\tilomega({\rm i}y)$ and 
Im$\tilomega({\rm i}y)$ are of even and odd 
$y$--parity respectively, and for fixed $x \geq 0$ the 
profiles Re$\tilomega(x+{\rm i}y)$ and 
Im$\tilomega(x+{\rm i}y)$ also have the same 
$y$--parity. Therefore, along the path of integration 
used to define the inner product, $\mathcal{W}(y)$ has 
the form $\mathcal{W}^{+}(y) + {\rm i} 
\mathcal{W}^{-}(y)$, 
where $\mathcal{W}^+$ is even and 
$\mathcal{W}^-$ is odd. 
Since $x+{\rm i}y$ is also of 
this form, it follows that all of the functions 
appearing in the list (\ref{listoffs}) are as well. Now, 
the integration measure $m_k(y)$ appearing in the inner 
product $\langle\, ,\rangle_k$ is of even $y$--parity, 
showing that $\langle\, ,\rangle_k$ is real--valued on the 
space (\ref{listoffs}). Whence the $c_{nj}$ are purely real. 
To see that $c_{nj} = 0$ for $j > 4$, first suppose that 
we have carried out the orthogonalization up to the 
function $g_{n-2}(y)$. In other words, we have $g_{n-2}(y)$ 
orthogonal to $g_{n-3}(y)$, $g_{n-4}(y)$, $\dots, g_1(y)$. 
Owing to the pattern in the list (\ref{listoffs}), 
multiplication of $g_{n-2}(y)$ by $x+{\rm i}y$ yields 
a new function $(\sigma g_{n-2})(y)$ which is not in the 
span of $g_{n-2}(y)$,$ g_{n-3}(y)$, $\dots, g_1(y)$. 
Moreover, $(\sigma g_{n-2})(y)$ is automatically 
orthogonal to $g_{n-j}(y)$ for $j>4$. Indeed, consider 
the identity 
\begin{equation}
(x+{\rm i}y)g_{n-2}(y) 
\overline{g}_{n-j}(y) = 
g_{n-2}(y) 
\overline{[2x g_{n-j}(y)-(x+{\rm i}y)g_{n-j}(y)]}
\end{equation}
which implies that
\begin{equation}
\big\langle (\sigma g_{n-2}),g_{n-j}\big\rangle_k
=  2x\big\langle g_{n-2},g_{n-j}\big\rangle_k
-\big\langle g_{n-2},(\sigma g_{n-j})\big\rangle_k\, .
\end{equation}
Again owing to the pattern in the list
(\ref{listoffs}), the function $(\sigma g_{n-j})(y) 
= (x+{\rm i}y)g_{n-j}(y)$ is in the span of $g_{n+2-j}(y)$,
$g_{n+1-j}(y)$, $\dots, g_1(y)$. Whence for $j>4$ we see 
that both terms on the {\sc rhs} of the last equation 
vanish, giving $c_{nj} = 0$.

\subsubsection{Numerical least--squares problem}\label{sbsbSec:3.3.3}
We sketch how 
we have handled three issues which arise in numerically 
implementing the least--squares solve.

{\em Numerical quadrature.}
To handle the integration necessary to solve (\ref{kthenergy})
via the described orthogonalization, we may change variables 
and write
\begin{equation}
\int^{\infty}_{-\infty} h(y){\rm d}y = \int^{\pi/2}_{-\pi/2}
h(\tan\theta) \sec^2\theta{\rm d}\theta\, ,
\label{quadra} 
\end{equation}
where $h(y)$ represents any relevant integrand. This integral is
numerically approximated as
\begin{equation}
\sum_{i=0}^{i_\mathrm{max}}\lambda_i h(\tan\theta_i)\sec^2\theta_i\, ,
\end{equation} 
where the $\theta_i$ and $\lambda_i$ are appropriate quadrature nodes and
weights.\footnote{Do not confuse the quadrature weights $\lambda_i$ 
with the $k$th weight function $m_k(y)$ defining $\langle\,,\rangle_k$. 
In (\ref{quadra}) $m_k(y)$ has been swept into $h(y)$.}Due to 
the sum--of--poles form of the {\sc fdrk}, all 
encountered integrands $h(\tan\theta)$ are periodic on 
$[-\pi/2,\pi/2]$. Moreover, all integrands are infinitely continuously
differentiable, save possibly at $\theta = 0$. By analogy with
integer--order Bessel functions, we expect regularity at $\theta = 0$
of order $2l+1$ \cite{AGH1}. Whence for high $l$ the trapezoid rule is 
effective and highly convergent. For low $l$ the profiles 
Re$\tilomega({\rm i}y)$ and Im$\tilomega({\rm i}y)$ are concentrated 
around and less regular at the $y$--origin. Therefore, for the first 
few $l$ we employ an adaptive trapezoid quadrature rule,
introducing more nodes around the origin.

{\em Evaluation of $P(\sigma)$, $Q(\sigma)$, 
and their derivatives.} Assume that we have produced a set 
of real constants $\{c_{nj}: n = 2,\dots,2d+1;\, 1\leq j \leq 4\}$ 
belonging to the $k$th linear least--squares problem 
(\ref{kthenergy}) described in {\sc Section}
\ref{sbsbSec:3.3.2}. Further, 
assume that the solution polynomials $P^{(k+1)}(\sigma)$ and 
$Q^{(k+1)}(\sigma)$ are (close enough to) the polynomials 
$P(\sigma)$ and $Q(\sigma)$ solving (\ref{leastsqr}). In other
words, assume that the iteration in $k$ has converged in some 
sense. In order to express the final approximating rational 
function as a pole sum (\ref{dpolesum}), 
we must first know how to evaluate these polynomials and their
derivatives at any $\sigma$ using only the $c_{nj}$ 
constants.\footnote{At each level in the iteration we 
must perform such an evaluation for 
$Q^{(k)}({\rm i}y)$, thereby determining $m_{k}(y)$ 
and in turn the $k$th inner product. The relevant $c_{nj}$ come from 
the previous level.} This may be done via the following algorithm.
Let $A_{1} = 0$, $B_{1} = 1$, $A_{2} = -1$, 
$B_{2} = - c_{21}$, $C_{1} = 0$, $D_{1} = 0$, $C_{2} = 0$,
and $D_{2} = 0$. Next, for $n = 3$ until $2d+1$, set
\begin{align}
& A_n  =  \sigma A_{n-2} - \sum_{j=1}^4
                           c_{nj} A_{n-j}\, ,
& C_n =  A_{n-2} + \sigma C_{n-2}
                  - \sum_{j=1}^4
                          c_{nj} C_{n-j}\, ,
\\
& B_n =  \sigma B_{n-2} - \sum_{j=1}^4
                           c_{nj} B_{n-j}\, ,
& D_n =  B_{n-2} + \sigma D_{n-2} 
                  - \sum_{j=1}^4
                          c_{nj} D_{n-j}\, .
\nonumber
\end{align}
Then $P(\sigma) = A_{2d+1}$,  $Q(\sigma) = B_{2d+1}$, 
$P'(\sigma) = C_{2d+1}$, $Q'(\sigma) = D_{2d+1}$. With the
ability to evaluate $Q(\sigma)$ and $Q'(\sigma)$ in hand, 
we may use Newton's method to obtain the zeros $\beta_n$ of 
$Q(\sigma)$ which are also the pole locations appearing in
(\ref{dpolesum}). The corresponding pole strengths are
then $\gamma_n = P(\beta_n)/Q'(\beta_n)$.

{\em Construction of the initial guess 
$Q^{(0)}(\sigma)$.} Recall that our sum--of--poles representation
of the {\sc fdrk} features both a discrete sector and a continuous 
sector. With this in mind, we consider the following two cases.
\begin{enumerate}
\item[(i)]{$d \geq N_l$. The number of approximating poles is equal 
to or greater than the number of actual poles from the discrete sector. 
Therefore, one approximating pole should correspond to each 
discrete--sector pole. Unless $d = N_l$, after each discrete--sector
pole has been thusly represented, some approximating poles 
will be ``left over.''  These should be placed on the negative 
Re$\sigma$ axis in order to mimic poles from the continuous sector 
(cut contribution). In case (i) we aim to naively mimic the
actual pole locations found in the exact sum--of--poles 
representation.}
\item[(ii)]{$d < N_l$. The number of approximating poles is
strictly less than number actual poles in the discrete sector,
and we do not attempt to naively mimic the pole locations found
in the exact sum--of--poles representation.} 
\end{enumerate}
For case (i) we set $Q^{(0)}(\sigma) = 
(\sigma - \varsigma_1)
(\sigma - \varsigma_2) \cdots 
(\sigma - \varsigma_d)$. 
Then for all $i\leq N_l$ we let $\varsigma_i = 
15\sigma_{l,i}(15)/\rho_B$, 
that is we set $\varsigma_i$ equal to the (numerically obtained) 
$i$th zero of $W_l(15\sigma;\sigma)$ times the ratio $15/\rho_B$
(we are also suppressing the $l$ which could be on the 
$\varsigma_i$).
If there are any, the remaining $d-N_l$ locations $\varsigma_i$ 
are chosen as follows. For each relevant $l$ (in practice for 
$l\leq 10$) we choose a fixed interval 
$[a,b]\subset\mathbb{R}_{\leq 0}$ 
lying on the negative Re$\sigma$ axis. This interval is 
``eyeballed'' to ensure that $[-b,-a]$ includes the essential 
support of the profile function $f_l(\chi;15)$. We then send 
$a\mapsto 15a/\rho_B$ and $b\mapsto 15b/\rho_B$. If $d = N_l+1$, 
we let $\varsigma_d = (a+b)/2$, and otherwise $\varsigma_i = 
a + (i-N_l-1)(b-a)/(d-N_l-1)$ with $i$ running from $N_l+1$ to 
$d$. 

For case (ii) we set $z = \sigma\rho_B$ and then
use the Bessel continued fraction expansion 
(\ref{continuedfraction}). As mentioned there, Lenz's method may be 
used to evaluate this continued fraction. This method is 
described in the appendix of Ref.~\cite{Thompson}. Via slight
modification it may also be used to return a partial denominator, 
which we use for the value $Q^{(0)}({\rm i}y)$. Indeed, even the 
most basic algorithm (see Ref.~\cite{Wall}) to 
compute continued fractions yields a partial denominator and 
may be used in this fashion.
%
%
\section{Evolution system and implementation 
of {\sc robc}}\label{Sec:4}
The goal of this section is to incorporate {\sc robc} into the 
MacCormack algorithm, the finite--difference scheme we have 
used to simulate wave propagation on Schwarzschild blackholes.  
The first subsection describes the spacetime foliation and associated 
first--order system of evolution equations on which our numerical 
work is based. The second reviews the MacCormack algorithm in the 
context of this system of {\sc pde}, and it also addresses the 
issue of inner boundary conditions at the horizon $H$. The third 
subsection covers implementation of {\sc robc}. The MacCormack 
algorithm is a simple, second--order--accurate, 
predictor--corrector scheme, and we show how {\sc robc} fit 
nicely within the framework of the interior prediction and 
correction. In this final subsection we also 
touch upon some memory issues relevant to our implementation.

\subsection{Spacetime foliation and evolution 
system}\label{sbSec:4.1}
\subsubsection{Chosen foliation of spacetime into spacelike 
slices}\label{sbsbSec:4.1.1}
Our numerical simulations of wave propagation on the Schwarzschild 
geometry feature physical coordinates. Moreover, for the time 
coordinate $t$ and its associated foliation of the spacetime into 
spacelike slices $\Sigma$ we have chosen one closely related 
to ingoing Eddington--Finkelstein coordinates. The letter 
$\Sigma$ is used both to denote the spacetime foliation and a 
generic slice of the foliation. There are several motivations for 
switching from the time $T$ (introduced in {\sc Section}
\ref{sbsbSec:1.1.1}) to $t$ 
(defined in a moment). Besides removing the coordinate degeneracy 
present in the diagonal line--element (\ref{dlesselement}), use of 
the time coordinate $t$ also makes the inner boundary conditions at 
$H$ particularly easy to set. Although we need not say more to 
motivate the upcoming discussion, further explanation may be 
found in {\sc Appendix} C of \cite{LauMathDiss}.

Let us introduce the ingoing 
Eddington--Finkelstein coordinate system, here in terms of
physical rather than dimensionless coordinates. We first pass to
physical coordinates by multiplying (\ref{dlesselement}) by
an overall factor of $4\mathrm{m}^2$ and subsequently sending
$4\mathrm{m}^2\mathrm{d}s^2\mapsto\mathrm{d}s^2$.
Next, we introduce the physical tortoise coordinate 
\begin{equation}
r_* = r + 2\mathrm{m}\log\big[r/(2\mathrm{m})-1\big]\, ,
\end{equation}
thereby reaching
\begin{equation}
\dee s^2 = F (-\dee T^2 + \dee r_*^2)
             + r^2 \dee \theta^2
             + r^2\sin^2\theta \dee \phi^2\, ,
\end{equation}
where again $F(r) = 1-2\mathrm{m}/r$.
Using the physical {\em advanced time} $v
= 2\mathrm{m}\nu = T + r_{*}$, we get 
the line--element in advanced Eddington--Finkelstein form,
\begin{equation}
\dee s^2 = - F \dee v^2
           + 2\dee v\dee r
             + r^2 \dee \theta^2
             + r^2\sin^2\theta \dee \phi^2\, .
\label{fourEFelementin}
\end{equation}
Like the retarded time discussed in {\sc Section} \ref{sbsbSec:1.1.1}, 
the advanced time $v$ also labels characteristic surfaces, but ones 
which are ingoing (cones which open up towards the past). From a
numerical standpoint, we do not want to work directly with a null
coordinate like $v$, rather a time coordinate $T_v$ obeying 
$v = T_v + r$. Since by definition
\begin{equation}
v = (T + r_* - r) + r\, ,
\end{equation}
the desired coordinate is
\begin{equation}
T_v = T + r_* - r = T + 2\mathrm{m}
                    \log\big[r/(2\mathrm{m})-1\big]\, ,
\label{EFtime}
\end{equation}
and we refer to $T_v = v - r$ as {\em ingoing
Eddington--Finkelstein time}, although as indicated above
it is the combination $T_v + r$ which labels ingoing
characteristics. Likewise, we could introduce a time 
$T_u = u + r$ based on the physical retarded time 
$u = 2\mathrm{m}\mu = T - r_*$, but such a time variable would 
be less suitable for numerical work (see {\sc Appendix} C of
\cite{LauMathDiss}). Let lowercase $t$ represent 
ingoing  Eddington--Finkelstein time shifted by a constant 
as follows:
\begin{equation}
t  =  T_v - 2\mathrm{m}\log\big[r_B/(2\mathrm{m})-1\big]
= T + 2\mathrm{m}\log
\left[\frac{r-2\mathrm{m}}{r_B-2\mathrm{m}}\right]\, . 
\label{inEFtime}
\end{equation}
The family of level--$t$ slices is the same as
the family of level--$T_v$ slices, and the combination 
$t+r$ also labels ingoing characteristics. However,
the $t$ coordinate has been tailored to ensure that 
$t = T$ {\em at the outer boundary} $r = r_B$. 

By construction
\begin{equation}
\dee v = \dee T_v + \dee r = \dee t + \dee r\, ,
\end{equation}
and using this differential relationship, we rewrite
(\ref{fourEFelementin}) as follows:
\begin{align}
\dee s^2 & =  - F \dee t^2
           + 2(1-F)\dee t\dee r
           + (2-F)\dee r^2
             + r^2 \dee \theta^2
             + r^2\sin^2\theta \dee \phi^2
\label{fourEFelementtee} \\
& =  -\left[\frac{1}{2-F}\right]\dee t^2
+ (2-F)\left[\dee r
+ \frac{(1-F)}{2-F}\dee t\right]^2
+ r^2 \dee \theta^2
             + r^2\sin^2\theta \dee \phi^2\, .
\nonumber
\end{align}
The last line can be written
\begin{equation}
\mathrm{d}s^2 = -N^2\mathrm{d}t^2 
           +  M^2(\mathrm{d}r + V^r\mathrm{d}t)^2
           +  r^2 \mathrm{d}\theta^2 
           + r^2\sin^2\theta \mathrm{d}\phi^2\, ,
\label{fourEFelement}
\end{equation}
where the {\em temporal lapse}, {\em radial lapse}, and 
{\em shift vector} are the following:
\begin{align}
N(r) & =  (1+2\mathrm{m}/r)^{-1/2}\, ,
\label{EFlapseandshift} \\
M(r) & =  (1+2\mathrm{m}/r)^{1/2}\, ,
\nonumber \\
V^r(r) & =  2\mathrm{m}/(2\mathrm{m}+r) \, .
\nonumber
\end{align}
In \cite{York} York discusses the geometric 
interpretations of such variables in a general setting. For 
now, the main point is that the line--element 
(\ref{fourEFelement}) is nondegenerate at the horizon $H$, 
since $N$, $M$, and $V^r$ are all well--behaved at $r = 2\mathrm{m}$. 

\subsubsection{Wave equation with respect to chosen 
foliation}\label{sbsbSec:4.1.2}
Let us derive the form of the wave equation in the 
Eddington--Finkelstein system $(t,r,\theta,\phi)$ of
coordinates. This can be achieved via coordinate 
transformation at the level of {\sc pde}, or by constructing
the d'Alembertian associated with (\ref{fourEFelement}). 
Choosing the first method, we start by rewriting 
(\ref{maineq2}) in terms of physical coordinates, in 
order to reach
\begin{equation}
\left(1-\frac{2\mathrm{m}}{r}\right)^{-1}
       \frac{\partial^2 \psi_l}{\partial T^2}
      -\frac{1}{r^2}\frac{\partial}{\partial r}
       \left[r^2\left(1-\frac{2\mathrm{m}}{r}\right)
       \frac{\partial\psi_l}{\partial r}
       \right]
      + \frac{l(l+1)\psi_l}{r^2}
      - \frac{2\mathrm{m}\jmath^2\psi_l}{r^3}
      = 0\, .
\label{physmaineq2}
\end{equation}
In performing the near trivial coordinate transformation
$(\tau,\rho) = (\tau(T),\rho(r))$, we have retained the 
symbol $\psi_l$ to denote the new function satisfying the 
slightly new form (\ref{physmaineq2}) of the {\sc pde}. Next, we 
introduce the function $U(t,r) = \psi_l(T(t),r)$, suppressing
from here on out the subscript $l$ which should be on $U$. Then, 
upon carrying out the coordinate transformation $(T,r) = (T(t),r)$ on 
(\ref{physmaineq2}), we find
\begin{eqnarray}
\lefteqn{\left(1+\frac{2\mathrm{m}}{r}\right)
\frac{\partial^2U}{\partial t^2} -
\frac{4\mathrm{m}}{r}
\frac{\partial^2U}{\partial t\partial r}-
\left(1-\frac{2\mathrm{m}}{r}\right)
\frac{\partial^2U}{\partial r^2}} 
& & \label{waveeqnEFin}\\
& & -
\frac{2\mathrm{m}}{r^2}
\frac{\partial U}{\partial t} +
\frac{2(\mathrm{m-r)}}{r^2}
\frac{\partial U}{\partial r}
+\frac{l(l+1)U}{r^2} -
\frac{2\mathrm{m}\jmath^2U}{r^3}
= 0
\nonumber
\end{eqnarray}
for the new form of the wave equation. Upon formal Laplace 
transformation on $t$, from this equation one recovers Leaver's 
normal form for the generalized spheroidal wave equation 
\cite{Leaver1}. In order to trade this second--order equation for 
a first--order system of equations, we introduce an outgoing 
characteristic derivative of $U$.

\subsubsection{Outgoing characteristic 
derivative}\label{sbsbSec:4.1.3}
\begin{figure}
\scalebox{0.70}{\includegraphics{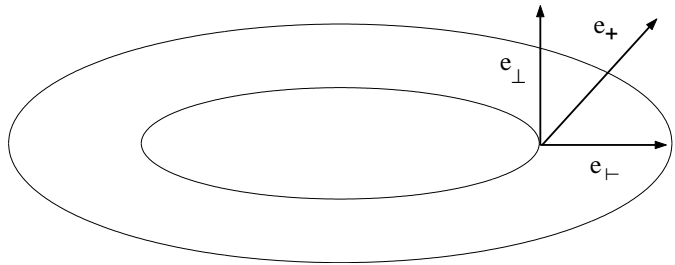}}
\caption{{\sc Concentric spheres in a $\Sigma$ slice.}   
The figure depicts two concentric spheres in a single
level--$t$ slice as well as the vectors discussed
in the text. The $\Sigma$ normal $e_\bot$ points out of
the slice into the future, while $e_\vdash$ points tangent
to the slice but normal to the round sphere. These arrows 
may be extended as vector fields over the whole exterior 
region.}
\label{fourFigure1}
\end{figure}

Construction of the characteristic derivative is based on
an understanding of the relevant normal vector fields 
associated with the $\Sigma$ foliation. Respectively,
the following unit vector fields
\begin{equation}
e_\bot = N^{-1}\big(\partial/\partial t
- V^r \partial/\partial r\big)\, ,\qquad
e_\vdash = M^{-1}\partial/\partial r
\label{normalvfs}
\end{equation}
are the future--pointing normal to the
$\Sigma$ foliation and the outward--pointing normal to 
concentric spheres within a given $\Sigma$ slice \cite{York}. 
These vectors are depicted in {\sc Fig.}~\ref{fourFigure1}.

With normal vector fields (\ref{normalvfs}), we 
introduce $e_{+} = e_\bot + e_\vdash$ which is also depicted in
{\sc Fig.}~\ref{fourFigure1}.  Then $X = e_{+}[U]$ is the 
characteristic derivative we seek, where $[\;]$ denotes the
standard operation of a vector field on a scalar function.
More explicitly,
\begin{equation}
 X = \frac{1}{N}
\frac{\partial U}{\partial t}   
+\left(\frac{1}{M}-\frac{V^r}{N}
\right)\frac{\partial U}{\partial r}\, .
\label{defofXU}
\end{equation}
We now take $(U, X)$ as our basic variables. Since $U$ and $X$ 
belong to a single spherical--harmonic mode, both
should carry $l$ and $m$ subscripts, but we 
continue to suppress these.

\subsubsection{First--order system of evolution 
equations}\label{sbsbSec:4.1.4}
To facilitate numerical implementation, we write 
(\ref{waveeqnEFin}) in a first--order form. The evolution 
equation for $U$,
\begin{equation}
\partial_t U = 
(V^r - M^{-1}N)\partial_r U + N X\, ,
\label{fourUdot}
\end{equation}
is just the definition (\ref{defofXU}) of $X$. 
Eq.~(\ref{waveeqnEFin}) now becomes the evolution equation for 
$X$. We write this equation in the general form
\begin{align}
\partial_t X = & 
(V^r + M^{-1}N)\partial_r X
- N\big(X^M_{-} + 2X^R_{+}\big)X 
+ R^2 X\partial_r\alpha
\label{fourXdot}
\\
&               
+ 2M^{-1}NX^R_{+}\partial_r U
- R^{-2} N l(l+1)U
+ 2\mathrm{m}NR^{-3}\jmath^2U\, ,
\nonumber
\end{align}
where $R$ is the areal radius, $X^M_{-}$ and $X^R_{+}$ are 
certain characteristic variables of the background geometry,
and $\alpha = N/(M R^2)$ is the ``dedensitized lapse.'' In 
this general notation one might denote our main characteristic 
variable (\ref{defofXU}) by $X^U_+$ rather than plain $X$, but 
there is no need to do so here. In terms of the round--sphere 
area $A$, the variable $R$ is defined geometrically as 
$[A/(4\pi)]^{1/2}$, and, therefore, $R = r$ for the 
Eddington--Finkelstein system. Moreover, in terms of the 
``Einstein--Christoffel variables'' $K_M$, $K_R$, $f_M$ and 
$f_R$ written in Eqs.~(2.2) and (2.3) of Ref.~\cite{Lau}, 
we have $X^M_{-} = - K_M - f_M$ and $X^R_{+} = -K_R + f_R$. 
For the evolution at hand all of these variables are fixed, 
and in the numerical implementation they amount to function 
calls given by 
\begin{align}
R(r) & =  r\, ,
\label{fcalls1}\\
\partial_r\alpha(r) & =  -\frac{2(\mathrm{m}
                      + r)}{(2\mathrm{m}r + r^2)^2}\, ,
\label{fcalls2}\\
X^R_{+}(r) & = 
\frac{r - 2\mathrm{m}}{2\mathrm{m}r + r^2}
\left[1+\frac{2\mathrm{m}}{r}\right]^{1/2}\, ,
\label{fcalls3}\\
X^M_{-}(r)  & = 
\frac{2\mathrm{m}^2 - 5\mathrm{m}r - 4 r^2}{r(2\mathrm{m} + r)^2}
\left[1+\frac{2\mathrm{m}}{r}\right]^{1/2} \, .
\label{fcalls4}
\end{align}
We remark that the variable $X^R_{+} = e_{+}[\log R]$, 
whence describes the rate of change of the logarithm of areal 
radius along the outgoing null direction. 
Note that $X^R_{+}(2\mathrm{m}) = 0$. 
Despite the somewhat general form (\ref{fourXdot}) we have 
chosen for the wave equation, it is nothing but a tedious 
exercise in elementary algebra to substitute the given 
expressions for $X$, $N$, $M$, $V^r$, $R$, $X^M_{-}$, $X^R_{+}$, 
and $\partial_r\alpha$ into (\ref{fourXdot}) 
in order to recover ($N$ times) Eq.~(\ref{waveeqnEFin}). 
In performing this exercise, note right away that 
\begin{equation}
V^r + M^{-1}N = 1
\label{velocityisone}
\end{equation} 
for the chosen coordinate system. Each of the evolution 
equations (\ref{fourUdot},\ref{fourXdot}) 
has the form of a forced advection equation, and in 
particular we can write (\ref{fourXdot}) as follows:
\begin{equation}
\partial_t X = \partial_r X + S(r,X,U,\partial_rU)\, ,
\label{forced} 
\end{equation}
where the source term $S$ is defined by comparison of this
equation with (\ref{fourXdot}). Note that $S$ also depends
on the parameters $l$, $\jmath$, and $\mathrm{m}$.

\subsection{Interior of the computational 
domain and horizon: numerical details}\label{sbSec:4.2}
\subsubsection{MacCormack predictor--corrector 
scheme} \label{sbsbSec:4.2.1}
Suppose that the radial mesh is the collection of $Q+1$ nodes 
$r_q$, where $r_0 = 2\mathrm{m}$, $r_{Q} = r_B = 2\mathrm{m}\rho_B$, 
and $r_q = 2\mathrm{m} + q\Delta r$. The spatial discretization 
step is $\Delta r = (r_B - 2\mathrm{m})/Q$. Once the number $Q$ of 
spatial subintervals is fixed, the temporal discretization step 
$\Delta t$ is determined by fixing the Courant number 
$\Delta t/\Delta r$. Let the temporal mesh nodes be 
$t_n = n\Delta t$, and define,
for example, the mesh function $U^n_q = U(t_n,r_q)$. To simplify 
the presentation, our analysis holds $U^n_q$ as the exact 
value of $U(t,r)$ at $(t_n,r_q)$. Our task of numerical 
evolution on the interior of the computational domain is then to 
produce approximations, $\widetilde{U}^{n+1}_q$ and 
$\widetilde{X}^{n+1}_q$, to the true values $U^{n+1}_q$ and 
$X^{n+1}_q$ (doing so for $0 < q < Q$), given in advance both 
$U^{n}_q$ and $X^{n}_q$
(for $0 \leq q \leq Q$) as well as the exact geometry function 
calls $N_{q}$, $M_q$, $V^r_{q}$, $R_q$, $[X^R_{+}]_{q}$, 
$[X^M_{-}]_{q}$, and $[\partial_r\alpha]_{q}$ listed in
Eqs.~(\ref{EFlapseandshift},\ref{fcalls1}--\ref{fcalls4}).
Here, for example, $M_q = M(r_q)$ and $[X^M_{-}]_{q} = 
X^M_{-}(r_q)$. As these functions calls are
time--independent, they need not carry the time superscript $n$. 
We choose the MacCormack scheme to accomplish our task. This 
scheme is in consistent conservation form \cite{LeVeque} as applied to a 
conservation law like the $1+1$ scalar wave equation. Since our 
primary focus is numerical implementation of {\sc robc}, it pays 
to keep the interior evolution simple in order to 
highlight the implementation. 

Let us quickly review the MacCormack scheme in the context of 
the variables at hand. First consider the following predicted 
variables which stem from straightforward differencing of 
(\ref{fourUdot},\ref{forced}):
\begin{align}
\bar{U}^{n+1}_q & =  U^n_q + \Delta t\big[\,(V^r - M^{-1}N)^{}_q\,
                      D_{-} U^n_q + N^{}_q \, X^n_q\big]\, ,
\label{interiorprediction} \\
\bar{X}^{n+1}_q & =  X^n_q + \Delta t\big[ 
                      D_{+} X^n_q + S^n_q\big]\, .
\nonumber
\end{align}
Here $D_{\mp}$ denotes either a downwind or upwind 
difference stencil,
\begin{equation}
D_{-}U^n_q = \big(U^n_q - U^n_{q-1}\big)\big/\Delta r\, ,
\quad
D_{+}X^n_q = \big(X^n_{q+1} - X^n_q\big)\big/\Delta r\, .
\end{equation}
To approximate the $\partial_rU$ appearing in the $S^n_q$ 
source in the second equation of (\ref{interiorprediction}), 
we use a centered difference stencil.
The predicted variables $\bar{U}^{n+1}_q$ and
$\bar{X}^{n+1}_q$ are not second--order--accurate 
approximations to the values $U^{n+1}_q$ and 
$X^{n+1}_q$ in question. However, the corrected variables
\begin{align}
\widetilde{U}^{n+1}_q & =  0.5
                  \big\{\bar{U}^{n+1}_q 
                + U^n_q
    + \Delta t\big[(V^r - M^{-1}N)^{}_{q}\,
                 D_+\bar{U}^{n+1}_q
                +  N^{}_q\, 
                   \bar{X}{}^{n+1}_q\big]\big\}\, ,
\label{interiorcorrection} \\
\widetilde{X}^{n+1}_q & =  0.5
                  \big\{\bar{X}^{n+1}_q
                + X^{n}_q
                + \Delta t\big[D_{-}\bar{X}^{n+1}_q
                + \bar{S}{}^{n+1}_q\big]\big\}\, ,
\nonumber 
\end{align}
are indeed second--order--accurate approximations. To approximate the 
advection derivative term in the correction phase (\ref{interiorcorrection}), 
we have incorporated stencils $D_{\pm}$ opposite to those $D_{\mp}$ used 
for the same term in the prediction phase (\ref{interiorprediction}). The 
source term $\bar{S}{}^{n+1}_q$ at the next time step is
built both with time--independent function calls and barred 
variables. We again use a centered stencil to handle the 
derivative term in the source, but now one
\begin{equation}
\overline{[\partial_r U]}{}^{n+1}_q = 0.5\big(\bar{U}^{n+1}_{q+1}
                                - \bar{U}^{n+1}_{q-1}
                             \big)\big/\Delta r
\end{equation} 
built with predicted values.

\subsubsection{Inner boundary conditions 
at $H$} \label{sbsbSec:4.2.2} From (\ref{fourXdot}) or (\ref{forced}) 
we see that $X$ propagates according to the cartoon $H\nwarrow B$, 
that is to say inward from $B$ towards $H$, and at unit 
speed with respect to the coordinate time axis $\partial/\partial t$. 
Eqs.~(\ref{fcalls1},\ref{fcalls2},\ref{velocityisone}) then determine
the following form for the $X$ evolution equation at the horizon:
\begin{eqnarray}
\lefteqn{\big[\partial_t X - \partial_r X\big]\big|_{r =2\mathrm{m}}
=} 
& & \label{fourdotXatH} \\
& &
-\big[N X^M_{-} X
- r^2 X\partial_r\alpha
+ r^{-2} N l(l+1)U
- 2\mathrm{m}Nr^{-3}\jmath^2U
\big]\big|_{r =2\mathrm{m}}\, ,
\nonumber
\end{eqnarray}
Notice the absence of the derivative term involving $\partial_r U$ 
on the {\sc rhs}. As $X^R_+(2\mathrm{m}) = 0$, it has been dropped.
Eq.~(\ref{fourdotXatH}) shows that $X$ propagates inward at $H$, 
whence there is no analytical issue of an inner boundary condition 
for $X$. 

The coordinate velocity appearing in (\ref{fourUdot}) is
\begin{equation}
-V^r + M^{-1}N = \frac{r - 2\mathrm{m}}{r + 2\mathrm{m}}\, ,
\label{Ucoordinatevelocity}
\end{equation} and it is monotonically increasing
on $[2\mathrm{m},\infty)$ with 
$0 \leq - V^r + M^{-1}N < 1$.\footnote{Based on 
(\ref{Ucoordinatevelocity}), the unit--speed propagation of $X$, 
and the well--known stability properties of the MacCormack 
algorithm \cite{LeVeque}, we can therefore expect numerical 
evolution stability for $\Delta t/\Delta r < 1$. We have $1$ on
the {\sc rhs} of this inequality, since $1$, the speed of $X$
everywhere, is larger than $(\rho_B-1)/(\rho_B+1)$, the maximum 
coordinate speed for $U$ over $[2\mathrm{m},r_B]$.} 
This shows that $U$ propagates according to the cartoon
$H\nearrow B$, at least for $r > 2\mathrm{m}$. More precisely, 
$U$ propagates outward at speed $|V^r - M^{-1}N|$ with respect 
to the coordinate time axis $\partial/\partial t$. Furthermore, 
at the horizon\footnote{In fact, the coordinate vector field
$\partial/\partial t$, a Killing direction, is null at the
horizon.}
\begin{equation}
\partial_t U\big|_{r = 2\mathrm{m}}
= NX\big|_{r = 2\mathrm{m}}\, ,
\label{horizoneq}
\end{equation}
since $-V^r + M^{-1}N$ vanishes for $r = 
2\mathrm{m}$. Therefore, $U$ propagates straight up the 
coordinate time axis at $H$, and there is no analytical issue 
of an inner boundary condition for $U$. Indeed, in 
tandem with the way (\ref{fourdotXatH}) the variable $X$ 
propagates, Eq.~(\ref{horizoneq}) implies that the region 
inside of $r = 2\mathrm{m}$ has no causal influence (insofar as 
the disturbances $U$ and $X$ are concerned) on the region 
outside of $r = 2\mathrm{m}$. Heuristically, $H$ acts as a 
one--way membrane.

As for {\em numerical} inner boundary conditions at $H$, 
we extend the radial mesh past the horizon by adding a 
ghost node $r_{-1} = 2\mathrm{m} - \Delta r$. 
At both the prediction stage and correction stage, values at the 
ghost node are obtained via a simple copy--over from its righthand 
neighbor $r_0 = 2\mathrm{m}$: 
$\bar{U}{}^{n+1}_{-1} = \bar{U}{}^{n+1}_{0}$, 
$\bar{X}{}^{n+1}_{-1} = \bar{X}{}^{n+1}_{0}$,
$\widetilde{U}{}^{n+1}_{-1} = \widetilde{U}{}^{n+1}_{0}$, 
and 
$\widetilde{X}{}^{n+1}_{-1} = \widetilde{X}{}^{n+1}_{0}$.
Of course,
these are not the correct inner boundary conditions, but any 
resulting numerical error is trapped within the region 
$r<2\mathrm{m}$ as shown by careful examination of the outlined 
MacCormack difference scheme at the point $r_0 = 2\mathrm{m}$. 
Indeed, owing to the aforementioned identity $[X^+_R]_0 = 0$
which kills the $\partial_r U$ term in (\ref{fourXdot}), the predicted 
variable $\bar{X}{}^{n+1}_0$ depends neither on $U^n_{-1}$ nor 
$X^n_{-1}$. Likewise, since $(V^r - M^{-1}N)_0 = 0$, the 
predicted variable $\bar{U}{}^{n+1}_0$ also depends neither on 
$U^n_{-1}$ nor $X^n_{-1}$. Next, because $\bar{X}{}^{n+1}_{-1}$ 
is a copy--over of $\bar{X}{}^{n+1}_0$ (so that 
$D_{-}\bar{X}{}^{n+1}_0 = 0$ holds), the corrected variable 
$\widetilde{X}{}^{n+1}_0$ does not depend on mesh values at 
$r_{-1}$. Therefore, we see that $\widetilde{U}{}^{n+1}_0$ 
also does not depend on values at $r_{-1}$, in parallel with
the fact that physically the region $r < 2\mathrm{m}$ cannot
influence $H$.

\subsection{Implementation of {\sc robc}}\label{sbSec:4.3}
\subsubsection{{\sc robc} in physical 
coordinates}\label{sbsbSec:4.3.1}
We may rewrite the convolution (\ref{ROBC}) in 
terms of the physical coordinates via division of the 
equation by an overall factor of $2\mathrm{m}$. The result is
\begin{equation}
{\sf X}_l(T,r_B) +
\frac{\psi_l(T,r_B)}{{\sf M}(r_B)r_B} =
r_B^{-1}{\sf N}(r_B)\int^T_0
\Omega_l(T-T';r_B)
\psi_l(T',r_B)\mathrm{d}T'\, ,
\label{physROBC}
\end{equation}
where we have defined the physical {\sc tdrk}
\begin{equation}
\Omega_l(T;r_B) = 
\frac{\omega_l(T/(2\mathrm{m});r_B/(2\mathrm{m}))}{2\mathrm{m}}\, .
\end{equation}
Also appearing in (\ref{physROBC}) is
\begin{equation}
{\sf X}_l = 
\frac{1}{{\sf N}}\frac{\partial\psi_l}{\partial T}
+\frac{1}{{\sf M}}\frac{\partial\psi_l}{\partial r}\, ,
\end{equation}
the characteristic derivative ${\sf e}_{+}[\psi_l]$ of
$\psi_l$ along a null direction ${\sf e}_{+}$ which points 
outward and towards the future. 

We note that the null vector
\begin{equation}
{\sf e}_+ = \frac{1}{{\sf N}}\frac{\partial}{\partial T}
+\frac{1}{{\sf M}}\frac{\partial}{\partial r}  
\end{equation}
points in the same direction as the one
\begin{equation}
e_+ = \frac{1}{N}\frac{\partial}{\partial t}
+\left(\frac{1}{M}-\frac{V^r}{N}\right)
\frac{\partial}{\partial r}
\end{equation}
considered earlier. However, they have different extents and are 
related by a pure boost,
\begin{equation}
{\sf e}_+ = e^{\vartheta} e_+\, ,
\label{boost}
\end{equation}
through a hyperbolic angle
\begin{equation}
\vartheta = {\textstyle \frac{1}{2}}
\log\left[\frac{1+N^{-1}MV^r}{1-N^{-1}MV^r}
\right]\, .
\end{equation}
Since the proper spatial velocity $MV^r/N = 2\mathrm{m}/r < 1$
for $r > 2\mathrm{m}$, the argument of the logarithm is positive on the
exterior region.

\subsubsection{{\sc robc} in Eddington--Finkelstein 
coordinates}\label{sbsbSec:4.3.2}
The physical convolution (\ref{physROBC}) is expressed with respect 
to spacelike slices which are level in the static 
time variable $T$. We now wish to express the convolution 
(\ref{physROBC}) with respect to the spacelike slices $\Sigma$
which are level in the ingoing Eddington-Finkelstein time $t$. 
Since we have arrange for $T = t$ at the outer boundary 
$r = r_B$, we may express the physical {\sc robc} given in 
(\ref{physROBC}) as follows:
\begin{equation}
e^{\vartheta_B} X(t,r_B) +
\frac{U(t,r_B)}{{\sf M}(r_B)r_B} =
r_B^{-1}{\sf N}(r_B)\int^t_0
\Omega(t-t';r_B)
U(t',r_B)\mathrm{d}t'\, .
\label{EFphysROBC}
\end{equation}
To reach this equation, drop the $l$ subscript on $\Omega_l$, 
enact a trivial change of variables, and appeal to the 
equalities
\begin{equation}
{\sf X}_l =  {\sf e}_{+}[\psi_l]
= e^{\vartheta} e_{+}[U] = e^{\vartheta} X\, .
\end{equation}
The middle equality stems from (\ref{boost}), and it of course 
holds at the outer boundary where we use the shorthand 
$\vartheta_B = \vartheta(r_B)$. In this middle equality we have
also dropped the subscript $l$. In (\ref{EFphysROBC}) we 
have chosen to retain ${\sf N}(r_B)$ and ${\sf M}(r_B)$, although 
they could be traded for $N(r_B)$ and $M(r_B)$ at the expense of 
introducing further boost factors. In any case, they are simply 
fixed constants, insofar as the {\sc robc} are concerned.

\subsubsection{Approximate time--domain radiation 
kernel}\label{sbsbSec:4.3.3}
From now until the end of this subsection we reinstate the policy of 
suppressing all factors of $l$. Moreover, for the most part we also 
suppress factors of $\rho_B$, or $r_B$ as the case may be.
The technique of kernel 
compression discussed in {\sc Section} \ref{sbSec:3.3} 
yields an approximate 
{\sc fdrk} $\tilxi(\sigma) = P(\sigma)/Q(\sigma)$ of the form 
(\ref{dpolesum}). The inverse Laplace transform 
$\mathcal{L}^{-1}[\tilxi](\tau)$ of this approximate {\sc fdrk}
is an approximate {\sc tdrk}
\begin{equation}
\xi(\tau) =  \sum_{n=1}^{d} 
\gamma_n \exp(\beta_n\tau)\, .
\end{equation}
Note that the $d$ pole locations $\beta_n = \beta^R_n + \mathrm{i}\beta^I_n$
and strengths $\gamma_n = \gamma^R_n + \mathrm{i}\gamma^I_n$ are generally
complex and the locations (are numerically observed to) lie in the 
lefthalf plane. In order to pass to a physical approximate {\sc tdrk} $\Xi(t)
= \xi(t/(2\mathrm{m}))/(2\mathrm{m})$, we set $\beta_n  = 2\mathrm{m}
b_n$ and $\gamma_n = 2\mathrm{m} c_n$, thereby reaching
\begin{equation}
\Xi(t) =  \sum_{n=1}^{d}
c_n \exp(b_n t)\, .
\end{equation}
We list a representative set of pole locations and strengths in
Table \ref{fourTable1}. Note that these must be scaled for use
in the physical kernel. For $\rho_B = 15$ as in the table and a 
blackhole of mass $\mathrm{m} = 2$, we would obtain the pole 
locations $b_n$ and strengths $c_n$ corresponding to $r_B
= 60$ upon dividing the entries in the table by $4 = 2\mathrm{m}$.
\begin{table}
\begin{center}
\begin{tabular}{|c|l|}
\hline
$\beta^R_1+\mathrm{i}\beta^I_1$ 
& $-0.20807160+\mathrm{i}0$\\
\hline
$\beta^R_2+\mathrm{i}\beta^I_2$ 
& $-0.18971337+\mathrm{i}0.06900399$\\
\hline
$\beta^R_3+\mathrm{i}\beta^I_3$ 
& $-0.13304751+\mathrm{i}0.17772749$\\
\hline
$\beta^R_4+\mathrm{i}\beta^I_4$ 
& $-0.18971337-\mathrm{i}0.06900399$\\
\hline
$\beta^R_5+\mathrm{i}\beta^I_5$ 
& $-0.13304751-\mathrm{i}0.17772749$\\
\hline
\hline
$\gamma^R_1+\mathrm{i}\gamma^I_1$ 
& $-0.07445512+\mathrm{i}0$\\
\hline
$\gamma^R_2+\mathrm{i}\gamma^I_2$ 
& $-0.17152322+\mathrm{i}0.07630724$\\
\hline
$\gamma^R_3+\mathrm{i}\gamma^I_3$ 
& $-0.12569363+\mathrm{i}0.17804157$\\
\hline
$\gamma^R_4+\mathrm{i}\gamma^I_4$ 
& $-0.17152322-\mathrm{i}0.07630724$\\
\hline
$\gamma^R_5+\mathrm{i}\gamma^I_5$ 
& $-0.12569363-\mathrm{i}0.17804157$\\
\hline
\end{tabular}
\end{center}
\vskip 2mm
\caption{{\sc Pole locations and strengths for 
a compressed kernel}. Here the compressed kernel
corresponds to $\jmath = 0$, $l = 4$, $\rho_B = 15$, and
$\varepsilon = 10^{-6}$. Note that the number of
poles is $d = 5$ in accord with Table 
\ref{threeTable2} from {\sc Section} \ref{sbsbSec:3.3.1}. 
Both the $0$ in $\beta_1$ and the $0$ 
in $\gamma_1$ correspond 
to output numbers from the compression algorithm
which are of order $10^{-47}$.}
\label{fourTable1}
\end{table}

Of the $d$ poles $d_\mathrm{sing}$ will lie on the negative real 
axis and there will be $d_\mathrm{pair}$ complex--conjugate pairs. 
So $d = d_\mathrm{sing} + 2d_\mathrm{pair}$ always holds. For the
compressed kernel shown in Table \ref{fourTable1}, we have
$d_\mathrm{sing} = 1$ and $d_\mathrm{pair} = 2$. We then break up 
$\Xi(t)$ as follows:
\begin{equation}
\Xi(t) = \sum_{i=1}^{d_\mathrm{sing}}H_i(t)
       + \sum_{j=1}^{d_\mathrm{pair}}G_j(t) \, .
\label{fourbreakupofkernel}
\end{equation}
In this decomposition
\begin{equation}
H_i(t) = \mu_i\exp(\kappa_i t)\, ,\;
G_j(t) = 
2\left[m^R_j\cos(k^I_j t) 
- m^I_j\sin(k^I_j t)
\right]\exp(k^R_j t)\, ,
\end{equation}
where $\kappa_i$ and $\mu_i$ respectively correspond 
to a pole location $b^R_n + \mathrm{i}0$ and strength 
$c^R_n + \mathrm{i}0$ (each lying on the negative real axis), 
while $k^R_j+\mathrm{i}k^I_j$ and $m^R_j + \mathrm{i}m^I_j$ 
respectively correspond to a pole location 
$b^R_n+\mathrm{i}b^I_n$ and 
strength $c^R_n+\mathrm{i}c^I_n$ (each lying properly 
in the second quadrant). One should not confuse
the $k^R_j+\mathrm{i}k^I_j$ with the zeros $k_{l,n}$ of the 
MacDonald function considered in {\sc Sections} 
\ref{sbsbSec:2.1.3} and \ref{sbsbSec:3.1.1}. 
For the compressed kernel
considered in Table \ref{fourTable1} we list these reordered
locations and strengths in Table \ref{fourTable2}. 
We also need to consider 
an auxiliary object, 
\begin{equation}
F_j(t) =
2\left[m^I_j\cos(k^I_j t)
+ m^R_j\sin(k^I_j t)  
\right]\exp(k^R_j t)\, ,
\end{equation}
not appearing directly in the kernel. We then use
\begin{equation}
(\Xi*U)(t) =
\sum_{i=1}^{d_\mathrm{sing}}\int^t_0      
H_i(t-t') U(t')\mathrm{d}t' +
\sum_{j=1}^{d_\mathrm{pair}}\int^t_0
G_j(t-t') U(t')\mathrm{d}t'\, ,
\label{fourapproximateconvol}
\end{equation}
as an approximation to the convolution $(\Omega * U)(t)$
in (\ref{EFphysROBC}).
\begin{table}
\begin{center}
\begin{tabular}{|l|l|}
\hline
$2\mathrm{m}\kappa_1$ & $-0.20807160$\\
\hline
$2\mathrm{m}(k^R_1 +\mathrm{i}k^I_1)$ 
& $-0.18971337+\mathrm{i}0.06900399$\\
\hline
$2\mathrm{m}(k^R_2 +\mathrm{i}k^I_2)$ 
& $-0.13304751+\mathrm{i}0.17772749$\\
\hline
\hline
$2\mathrm{m}\mu_1$ & $-0.07445512$\\
\hline
$2\mathrm{m}(m^R_1 +\mathrm{i}m^I_1)$ 
& $-0.17152322+\mathrm{i}0.07630724$\\
\hline
$2\mathrm{m}(m^R_2 +\mathrm{i}m^I_2)$ 
& $-0.12569363+\mathrm{i}0.17804157$\\
\hline
\end{tabular}
\end{center}
\vskip 2mm
\caption{{\sc Reordered pole locations and 
strengths for a compressed 
kernel}. Here we show the reordered locations
and strengths which render the compressed kernel
in Table \ref{fourTable1} manifestly real. 
Compare these numbers with those in Table
\ref{fourTable1}. A choice of mass must be made 
to determine these values. In this table and the
text one should make a distinction between the 
mass (plain $\mathrm{m}$) and the numerical pole 
strength $m_j$ (italic $m$ and with a subscript).}
\label{fourTable2} 
\end{table}

The update $(\Xi*U)(t+\Delta t)$ plays a central role in our 
implementation of {\sc robc}. We write it as follows:
\begin{equation}
(\Xi*U)(t+\Delta t)  = 
\int_0^t \Xi(t+\Delta t -t') U(t')\mathrm{d}t'
+ \int_t^{t+\Delta t}\Xi(t+\Delta t -t') U(t')\mathrm{d}t'\, ,
  \label{splitofXistarU}
\end{equation}
respectively referring to the first and second integrals on 
the {\sc rhs} as the 
{\em history part} ${}_\mathtt{H}(\Xi*U)(t+\Delta t)$ 
and the {\em local part} ${}_\mathtt{L}(\Xi*U)(t+\Delta t)$ 
of the updated convolution. 
In the same fashion, $(H_i*U)(t+\Delta t)$, 
$(F_j*U)(t+\Delta t)$, and $(G_j*U)(t+\Delta t)$ can each be
split into history and local parts. Computing the update 
$(\Xi*U)(t+\Delta t)$ then amounts to computing the history 
and local parts of both $(G_j*U)(t+\Delta t)$ and 
$(H_i*U)(t+\Delta t)$. We discuss the local parts of these 
convolutions below, but here note that the history parts may 
be computed via the following {\em exact} identities: 
\begin{align}
& \label{HGFidentities}
\\
{}_\mathtt{H}(H_i*U)(t+\Delta t) & =  
\exp(\kappa_i\Delta t) (H_i*U)(t)\, ,
\nonumber	
\\
{}_\mathtt{H}(G_j*U)(t+\Delta t) & = 
\exp(k^R_j\Delta t)  
\big[\cos(k^I_j\Delta t)(G_j*U)(t)
-\sin(k^I_j\Delta t)(F_j*U)(t)\big]\, ,
\nonumber\\
{}_\mathtt{H}(F_j*U)(t+\Delta t) & = 
\exp(k^R_j\Delta t)
\big[\cos(k^I_j\Delta t)(F_j*U)(t)
+\sin(k^I_j\Delta t)(G_j*U)(t)\big]\, .
\nonumber
\end{align}
These follow from elementary exponential and 
trigonometric identities. The
{\em recursive evaluation} of the history dependence afforded by
identities (\ref{HGFidentities}) leads to a considerable reduction
in computational storage cost. A direct evaluation of the boundary
convolution would require $O(t/\Delta t)$ memory locations to store
the whole history of $U$ along the timelike cylinder ${}^3\!B$ (see 
{\sc Figure} \ref{introFigure1} in the introduction). Our algorithm 
requires only the memory needed to store (the constituent pieces for) 
$(\Xi*U)^n$ as well as the numerical pole strengths and locations ($2d 
= 2d_\mathrm{sing} + 4d_\mathrm{pair}$ real numbers).

\subsubsection{Incorporation of {\sc robc} into predictor--corrector 
algorithm}\label{sbsbSec:4.3.4}
At $t = t_n$ suppose that we are given
(i) $U^n_q$ and $X^n_q$ for $0\leq q\leq Q$, (ii) 
$(\Xi*U)^n = (\Xi*U)(t_n;r_B)$, and (iii) the necessary procedures 
to perform interior update as described in
{\sc Section} \ref{sbsbSec:4.2.1}. For the second assumption, we 
actually require 
$\forall i, j$ that $(H_i*U)^n$, $(G_j*U)^n$, and $(F_j*U)^n$ 
are individually given. Again, for ease of presentation we 
assume that all of these present time--step expressions are exact. 
Our implementation of {\sc robc} then amounts to the following 
task: use (\ref{EFphysROBC}) to produce second--order accurate
approximations 
\begin{equation}
\widetilde{U}^{n+1}_Q\, ,\quad 
\widetilde{X}^{n+1}_Q\, ,\quad
\widetilde{(\Xi*U)}{}^{n+1}
\end{equation}
to the true values $U^{n+1}_{Q}$, $X^{n+1}_{Q}$,
and $(\Xi*U)^{n+1}$. The notation $\widetilde{(\Xi*U)}{}^{n+1}$ is 
shorthand for the pieces
\begin{equation}
\widetilde{(H_i*U)}{}^{n+1}\, ,\quad
\widetilde{(G_j*U)}{}^{n+1}\, ,\quad\widetilde{(F_j*U)}{}^{n+1}\, .
\end{equation}
We emphasize that $U^{n+1}_{Q}$ and $X^{n+1}_{Q}$ are the true 
values associated with a field freely radiating on a larger 
domain. 

We accomplish the task using the framework of the interior
prediction--correction algorithm. Before entering the 
prediction phase, let us see what can be precomputed and 
stored for use during the course of the algorithm. First,
we have
\begin{equation}
\Xi(0) =  \sum_{i=1}^{d_{\mathrm{sing}}}H_i(0)
                 +    \sum_{j=1}^{d_{\mathrm{pair}}}G_j(0)\, ,
\qquad
\Xi(\Delta t)  =  \sum_{i=1}^{d_{\mathrm{sing}}}H_i(\Delta t)
                 +    \sum_{j=1}^{d_{\mathrm{pair}}}G_j(\Delta t)
\end{equation}
at our disposal from $t = t_0 = 0$ onwards. Second, at the 
time step $t = t_n$, we may appeal to the identities 
(\ref{HGFidentities}) in order to exactly calculate and store 
the history parts 
${}_\mathtt{H}(H_i*U)^{n+1}$, 
${}_\mathtt{H}(G_j*U)^{n+1}$, and 
${}_\mathtt{H}(F_j*U)^{n+1}$ $\forall i, j$. With these we then
exactly calculate the history part ${}_\mathtt{H}(\Xi*U)^{n+1}$ 
of the updated total convolution $(\Xi*U)^{n+1}$.

Let us now describe the prediction phase. Due to the way we 
have chosen the difference stencils in 
(\ref{interiorprediction}), the boundary prediction 
$\bar{U}^{n+1}_{Q}$ may be calculated from the same formula, 
\begin{equation}
\bar{U}^{n+1}_{Q} = U^{n}_{Q}
                + \Delta t\big[(V^r - M^{-1}N)^{}_{Q}\, 
                  (U^n_Q-U^n_{Q-1})\big/\Delta r
                + N^{}_{Q}\, X^n_Q\big]\, ,
\end{equation}
used in the interior prediction. We then substitute $\bar{U}^{n+1}_Q$ 
into the approximation to (\ref{EFphysROBC}) obtained by 
replacing $\Omega$ with $\Xi$, thereby obtaining the ``predicted''
characteristic variable at the next time step. Namely, 
\begin{equation}
\bar{X}^{n+1}_Q =
r_B^{-1}{\sf N}_B e^{-\vartheta_B}
\left[{}_\mathtt{H}(\Xi*U)^{n+1}
+ {}_\mathtt{L}\overline{(\Xi*U)}{}^{n+1}
- \bar{U}^{n+1}_Q\right]\, ,
\label{firstROBC}
\end{equation}
where we use the shorthands ${\sf N}_B = {\sf N}(r_B)
= 1/{\sf M}(r_B)$ and
\begin{equation}
{}_\mathtt{L}\overline{(\Xi*U)}{}^{n+1}
 = 0.5\Delta t\big[\Xi(\Delta t) U^{n}_{Q} + 
\Xi(0)\bar{U}^{n+1}_{Q}\big]\, .
\label{predictlocalpart}
\end{equation}
The last shorthand ${}_\mathtt{L}\overline{(\Xi*U)}{}^{n+1}$ is a 
prediction for the local part ${}_\mathtt{L}(\Xi*U)^{n+1}$ of 
the updated total convolution $(\Xi*U)^{n+1}$. This prediction
is based on trapezoidal approximation of the second integral appearing 
on the {\sc rhs} of (\ref{splitofXistarU}). In {\sc Section}
\ref{sbsbSec:5.1.2} we also consider what results from both 
rectangular and parabolic approximations as well as the
trapezoidal approximation at hand.

Turning to the correction phase, we first obtain
\begin{align}
\widetilde{U}{}^{n+1}_{Q} = &   0.5
                  \Big\{\bar{U}{}^{n+1}_Q
                  + U^{n}_{Q} \\
                &   + \Delta t\Big[ (V^r - M^{-1}N)^{}_{Q}\, 
                  (\bar{U}{}^{n+1}_{Q+1}-\bar{U}{}^{n+1}_{Q})
                  \big/\Delta r
                + N^{}_{Q}\, 
                  \bar{X}{}^{n+1}_{Q}\Big]
                  \Big\}\, .
\nonumber 
\end{align}
In this formula $\bar{U}^{n+1}_{Q+1}$ is not available from 
the prediction phase, whence it is computed via the extrapolation
\begin{equation}
\bar{U}{}^{n+1}_{Q+1} = 3\bar{U}{}^{n+1}_{Q} - 3\bar{U}{}^{n+1}_{Q-1}
                + \bar{U}{}^{n+1}_{Q-2}\, .
\label{fourextrap}
\end{equation}
We discuss this detail of our implementation in the
{\sc Appendix}. Next, 
we compute the ``corrected'' variable
\begin{equation}
\widetilde{X}{}^{n+1}_{Q} = r_B^{-1}
{\sf N}_B e^{-\vartheta_B}
\left[{}_\mathtt{H}(\Xi*U){}^{n+1}
+ {}_\mathtt{L}\widetilde{(\Xi*U)}{}^{n+1}
- \widetilde{U}{}^{n+1}_{Q}\right]\, ,
\label{secondROBC}
\end{equation}
where as before
\begin{equation}
{}_\mathtt{L}\widetilde{(\Xi*U)}{}^{n+1}
 = 0.5\Delta t\big[\Xi(\Delta t) U^{n}_{Q} +
\Xi(0)\widetilde{U}{}^{n+1}_{Q}\big]\, .
\label{correctlocalpart}
\end{equation}
To end the correction phase, we similarly obtain 
$\forall i,j$
\begin{equation}
{}_\mathtt{L}\widetilde{(F_j*U)}{}^{n+1}\, ,\qquad
{}_\mathtt{L}\widetilde{(G_j*U)}{}^{n+1}\, ,\qquad
{}_\mathtt{L}\widetilde{(H_i*U)}{}^{n+1}\, ,
\end{equation}
and use these to update the constituent pieces of the
convolution. For example, we take
\begin{equation}
  \widetilde{(G_j*U)}{}^{n+1}
= {}_\mathtt{H}(G_j*U){}^{n+1}
+ {}_\mathtt{L}\widetilde{(G_j*U)}{}^{n+1}\, ,
\end{equation}
as the desired approximation to the updated total 
convolution $(G_j*U)^{n+1}$.
%
%
\section{Numerical tests}\label{Sec:5}
This section documents the results of several numerical
tests of our implementation of {\sc robc}. Throughout, we 
consider a blackhole of mass $\mathrm{m} = 2$ enclosed   
within an outer boundary of radius $r_B = 60$ so that the
horizon is located at $2\mathrm{m} = 4$ and $\rho_B = 15$.
For this choice of $r_B$ the outer boundary $B$ lies in the
weak--field region (see the second paragraph of {\sc Section} 
\ref{Sec:3}). As mentioned in the introduction, we have 
chosen $\mathrm{m} = 2$ only to have an example for which the 
mass is neither $1$ nor $\frac{1}{2}$. Our numerical tests 
cover both $\jmath = 0,2$ cases as well as various values of 
$l$. Results carried out with $\rho_B = 20$ are similar, but 
not reported. A more detailed investigation of these boundary 
conditions is forthcoming \cite{EvansLau}.

\subsection{One--dimensional radial 
evolutions}\label{sbSec:5.1} Tests based on 
one--dimensional evolutions most effectively examine 
the issues of convergence and error for our {\sc robc}, 
and this subsection is devoted to them. We first 
examine both  issues in the context of a short--time 
numerical evolution. Subsequently, we further examine
the issue of numerical error for our {\sc robc} by 
considering two long--time evolutions. We also consider
a third long--time evolution, although not in the
context of an error analysis.

\subsubsection{Short--time evolution} \label{sbsbSec:5.1.1}
\begin{figure}
\scalebox{0.70}{\includegraphics{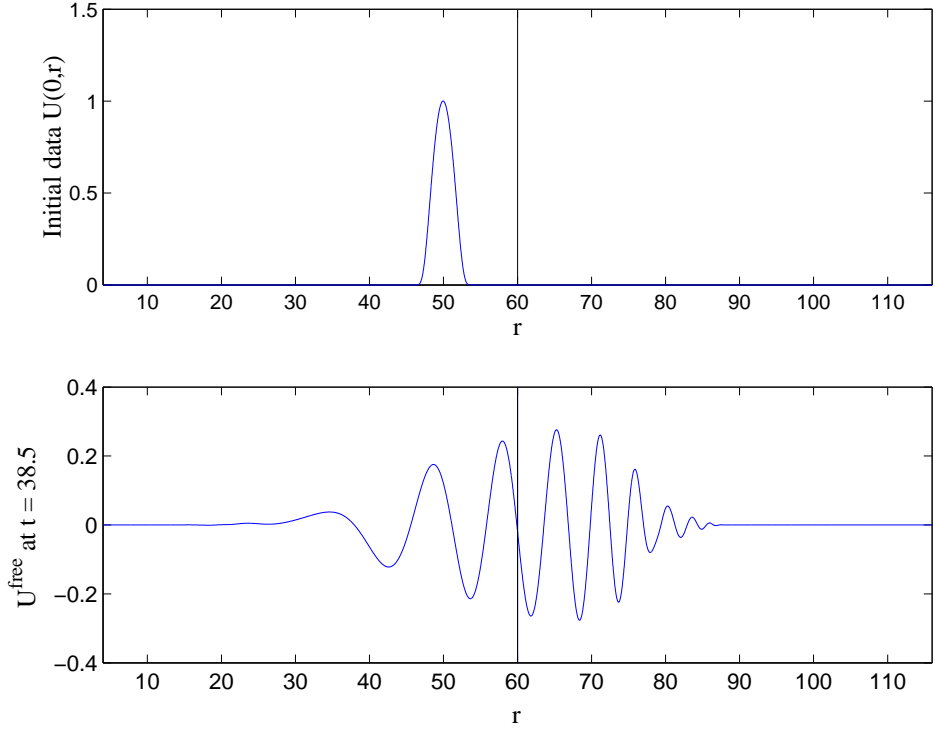}}
\caption{{\sc Free evolution on a
larger domain.} The parameters for the evolution are
$\Delta t \big/\Delta r = 0.5$,
$\mathrm{m} = 2$, $l = 64$, and $\jmath = 2$.
There are $32768$ subintervals dividing $[4,116]$
and 16384 subintervals dividing $[4,60]$.
\label{fiveFigure1}}
\end{figure}

Given fixed values for $l$ and $\jmath$, we have taken the 
initial data $U(0,r)$ shown in the top plot of 
{\sc Fig}.~\ref{fiveFigure1}. We have set $U(0,r) = f(r)/r$, 
where
\begin{equation}
f(r) = A\exp\left[\frac{8\delta}{(b-a)^{2}}\right]
       \left/\exp\left[\frac{\delta}{(r-a)^{2}}
                 + \frac{\delta}{(b-r)^{2}}\right]
                                           \right.
\label{bumpfunction}
\end{equation}
for $A = 50$, $a = 45$, $b = 55$, and $\delta = 20$. The 
profile $U(0,r)$ is then compactly supported on $[45,55]$
and of $C^\infty$ class. We choose $X(0,r) = -U(0,r)/r$ 
as the other initial condition. The resulting initial data set 
would be a pure outgoing were we considering $l=0$ and flat 
Minkowski spacetime. For the case at hand, whatever the fixed 
values $l$ and $\jmath$ are, the initial data is not purely 
outgoing but will reach the outer boundary $r_B = 60$ in a 
time (about 5) which is short relative to the crossing time of 
the domain $[4,60]$. Recall that the spatial step size is 
$\Delta r = (r_B - 2\mathrm{m})/Q$ and the radial mesh points 
are $r_q = 2\mathrm{m} + q\Delta r$. Now we have the 
particular values $r_0 = 2\mathrm{m} = 4$ and $r_Q = r_B = 
60$. We arrange for a spatial discretization such that 
$r_{2Q} = 116$. In particular, if $r_{16384} = 60$, then 
we have $r_{32768} = 116$. Notice that $112 = 116-4$ is twice 
$56 = 60-4$, and so $[4,116]$ is twice as large as $[4,60]$. 

In order to generate an accurate reference solution, we 
{\em numerically} evolve the initial data on the larger domain 
$[4,116]$ until $t = 38.5$. For this evolution (and all others 
in this subsection) we choose a Courant factor 
$\Delta t \big/\Delta r = 0.5$, and at $r_{2Q} = 116$ we use 
simple copy--over boundary conditions, since the disturbance 
does not reach this farther outer boundary in the short time 
$t = 38.5$. The result of our numerical evolution on $[4,116]$ 
is a numerical solution $U^{\mathrm{free}}$, described as 
``free'' since it arises from a free evolution. Any numerical 
error in $U^{\mathrm{free}}$ stems solely from the finite 
difference scheme and not from boundary conditions. The bottom 
plot in {\sc Fig}.~\ref{fiveFigure1} depicts the result of such 
an evolution for the case $l=64$ and $\jmath = 2$ (gravitational 
radiation). In both this plot and the plot of the initial 
data above it, we have indicated $r_B = 60$ by a vertical line.
The free solution that we compute corresponds to $2^{14}$ 
subintervals of $[4,60]$ and so $2^{15}$ subintervals of 
$[4,116]$. Over the smaller domain $[4,60]$, we have estimated
that
\begin{equation}
\|\Delta U^{\mathrm{free}}\|_\infty =
\mathrm{sup}\big\{|U^{\mathrm{free}}_q
-U^{\mathrm{true}}_q|: q =0,\dots,16384\big\} \simeq 
3.6\times 10^{-6}\, ,
\end{equation}
where the $U^{\mathrm{true}}_q$ are exact mesh--point 
values of the true solution.
Notice that this supremum error measure is computed with 
truncated arrays (the full range of $q$ is $0$ to $32768$). 
We have obtained this estimate by a suitable convergence 
analysis of the truncated free solution and Richardson 
extrapolation \cite{Atkinson}. 
\begin{figure}
\scalebox{0.70}{\includegraphics{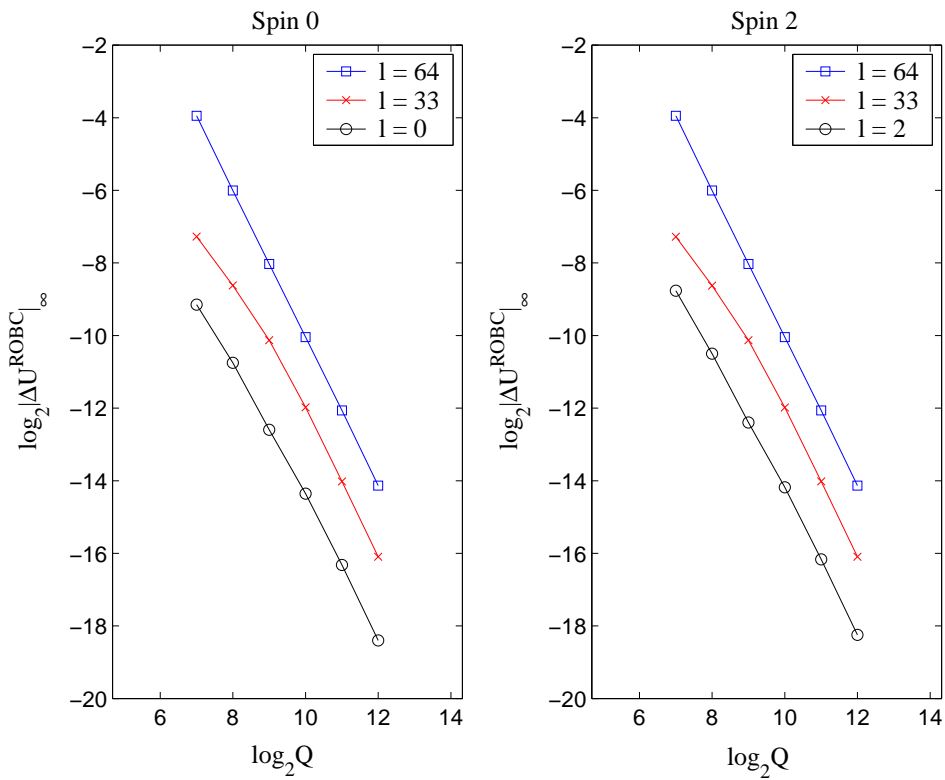}}
\caption{{\sc Error of the solution $U^{\mathrm{ROBC}}$.}
We plot the absolute supremum error of the {\sc robc} solution
for various $j$, $l$, and $Q$. In all cases, these errors have
been computed with the free solution corresponding to $2^{14}$
subintervals of $[4,60]$ in place of the true solution.}
\label{fiveFigure2}
\end{figure}

Using our $\varepsilon = 10^{-10}$ {\sc robc}, we may also evolve the 
initial data to $t = 38.5$ directly on the smaller domain $[4,60]$, 
thereby generating another numerical solution $U^{\mathrm{ROBC}}$. 
Furthermore, we may generate yet another numerical solution 
$U^{\mathrm{SOBC}}$ by using {\em Sommerfeld outer boundary conditions} 
({\sc sobc}) in place of our {\sc robc}. By {\sc sobc} we mean the local 
boundary conditions defined by setting the numerical kernel $\Xi_l = 0$ 
throughout the evolution. The short--time numerical tests at hand
then amount to numerical examinations of the convergence and accuracy 
of both $U^{\mathrm{ROBC}}$ and $U^{\mathrm{SOBC}}$ relative to the free 
solution $U^{\mathrm{free}}$ (here serving in lieu of the true solution). 
Let us first consider the issue of convergence in the next paragraph, 
and afterwards turn to accuracy.
\begin{figure}  
\scalebox{0.70}{\includegraphics{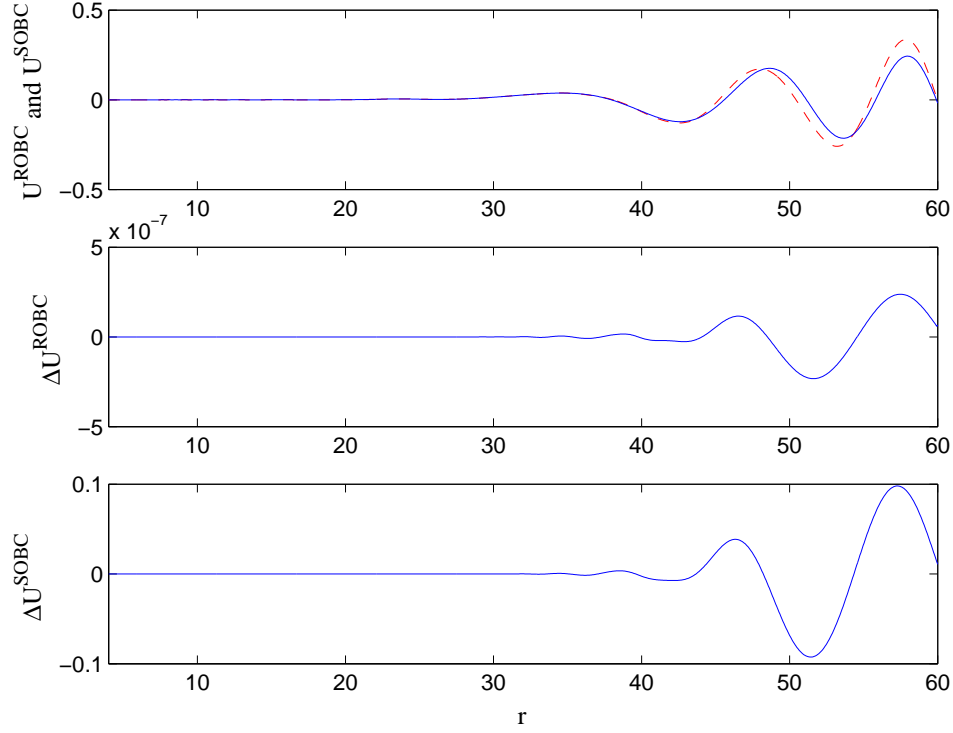}}
\caption{{\sc Comparison of robc and sobc.}
Here $\jmath = 2$, $l = 64$, and all profiles have    
been computed with 16384 subintervals of $[4,60]$. In the
top plot the $U^{\mathrm{SOBC}}$ solution is the dashed line.
\label{fiveFigure3}}
\end{figure}

Suppose $U^{\mathrm{ROBC}}$ is obtained for $Q = 2^8$, with the 
supremum error $\|\Delta U^{\mathrm{ROBC}}\|_\infty$ between 
$U^{\mathrm{ROBC}}$ and $U^{\mathrm{free}}$ is given by the formula
\begin{equation}
\|\Delta U^{\mathrm{ROBC}}\|_\infty =
\mathrm{sup}\big\{|U^{\mathrm{ROBC}}_q
-U^{\mathrm{free}}_{64q}|: q =0,\dots,256\big\}\, ,
\label{fivesuperror}
\end{equation}
where we note that $64 = 2^{14}/2^8$. In general, a 
stride of $2^{14}/2^k$ must be used to evaluate $\|\Delta 
U^{\mathrm{ROBC}}\|_\infty$, when the array $U^{\mathrm{ROBC}}$ corresponds 
to $Q = 2^k$ for $k \leq 14$ (and $U^{\mathrm{free}}$ corresponds to $2^{14}$ 
subintervals of $[4,60]$ as set up above).
Let us switch to {\sc matlab} notation for such norms, so that
(\ref{fivesuperror}) becomes\footnote{We have slightly
abused this notation, since {\sc matlab} requires that the first
entry of an array is labeled by $1$.} 
\begin{equation}
\|\Delta U^{\mathrm{ROBC}}\|_\infty =
\mathrm{norm}\big(U^{\mathrm{ROBC}}(0:256)
-U^{\mathrm{free}}(0:64:16384),\mathrm{inf}\big)\, .  
\end{equation}
As we increase the number of subintervals used to compute 
$U^{\mathrm{ROBC}}$, we can check to see if the {\sc robc} numerical 
solution initially converges to the free one. The plots in 
{\sc Fig}.~\ref{fiveFigure2} document this convergence for a small
sampling of $\jmath$ and $l$ values. Due to the equal scales of their
horizontal and vertical axes, these plots indicate that our 
implementation of {\sc robc} is second--order accurate as expected. 
Using L2 errors instead of supremum errors, we obtain the same 
convergence rates as those indicated. Plots analogous to those in 
{\sc Fig}.~\ref{fiveFigure2} indicate that for any choice of 
$\jmath$ and $l$ values in our sampling the corresponding 
numerical solution $U^{\mathrm{SOBC}}$ does not converge to 
the free solution.
\begin{figure}
\scalebox{0.68}{\includegraphics{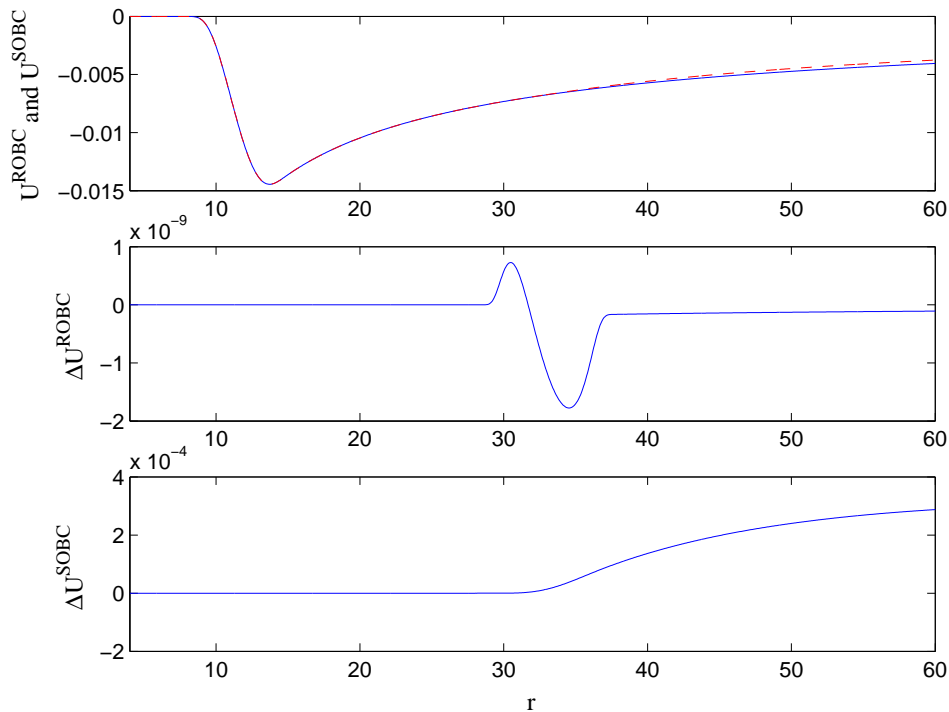}}
\caption{{\sc Comparison of robc and sobc.}
Here $\jmath = 0$, $l = 0$, and all profiles have
been computed with 16384 subintervals of $[4,60]$. In the
top plot the $U^{\mathrm{SOBC}}$ solution is the dashed line.
\label{fiveFigure4}}
\end{figure}

Let us now consider the accuracy of both the {\sc robc} and 
{\sc sobc} solutions, fixing for both a spatial resolution of 
$2^{14}$ subintervals of $[4,60]$, the same resolution as for 
$U^{\mathrm{free}}$. For the choices $\jmath = 2$ and $l = 64$, 
{\sc Fig}.~\ref{fiveFigure3} depicts the numerical solutions 
$U^{\mathrm{ROBC}}$ and $U^{\mathrm{SOBC}}$ as well as the errors 
$\Delta U^{\mathrm{ROBC}} = U^{\mathrm{ROBC}}-U^{\mathrm{free}}$ and 
$\Delta U^{\mathrm{SOBC}} = U^{\mathrm{SOBC}}-U^{\mathrm{free}}$. 
{\sc Fig.}~\ref{fiveFigure4} depicts the same plots, but for 
$\jmath =  0$ and $l = 0$. For the sake of comparison, the top plots 
in these figures superimpose $U^{\mathrm{ROBC}}$ and 
$U^{\mathrm{SOBC}}$. That they differ is even evident to the eye, and 
a glance at the middle and bottom plots in the figures shows that the 
{\sc sobc} errors are many orders of magnitude larger than the 
{\sc robc} errors. For the $\jmath = 2$ and $l = 64$ case shown in 
{\sc Fig.}~\ref{fiveFigure3}, we note that 
$\|\Delta U^{\mathrm{ROBC}}\|_\infty$ (now computed with a stride 
of one) is about $2.38\times10^{-7}$. This value is smaller than 
$3.6\times 10^{-6}$, our aforementioned estimate for the supremum 
error
\begin{equation}
\|\Delta U^{\mathrm{free}}\|_\infty =
\mathrm{norm}\big(U^{\mathrm{free}}(0:16384)
-U^{\mathrm{true}}(0:16384),\mathrm{inf}\big)
\end{equation}
in the free solution. In Table \ref{fiveTable1} we list these 
error measures for the other values of $\jmath$ and $l$ in our small 
sampling. In all cases the {\sc robc} and free numerical solutions 
agree to within our confidence in the free solution. On the contrary,
the {\sc sobc} and free numerical solutions do not agree to within
this confidence, as is evident from the errors 
$\|\Delta U^{\mathrm{SOBC}}\|_\infty$ also listed in the table.
\begin{table}\begin{center}
\begin{tabular}{|c||c|c|c|c|}
\hline
$l$   & $\|\Delta U^{\mathrm{free}}\|_\infty$ & 
        $\|\Delta U^{\mathrm{ROBC}}\|_\infty$ &
        $\|\Delta U^{\mathrm{ROBC}}\|_\infty$ &
        $\|\Delta U^{\mathrm{SOBC}}\|_\infty$ \\ 
\hline
$64$ & $3.6  \times 10^{-6}$ 
     & $2.46 \times 10^{-7}$ 
     & $2.38 \times 10^{-7}$ 
     & $9.80 \times 10^{-2}$ \\
\hline
$33$ & $9.7  \times 10^{-7}$ 
     & $5.97 \times 10^{-8}$ 
     & $5.65 \times 10^{-8}$ 
     & $7.21 \times 10^{-2}$ \\
\hline
$0$  & $1.9  \times 10^{-7}$  
     & $1.77 \times 10^{-9}$ 
     & $1.78 \times 10^{-9}$ 
     & $2.88 \times 10^{-4}$ \\
\hline
\end{tabular}
\end{center}
\vskip 2mm
\caption{{\sc Error measures for short--time test.} The 
table lists various supremum error measures for the 
$\jmath = 0$ case (the corresponding errors for $\jmath
= 2$ look the same). The first column of values for
$\|\Delta U^{\mathrm{free}}\|_\infty$ are estimates for
the error in the free solution, and we have obtained
these estimates via a convergence analysis of the free
evolution. In the last column, for example, the values
$\|\Delta U^{\mathrm{SOBC}}\|_\infty$ measure the difference
between the free and {\sc sobc} solutions when both are
computed with a spatial resolution of 16384 subintervals.
We list two columns for $\|\Delta U^{\mathrm{ROBC}}\|_\infty$.
The first corresponds to $\varepsilon = 10^{-6}$ and the
second to $\varepsilon = 10^{-10}$ {\sc robc}. We see 
essentially no difference between these at the resolution
of this test.}
\label{fiveTable1}
\end{table}

\subsubsection{Long--time evolutions} \label{sbsbSec:5.1.2}
We consider three long--time evolutions, taking for the 
first two $\jmath = 2$, $l = 2$, and $\Delta t \big/\Delta r = 0.5$ 
for all runs. Moreover, for the first two we choose initial data 
$U(0,r) = f(r)/r$ with $f(r)$ as before 
in (\ref{bumpfunction}), but now with for $A = 10$, $a = 5$, $b = 15$, 
and  $\delta = 20$. The profile $U(0,r)$ is then a compactly supported 
bump function of unit height on $[5,15]$, an interval overlapping 
the circular photon orbit at $3\mathrm{m} = 6$ \cite{MTW}. We again 
set $X(0,r) = -U(0,r)/r$ in order to crudely mimic outgoing initial data.
Such data describes a wave packet that, although initially close 
to the horizon, somewhat escapes from the blackhole. For both 
evolutions, we consider a numerical test of our {\sc robc}. Each one 
amounts to comparing a numerical solution $U^\mathrm{ROBC}$ generated 
with our {\sc robc} to a numerical solution $U^\mathrm{free}$ generated 
on a larger domain, now either four or sixteen times the size of $[4,60]$.

These long--time tests show that our {\sc robc} yield accurate 
numerical solutions corresponding to certain celebrated physical 
phenomena. In addition, the first long--time test further 
elucidates the error properties of the {\sc robc} solution 
relative to the free one. Assuming that $U^\mathrm{ROBC}$ and 
$U^\mathrm{free}$ are obtained at the same spatial resolution, 
any difference 
$\|U^\mathrm{ROBC} - U^\mathrm{free}\|_\infty$ between them stems
from three possible sources: (i) use of an approximate kernel in 
the integral convolution at the boundary, (ii) the integration 
rule used to evaluate the integral convolution, and (iii) error 
in the boundary values of the solution $U$ itself (which are fed 
into the convolution). Source (iii) has the potential for 
feedback. Our first long--time evolution and experiment examines the 
competing influence of these sources. Based on this examination, we will
conclude (to no surprise) that, given our second--order--accurate 
scheme for interior update, our trapezoidal implementation of the 
{\sc robc} is optimal.
\begin{figure}
\scalebox{0.70}{\includegraphics{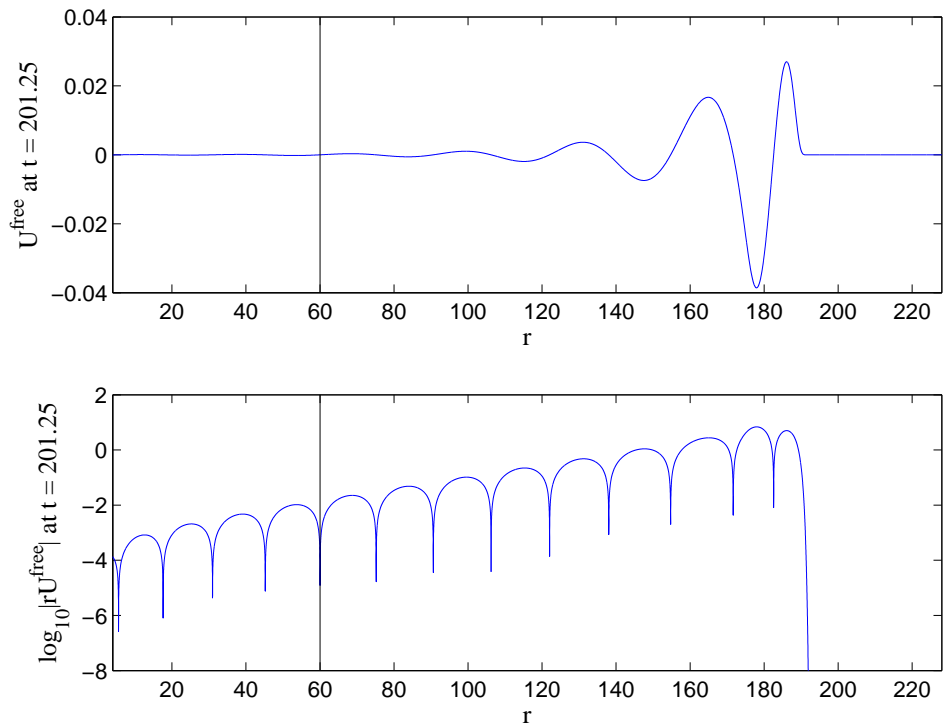}}
\caption{{\sc Free evolution on $[4,228]$ at $t = 201.25$.}
\label{fiveFigure5}}
\end{figure}

\begin{figure}
\scalebox{0.68}{\includegraphics{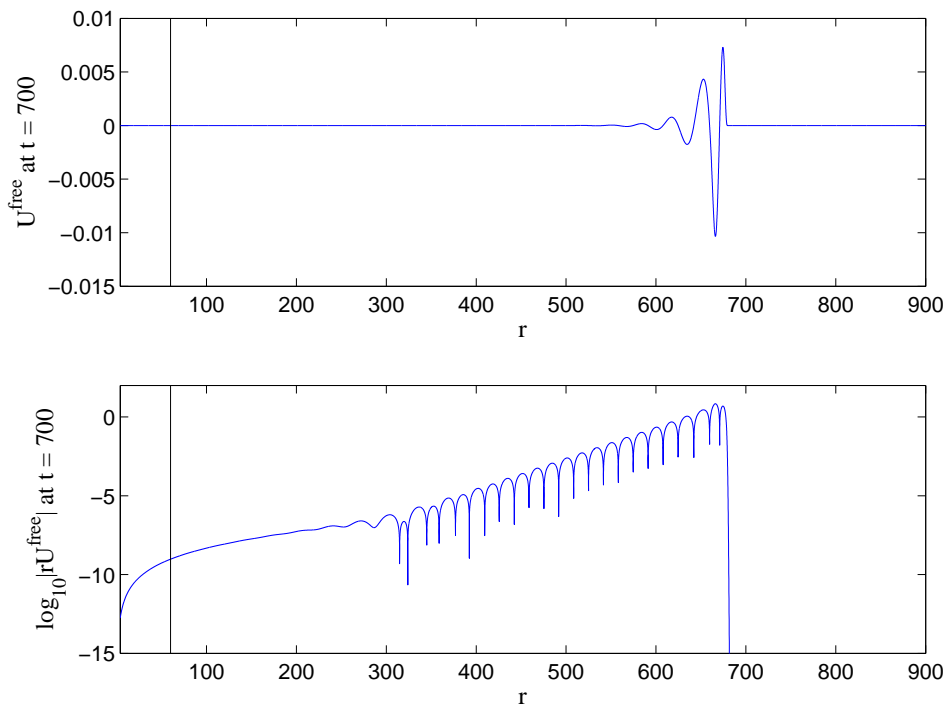}}
\caption{{\sc Free evolution on $[4,900]$ at
$t = 700$.}
\label{fiveFigure6}}
\end{figure}

\begin{figure}
\scalebox{0.75}{\includegraphics{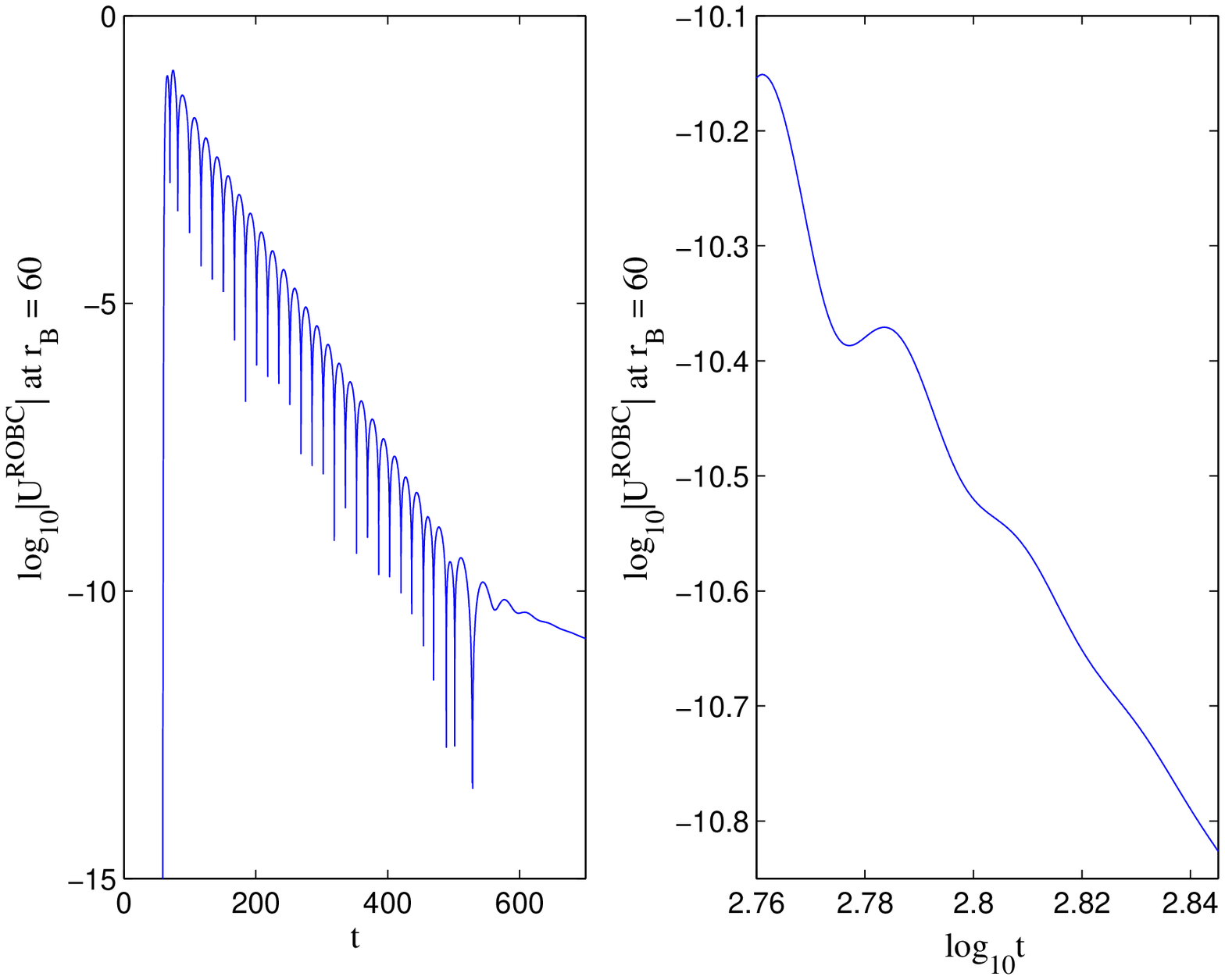}}
\caption{{\sc Quasinormal ringing and decay tail.}
For the described second long--time evolution, we 
record the time history of the field at the fixed position
$r_B = 60$ (the vertical line in {\sc Fig}.~\ref{fiveFigure6}). 
The lefthand plot depicts quasinormal ringing. The righthand 
plot, a zoom--in of the late--time history of the field, 
features $\log_{10}t$ to highlight the onset of the $t^{-7}$ 
power--law decay. Each plot is a history of the numerical 
boundary value $U^\mathrm{ROBC}_Q$.
\label{fiveFigure7}}
\end{figure}

{\em First long--time evolution: quasinormal ringing.} 
For the first test, we evolve the data on $[4,228]$ until 
$t = 201.25$, thereby obtaining a new free solution. The top plot 
in {\sc Fig}.~\ref{fiveFigure5} depicts this free solution. Notice 
that the leading edge of the front has advanced from $r = 15$ to 
about $r = 200$, and that overall the disturbance falls off rapidly 
with decreasing $r$. However, the bottom plot, a graph of the 
base--10 logarithm of the absolute value of $rU^\mathrm{free}$ (with
multiplication by $r$ correcting for $1/r$ fall--off in
amplitude), exhibits the behavior called {\em quasinormal ringing} in 
the physics literature. This behavior is also depicted in 
{\sc Fig.}~\ref{fiveFigure6} which shows the free evolution of the 
same data on $[4,900]$ until $t = 700$. As the leading front 
advances, it leaves exponentially decaying oscillations in its 
wake. The phenomenon is better described by examining the time 
history of the solution at a fixed radius, as is done in 
{\sc Fig}.~\ref{fiveFigure7}. The complex frequency 
corresponding to this behavior is the least--damped (or fundamental) 
$l = 2$ quasinormal mode of the $\mathrm{m} = 2$ blackhole 
in question. Kokkotas and Schmidt record a value of 
$0.37367-{\rm i}0.08896$ for the product of this frequency and the 
blackhole mass \cite{KokkotasSchmidt}. Dividing this value by 2, we 
find $0.186835-\mathrm{i}0.04448$ as the least--damped $l = 2$ 
quasinormal mode of our black hole. The estimate $\pi/0.186835 \simeq 
16.8$ indeed corresponds to about half a wavelength in the lefthand 
plot of {\sc Fig.}~\ref{fiveFigure7} (as measured from one spike to 
the next), and the damping coefficient $0.04448/\ln 10 \simeq 0.019$ 
matches minus the slope of the decaying amplitude in the same plot.

To further examine the error properties of the {\sc robc} solution, 
we generate the numerical solution $U^{\mathrm{ROBC}}$ on $[4,60]$ 
for $t = 201.25$, that we may compare it with $U^{\mathrm{free}}$ 
restricted to this smaller interval. The outer radius $r_B = 60$ is 
marked by a black line in {\sc Fig.}~\ref{fiveFigure5}. Using 
{\sc matlab} notation, for $Q$ subdivisions of $[4,60]$ we have the 
numerical solution $U^{\mathrm{ROBC}}(0:Q)$, and for $4Q$ 
subdivisions of $[4,228]$ we have the numerical solution 
$U^{\mathrm{free}}(0:4Q)$. We may then compare the truncated solution 
$U^{\mathrm{free}}(0:Q)$ directly with $U^{\mathrm{ROBC}}(0:Q)$ by 
computing the supremum error between the two. Doing so for a sequence 
of spatial resolutions, we construct the column labeled
{\em trapezoidal} in Table \ref{fiveTable2}. By changing the spatial 
resolution with $\Delta t\big/\Delta r = 0.5$ fixed, we also change the 
number of temporal steps taken in generating each solution; however, we 
always compare {\sc robc} and free solutions generated with the same 
number of temporal steps. Taking the base--2 logarithm of successive 
ratios of the trapezoidal errors in Table \ref{fiveTable2}, we see a 
second--order convergence. Such convergence indicates that source (i) 
from the paragraph before last is not the dominant contribution to
$\|U^\mathrm{ROBC} - U^\mathrm{free}\|_\infty$. Indeed, we have the 
same approximating numerical kernel regardless of mesh resolution. 
Therefore, for the numerical experiment at hand (which is based on 
our trapezoidal implementation of the {\sc robc}) this error measure 
stems chiefly from the aforementioned sources (ii) and (iii).
\begin{table}\begin{center}
\begin{tabular}{|c||c|c|c|}\hline
Q & rectangular & trapezoidal & parabolic \\
\hline\hline
256   & $1.6571\times 10^{-5}$ & $7.2426\times 10^{-7}$  & $6.8587\times 10^{-7}$
\\\hline
512   & $8.3088\times 10^{-6}$ & $1.7819\times 10^{-7}$  & $1.6877\times 10^{-7}$
\\\hline
1024  & $4.1663\times 10^{-6}$ & $4.3814\times 10^{-8}$  & $4.1495\times 10^{-8}$
\\\hline
2048  & $2.0843\times 10^{-6}$ & $1.0856\times 10^{-8}$  & $1.0278\times 10^{-8}$
\\\hline
4096  & $1.0416\times 10^{-6}$ & $2.7074\times 10^{-9}$  & $2.5627\times 10^{-9}$
\\\hline
8192  & $5.2053\times 10^{-7}$ & $6.7626\times 10^{-10}$ & $6.4011\times 10^{-10}$
\\\hline
16384 & $2.6023\times 10^{-7}$ & $1.6888\times 10^{-10}$ & $1.5985\times 10^{-10}$
\\\hline
32768 & $1.3011\times 10^{-7}$ & $4.2149\times 10^{-11}$ & $3.9891\times 10^{-11}$
\\\hline
\end{tabular}
\end{center}
\vskip 2mm
\caption{{\sc Supremum errors for first long--time
test.} We list errors
$
\mathrm{norm}\big(
U^\mathrm{ROBC}(0:Q) -
U^\mathrm{free}(0:Q),\mathrm{inf}\big)
$
corresponding to (forward) rectangular, trapezoidal, and
parabolic implementations of the {\sc robc}.
\label{fiveTable2}}
\end{table}
%
%
%
%
%
%
%
%
%
%
%

To further resolve the influence of these error sources, 
we have carried out the experiment using the rectangle rule 
(both forward and backward cases), rather than the trapezoid 
rule as in (\ref{predictlocalpart}), to evaluate the local part 
of the boundary Laplace convolution. The corresponding errors 
for either case, with those for the forward case listed in the 
first column of Table \ref{fiveTable2}, decrease at a first--order 
rate. Therefore, we conclude that for a rectangular implementation 
the dominant contribution to 
$\|U^{\mathrm{ROBC}}-U^\mathrm{free}\|_\infty$ is source (ii). 
We have also carried out the experiment using Simpson's rule 
(parabolic rule) to approximate the local part of the convolution. 
Doing so involves a certain parabolic interpolation to get 
half--time--step numerical values for the field $U$. In this case, 
the corresponding errors, listed in the third column of Table 
\ref{fiveTable2}, again decrease at a second--order rate, although 
these errors are slightly smaller than their trapezoid counterparts. 
If (ii) were the dominant contribution to 
$\|U^{\mathrm{ROBC}}-U^\mathrm{free}\|_\infty$ for our trapezoidal 
implementation of the {\sc robc}, then switching to a parabolic 
implementation should lead to error decrease at a fourth--order rate.
Since it does not, we conclude that our trapezoidal {\sc robc} are 
not the dominant source of error (assuming a second--order 
interior scheme).
These considerations suggest that there is little point in 
improving the integration rule used for the Laplace convolution, 
unless one also implements a higher order scheme for the interior. 

\begin{figure}
\scalebox{0.70}{\includegraphics{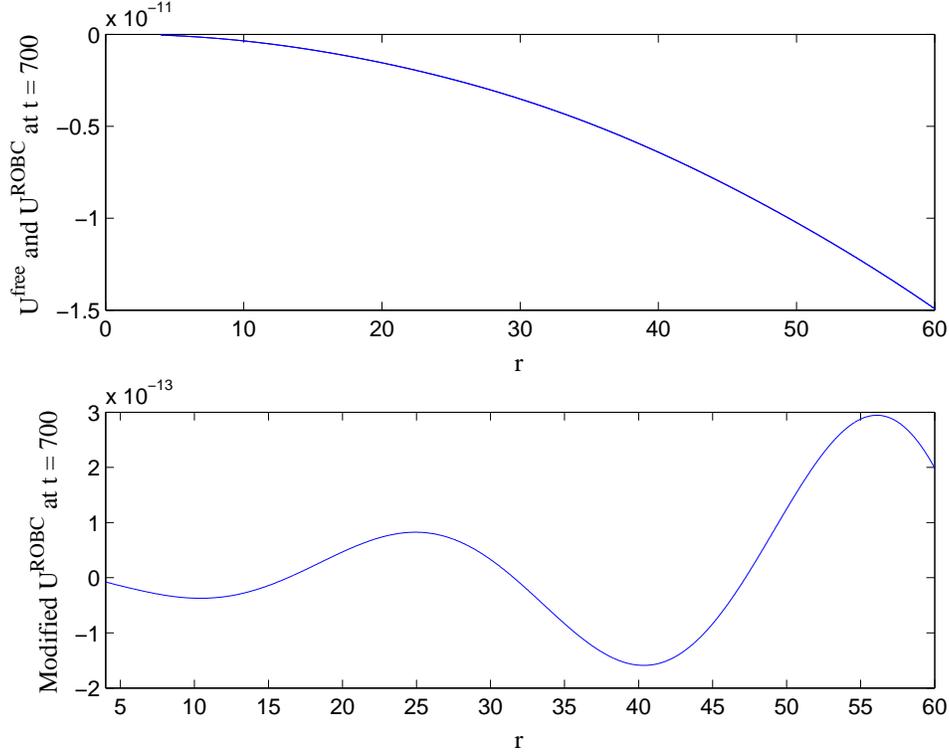}}
\caption{{\sc $U^{\mathrm{free}}$ and $U^{\mathrm{ROBC}}$ at
$t = 700$.} The modified $U^{\mathrm{ROBC}}$ solution has been
obtained by removing the cut contribution to the {\sc tdrk} as
described in the text.
\label{fiveFigure8}}
\end{figure}

{\em Second long--time evolution: decay tails.} 
Our second long--time, one--dimensional, test evolution shows 
that the {\sc robc} yield a reasonably accurate numerical solution 
even for very late times 
relative to the crossing time of the domain. After the quasinormal 
ringing dies down, the solution at a fixed radius $r$ is known to 
behave as order $t^{-(2l+3)}$, provided the initial field configuration 
is not time--symmetric \cite{KokkotasSchmidt}, and this behavior is
seen in the righthand plot of {\sc Fig}.~\ref{fiveFigure7}. 
Heuristically, this late--time {\em decay tail} stems from the 
back--scattering of outgoing waves off of the background curvature. 
{\sc Fig}.~\ref{fiveFigure6} depicts the solution at $t = 700$, 
and the bottom plot shows that by this late time a sizable region 
has entered the power--law decay regime. Note that $r_B = 60$ is again 
marked by a black line in the figure. For $Q$ subdivisions of 
$[4,60]$, there are $16Q$ subdivisions of $[4,900]$. Therefore, 
we truncate the full numerical solution 
$U^{\mathrm{free}}(0:16Q)$, and like before compare
$U^{\mathrm{free}}(0:Q)$ directly with $U^{\mathrm{ROBC}}(0:Q)$.
Doing so for a sequence of spatial resolutions, we construct 
Table \ref{fiveTable3}. Due to round--off error, these errors do 
not decrease at a second--order rate. However, with the spatial 
resolution $4096$ indicated in the table's fifth row as an example, 
we note that the supremum relative error\footnote{The $./$ is 
{\sc matlab} notation for component--by--component division.}
\begin{equation}
\mathrm{norm}((U^\mathrm{free}(0:4096)
-U^\mathrm{ROBC}(0:4096))./U^\mathrm{free}(0:4096),\mathrm{inf})
\end{equation}
is the reasonably accurate value $2.1019 \times 10^{-4}$. For 
{\sc sobc}, rather than $6.0051\times 10^{-16}$ for the absolute error 
and 
$2.1019 \times 10^{-4}$ for the relative error, we get the numbers 
$2.9569\times 10^{-9}$ and $5.2668\times 10^3$.
{\sc Fig.}~\ref{fiveFigure8} depicts the solution at 
$t = 700$ on $[4,60]$. The top plot superimposes 
both $U^{\mathrm{free}}(0:4096)$ and $U^{\mathrm{ROBC}}(0:4096)$
to show that they match to the eye. 

In the bottom plot of {\sc Fig.}~\ref{fiveFigure8} we show a   
modified $U^{\mathrm{ROBC}}(0:4096)$. This profile has been
generated by ``turning off'' the cut contribution to the kernel.
More precisely, to obtain this profile, in
(\ref{fourapproximateconvol}) we have by hand set $H_i(t-t') = 0$,
$\forall i$ throughout the evolution. For this $l = 2$  case, the
kernel $\Xi(t) = G_1(t)$ in
(\ref{fourbreakupofkernel}) then corresponds to a single conjugate
pair of poles. From this experiment, we might infer that the  
phenomenon of decay tails is handled by the cut contribution to the
kernel. As seen in {\sc Sections} \ref{sbSec:3.1} and \ref{sbSec:3.2}
(but there for $\jmath = 0$), this contribution stems from a continuous
distribution of poles.
\begin{table}\begin{center}
\begin{tabular}{|l||c|}\hline
$\mathrm{norm}\big(
U^\mathrm{ROBC}(0:256) -
U^\mathrm{free}(0:256),
\mathrm{inf}\big)$
&  $1.8166\times 10^{-15}$\\
\hline
$\mathrm{norm}\big(
U^\mathrm{ROBC}(0:512) -
U^\mathrm{free}(0:512),\mathrm{inf}\big)$
&  $1.1427\times 10^{-15}$\\
\hline
$\mathrm{norm}\big(
U^\mathrm{ROBC}(0:1024) -
U^\mathrm{free}(0:1024),\mathrm{inf}\big)$
&  $7.6272\times 10^{-16}$\\
\hline
$\mathrm{norm}\big(
U^\mathrm{ROBC}(0:2048) -
U^\mathrm{free}(0:2048),\mathrm{inf}\big)$
& $6.2889\times 10^{-16}$\\
\hline
$\mathrm{norm}\big(
U^\mathrm{ROBC}(0:4096) -
U^\mathrm{free}(0:4096),\mathrm{inf}\big)$
& $6.0051\times 10^{-16}$\\
\hline
$\mathrm{norm}\big(
U^\mathrm{ROBC}(0:8192) -
U^\mathrm{free}(0:8192),\mathrm{inf}\big)$
& $5.9483\times 10^{-16}$\\
\hline
\end{tabular}
\end{center}
\vskip 2mm
\caption{{\sc Supremum errors for second long--time
test.} Here we are using {\sc matlab} notation for the
supremum error as explained in the text.
\label{fiveTable3}}
\end{table}

\begin{figure}
\scalebox{0.70}{\includegraphics{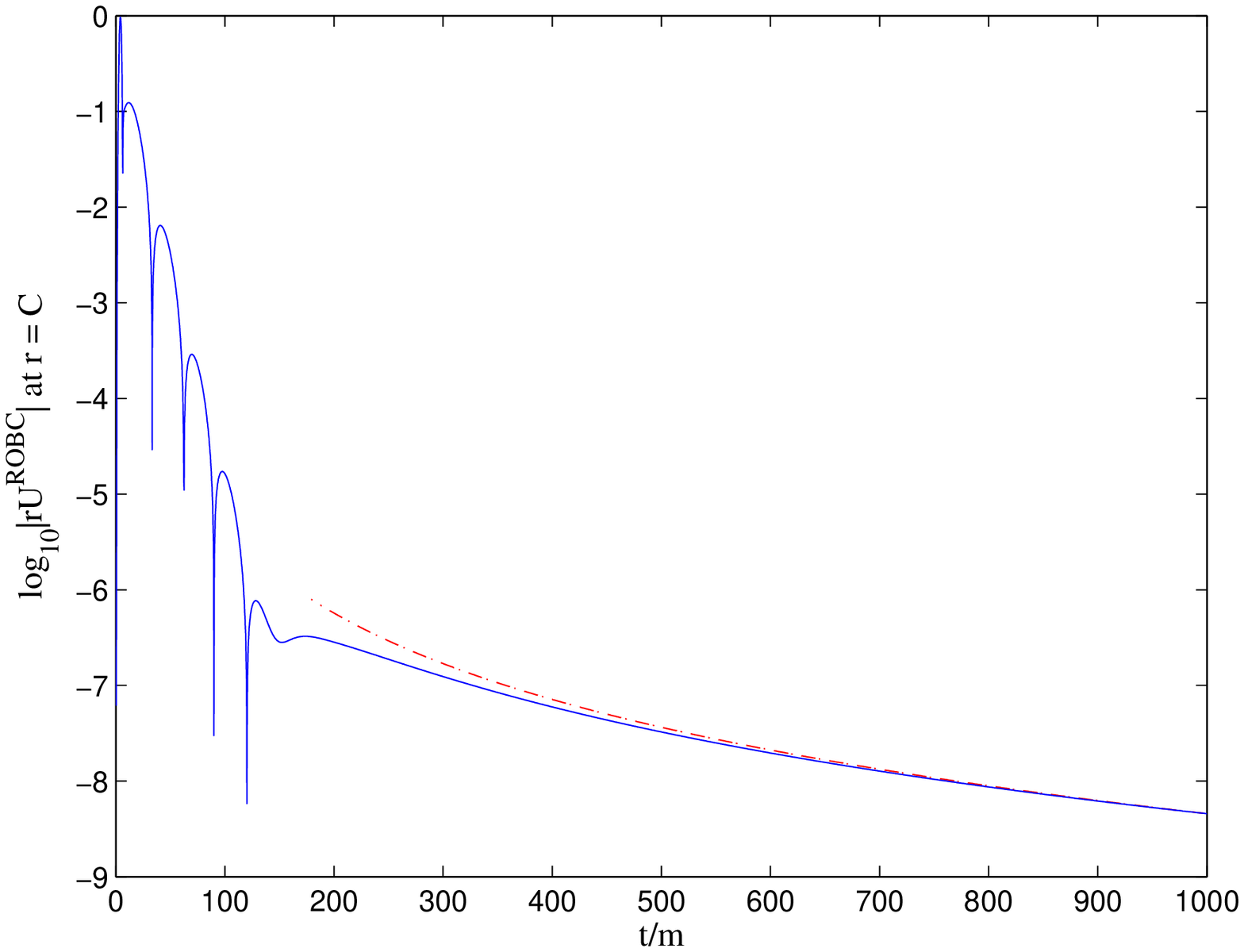}}
\caption{{\sc Long--time history of the field
$rU^{\mathrm{ROBC}}$ at $r = C \simeq 2.5569\mathrm{m}$.}
Here we show the result of the third long--time experiment
with $\jmath,l=0$ and initial data inspired by \cite{ABBP}.
$C$ is the radial value corresponding to vanishing 
tortoise coordinate. More precisely, here we consider the history
at the radial mesh point $\simeq 2.5572\mathrm{m}$ closest to 
$C$. This plot may be compared with {\sc Figure} 2 in \cite{ABBP}.
The dash--dot line is the base--10 logarithm of
$4.57(\mathrm{m}/t)^3$, similar to the matching curve 
considered in \cite{ABBP}. This plot exhibits the power--law
decay of the field. We stress that this plot has
been obtained with the outer boundary radius $r_B
= 30\mathrm{m}$. In carrying out the experiment,
we set $\mathrm{m} = 2$.
\label{newfiveFigure}}   
\end{figure}

{\em Third long--time evolution: decay tails revisited.}
Here we consider an experiment described by Allen, Buckmiller,
Burko, and Price \cite{ABBP}, in order to further demonstrate
that our {\sc robc} capture the phenomenon of decay tails. In 
terms of the standard retarded time $u = T - r_*$, we introduce a 
function $f(u) = \big[u(u-8\mathrm{m})/(16\mathrm{m}^2)\big]{}^8$ 
for $0\leq u \leq 8\mathrm{m}$, with $f(u) = 0$ otherwise. 
Since we work with Eddington--Finkelstein coordinates, whereas
\cite{ABBP} worked with static time coordinates, we set $u(t,r) 
= t + r - 2 r_* - C$. Here $C$ is a constant, chosen to ensure 
that $C + 2\mathrm{m}\log[-1+C/(2\mathrm{m})] = 0$, and in turn 
$u(t,C) = t$. [Before in (\ref{inEFtime}) and the analysis
following that equation, $t$ denoted a time variable which 
differs from this one by a constant, but no matter.] We obtain 
an initial $\jmath, l = 0$ scalar wave packet as 
$U(0,r) = f(u(0,r))/r$, and also complete the data accordingly 
with $X(0,r)$ obtained from (\ref{defofXU}). That is, we get 
$X(0,r)$ by computing $e_+[f(u(t,r))/r]$ and then setting $t=0$. 
Therefore, the data describes an essentially outgoing pulse 
which at $t = 0$ is of unit height and roughly supported on 
$[2\mathrm{m},C]$. Next, on a numerical domain 
$[2\mathrm{m},30\mathrm{m}]$, we evolve the data until $t = 
1000\mathrm{m}$ with our $\varepsilon = 10^{-10}$ {\sc robc}, 
along the way recording the time history of the product
$r U^\mathrm{ROBC}$ of radius and numerical field 
at the fixed radius $C$ (more precisely, at the radial 
mesh point closest to $C$). Here we multiply by radius 
(whereas in {\sc Fig}.~\ref{fiveFigure7} we did not) in order
to have better agreement with \cite{ABBP}. The resulting
history is shown in {\sc Fig}.~\ref{newfiveFigure}. We stress 
that our domain size is more than 16 times smaller than the one 
considered in \cite{ABBP}.

We believe this experiment shows that our {\sc robc} indeed
resolve decay tails. This is not to say that the results of
\cite{ABBP} are incorrect. Those results are applicable 
for the type of boundary conditions considered, namely 
approximate conditions which are not history dependent. We intend 
to carry out other more careful tests of our {\sc robc} along the 
lines spelled out in \cite{ABBP}. Results will be presented in a 
forthcoming work \cite{EvansLau}.

\subsection{Three--dimensional evolution} \label{sbSec:5.2}
Our final test is three--dimensional, and 
considers the scenario of a wave--packet striking the 
computational boundary at a shallow angle (or at least not 
on a perpendicular). Such a scenario is known to be difficult 
test case for radiation boundary conditions in flatspace. 
We carry out this test only for the scalar $\jmath = 0$ case.

\subsubsection{Description and set--up} \label{sbsbSec:5.2.1}
Consider the bump function
\begin{equation}
f(t,x,y,z) = 
A\exp\left(\frac{2\delta}{b^{2}}\right)
       \left/\exp\left[\frac{\delta}{\big(\big|\mathbf{r} 
                 - \mathbf{v}t -\mathbf{r}_{0}\big|+b\big){}^{2}}  
                 + \frac{\delta}{\big(b-
                 \big|\mathbf{r} - \mathbf{v}t 
                 -\mathbf{r}_{0}\big|\big){}^{2}}\right]
                                           \right.\, ,
\label{bumpfunction3}
\end{equation}
where
\begin{equation}
\big|\mathbf{r} - \mathbf{v}t -\mathbf{r}_{0}\big| 
= \big[(x - v_x t - x_0)^2  
        + (y - v_y t - y_0)^2
        + (z - v_z t - z_0)^2\big]^{1/2}\, .
\end{equation}
Were we considering wave propagation on flat Minkowski
spacetime, $f$ would be an exact solution. Were this the case, 
the bump function would have center $\mathbf{r}_0 = 
(x_0,y_0,z_0)$ at $t=0$ and would travel without distortion 
and at unit speed in the direction $\mathbf{v} = (v_x,v_y,v_z)$. 
Here we are assuming that $v_x^2 + v_y^2 + v_z^2 = 1$. 

Let us interpret $(x,y,z)$ as the ``Cartesian coordinates''
belonging to the Edding\-ton--Finkelstein coordinates 
$(r,\theta,\phi)$ from {\sc Section} \ref{sbsbSec:4.1.1} (the 
same {\em spatial} coordinates as in {\sc Section} 
\ref{sbsbSec:1.1.1}). That is to say, $x = r\sin\theta\cos\phi$, 
and so on.\footnote{These are not 
the well--known Schwarzschild system of {\em isotropic 
coordinates} \cite{MTW}.} Doing so gives us a 
diffeomorphism between the initial Schwarzschild time slice 
$\Sigma$ and three--dimensional Euclidean space $\mathbb{E}^3$, 
itself viewed as the initial surface of flat spacetime. On 
$\mathbb{E}^3$ we obtain the data $f|_{t=0}$ and 
$\partial_t f|_{t=0}$ which is then taken over as initial data
on $\Sigma$. The Schwarzschild evolution of this data is not 
the spacetime function (\ref{bumpfunction3}). Nevertheless, on 
physical grounds, one expects this evolution to be 
qualitatively similar to the flat--spacetime evolution 
(\ref{bumpfunction3}) of the same data, provided that (i)
the center $(x_0,y_0,z_0)$ is of large enough radius 
$(x_0^2 + y_0^2 + z_0^2)^{1/2}$ and (ii) the support $b$ 
relative to that center is small enough.

We have written a three--dimensional spectral code to evolve
such data. The basic idea, implemented numerically,
is to obtain a data set, $U_{lm}(0,r)$ and $X_{lm}(0,r)$,
for each spherical mode $(l,m)$, a set we then evolve via the 
one--dimensional purely radial scheme described in {\sc Section} 
\ref{Sec:4} (although throughout that section we 
suppressed the $lm$ subscripts on these mode variables). 
The exact data for each mode are obtained by 
harmonic {\em analysis} as follows:
\begin{equation}
U_{lm}(0,r) = \int_0^{2\pi}\int_{0}^{\pi} 
f(0,x(r,\theta,\phi),y(r,\theta,\phi),z(r,\theta,\phi))
\overline{Y}_{lm}(\theta,\phi)\sin^2\theta 
{\rm d}\theta{\rm d}\phi ,
\end{equation}
and  
\begin{equation}
X_{lm}(0,r) = \int_0^{2\pi}\int_{0}^{\pi}
g(0,x(r,\theta,\phi),y(r,\theta,\phi),z(r,\theta,\phi))
\overline{Y}_{lm}(\theta,\phi)\sin^2\theta
{\rm d}\theta{\rm d}\phi ,
\end{equation}
where $g = \partial_t f 
         + r^{-1}(x\partial_x f + y\partial_y f + z\partial_z f)$.
After all one--dimensional evolutions have been carried out,
we have a collection of $U_{lm}(t,r)$ modes (here viewed as 
exact for ease of presentation) with which we can construct 
the three--dimensional solution
\begin{equation}
U(t,r,\theta,\phi) = \sum_{l=0}^\infty\sum_{m=-l}^l
U_{lm}(t,r)Y_{lm}(\theta,\phi)
\end{equation}
via harmonic {\em synthesis}. 
Numerically, we obtain $f|_{t=0}$ and $g|_{t=0}$ on a 
spherical--polar grid 
\begin{equation}
\big\{(r_q,\theta_i,\phi_j): 0\leq q \leq Q, 1\leq i \leq I, 
1\leq j \leq J\big\}
\end{equation}
by exact function calls. For each $r_q$
we then perform a harmonic analysis on the restriction of the
initial data to the two--dimensional spherical grid 
$(\theta_i,\phi_j)$. We have used Adams and Swarztrauber's
{\sc spherepack 3.0} to numerically perform all harmonic 
analysis and synthesis, and in particular those subroutines 
({\tt shagci}, {\tt shagc}, {\tt shsgci}, and {\tt shsgc}) 
from {\sc spherepack} which take the $\theta_i$ as Gaussian 
points in colatitude.
\begin{figure}
\scalebox{0.70}{\includegraphics{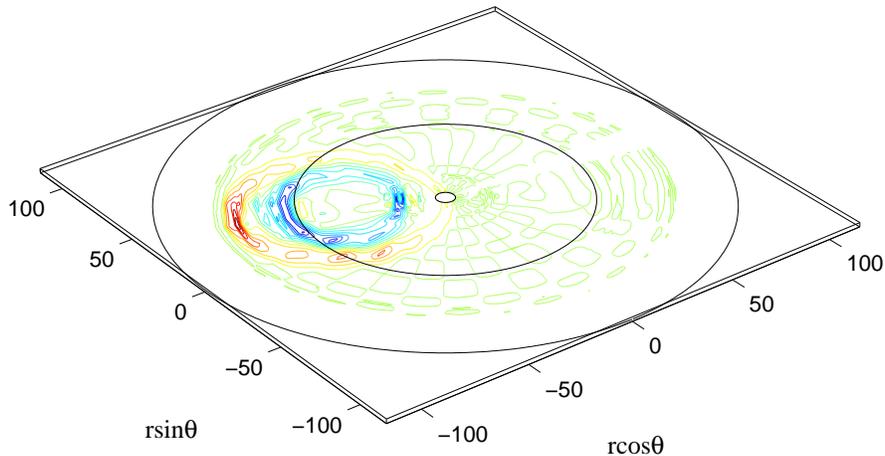}}
\caption{{\sc Equatorial cross section of $U^\mathrm{free}$ at
$t = 38.5$.} The initial data for this evolution is described
in the text. The outer, middle, and inner circles are $r = 116$,
$60$, and $4$. For this plot $(Q,I,J) = (2048,33,32)$, so that
the shown spherical domain (determined
by $4 \leq r \leq 116$) corresponds to a $(4096,33,32)$
discretization. The oscillations spread out over the whole
angular domain result from the harmonic truncation of the initial
data.
\label{fiveFigure9}}
\end{figure}

We work with an angular grid corresponding to $(I,J) = (33,32)$,
with the first number odd so that the equator is a Gaussian point.
The spectral truncation in {\sc spherepack} is triangular, and 
with this angular resolution we only sample harmonic modes for 
$l \leq 32$. Prior to the harmonic analysis described above, we 
perform a preliminary analysis followed immediately by synthesis. 
This initial procedure corresponds to truncation of the full 
harmonic series for $f|_{t=0}$ and $g|_{t=0}$, and it does 
modify of the described initial data. The resulting data is now 
spread out over the whole angular domain, although it remains 
concentrated at $(x_0,y_0,z_0)$.
 
\subsubsection{Short--time three--dimensional 
evolution} \label{sbsbSec:5.2.2}
The three--dimensional numerical test we now describe is somewhat 
analogous to the one--dimensional short--time test described in 
{\sc Section} \ref{sbsbSec:5.1.1}. We consider a spherical domain 
bounded internally by $r=4$ and externally by $r=116$. On this 
large domain, we choose initial data as described in the last 
subsubsection. That is to say, in addition to the discretization 
sizes $(Q,I=33,J=32)$ of our three--dimensional domain, 
in (\ref{bumpfunction3}) we also choose the vectors $(x_0,y_0,z_0)$ 
and $(v_x,v_y,v_z)$ as well as the values of $A$, $\delta$, and $b$.
We then numerically evolve 
this data until $t = 38.5$. In setting up the initial data, due 
care must be taken to ensure that (i) the support of the data is 
contained within $r < 60$ and (ii) in the short--time $t = 38.5$ 
the propagating solution spills over $r = 60$ but does not reach 
$r = 116$. To satisfy these 
criteria, we have chosen a bump function (\ref{bumpfunction3}) 
centered at $(x_0,y_0,z_0) = (-40,20,14)$, with initial velocity 
$(v_x,v_y,v_z) = (-0.87,-0.4,-0.3)$. Moreover, we have chosen
$A = 1$, $b = 12$, and $\delta = 0.1$.

The byproduct of our evolution is a three--dimensional
free solution $U^{\mathrm{free}}$ at $t = 38.5$.
As before, any numerical error in $U^{\mathrm{free}}$ stems solely 
from the finite--difference scheme and not the boundary 
conditions. Having generated such a free solution from the 
initial data described above, we show the result in 
{\sc Fig.}~\ref{fiveFigure9}, a depiction of the equatorial 
cross section of $U^{\mathrm{free}}$. As mentioned, the number 
$I=33$ of $\theta_i$ points ensures that $\theta_{17} = \pi/2$, 
allowing for such a figure. Our 
initial data has been designed to yield an evolution with an 
interesting cross section.
\begin{table}\begin{center}
\begin{tabular}{|c||c|c|c|c|}
\hline
$Q$   & $\|\Delta U^{\mathrm{ROBC}}\|_\infty$ &
$\|\Delta U^{\mathrm{ROBC}}\|_2$ &
$\|\Delta U^{\mathrm{SOBC}}\|_\infty$ &
$\|\Delta U^{\mathrm{SOBC}}\|_2$ \\
\hline
$256$  & $1.7593\times 10^{-4}$
       & $4.9594\times 10^{-6}$
       & $7.3416\times 10^{-2}$ 
       & $8.3218\times 10^{-3}$ \\
\hline
$512$  & $1.6646\times 10^{-5}$
       & $1.1387\times 10^{-6}$
       & $7.3137\times 10^{-2}$
       & $8.2895\times 10^{-3}$ \\
\hline
$1024$ & $3.4185\times 10^{-6}$
       & $2.7872\times 10^{-7}$
       & $7.3131\times 10^{-2}$
       & $8.2724\times 10^{-3}$ \\
\hline
$2048$ & $7.9344\times 10^{-7}$
       & $6.8970\times 10^{-8}$
       & $7.3131\times 10^{-2}$
       & $8.2656\times 10^{-3}$ \\
\hline
\end{tabular}
\end{center}
\vskip 2mm
\caption{{\sc Error measures for the three--dimensional test.}}
\label{fiveTable4}
\end{table}

The numerical test consists of again evolving the data, but
now on the smaller spherical domain determined by 
$4 \leq r \leq 60$ and using the $\varepsilon = 10^{-10}$
{\sc robc}. As a result, we
get another three--dimensional array $U^{\mathrm{ROBC}}$ which
we compare with a suitably truncated $U^{\mathrm{free}}$. 
Note that now both $U^{\mathrm{ROBC}}$ and 
$U^{\mathrm{free}}$ are three--dimensional arrays of physical 
values, whereas in the first subsection they stood for one--dimensional 
arrays of spectral values. We consider 
two error measures corresponding to the difference 
$\Delta U^{\mathrm{ROBC}} = U^{\mathrm{ROBC}} - 
U^{\mathrm{free}}$. The first is simply the absolute supremum error 
\begin{equation}
\|\Delta U^{\mathrm{ROBC}}\|_\infty
= \mathrm{sup}\big\{
\big|U^{\mathrm{ROBC}}_{q,i,j} - 
U^{\mathrm{free}}_{q,i,j}\big|: 0\leq q\leq Q,  
1\leq i\leq I,  1\leq j\leq J\big\}\, ,
\label{3dsuperr}
\end{equation}
while the second is a discretization of the L2 error
\begin{equation}
\|\Delta U^{\mathrm{ROBC}}\|_2 =
\frac{1}{\mathrm{vol}(\Sigma)}
\left[\sum_{j=1}^{J}\sum_{i=1}^I \sum_{q=1}^Q
\big|U^{\mathrm{ROBC}}_{q,i,j} 
- U^{\mathrm{free}}_{q,i,j}\big|{}^2 M^{}_q r^2_q\sin\theta_i
\Delta r \Delta \theta_i \Delta\phi\right]^{1/2}\, ,
\label{3dl2err}
\end{equation}
where the radial lapse factor $M_q = (1+4/r_q)^{1/2}$ ensures
that this discretization corresponds to proper integration. 
The volume of the computational domain is 
\begin{equation}
\mathrm{vol}(\Sigma) = 4\pi\int_4^{60} (1+4/r)^{1/2} r^2{\rm d}r \simeq
9.48255 \times 10^5.
\end{equation}
Since the $\theta$--mesh is comprised of Gaussian points 
$\theta_i$, we must also include the appropriate Gaussian weights
$\Delta \theta_i$ in the sum (both the $\theta_i$ and 
$\Delta \theta_i$ are obtained with the routine {\tt gaqd} 
from {\sc spherepack}). We emphasize that 
unlike the situation in the first subsection where all error 
measures have been computed in the spectral space of spherical 
harmonics, these error measures are computed in physical 
space. 

In Table \ref{fiveTable4} we list errors for the scenario shown in 
{\sc Fig}.~\ref{fiveFigure9}, doing so for radial 
discretizations corresponding to $Q=256$, $512$, $1024$ and $2048$ (so 
that the free solution has been computed with radial
discretizations corresponding to $2Q = 512$, $1024$, $2048$ and $4096$). 
In the table we have also listed the corresponding Sommerfeld 
errors, which have been computed as in (\ref{3dsuperr}) and 
(\ref{3dl2err}) but with $U^{\mathrm{SOBC}}$ in place of 
$U^{\mathrm{ROBC}}$. Notice that the {\sc robc} give rise to perfect
second--order convergence, whereas the {\sc sobc} do not. Moreover,
the {\sc robc} numerical solutions are several orders of magnitude
more accurate than their {\sc sobc} counterparts. It is also of 
interest to consider the relative computational cost of both {\sc robc} 
and {\sc sobc}. As {\sc sobc} are purely local boundary conditions, they 
are of course less expensive than our {\sc robc}. However,
despite being strikingly more accurate, we find our {\sc robc} to be 
nearly as cheap as {\sc sobc}, as is evident by the relative timings
listed in Table \ref{fiveTable5}.
\begin{table}\begin{center}
\begin{tabular}{|c||c|c|}  
\hline
$Q$ & {\sc robc} & {\sc sobc} \\
\hline
\hline
256  & 50.00 & 46.46 \\
\hline
512  & 49.45 & 48.15 \\
\hline
1024 & 43.21 & 41.85 \\
\hline
2048 & 19.55 & 18.18 \\
\hline
\end{tabular}
\end{center} 
\vskip 2mm
\caption{{\sc Relative times for three--dimensional evolutions.} Here we
list the relative run times as percentages for the various $Q$ values.  
With $Q = 1024$, for example, we generated the free solution on the larger
domain via a run of 1817.26{\tt s}. The corresponding {\sc robc} and
{\sc sobc} runs on the smaller domain took 785.18{\tt s} and 760.61{\tt s}
respectively. The entry 43.21 in the table corresponds to $100\times
785.18/1817.26$, and so on.}
\label{fiveTable5}
\end{table}
%
%
\section{Discussion}
We start this final section by comparing our results with 
those of our main reference \cite{AGH1}, thereby gaining 
some perspective on what we have achieved. After that
comparison we discuss possible applications and extensions.

\subsection{Comparison with the work of Alpert, Greengard, 
and Hagstrom}
Our rapid implementation of {\sc robc} on the Schwarzschild 
geometry has been inspired by Alpert, Greengard, and 
Hagstrom's implementation on flatspace. Although we have followed 
their approach closely, our implementation falls somewhat short 
of theirs. Let us point out why this is the case by comparing both 
implementations from theoretical and numerical standpoints.

From a theoretical perspective our work is less satisfactory
than that of {\sc agh}, chiefly because  ---as mentioned in 
the introduction--- we are unable to give a rigorous 
asymptotic analysis of our implementation. For each angular $l$ 
index {\sc agh} deal with a Bessel {\sc fdrk} 
which is theoretically known to admit a sum--of--poles 
representation, one corresponding to a purely discrete sum in 
the case of 3+1 wave propagation and to a discrete and continuous 
sum in the 2+1 case. Moreover, due to the exhaustively studied 
properties of Bessel functions, 
they also start with a wealth of useful information about how 
the pole locations, pole strengths, and (in the 2+1 case) cut
profile for the representation behave in all conceivable 
asymptotic regimes (which include certain scaling properties 
as $l$ becomes large). With tight control over where the 
physical poles accumulate and the behavior of cut profile, they 
are able to borrow ideas from the fast--multipole method in 
order to replace (discrete and continuous) physical pole 
sums with (fully discrete) approximate pole sums of fewer terms. 
As a result, they rigorously prove that the sum--of--poles 
representation for the Bessel {\sc fdrk} admits a rational 
approximation, one uniformly valid in the righthalf frequency 
plane and exhibiting exponential convergence as the number of 
approximating poles is increased. For the scenario
we have considered, such an analysis would seem out of the 
question. Indeed, although we have provided extremely convincing 
numerical evidence that the Heun {\sc fdrk} admits a sum--of--poles 
representation, one strikingly similar to the 2+1 Bessel case,
even this is conjecture from a theoretical standpoint. The 
battery of asymptotics needed to theoretically prove that such 
a representation admits a rational approximation in the style
of {\sc agh} is certainly well beyond the author's knowledge 
of special functions.

Our numerical implementation of {\sc robc} also falls somewhat
short of {\sc agh}'s, in that they considered a higher bandwidth 
(their $l \leq 1024$ compared to our $l \leq 64$) and a smaller 
best error tolerance (their $\varepsilon = 10^{-15}$ compared 
to our $\varepsilon = 10^{-10}$). However, we believe that on 
this count our work is closer to being on the same footing 
with theirs. Indeed, insofar as {\em numerical} implementation 
of flatspace {\sc robc}
is concerned, their elegant asymptotic analysis proving the 
existence of rational approximations is somewhat beside the point. 
Ultimately, their compression algorithm (which yields the desired 
rational approximations) relies only on the ability to evaluate 
the Bessel {\sc fdrk} along the imaginary axis of the frequency 
plane. For Bessel functions, which obey certain order recursion 
relations, such evaluation can be efficiently done via the 
continued fraction expression (\ref{continuedfraction}) 
following from such relations.
We stress that such evaluation requires no knowledge of the 
sum--of--poles representation for the kernel. Although we have 
no such continued fraction expression with which to evaluate the 
Heun {\sc fdrk}, we have seen that our integration method 
is almost as accurate (in the sense spelled out by
{\sc Section} \ref{sbsbSec:2.2.4}). 
For Bessel kernels we have observed that 
the numerical path integration required by our method is more 
expensive than continued fraction evaluation, and all the more so 
as the order $l+1/2$ gets large, although we 
have not made a systematic comparison of the two methods. However, 
the cost of evaluation is almost beside the point, since in 
principle any extra cost associated with our method need only be 
incurred once. Of true importance is accuracy, and through the 
use of extended precision we believe it possible to build Heun 
kernels corresponding to $\varepsilon = 10^{-15}$ and through 
a bandwidth of $1024$.\footnote{Of course, had {\sc agh} taken
advantage of extended precision with the continued fraction method, 
they presumably could have pushed beyond even their reported 
numbers.} We point out that, insofar as both gravitational wave 
astronomy \cite{Schutz} and the post--Newtonian approximation 
\cite{Thorne,Blanchet} of the gravitational field are concerned, 
$l$ values well below $64$ are the ones primarily relevant to 
gravitational wave observation.

Another numerical shortcoming of this work
is that we have mainly considered certain discrete values of 
the outer boundary radius $\rho_B$, whereas {\sc agh} allow for 
any value.\footnote{They are easily able to allow for 
any value for the following reason: The flatspace radial wave equation 
arising from Laplace and spherical--harmonic transformation 
(essentially the modified Bessel equation) can be expressed solely 
in terms of $sr$, the product of Laplace frequency and radius.} 
We hope to extend our results to a continuous interval in 
$\rho_B$, say $[15,25]$, via construction of our rational 
approximations at Chebyshev nodes in $1/\rho_B$ and subsequent 
interpolation. The only reason we have not done this 
so far is that the straight compression algorithm described in 
{\sc Section} \ref{sbSec:3.3} yields rational approximations which do not 
appear to vary that smoothly with $1/\rho_B$ (it is a nonlinear 
minimization after all). We anticipate that this problem can be 
mitigated. Indeed, using the described compression, one could 
construct a rational approximation at a Chebyshev node $\xi^1_B = 
1/\rho_B^1$ near 1/15, and as a result obtain a number of pole 
locations [the $Q$ in (\ref{leastsqr})]. For the other nodes, 
one could simply fix a $Q$ by suitable scaling of the collected 
locations for $\xi^1_B$, and then solve a straight least squares 
problem for $P$ in (\ref{leastsqr}). This straight least--square 
solve would be the same problem as (\ref{leastsqr}) but with 
$Q$ fixed. In any case, we do plan to build up some form of 
our {\sc robc} which is valid over a continuous $\rho_B$ 
interval.

\subsection{Potential applications and extensions}

We briefly describe several potential applications of our 
{\sc robc}. The first arises in the theory of non--spherical stellar 
collapse \cite{MTW}. A massive star, having expended its nuclear fuel, 
will collapse into a blackhole, with gravitational waves radiated in 
the process. Provided that the initial configuration of the collapsing 
system is not highly non--spherical, the evolution of such waves 
are governed by the Regge--Wheeler and Zerilli equations. Our 
{\sc robc} are therefore applicable to such scenarios. The classic 
work on perturbed Oppenheimer--Snyder collapse is that of Cunningham, 
Moncrief, and Price \cite{Cunningetal1,Cunningetal2}, but we hope to
carry out further numerical investigations with our {\sc robc}. Another
field of recent interest where our {\sc robc} should find application
is the theory of stellar perturbations, in particular oscillation modes 
of stars as sources for gravitational radiation (see for 
example \cite{Anderssonetal1,Anderssonetal2,Allenetal}). As a third 
application, we anticipate that our {\sc robc} can by applied to the 
Cauchy--perturbative matching scheme of Rupright, Abrahams, and 
Rezzolla \cite{Rupright,Rezzolla}, 
a numerical approach to {\sc robc} for full general relativity. 
The idea is to match the Cauchy evolution of the full Einstein 
equations to a set of one--dimensional, linear, radial 
evolutions of perturbative modes on Schwarzschild, with the 
matching taking place at an extraction two--surface 
large enough to ensure the validity of perturbation theory. In 
this approach, each of the modes is separately evolved on a 
large radial domain until the next time--step, allowing for
both wave--form extraction and updating of the interior solution 
at the extraction surface. The equation governing the evolution
of such modes is related to the one we have considered.
Rupright {\em et al} employ Sommerfeld outer boundary conditions 
for their radial evolutions. We expect that our {\sc robc} could 
improve their method, perhaps even doing away with the separate 
radial evolutions altogether (the idea here would be to impose 
our {\sc robc} directly at the extraction surface,
provided it indeed sits well within the perturbative region).

Let us mention several other arenas where the issue of
{\sc robc} might be examined along the lines laid down here. 
The most obvious would be {\sc robc} for
time--domain wave propagation on more complicated 
blackholes: the Reissner--Nordstr{\o}m solution 
(charged Schwarzschild), the rotating Kerr solution, and 
the charged Kerr solution (also known as Kerr--Newman) \cite{MTW}. 
While tackling the problem of {\sc robc} for a metric as 
complicated as the Kerr solution would be very difficult, one might 
first obtain {\sc robc} for an approximate solution (to the Einstein 
equations) which exhibits rotation, such as the Brill--Cohen metric 
(see p.~699 of \cite{MTW} and references therein).
On a different track, {\sc robc} might be investigated for other 
types of {\sc pde} on the Schwarzschild geometry, such as 
curved--space versions of the Klein--Gordon or Euler equations. 
Implementation of {\sc robc} for 
fluid equations on blackholes would have application in 
realistic studies of blackhole accretion. The methods of this
work could be extended to the Schr\"{o}dinger equation
with a potential, like the Coulomb potential, which is not of 
compact support. (In his dissertation \cite{Jiang} 
Jiang studied {\sc robc} for 
the Schr\"{o}dinger equation via the {\sc agh} approach, but only 
considered compactly supported potentials.) Implementation
of {\sc robc} in such a setting would involve the special function
theory of Coulomb wave functions or the related Whittaker 
functions (both incarnations of confluent hypergeometric 
functions)\cite{Slater,Curtis,Thompson}. Finally, we 
remark on a more practical possibility,
namely, extension of these methods to the ordinary wave 
equation but expressed in prolate or oblate spheroidal 
coordinates \cite{MorFesh}. Ref.~\cite{SinhaMacPhie} 
addresses computational issues associated with evaluating 
spheroidal wave functions. The issue of {\sc robc} in this 
setting would be not unlike the issue for Kerr blackholes. 
In any case, the spheroidal wave equation is essentially 
the confluent Heun 
equation, so one might expect our methods to be applicable.
Radiative boundary conditions which are tied to spheroidal
surfaces might prove useful in modeling phenomena with 
slender geometries, such as those arising in antenna 
design (see Ref.~\cite{antenna} and references
therein).

Yet another extension of our work would involve coupling {\sc robc} 
to high--order interior schemes, such as the one described in
\cite{AGH3} and \cite{Kress}, but without reduction of accuracy. This 
is a delicate issue pertinent to the simulation of waves on flatspace 
as well as on blackholes. We have presented a second--order accurate 
implementation of {\sc robc}, but it would be beneficial to achieve 
a fourth or higher--order implementation. Even with the exact {\sc robc} 
in hand, it is a nontrivial problem to couple them to a high--order 
interior scheme, as a naive coupling gives rise to numerical boundary 
layers which ---upon spatial differencing--- spoil the order of accuracy. 
Indeed, such problems occur when trying to couple exact Drichlet or 
Neumann boundary conditions to well--known high--order schemes, such 
as Runga--Kutta schemes, and they have a long history 
\cite{gustafsson:1975}--\cite{pathria:1997}. For achieving a successful 
high--order implementation of {\sc robc}, a strategy based upon 
Picard integral 
deferred correction would seem promising (see \cite{KressGust} as well
as comments made by Minion \cite{Minion2003}). 
%
%
\section*{Acknowledgments}
The material presented here has been based on \cite{LauMathDiss},
a 2003 Ph.D.~dissertation in applied mathematics. I am grateful for
the generous guidance of my academic advisor, Professor M.~L.~Minion.
Beyond frequently helping me throughout the course of this project,
he also carefully read multiple earlier drafts of my dissertation,
offering many invaluable criticisms and suggestions for improvement.

I also thank the other members of my Ph.D. dissertation committee:
Professors C.~R. Evans (UNC Physics \& Astronomy), M.~G.~Forest,
S.~Mitran, and
M.~E.~Taylor. Thanks also to both Doctors L.~Lee and R.~Zhou for helpful
conversations and Professor T.~Hagstrom (University of New Mexico)
for email correspondence.

I want to especially thank Professor C.~R.~Evans for crucial discussions
(particularly early on) both about this problem and wave simulation on
blackholes in general. Several key ideas in the work were developed
during these discussions, and joint work with Professor Evans on this
subject will appear elsewhere. I also acknowledge the influence of
Professor M.~E.~Taylor, as well as encouragement and general support
from both Professor Taylor and Professor J.~M.~Hawkins.
                                                                         
For collaborations on other projects affecting this one, I thank
Professor W.~Kummer (Vienna Technical University), Professor
A.~N.~Petrov (Moscow State University), and especially both
Professors J.~D.~Brown (North Carolina State University) and
J.~W.~York (Cornell University). I make special note of the
lasting influence and inspiration of Professor York, under whose
supervision I received a 1994 physics Ph.D.~in general relativity.
%
%
\begin{appendix}
\section{Modified MacCormack scheme}\label{Sec:A}
In {\sc Section} \ref{sbsbSec:4.3.4} we use the extrapolation 
(\ref{fourextrap}) in lieu of the predicted variable 
$\bar{U}^{n+1}_{Q+1}$ which is not yet available, and this is 
a salient feature of our implementation of {\sc robc}. We stress 
that this extrapolation takes place only at the outermost mesh 
point $r_Q = r_B$. Nevertheless, in order to investigate its 
validity, we ask whether or not it may be extended to a valid 
numerical scheme over the whole interior computational domain. 

\begin{figure}[h]
\scalebox{0.75}{\includegraphics{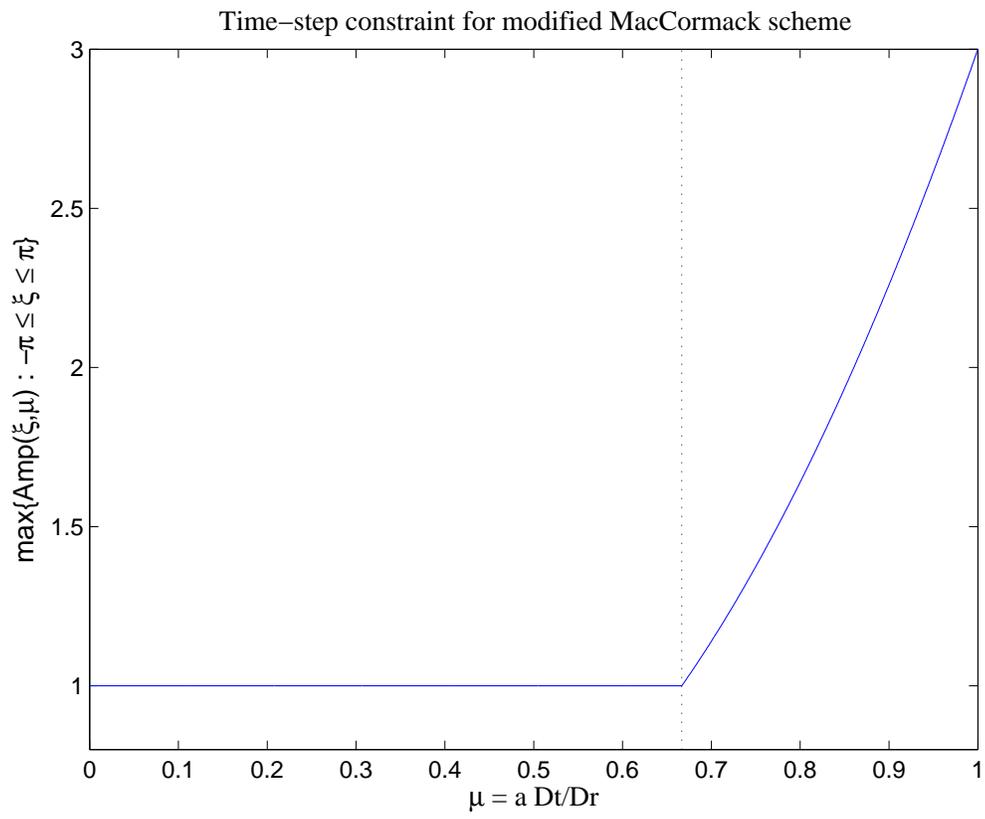}}
\caption{{\sc First output from the given matlab code.}
\label{appendixDFigure1}}
\end{figure}

\begin{figure}[t]
\scalebox{0.75}{\includegraphics{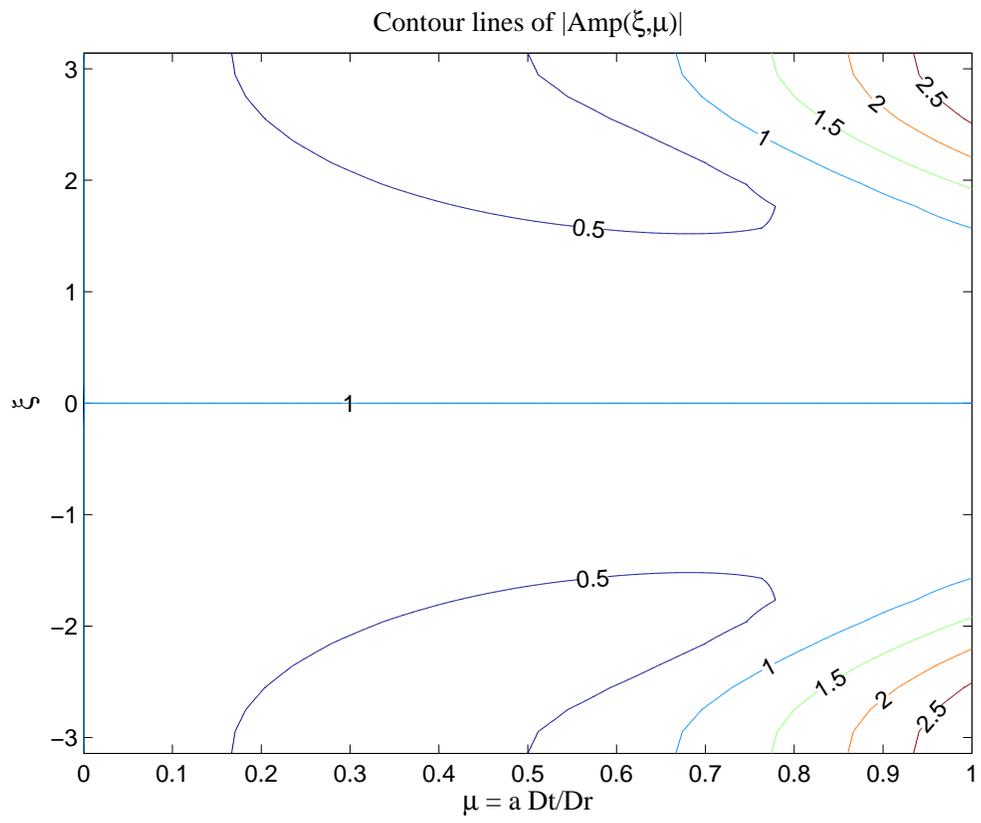}}
\caption{{\sc Second output from the given matlab code.}
\label{appendixDFigure2}}
\end{figure}

For simplicity and in order to make precise statements 
about the scheme to be considered, let us focus attention 
on the simplest conservation law
\begin{equation}
\frac{\partial U}{\partial t} 
+ a \frac{\partial U}{\partial r} = 0\, ,
\label{Dmodel}
\end{equation}
with $a > 0 $, as a model of Eqs.~(\ref{fourUdot}) and
(\ref{Ucoordinatevelocity}).
Now consider the following modified MacCormack scheme.
The prediction phase is unchanged. Namely, we take
\begin{equation}
\bar{U}{}^{n+1}_{q} = U^{n}_{q} 
                    - \mu
                      \big(U^n_q - U^n_{q-1}\big)\, ,
\end{equation}
with $\mu = a\Delta t\big/\Delta r$ (not to be confused with
the retarded time coordinate).
However, for the corrected variable we set 
\begin{equation}
\widetilde{U}{}^{n+1}_{q}
= \big[\bar{U}{}^{n+1}_q + U^n_q - 
\mu
\big(2\bar{U}{}^{n+1}_{q} - 3\bar{U}^{n+1}_{q-1}
+\bar{U}^{n+1}_{q-2}\big)\big]\, .
\end{equation}
The three terms within the round parenthesis 
correspond to $(\bar{U}{}^{n+1}_{q+1} - \bar{U}{}^{n+1}_q)$ in 
the straight MacCormack scheme. However, we use the 
extrapolation
\begin{equation}  
3\bar{U}{}^{n+1}_{q} - 3\bar{U}{}^{n+1}_{q-1} + 
\bar{U}{}^{n+1}_{q-2}
\end{equation}
in place of $\bar{U}{}^{n+1}_{q+1}$. This modified scheme is
second--order accurate, as can be verified by a calculation
based on Taylor series. Moreover, it can be written as
\begin{equation}
\widetilde{U}{}^{n+1}_{q} = U^n_q 
- \mu
  \big[F(U^n_q,U^n_{q-1},U^n_{q-2})
-      F(U^n_{q-1},U^n_{q-2},U^n_{q-3})\big]\, ,
\end{equation}
in terms of the flux function $F(U,V,W)$ determined by
\begin{equation}
2F(U,V,W) = 3 U - V
         - \mu\big(2U-3V+W\big)\, .
\end{equation}
Since $F(U,U,U) = U$, the scheme is in {\em consistent 
conservation form} as applied to a conservation law like
the $1+1$ flatspace advection equation \cite{LeVeque}. 
Following Minion's description \cite{Minion1996} of a standard 
stability analysis, we have written a short {\sc matlab} code 
(given below) which demonstrates that the modified scheme is stable 
for the time--step constraint $a\Delta t\big/\Delta r < 2/3$ (or a 
number shockingly close to $2/3$). The output plots for this code 
are depicted in {\sc Figs.}~\ref{appendixDFigure1} and 
\ref{appendixDFigure2}. The second plot shows that instability in 
the scheme stems from the highest modes.

For the straight interior MacCormack scheme as applied to the 
{\em system} (\ref{fourUdot}) and (\ref{fourXdot}) of evolution 
equations, we have remarked in {\sc Section} \ref{sbsbSec:4.2.2} 
that we expect evolution stability for $\Delta t\big/\Delta r < 1$. 
Recall that this is the expectation because the variable $X$ 
propagates everywhere with speed $1$, while $(\rho_B-1)/(\rho_B+1) 
< 1$ is the maximum speed of $U$ over the domain $[2\mathrm{m},r_B]$, 
as is evident from (\ref{Ucoordinatevelocity}). We now assume that 
our stability analysis for the modified MacCormack scheme ---carried 
out using the model equation (\ref{Dmodel})--- also pertains to
(\ref{fourUdot}). Since the modified scheme is only used at the 
outermost boundary point, we let $a = (\rho_B-1)/(\rho_B+1)$. 
Then for $\rho_B = 15$, we get $a = 14/16$, and in turn the 
constraint $\Delta t\big/\Delta r < 16/21 \simeq 0.7619$ associated 
with the modified scheme. Since the {\sc rhs} of this constraint is 
less than $1$, the issue at hand is whether it is $16/21$ or $1$
which limits evolution stability. From numerical experiments, 
such as those in {\sc Section} \ref{Sec:5}, 
we find that stability is ensured 
so long as $\Delta t\big/\Delta r < 1$. Although the modified 
MacCormack time--step constraint is more restrictive, it is 
not a limitation for our numerical evolutions, presumably because 
the modified scheme is implemented only at a single point.

\noindent
\small
\begin{verbatim}
%
%  S R Lau
%  Applied Mathematics Group
%  Department of Mathematics
%  University of North Carolina
%  Chapel Hill, NC 27599-3250
%  USA
%
%  26 October 2003
%
%  Matlab code: ModifiedMacCormack.m
%
%  Code produces two plots, each of which elucidates the time-step 
%  stability associated with applying a modified MacCormack scheme 
%  to the 1+1 linear advection equation,
%
%  U_t + a U_r = 0,
%
%  where a > 0 is a constant velocity. Our method of analysis 
%  follows M. L. Minion, "On the Stability of Godunov-Projection
%  Methods for Incompressible Flow," J. Comp. Phys. {\bf 123}, 435 
%  (1996). 
%
%  The modified MacCormack scheme is as follows. In terms of the 
%  Courant-Friedrich-Levy factor 
%
%  mu = a Dt/Dr, 
%
%  we obtain the predicted variable
%
%  barU(n+1,q) = U(n,q)  - mu [U(n,q) - U(n,q-1)].             (A)
%
%  Next, we construct the corrected variable
%
%  tilU(n+1,q) = 0.5{barU(n+1,q) + U(n,q) - mu
%            [2barU(n+1,q) - 3barU(n+1,q-1) + barU(n+1,q-2)]}. (B)
%
%  The three terms within the square parenthesis correspond
%  to [barU(n+1,q+1) - barU(n+1,q)] in the straight MacCormack
%  scheme. However, here we use the extrapolation
%
%  3barU(n+1,q) - 3barU(n+1,q-1) + barU(n+1,q-2)
%
%  in place of barU(n+1,q+1). This scheme is inspired by how
%  ROBC have been implemented within the MacCormack scheme in
%  S  R  Lau, "Rapid Evaluation of Radiation Boundary Kernels for
%  Time-Domain Wave Propagation on Blackholes." UNC-Chapel Hill
%  Ph.D. dissertation (2003). See the end of Chapter 4.
%
%  To perform a standard Von Neumann stability analysis, suppose 
%  that U is periodic on [0,2pi], and represented by the discrete
%  Fourier series
%
%           Q/2-1
%  U(j) =    Sum    hatU(p) exp(j p Dr),                       (C)
%          p = -Q/2
%
%  where Dr = 2pi/Q and hatU is the Fourier transform of U. Then 
%  the range of xi = p Dr lies in [-pi,pi]. To define Amp(xi,mu), 
%  the amplification factor, we substitute the spectral form (C) 
%  into equations (A) and (B) above, in order to get the symbol 
%  Amp for one step of the method:
%
%  hatU(n+1,p) = Amp(xi,mu) hatU(n,p).
%
%  Having given the relevant background, let us turn to the code.
%
%  Step 1. Build up a two-by-two mesh grid with CFL factor mu 
%  along x-axis and angle xi = p Dr along y-axis. The argument 
%  of each trigonometric function which makes up the amplification 
%  factor Amp involves p Dr = 2 pi p/Q, so to test Amp we sample 
%  xi from -pi to pi at regular intervals (continuity of Amp 
%  ensures that we need not sample every xi).
%
%  Choose discretization sizes for mu and xi arrays, and build the 
%  mesh grid.
%
Dmu = 0.001;
Dxi = pi/16;
[xi,mu] = meshgrid(-pi:Dxi:pi,0:Dmu:1); 
%
%  Step 2. Build corresponding two-by-two array of values for 
%  the Amp symbol particular to the modified MacCormack scheme.
%
Exp1  = exp(-i*xi);
Exp2  = exp(-2*i*xi);
Exp3  = exp(-3*i*xi);
term1 = 3-4*Exp1+Exp2;
term2 = 2-5*Exp1+4*Exp2-Exp3;
Amp   = 1-0.5*mu.*term1+0.5*mu.*mu.*term2;
%
%  Step 3. Make the first plot. For each point in the mu array, 
%  we compute the maximum of |Amp(xi,mu)| over all xi. Note mu = 0 
%  always gives an amplification factor of unity, so maxAmpMod > 1 
%  marks instability.
%
[J K] = size(Amp);
for j = 1:J
maxAmpMod(j)  = max(abs(Amp(j,:)));
end
figure(1)
hold off
plot(mu(1:J,1),maxAmpMod)
hold on
xlabel('\mu = a Dt/Dr')
ylabel('max\{Amp(\xi,\mu) : -\pi \leq \xi \leq \pi\}')
title('Time--step constraint for modified MacCormack scheme')
plot((2/3)*ones(size([0.8:0.1:3])),[0.8:0.1:3],'k:')
axis tight
%
%  Step 4. Make the second plot. Here we plot contour lines of 
%  the modulus of the Amplification factor in order to see which 
%  xi are the most amplified.
%
figure(2)
hold off
[CS H] = contour(mu,xi,abs(Amp));
xlabel('\mu = a Dt/Dr')
ylabel('\xi')
title('Contour lines of |Amp(\xi,\mu)|')
clabel(CS,H,'manual')
%
%  Press "return" to exit manual labeling.
%
\end{verbatim}
\clearpage
\section{Numerical Tables for Compressed Kernels}\label{Sec:B}
We list the pole locations $\beta_{n}$ and 
strengths $\gamma_{n}$ for a few compressed kernels 
used in the numerical tests of {\sc Section}~\ref{Sec:5}. 
A table entry of $0$ indicates an output number from the 
compression algorithm which is less than $10^{-30}$ in 
absolute value. For those $\beta_{n}$ and $\gamma_{n}$ 
appearing in conjugate pairs, we list only the one with 
positive imaginary part.
 \begin{table}[h]
 \begin{center}
 \begin{tabular}{|c||l|l|}
 \hline
 &\hspace{4cm} & \hspace{1cm}\\
 $n$ & $\beta^R_n$ & $\beta^I_n$
 \\ \hline
 1 & $-$0.407381913421500E+00 & 0
 \\ \hline
 2 & $-$0.292310786462780E+00 & 0
 \\ \hline
 3 & $-$0.212438334101443E+00 & 0
 \\ \hline
 4 & $-$0.154351731119108E+00 & 0
 \\ \hline
 5 & $-$0.111621522695175E+00 & 0
 \\ \hline
 6 & $-$0.801606769182339E$-$01 & 0
 \\ \hline
 7 & $-$0.570590618998134E$-$01 & 0
 \\ \hline
 8 & $-$0.401964994812843E$-$01 & 0
 \\ \hline
 9 & $-$0.280020468146328E$-$01 & 0
 \\ \hline
10 & $-$0.192813689223556E$-$01 & 0
 \\ \hline
11 & $-$0.131170606343391E$-$01 & 0
 \\ \hline
12 & $-$0.881053984705117E$-$02 & 0
 \\ \hline
13 & $-$0.583815623973656E$-$02 & 0
 \\ \hline
14 & $-$0.381311155357965E$-$02 & 0
 \\ \hline
15 & $-$0.245325710136527E$-$02 & 0
 \\ \hline
16 & $-$0.155499824057403E$-$02 & 0
 \\ \hline
17 & $-$0.970762744563191E$-$03 & 0
 \\ \hline
18 & $-$0.585846647089220E$-$03 & 0
 \\ \hline
19 & $-$0.373076210020314E$-$03 & 0
 \\ \hline
20 & $-$0.183147313598350E$-$03 & 0
 \\ \hline
 \end{tabular}
 \begin{tabular}{|c||l|l|}
 \hline
 &\hspace{4cm} & \hspace{1cm}\\
 $n$ & $\gamma^R_n$ & $\gamma^I_n$
 \\ \hline
 1 & $-$0.560438492794723E$-$06 & 0
 \\ \hline
 2 & $-$0.149307319758660E$-$04 & 0
 \\ \hline
 3 & $-$0.100427245970089E$-$03 & 0
 \\ \hline
 4 & $-$0.294347741100713E$-$03 & 0
 \\ \hline
 5 & $-$0.482077412607607E$-$03 & 0
 \\ \hline
 6 & $-$0.512454401482300E$-$03 & 0
 \\ \hline
 7 & $-$0.391967225466562E$-$03 & 0
 \\ \hline
 8 & $-$0.233213890995425E$-$03 & 0
 \\ \hline
 9 & $-$0.114815532845651E$-$03 & 0
 \\ \hline
10 & $-$0.490292206150649E$-$04 & 0
 \\ \hline
11 & $-$0.187703451634142E$-$04 & 0
 \\ \hline
12 & $-$0.658176657111368E$-$05 & 0
 \\ \hline
13 & $-$0.214158551560558E$-$05 & 0
 \\ \hline
14 & $-$0.651562115228849E$-$06 & 0
 \\ \hline
15 & $-$0.186157771876132E$-$06 & 0
 \\ \hline
16 & $-$0.501212404555003E$-$07 & 0
 \\ \hline
17 & $-$0.128840007496407E$-$07 & 0
 \\ \hline
18 & $-$0.314418019702290E$-$08 & 0
 \\ \hline
19 & $-$0.569152368353698E$-$09 & 0
 \\ \hline
20 & $-$0.244872495848780E$-$09 & 0
 \\ \hline
 \end{tabular}
 \\[5mm]
 \caption{
Kernel for $\jmath =  0$, $l =  0$, 
$\rho_B =  15$, 
$\varepsilon = 10^{-10}$.}
 \end{center}
 \end{table}
 \begin{table}[t]
 \begin{center}
 \begin{tabular}{|c||l|l|}
 \hline
 & \hspace{3.5cm} & \hspace{3.5cm} \\
 $n$ & $\beta^R_n$ & $\beta^I_n$
 \\ \hline
 1 & $-$0.347849789139467E+01 &  0
 \\ \hline
 2 & $-$0.328013420888283E+01 &  0.122459742396181E+01
 \\ \hline
 3 & $-$0.277834690894067E+01 &  0.221807916777226E+01
 \\ \hline
 4 & $-$0.216780239221729E+01 &  0.288987484597755E+01
 \\ \hline
 5 & $-$0.160611143055369E+01 &  0.328338093803925E+01
 \\ \hline
 6 & $-$0.116571976808040E+01 &  0.348766466680969E+01
 \\ \hline
 7 & $-$0.870203833982946E+00 &  0.359360642350047E+01
 \\ \hline
 8 & $-$0.654574374170026E+00 &  0.373250677729775E+01
 \\ \hline
 9 & $-$0.383775970479126E+00 &  0.391836743018780E+01
 \\ \hline
 \end{tabular}
 \begin{tabular}{|c||l|l|}
 \hline
 & \hspace{3.5cm} & \hspace{3.5cm} \\
 $n$ & $\gamma^R_n$ & $\gamma^I_n$
 \\ \hline
 1 & $-$0.351882332982489E+02 &  0
\\ \hline
 2 & $-$0.288247787077508E+02 &  0.145963809805392E+02
\\ \hline
 3 & $-$0.159408675098397E+02 &  0.197009045186916E+02
\\ \hline
 4 & $-$0.579079784769916E+01 &  0.165503367808947E+02
\\ \hline
 5 & $-$0.888910445944541E+00 &  0.110514097610705E+02
\\ \hline 
 6 &  +0.617612362898097E+00 &  0.649132105745073E+01
\\ \hline
 7 &  +0.488858645232752E$-$01 &  0.363187756402738E+01
\\ \hline
 8 & $-$0.604551383223953E+00 &  0.358904868249248E+01
\\ \hline
 9 & $-$0.352475685403885E+00 &  0.379107013210578E+01
 \\ \hline
 \end{tabular}
 \\[5mm]
 \caption{
 Kernel for $\jmath =  2$, $l =  64$, 
$\rho_B =  15$, $\varepsilon = 10^{-10}$. Complex
conjugation of entries 2--9 gives entries 10--17.}
 \end{center}
 \end{table}
 \begin{table}[b]
 \begin{center}
 \begin{tabular}{|c||l|l|}
 \hline
 & \hspace{3.75cm} & \hspace{3.5cm} \\
 $n$ & $\beta^R_n$ & $\beta^I_n$
 \\ \hline
 1 & $-$0.375616176922446E+00 & 0
 \\ \hline
 2 & $-$0.252285897920276E+00 & 0
 \\ \hline
 3 & $-$0.171458781191188E+00 & 0
 \\ \hline
 4 & $-$0.116562490243471E+00 & 0
 \\ \hline
 5 & $-$0.764331990276028E$-$01 & 0
 \\ \hline
 6 & $-$0.468091057981411E$-$01 & 0
 \\ \hline
 7 & $-$0.263730137927373E$-$01 & 0
 \\ \hline
 8 & $-$0.125652994567027E$-$01 & 0
 \\ \hline
 9 & $-$0.947795178946719E$-$01 &  0.599312024947409E$-$01
 \\ \hline
 \end{tabular}
 \begin{tabular}{|c||l|l|}
 \hline
 & \hspace{3.75cm} & \hspace{3.5cm} \\
 $n$ & $\gamma^R_n$ & $\gamma^I_n$
 \\ \hline
 1 & $-$0.942815440763951E$-$05 & 0
 \\ \hline
 2 & $-$0.366046310052021E$-$03 & 0
 \\ \hline
 3 & $-$0.374027383588918E$-$02 & 0
 \\ \hline
 4 & $-$0.872734265927278E$-$02 & 0
 \\ \hline
 5 & $-$0.147189136342128E$-$02 & 0
 \\ \hline
 6 & $-$0.501356988668342E$-$04 & 0
 \\ \hline
 7 & $-$0.973423621068051E$-$06 & 0
 \\ \hline
 8 & $-$0.728807025058112E$-$08 & 0
 \\ \hline
 9 & $-$0.894836172990982E$-$01 &  0.620643548936884E$-$01
 \\ \hline
\end{tabular}
\\[5mm]
\caption{
Kernel for $\jmath =  2 $, $l =  2$, 
$\rho_B =  15$, $\varepsilon = 10^{-10}$. 
Complex conjugation of entry 9 gives entry 10.}
 \end{center}
 \end{table}
\end{appendix}

\end{document}